\documentclass[seceqn]{elsart}
\usepackage{setspace}
\usepackage{amssymb}
\usepackage{amsmath}
\usepackage{graphicx}
\usepackage{amsfonts}
\makeindex

\newcommand{\s}{\sigma}

\newcommand{\vp}{V_\phi(\phi)}
\newcommand{\vs}{V_\sigma(\sigma)}

\newcommand{\be}{\begin{equation}}
\newcommand{\ee}{\end{equation}}

\begin{document}

\begin{frontmatter}

\title{Quintom Cosmology: theoretical implications and observations}

\author[ad1]{Yi-Fu Cai},
\ead{caiyf@mail.ihep.ac.cn}
\author[ad2]{Emmanuel N. Saridakis},
\ead{msaridak@phys.uoa.gr}
\author[ad3,ad4]{Mohammad R. Setare} and
\ead{rezakord@ipm.ir}
\author[ad5] {Jun-Qing Xia}
\ead{xia@sissa.it}
\address[ad1]{Institute of High Energy Physics, Chinese Academy of
 Sciences, P.O. Box 918-4,
 Beijing 100049, P.R. China}
\address[ad2]{Department of Physics, University of Athens,
 GR-15771 Athens, Greece}
\address[ad3]{Department of Science, Payame Noor University, Bijar,
 Iran}
\address[ad4]{Research Institute for Astronomy and
 Astrophysics of Maragha, P. O. Box 55134-441, Maragha, Iran}
\address[ad5]{Scuola Internazionale Superiore di Studi Avanzati,
 Via Bonomea 265, 34136 Trieste, Italy}

\begin{abstract}
We review the paradigm of quintom cosmology. This scenario is
motivated by the observational indications that the equation of
state of dark energy across the cosmological constant boundary is
mildly favored, although the data are still far from being
conclusive. As a theoretical setup we introduce a no-go theorem
existing in quintom cosmology, and based on it we discuss the
conditions for the equation of state of dark energy realizing the
quintom scenario. The simplest quintom model can be achieved by
introducing two scalar fields with one being quintessence and the
other phantom. Based on the double-field quintom model we perform
a detailed analysis of dark energy perturbations and we discuss
their effects on current observations. This type of scenarios
usually suffer from a manifest problem due to the existence of a
ghost degree of freedom, and thus we review various alternative
realizations of the quintom paradigm. The developments in particle
physics and string theory provide potential clues indicating that
a quintom scenario may be obtained from scalar systems with higher
derivative terms, as well as from non-scalar systems.
Additionally, we construct a  quintom realization in the framework
of braneworld cosmology, where the cosmic acceleration and the
phantom divide crossing result from the combined effects of the
field evolution on the brane and the competition between four and
five dimensional gravity. Finally, we study the outsets and fates
of a universe in quintom cosmology. In a scenario with null energy
condition violation one may obtain a bouncing solution at early
times and therefore avoid the Big Bang singularity. Furthermore,
if this occurs periodically, we obtain a realization of an
oscillating universe. Lastly, we comment on several open issues in
quintom cosmology and their connection to future investigations.
\end{abstract}

\end{frontmatter}

\tableofcontents

\pagebreak

  \maketitle

\pagenumbering{roman}

\setcounter{page}{5}

\pagenumbering{arabic} \setcounter{page}{1}

\section{Introduction}\label{sec:introduction}

Accompanied by the recent developments on distance detection
techniques, such as balloons, telescopes and satellites, our
knowledge about cosmology has been greatly enriched. These new
discoveries in astrophysical experiments have brought many
challenges for the current theory of cosmology. The most
distinguished event is that two independent observational signals
on distant Type Ia supernovae (SNIa) in 1998 have revealed the
speeding up of our universe \cite{Riess:1998cb,Perlmutter:1998np}.
This acceleration implies that if the theory of Einstein's gravity
is reliable on cosmological scales, then our universe is dominated
by a mysterious form of matter. This unknown component possesses
some remarkable features, for instance it is not clustered on
large length scales and its pressure must be negative in order to
be able to drive the current acceleration of the universe. This
matter content is called ``dark energy'' (DE). Observations show
that the energy density of DE occupies about $70\%$ of today's
universe. However, at early cosmological epochs DE could not have
dominated since it would have destroyed the formation of the
observed large scale structure. These features have significantly
challenged our thoughts about Nature. People begin to ask
questions like: What is the constitution of DE? Why it dominates
the evolution of our universe today? What is the relation among
DE, dark matter and particle physics, which is successfully
constructed?

The simplest solution to the above questions is a cosmological
constant $\Lambda$ \cite{Weinberg:1988cp, Carroll:1991mt,
Krauss:1995yb, Huey:1998se}. As required by observations, the
energy density of this constant has to be
$\rho_{\Lambda}\sim(10^{-3}eV)^4$, which seems un-physically small
comparing to other physical constants in Einstein's gravity. At
the classical level this value does not suffer from any problems
and we can measure it with progressively higher accuracy by
accumulated observational data. However, questioning about the
origin of a cosmological constant, given by the energy stored in
the vacuum, does not lead to a reasonable answer. Since in
particle physics the vacuum-energy is associated with phase
transitions and symmetry breaking, the vacuum of quantum
electrodynamics for instance implies a $\rho_{\Lambda}$ about 120
orders of magnitude larger than what has been observed. This is
the worst fine-tuning problem of physics.

Since the fundamental theory of nature that could explain the
microscopic physics of DE is unknown at present, phenomenologists
take delight in constructing various models based on its
macroscopic behavior. There have been a number of review articles
on theoretical developments and phenomenological studies of dark
energy and acceleration and here we would like to refer to Refs.
\cite{Peebles:2002gy, Padmanabhan:2002ji, Copeland:2006wr,
Albrecht:2006um, Linder:2008pp, Frieman:2008sn, Caldwell:2009ix,
Silvestri:2009hh} as the background for the current paper. Note
that, the most powerful quantity of DE is its equation of state
(EoS) effectively defined as $w_{DE}\equiv p_{DE}/\rho_{DE}$,
where $p_{DE}$ and $\rho_{DE}$ are the pressure and energy density
respectively. If we restrict ourselves in four dimensional
Einstein's gravity, nearly all DE models can be classified by the
behaviors of equations of state as following:
\begin{itemize}

\item \textsl{Cosmological constant:} its EoS is exactly equal to
$w_{\Lambda}=-1$.

\item \textsl{Quintessence:} its EoS remains above the
cosmological constant boundary, that is $w_{Q}\geq-1$
\cite{Ratra:1987rm,Wetterich:1987fm}.

\item \textsl{Phantom:} its EoS lies below the cosmological
constant boundary, that is $w_{P}\leq-1$ \cite{Caldwell:1999ew,
Caldwell:2003vq}.

\item \textsl{Quintom:} its EoS is able to evolve across the
cosmological constant boundary \cite{Feng:2004ad}.

\end{itemize}

With the accumulated observational data, such as SNIa, Wilkinson
Microwave Anisotropy Probe observations (WMAP), Sloan Digital Sky
Survey (SDSS) and forthcoming Planck etc., it becomes possible in
the recent and coming years to probe the dynamics of DE by using
parameterizations of its EoS, constraining the corresponding
models. Although the recent data-fits show a remarkable agreement
with the cosmological constant and the general belief is that the
data are far from being conclusive, it is worth noting that some
data analyses suggest the cosmological constant boundary (or
phantom divide) is crossed\cite{Feng:2004ad,Huterer:2004ch}, which
corresponds to a class of dynamical models with EoS across $-1$,
dubbed {\it quintom}. This potential experimental signature
introduced an additional big challenge to theoretical cosmology.
As far as we know all consistent theories in physics satisfy the
so-called Null Energy Condition (NEC), which requires the EoS of
normal matter not to be smaller than the cosmological constant
boundary, otherwise the theory might be unstable and unbounded.
Therefore, the construction of the quintom paradigm is a very hard
task theoretically. As first pointed out in Ref.
\cite{Feng:2004ad}, and later proven in Ref.  \cite{Xia:2007km}
(see also Refs. \cite{Vikman:2004dc, Hu:2004kh, Caldwell:2005ai,
Zhao:2005vj, Kunz:2006wc}),  for a single fluid or a single scalar
field with a generic lagrangian of form ${\cal
L}(\phi,\partial_\mu\phi\partial^\mu\phi)$ there exists in general
a no-go theorem forbidding the EoS crossing over the cosmological
constant. Hence, at this level, a quintom scenario of DE is
designed to enlighten the nature of NEC violation. Due to this
unique feature, quintom cosmology differs from any other paradigm
in the determination of the cosmological evolution and the fate of
the universe.

This review is primarily intended to present the current status of
research on quintom cosmology, including theoretical constructions
of quintom models, its perturbation theory and predictions on
observations. Moreover, we include the discussions about quintom
cosmology and NEC, in order to make the nature of DE more
transparent. Finally, we examine the application of quintom in the
early universe, which leads to a nonsingular bouncing solution.

This work is organized as follows. In Section \ref{sec:basic} we
begin with the basics of Friedmann-Robertson-Walker cosmology and
we introduce a concordant model of $\Lambda$CDM, referring briefly
to scenarios beyond $\Lambda$CDM. In Section
\ref{sec:quintombasic} we present the theoretical setup of quintom
cosmology and we discuss the conditions for the DE EoS crossing
$-1$. In Section \ref{sec:quintomsimple} we introduce the simplest
quintom scenario, which involves two scalar fields, and we extract
its basic properties. Section \ref{sec:perturbation} is devoted to
the discussion of the perturbation theory in quintom cosmology and
to the examination of its potential signatures on cosmological
observations. Due to the existence in quintom cosmology of a
degree-of-freedom violating NEC, the aforementioned simplest model
often suffers from a quantum instability inherited from phantom
behavior. Therefore, in Section \ref{sec:quithd} we extend to a
class of quintom models involving higher derivative terms, since
these constructions might be inspired by fundamental theories such
as string theory. In Section \ref{sec:noscalar} we present the
constructions of quintom behavior in non-scalar models, such are
cosmological systems driven by a spinor or vector field, while in
Section \ref{sec:brane} we turn to the discussion of
quintom-realizations in modified (or extended) Einstein's gravity.
In Section \ref{sec:quintomEC} we examine the violation of NEC in
quintom cosmology. Additionally, we apply it to the early
universe, obtaining a nonsingular bouncing solution in four
dimensional Einstein's gravity, and we further give an example of
an exactly cyclic solution in quintom cosmology which is
completely free of spacetime singularities. Finally, in Section
\ref{sec:conclu} we conclude, summarize and outline future
prospects of quintom cosmology. Lastly, we finish our work by
addressing some unsettled issues of quintom cosmology. Throughout
the review we use the normalization of natural units $c=\hbar=1$
and we define $\kappa^2=8\pi G=M_p^{-2}$.


%

\section{Basic cosmology}\label{sec:basic}

Modern cosmology is based on the assumptions of large-scale
homogeneity and isotropy of the universe, associated with the
assumption of Einstein's general relativity validity on
cosmological scales. In this section we briefly report on current
observational status of our universe and in particular of DE, and
we review the elements of FRW cosmology.

\subsection{Observational evidence for DE}

With the accumulation of observational data from Cosmic Microwave
Background measurements (CMB), Large Scale Structure surveys (LSS)
and Supernovae observations, and the improvement of the data
quality, the cosmological observations play a crucial role in our
understanding of the universe. In this subsection, we briefly
review the observational evidence for DE.

\subsubsection{Supernovae Ia}

In $1998$ two groups  \cite{Riess:1998cb,Perlmutter:1998np}
independently discovered the accelerating expansion of our current
universe, based on the analysis of SNIa observations of the
redshift-distance relations. The luminosity distance $d_L$, which
is very important in astronomy, is related to the apparent
magnitude $m$ of the source with an absolute magnitude $M$ through
\begin{equation}
\mu\equiv m-M=5\log_{10}\left(\frac{d_L}{\rm Mpc}\right)+25~,
\end{equation}
where $\mu$ is the distance module. In  flat FRW universe the
luminosity distance is given by:
\begin{equation}
d_L=\frac{1+z}{H_0}\int^z_0{\frac{dz'}{E(z')}},
\end{equation}
with
\begin{equation}
E(z)\equiv\frac{H(z)}{H_0}=\left\{\Omega_{m0}(1+z)^3+\Omega_{r{0}}(1+z)^4
+\Omega_{\Lambda{0}}\exp\left[{3\int^z_0{\frac{1+w(z')}{1+z'}dz'}}\right]\right\}^{1/2}~,
\end{equation}
where $H_0$ is the Hubble constant, $w$ is the EoS of DE
component, $z$ is the redshift, and
$\Omega_{i0}=8\pi{G}\rho_{i0}/(3H_0^2)$ is the density parameter
of each component (with
$\Omega_{m0}+\Omega_{r0}+\Omega_{\Lambda0}=1$), i.e its energy
density divided by the critical density at the present epoch.

\begin{figure}[tb]
\begin{center}
\includegraphics[
width=4.3in] {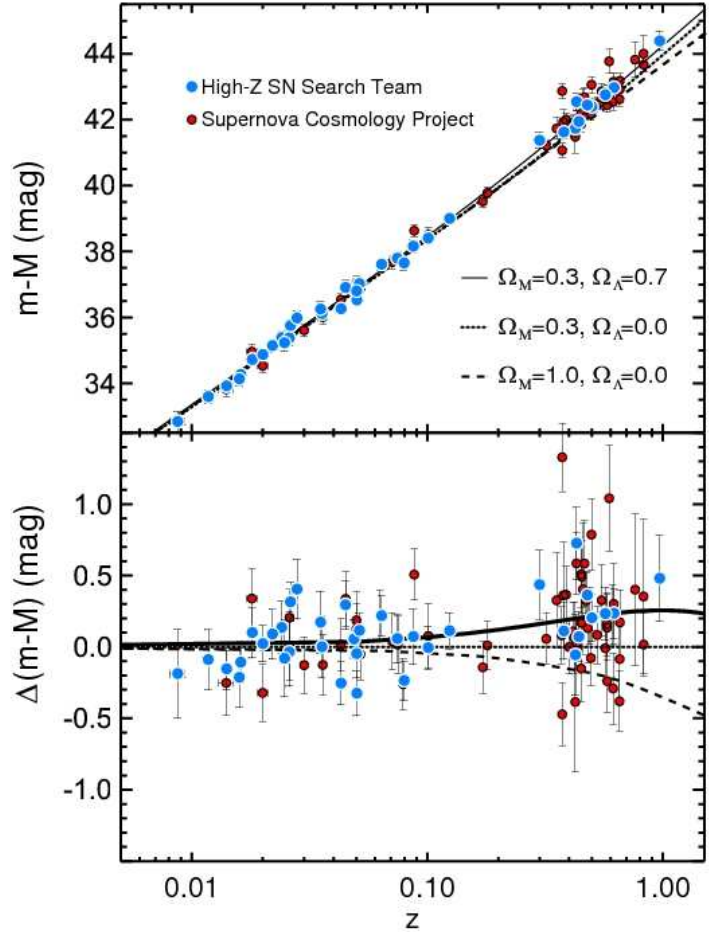} \caption{(Color online) {\it  Hubble
diagram of SNIa measured by SCP and HSST groups. In the bottom
graph, magnitudes relative to a universe with
($\Omega_{m0}=0.3$,$\Omega_{\Lambda{0}}=0$) are shown. From Ref.
\cite{Perlmutter:2003kf}.} } \label{fig1}
\end{center}
\end{figure}

In $1998$ both the the high-z supernova team (HSST)
 \cite{Riess:1998cb} and the supernova cosmology project (SCP)
 \cite{Perlmutter:1998np}, had found that distant SNIa are dimmer than
they would be in dark-matter dominated, decelerating universe.
When assuming the flat FRW cosmology
($\Omega_{m0}+\Omega_{\Lambda{0}}=1$), Perlmutter {\it et al.}
found that at present the energy density of dark-matter component
is:
\begin{equation}
\Omega^{\rm flat}_{m0} = 0.28 ~~~~~^{+0.09}_{-0.08}~(1\sigma~{\rm
statistical}) ~~~~^{+0.05}_{-0.04}~({\rm
identified~systematics})~,
\end{equation}
based on the analysis of $42$ SNIa at redshifts between $0.18$ and
$0.83$ \cite{Perlmutter:1998np}. In Fig.\ref{fig1} we present the
Hubble diagram of SNIa measured by SCP and HSST groups, together
with the theoretical curves of three cosmological models. One can
observe that the data are inconsistent with a $\Lambda=0$ flat
universe and strongly support a nonzero and positive cosmological
constant at greater than $99\%$ confidence. See Ref.
\cite{Riess:2004nr, Astier:2005qq, Riess:2006fw, Miknaitis:2007jd,
Kowalski:2008ez, Hicken:2009dk} for recent progress on the SNIa
data analysis.

\subsubsection{CMB and LSS}

Besides the SNIa observations, the CMB and LSS measurements also
provide evidences for a dark-energy dominated universe. The recent
WMAP data  \cite{Komatsu:2008hk} are in good agreement with a
Gaussian, adiabatic, and scale-invariant primordial spectrum,
which are consistent with single-field slow-roll inflation
predictions. Moreover, the positions and amplitudes of the
acoustic peaks indicate that the universe is spatially flat, with
$-0.0179 < \Omega_{K0} < 0.0081 (95\% {\rm CL})$.

\begin{figure}[tb]
\begin{center}
\includegraphics[
width=6.2in] {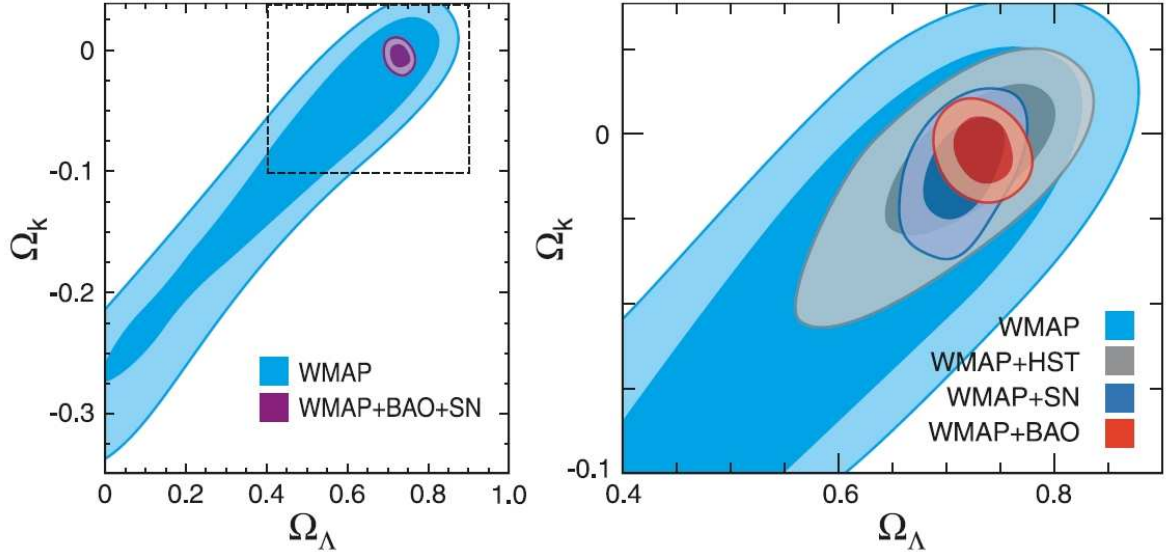} \caption{(Color online) {\it
Two-dimensional contours of the vacuum energy density,
$\Omega_{\Lambda}$, and the spatial curvature parameter
$\Omega_K$. From Ref. \cite{Komatsu:2008hk}.} } \label{fig2}
\end{center}
\end{figure}

Under the assumption of DE EoS being  $w\equiv-1$, the so called
$\Lambda$CDM model, the 5-year WMAP data, when they combine with
the distance measurements from the SNIa and the Baryon Acoustic
Oscillations (BAO) in the distribution of galaxies, imply that the
DE density parameter is $\Omega_{\Lambda0}=0.726\pm0.015$ at
present. In Fig.\ref{fig2} we illustrate the two dimensional
contour plots of the vacuum energy density $\Omega_{\Lambda}$ and
the spatial curvature parameter $\Omega_K$. One can see that the
flat universe without the cosmological constant is obviously ruled
out. Furthermore, the LSS data of the SDSS
 \cite{Tegmark:2003ud, Tegmark:2006az, Seljak:2004xh} agree with the WMAP data in favoring the
flat universe dominated by the DE component.

\subsubsection{The age of the universe}

An additional evidence arises from the comparison of the age of
the universe with other age-independent estimations of the oldest
stellar objects. In  flat FRW cosmology the age of the universe is
given by:
\begin{equation}
t_u=\int^{t_u}_0dt'=\int^{\infty}_0\frac{dz}{H(1+z)}=\int^{\infty}_0\frac{dz}{H_0(1+z)E(z)}~\label{ageeq}.
\end{equation}
In the dark-matter dominated flat universe ($\Omega_K=0$,
$\Omega_{m0}=1$), relation (\ref{ageeq}) leads to a universe-age:
\begin{equation}
t_u=\int^{\infty}_0\frac{dz}{H_0(1+z)\sqrt{\Omega_{m0}(1+z)^3}}=\frac{2}{3H_0}~.
\end{equation}
We have neglected the radiation contribution, setting
$\Omega_{\gamma0}=0$, since the radiation dominated period is much
shorter than the total age of the universe. Combining with the
present constraints on the Hubble constant from the Hubble Space
Telescope Key project (HST)  \cite{Freedman:2000cf},
$h=0.72\pm0.08$, the age of universe is $7.4 {\rm Gyr} < t_u <
10.1 {\rm Gyr}$.

On the other hand, many groups have independently measured the
oldest stellar objects and have extracted the corresponding
constraints the age of the universe: $11 {\rm Gyr} < t_u < 15 {\rm
Gyr}$  \cite{Jimenez:1996at, Richer:2002tg, Hansen:2002ij,
Krauss:2003em}. Moreover, the present 5-year WMAP data provide the
age-limit of $t_u=13.72\pm0.12 {\rm Gyr}$, when assuming the
$\Lambda$CDM model. Thus, one can conclude that the age of a flat
universe dominated by dark matter is inconsistent with these age
limits.

\begin{figure}[tb]
\begin{center}
\includegraphics[
width=4.3in] {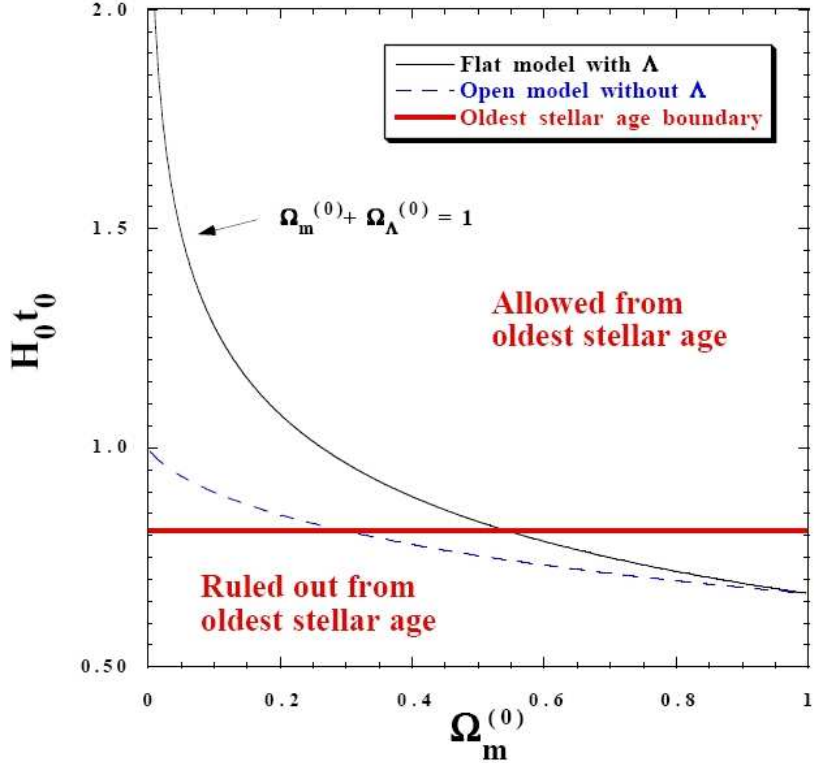} \caption{(Color online) {\it The age of the
universe $t_u$ (in units of $H^{-1}_0$) versus $\Omega_{m0}$. From
Ref. \cite{Copeland:2006wr}.} } \label{fig3}
\end{center}
\end{figure}

The above contradiction can be easily resolved by the flat
universe with a cosmological constant. Including the DE component,
the age of the universe reads:
\begin{equation}
t_u=\int^{\infty}_0\frac{dz}{H_0(1+z)\sqrt{\Omega_{m0}(1+z)^3+\Omega_{\Lambda0}}}~.
\end{equation}
Therefore, the age of universe will increase when
$\Omega_{\Lambda0}$ becomes large. In Fig.\ref{fig3} we depict the
 universe age  $t_u$ (in units of $H^{-1}_0$) versus $\Omega_{m0}$.
In this case, $t_u=1/H_0\approx 13.58 \,{\rm Gyr}$ when
$\Omega_{m0}=0.26$ and $\Omega_{\Lambda0}=0.74$. As we observe,
the age limits from the oldest stars are safely satisfied.

\subsection{A concordant model: $\Lambda$CDM}

From observations such as the large scale distribution of galaxies
and the CMB near-uniform temperature, we have deduced that our
universe is nearly homogeneous and isotropic. Within this
assumption we can express the background metric in the FRW form,
\begin{equation}
ds^2=dt^2-a^2(t)\left[\frac{dr^2}{1-kr^2}+r^2d\Omega_2^2\right]~,
\end{equation}
where $t$ is the cosmic time, $r$ is the spatial radius
coordinate, $\Omega_2$ is the 2-dimensional unit sphere volume,
and the quantity $k$ characterizes the curvature of 3-dimensional
space of which $k=-1,0,1$ corresponds to open, flat and closed
universe respectively. Finally, as usual, $a(t)$ is the scale
factor, describing the expansion of the universe.

A concordant model only involves radiation $\rho_r$, baryon matter
$\rho_{b}$, cold dark matter $\rho_{dm}$ and a cosmological
constant $\Lambda$. Thus, this class consists the so-called
$\Lambda$CDM models. In the frame of standard Einstein's gravity,
the background evolution is determined by the Friedmann equation
\begin{equation}
H^2=\frac{8\pi
G}{3}\Big(\rho_r+\rho_b+\rho_{dm}+\rho_k+\rho_{\Lambda}\Big)~,
\end{equation}
where $H\equiv\frac{\dot a}{a}$ is the Hubble parameter, and we
have defined the effective energy densities of spatial curvature
and cosmological constant $\rho_k\equiv-\frac{3k}{8\pi Ga^2}$ and
$\rho_{\Lambda}=\frac{\Lambda}{8\pi G}$, respectively.
Furthermore,  the continuity equations for the various matter
components write as
\begin{equation}
\dot\rho_i+3H(1+w_i)\rho_i=0~,
\end{equation}
where the equation-of-state parameters $w_i\equiv
\frac{p_i}{\rho_i}$ are defined as the ratio of pressure to energy
density. In particular, they read $w_r=\frac{1}{3}$ for radiation,
$w_b=0$ for baryon matter, $w_{dm}=0$ for cold dark matter,
$w_k=-\frac{1}{3}$ for spatial curvature, and $w_\Lambda=-1$ for
cosmological constant. One can generalize the EoS of the $i$-th
component as a function of the redshift $w_i(z)$, with the
redshift given as $1+z=\frac{1}{a}$. Therefore, the evolutions of
the various energy densities are given by
\begin{equation}
\rho_i=\rho_{i0}~\exp\left\{{3\int_0^z[1+w_i(\tilde
z)]d\ln(1+\tilde z)}\right\}~.
\end{equation}

From observations we deduce that $\Lambda$ is of the order of the
present Hubble parameter $H_0$, of which the energy density is
given by
\begin{equation}
\rho_\Lambda \sim 10^{-47}{\rm GeV}^4~.
\end{equation}
This provides a critical energy scale
\begin{equation}
M_{\Lambda} \sim \rho_{\Lambda}^{\frac{1}{4}} \sim 10^{-3}{\rm
eV}~.
\end{equation}
As far as we know, this energy scale is far below any cut-off or
symmetry-breaking scales in quantum field theory. Therefore, if
the cosmological constant originates from a vacuum energy density
in quantum field theory, we need to find another constant to
cancel this vacuum energy density but leave the rest slightly
deviated from vanishing. This is a well-known fine-tuning problem
in $\Lambda$CDM cosmology (for example see Ref.
\cite{Weinberg:2000yb} for a comprehensive discussion).

Attempts on alleviating the fine-tuning problem of $\Lambda$CDM
have been intensively addressed in the frame of string theory,
namely, Bousso-Polchinski (BP) mechanism \cite{Bousso:2000xa},
Kachru-Kallosh-Linde-Trivedi (KKLT) scenario \cite{Kachru:2003aw},
and anthropic selection in multiverse \cite{Weinberg:1987dv}.

\subsection{Beyond $\Lambda$CDM}

DE scenario constructed by a cosmological constant corresponds to
a perfect fluid with EoS $w_{\Lambda}=-1$. Phenomenologically, one
can construct a model of DE with a dynamical component, such as
the quintessence, phantom, K-essence, or quintom. With accumulated
astronomical data of higher precision, it becomes possible in
recent years to probe the current and even the early behavior of
DE, by using parameterizations for its EoS, and additionally to
constrain its dynamical behavior. In particular, the new released
5-year WMAP data have given  the most precise probe on the CMB
radiations so far. The recent data-fits of the combination of
5-year WMAP with other cosmological observational data, remarkably
show the consistency of the cosmological constant scenario.
However, it is worth noting that dynamical DE models are still
allowed, and especially the subclass of dynamical models with EoS
across $-1$.

Quintessence is regarded as a DE scenario with EoS larger than
$-1$. One can use a canonical scalar field to construct such
models, which action is written as  \cite{Ratra:1987rm,
Peebles:1987ek, Wetterich:1994bg, Zlatev:1998tr},
\begin{equation}
S_Q = \int d^4x \sqrt{-g} \bigg[
\frac{1}{2}\partial_\mu\phi\partial^{\mu}\phi-V(\phi) \bigg]~.
\end{equation}
By varying the action with respect to the metric, one gets the
energy density and pressure of quintessence,
\begin{equation}
\rho_Q=\frac{1}{2}\dot\phi^2+V(\phi)~,~~p_Q=\frac{1}{2}\dot\phi^2-V(\phi)~,
\end{equation}
and correspondingly the EoS is given by
\begin{equation}
w_Q=\frac{\dot\phi^2-2V}{\dot\phi^2+2V}~.
\end{equation}

However, recent observation data indicate that the contour of the
EoS for DE includes a regime with $w<-1$. The simplest scenario
extending into this regime uses a scalar field with a negative
kinetic term, which is also referred to be a ghost
\cite{Caldwell:1999ew}. Its action takes the form
\begin{equation}
S_P = \int d^4x \sqrt{-g} \bigg[
-\frac{1}{2}\partial_\mu\phi\partial^{\mu}\phi-V(\phi) \bigg]~,
\end{equation}
and thus the DE EoS is
\begin{equation}
w_P=\frac{\dot\phi^2+2V}{\dot\phi^2-2V}~.
\end{equation}
Although this model can realize the EoS to be below the
cosmological constant boundary $w=-1$, it suffers from the problem
of quantum instability with its energy state unbounded from below
\cite{Carroll:2003st, Cline:2003gs}. Moreover, if there is no
maximal value of its potential, this scenario is even unstable at
the classical level, which is referred as a Big Rip
singularity\cite{Caldwell:2003vq}.

Another class of dynamical DE model is K-essence
\cite{Chiba:1999ka, ArmendarizPicon:2000dh,
ArmendarizPicon:2000ah}, with its Lagrangian $P$ being a general
function of the kinetic term \cite{ArmendarizPicon:1999rj,
Garriga:1999vw}. One can define the kinetic variable
$X\equiv\frac{1}{2}\partial_\mu\phi\partial^\mu\phi$, obtaining
\begin{equation}
S_K = \int d^4x \sqrt{-g} P(\phi, X)~,
\end{equation}
and  the DE energy density and pressure are given by,
\begin{equation}
\rho_K=2X\frac{\partial P}{\partial X}-P~,~~p=P(\phi, X)~.
\end{equation}
 Consequently, its EoS is expressed as
\begin{equation}
w_K=-1+\frac{2XP_{,X}}{2XP_{,X}-P}~,
\end{equation}
where the subscript ``$,X$" denotes the derivative with respect to
$X$. By requiring a positive energy density, this model can
realize either $w>-1$ or $w<-1$, but cannot provide a consistent
realization of the $-1$-crossing, as will be discussed in the next
section.

Regarding the above three classes of DE scenarios together with
the cosmological constant model, there is a common question needed
to be answered, namely why the universe enters a period of cosmic
acceleration around today, in which the DE energy density is still
comparable to that of the rest components. In particular, if its
domination was stronger than  observed  then the cosmic
acceleration would begin earlier and thus large-scale structures
such as galaxies would never have the time to form. Only a few
dynamical DE models present the so-called tracker
behavior\cite{Zlatev:1998tr, Steinhardt:1999nw, Copeland:2006wr},
which avoids this coincidence problem. In these models, DE
presents an energy density which initially closely tracks the
radiation energy density until matter-radiation equality, it then
tracks the dark matter density until recently, and finally it
behaves as the observed DE.

\subsection{Observational Evidence for Quintom DE Scenario}

In this subsection we briefly review the recent observational
evidence that mildly favor the quintom DE scenario.

In $2004$, with the accumulation of supernovae Ia data, the time
variation of DE EoS   was allowed to be constrained. In Ref.
\cite{Huterer:2004ch} the authors produced uncorrelated and nearly
model independent band power estimates (basing on the principal
component analysis  \cite{Huterer:2002hy}) of the EoS of DE and
its density as a function of redshift, by fitting to the SNIa
data. They found marginal ($2\sigma$) evidence for $w(z)<-1$ at $z
< 0.2$, which is consistent with other results in the literature
\cite{Wang:2004py, Alam:2004jy, Wang:2003gz, Alam:2003fg,
Padmanabhan:2002vv, Zhu:2004cu}.

This result implied that the EoS of DE could vary with time. Two
type of parameterizations for $w_{\rm DE}$ are usually considered.
One form (Model A) is:
\begin{equation}
w_{\rm DE}=w_0+w'z~,
\end{equation}
where $w_0$ is the EoS at present and $w'$ characterizes the
running of $w_{\rm DE}$. However, this parametrization is only
valid in the low redshift,  suffering from a severe divergence
problem at high redshift, such as the last scattering surface
$z\sim1100$. Therefore, an alternative form (Model B) was proposed
by Ref. \cite{Chevallier:2000qy, Linder:2002et}:
\begin{equation}
w_{\rm DE}=w_0+w_1(1-a)=w_0+w_1\frac{z}{1+z}~,
\end{equation}
where $a$ is the scale factor and $w_1=-dw/da$. This
parametrization exhibits a very good behavior at high redshifts.

\begin{figure}[tb]
\begin{center}
\includegraphics[
width=6.in] {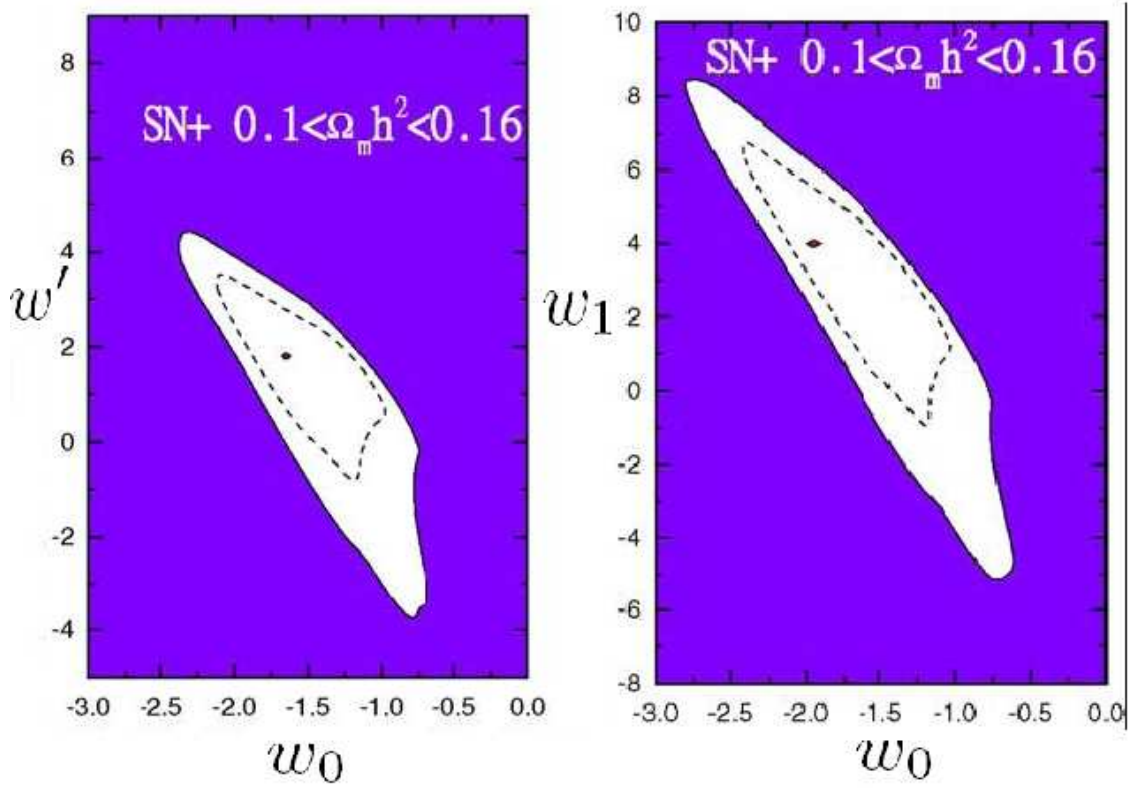} \caption{(Color online) {\it
Two-dimensional constraints on the DE parameters in two different
parameterizations. The left panel and right panel correspond to
Models A and B respectively (see text). From Ref.
 \cite{Feng:2004ad}.} } \label{fig4}
\end{center}
\end{figure}

In Ref. \cite{Feng:2004ad} the authors used the ``gold" sample of
157 SNIa, the low limit of cosmic ages and the HST prior, as well
as the uniform weak prior on $\Omega_Mh^2$, to constrain the free
parameters of the aforementioned two DE parameterizations. As
shown in Fig.\ref{fig4}, they found that the data seem to favor an
evolving DE with EoS being below $-1$ at present, evolved from
$w>-1$ in the past. The best fit value of the current EoS is
$w_0<-1$, with its running being larger than $0$.

\begin{figure}[tb]
\begin{center}
\includegraphics[
width=4.3in] {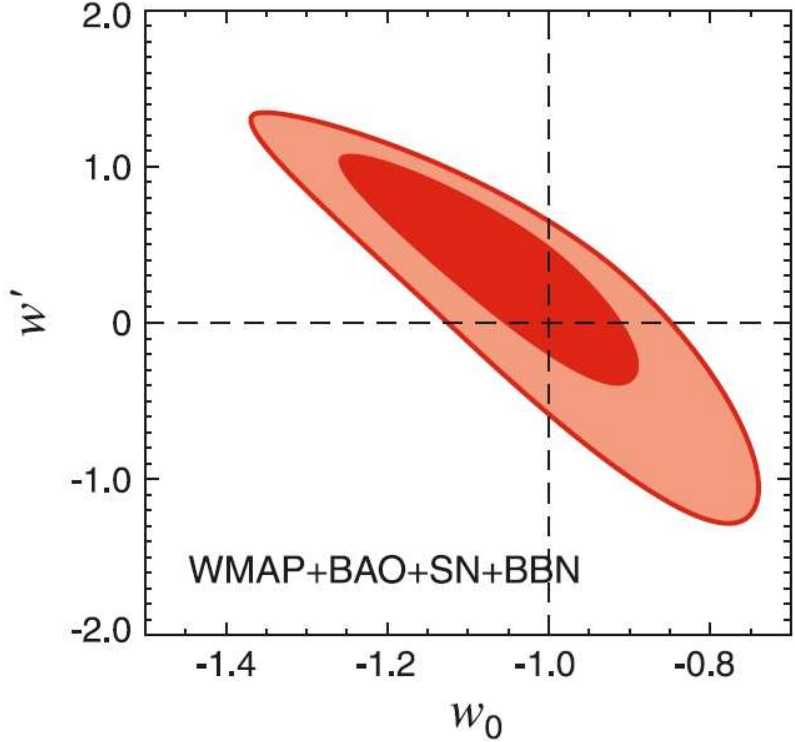} \caption{(Color online) {\it
Two-dimensional constraints on the DE parameters ($w_0$,$w'$) of
Model A (see text). From Ref. \cite{Komatsu:2008hk}.} }
\label{fig5}
\end{center}
\end{figure}

Apart from the SNIa data, CMB and LSS data  can be also used to
study the variation of DE EoS. In Ref. \cite{Hannestad:2004cb},
the authors used the first year WMAP, SDSS and 2dFGRS data to
constrain the DE models, and they found that  the data evidently
favor a strongly time-dependent $w_{\rm DE}$ at present, which is
consistent with other results in the literature \cite{Xia:2004rw,
Xia:2005ge, Xia:2006cr, Zhao:2006bt, Xia:2006rr, Xia:2006wd,
Zhao:2006qg, Wang:2007mza,Wright:2007vr,Li:2008cj}. As observed in
Fig.\ref{fig5}, using the latest 5-year WMAP data, combined with
SNIa and BAO data, the constraints on the DE parameters of Model B
are: $w_0=-1.06\pm0.14$ and $w_1=0.36\pm0.62$
\cite{Komatsu:2008hk, Xia:2008ex, Li:2008vf}. Thus, one deduces
that current observational data mildly favor $w_{\rm DE}$ crossing
the phantom divide during the evolution of universe. However, the
$\Lambda$CDM model still fits the data in great agreement.


%

\section{Quintom cosmology: Theoretical basics}\label{sec:quintombasic}

The scenario that the EoS of DE crosses the cosmological constant
boundary is referred as a ``Quintom" scenario. Its appearance has
brought another question, namely why does the universe  enter a
period of cosmic super-acceleration just today. The discussion of
this second coincidence problem has been carried out extensively
in a number of works \cite{Feng:2004ad, Feng:2004ff, Guo:2004fq,
Zhang:2005eg}. However, the explicit construction of Quintom
scenario is more difficult than other dynamical DE models, due to
a no-go theorem.

\subsection{A No-Go theorem}\label{sec:nogo}

In this subsection, we proceed to the detailed presentation and
proof of the ``No-Go" theorem which forbids the EoS parameter of a
single perfect fluid or a single scalar field to cross the $-1$
boundary. This theorem states that in a DE theory described by a
single perfect fluid, or a single scalar field $\phi$ minimally
coupled to Einstein Gravity with a lagrangian ${\cal L}=P(\phi,
X)$, in  FRW geometry, the DE EoS $w$ cannot cross over the
cosmological constant boundary. It has been proved or discussed in
various approaches in the literature
\cite{Xia:2007km,Vikman:2004dc, Hu:2004kh, Caldwell:2005ai,
Zhao:2005vj, Kunz:2006wc}, and in the following we would desire to
present a proof from the viewpoint of stability of DE
perturbations.

Let us first consider the fluid-case. Generally, a perfect fluid,
without viscosity and conduct heat, can be described by parameters
such as pressure $p$, energy density $\rho$ and entropy $S$,
satisfying a general EoS $p=p(\rho, S)$. According to the fluid
properties, such a perfect fluid can be classified into two
classes, namely barotropic and non-barotropic ones.

Working in the conformal Newtonian gauge\footnote{We refer to Ref.
\cite{Ma:1995ey} for a comprehensive study on cosmological
perturbation theory.}, one can describe the DE perturbations in
Fourier space as
\begin{eqnarray}
\label{dotdelta} \delta'&=&-(1+w)(\theta-3\Phi')-3{\cal H}(c_s^2-w)\delta\\
\label{dottheta} \theta'&=&-{\cal
H}(1-3w)\theta-\frac{w'}{1+w}\theta+k^2\left(\frac{c_s^2\delta}{1+w}+\Psi\right)~,
\end{eqnarray}
where the prime denotes the derivative with respect to conformal
time defined as $\eta\equiv\int dt/a$, ${\cal H}$ is the conformal
Hubble parameter, $c_s^2\equiv\delta p/\delta\rho$ is the sound
speed and
\begin{eqnarray}\label{adisoseq}
\delta\equiv\delta\rho/\rho~,~~\theta\equiv\frac{ik^j\delta
T^0_j}{\rho+p}~,
\end{eqnarray}
are the relative density and velocity perturbations respectively.
Finally, as usual the variables $\Phi$ and $\Psi$ represent metric
perturbations of scalar type.

If the fluid is barotropic, the iso-pressure surface is identical
with the iso-density surface, thus the pressure  depends only on
its density, namely $p=p(\rho)$. The adiabatic sound speed is
determined by
\begin{eqnarray}\label{soundspadiab}
c_a^2\equiv c_s^2|_{\rm
adiabatic}=\frac{p'}{\rho'}=w-\frac{w'}{3{\cal H}(1+w)}~.
\end{eqnarray}
From this relation we can see that the sound speed of a single
perfect fluid is apparently divergent when $w$ crosses $-1$, which
leads to instability in DE perturbation.

If the fluid is non-barotropic, the pressure  depends generally on
both its density and entropy, $p=p(\rho, S)$. The simple form of
the sound speed given in Eq.(\ref{adisoseq}) is not well-defined.
From the fundamental definition of the sound speed, that is taking
gravitational gauge invariance into consideration, we can obtain a
more general relationship between the pressure and the energy
density as follows:
\begin{eqnarray}
\label{gauge sound
speed}
 \delta\hat p=\hat c_s^2\delta\hat\rho~,
\end{eqnarray} with
the hat denoting gauge invariance. A gauge-invariant form of
density fluctuation can be written as
\begin{eqnarray}\label{gaugerho}
\delta\hat\rho=\delta\rho+3{\cal H}\left(\rho+
p\right)\frac{\theta}{k^2}~,\end{eqnarray} and correspondingly a
gauge-invariant perturbation of pressure writes
\begin{eqnarray}\label{gaugep}
\delta\hat p=\delta p+3{\cal H}c_a^2\left(\rho+
p\right)\frac{\theta}{k^2}~.
\end{eqnarray}
Finally, the gauge-invariant intrinsic entropy perturbation
$\Gamma$ can be described as
\begin{eqnarray}
\Gamma=\frac{1}{w\rho}(\delta
p-c_a^2\delta\rho)=\frac{1}{w\rho}\left(\delta\hat
p-c_a^2\delta\hat\rho\right)~.
\end{eqnarray}
Combining Eqs.(\ref{gauge sound speed})-(\ref{gaugep}), we obtain
the following expression, \label{deltap}
\begin{eqnarray} \delta
p&=&\hat{c}_s^2\delta\rho+\frac{3{\cal
H}\rho\theta(1+w)(\hat{c}_s^2-c_a^2)}{k^2}\nonumber\\
&=&\hat{c}_s^2\delta\rho+\hat{c}_s^2\frac{3{\cal
H}\rho\theta(1+w)}{k^2}-\frac{3{\cal
H}\rho\theta(1+w)}{k^2}w+\frac{\rho\theta w'}{k^2}.\ \ \ \ \
\end{eqnarray}
From Eq.(\ref{adisoseq}), one can see that $\theta$ will be
divergent when $w$ crosses $-1$, unless it satisfies the condition
\begin{eqnarray}\label{conditiontheta}
\theta w'=k^2\frac{\delta
p}{\bar\rho}~.
\end{eqnarray}

Substituting the definition of adiabatic sound speed $c^2_a$
(relation (\ref{soundspadiab})) and the condition
(\ref{conditiontheta}) into (\ref{gaugep}), we obtain $\delta\hat
p=0$. Since $\Gamma=\frac{1}{w\bar\rho}(\delta\hat
p-c_a^2\delta\hat\rho)$, it is obvious that due to the divergence
of $c_a^2$ at the crossing point, we have to require
$\delta\hat\rho=0$ to maintain a finite $\Gamma$. Thus, we deduce
the last possibility, that is $\delta\hat p=0$ and
$\delta\hat\rho=0$. From relations (\ref{gaugerho}),(\ref{gaugep})
this case requires that $\delta p=-c_a^2\frac{3{\cal
H}\bar\rho\theta(1+w)}{k^2}$ and $\delta\rho=-\frac{3{\cal
H}\bar\rho\theta(1+w)}{k^2}$, and thus $\delta p=c_a^2\delta\rho$.
Therefore, it returns to the case of adiabatic perturbation, which
is divergent as mentioned above.

In conclusion, from classical stability analysis  we demonstrated
that there is no possibility for a single perfect fluid to realize
$w$ crossing $-1$. For other proofs, see Refs. \cite{Hu:2004kh,
Kunz:2006wc}.

Let us now discuss the case of a general single scalar field. The
analysis is an extension of the discussion in Ref.
\cite{Zhao:2005vj}. The action of the field is given by
\begin{eqnarray}
S=\int d^4 x \sqrt{-g}P(\phi, X)~,
\end{eqnarray}
where $g$ is the determinant of the metric $g_{\mu\nu}$. In order
to study the DE EoS we first write down its energy-momentum
tensor. By definition  $\delta_{g_{\mu\nu}}S=-\int d^4 x
\frac{\sqrt{-g}}{2}T^{\mu\nu}\delta g_{\mu\nu}$ and comparing with
the fluid definition in the FRW universe one obtains:
\begin{eqnarray}
p&=&-T^{i}_{i}={\cal L}~,\\
\rho&=& T^{0}_{0}=2 X p_{,X}-p~,
\end{eqnarray}
where $``_{,X}"$ stands for $``\frac{\partial}{\partial X}"$.
Using the formulae above, the EoS is given by
\begin{eqnarray}
w=\frac{p}{\rho}=\frac{p}{2 X
p_{,X}-p}=-1+\frac{2Xp_{,X}}{2Xp_{,X}-p}~.
\end{eqnarray}
This means that, at the crossing point $t^*$, $Xp_{,X}|_{t^*}=0$.
Since $w$ needs to cross $-1$, it is required that $Xp_{,X}$
changes sign before and after the crossing point. That is, in the
neighborhood of $t^*$, $(t^*-\epsilon,t^*+\epsilon)$, we have
\begin{eqnarray}\label{crosscondition}
Xp_{,X}|_{t^*-\epsilon}\cdot Xp_{,X}|_{t^*+\epsilon}<0~.
\end{eqnarray}
Since $X=\frac{1}{2}\dot\phi^2$ is non-negative, relation
(\ref{crosscondition}) can be simplified as
$p_{,X}|_{t^*-\epsilon}\cdot p_{,X}|_{t^*+\epsilon}<0$. Thus, due
to the continuity of perturbation during the crossing epoch, we
acquire $p_{,X}|_{t^*}=0$.

Let us now consider the perturbations of the field. We calculate
the perturbation equation with respect to conformal time $\eta$ as
\begin{eqnarray}
u''-c_s^2\nabla^2 u-\left[\frac{z''}{z}+3c_s^2({\cal H}'-{\cal
H}^2)\right]u=0~,
\end{eqnarray} where we have defined
\begin{eqnarray}
u\equiv az\frac{\delta\phi}{\phi'}~,~~z\equiv
\sqrt{\phi'^2|\rho_{,X}|}~.
\end{eqnarray}
A Fourier expansion of the perturbation function $u$ leads to the
dispersion relation:
\begin{eqnarray} \label{omegadeee}
\omega^2=c_s^2k^2-\frac{z''}{z}-3c_s^2({\cal H}'-{\cal H}^2)~,
\end{eqnarray}
with $c_s^2$ defined as $\frac{p_{,X}}{\rho_{,X}}$. Stability
requires $c_s^2>0$. Since at the crossing point $p_{,X}=0$ and
$p_{,X}$  changes  sign, one can always find a small region where
$c_s^2<0$, unless $\rho_{,X}$ also becomes zero with a similar
behavior as $p_{,X}$. Therefore, the parameter $z$ vanishes at the
crossing and thus  ${\cal H}'-{\cal H}^2$ in (\ref{omegadeee}) is
finite. Hence, when assuming that the universe is fulfilled by
that scalar field it turns out to be zero, since at the crossing
point we have $\rho_{,X}=0$, $z=0$. Consequently, if $z''\neq0$ at
the crossing point the term $\frac{z''}{z}$ will be divergent.
Note that even if $z''=0$ this conclusion is still valid, since in
this case $\frac{z''}{z}=\frac{z^{'''}}{z^{'}}$. But $z$ is a
non-negative parameter  with $z=0$ being its minimum and thus $z'$
must vanish, therefore $\frac{z^{'''}}{z^{'}}$ is either divergent
or equal to $\frac{z^{(4)}}{z''}$, where $z''$ is also equal to
zero as  discussed above. Along this way, if we assume that the
first $(n-1)$-th derivatives of $z$ with respect to $\eta$ vanish
at the crossing point and that $z^{(n)}\neq 0$,  we can always use
the L'Hospital theorem until we find that
$\frac{z''}{z}=\frac{z^{(n)}}{z^{(n-2)}}$, which will still be
divergent. Therefore, the dispersion relation will be divergent at
the crossing point as well, and hence the perturbation will also
be unstable.

In summary, we have analyzed the most general case of a single
scalar field described by a lagrangian of the form ${\cal L}={\cal
L}(\phi, \partial_{\mu}\phi\partial^{\mu}\phi)$ and we have
studied different possibilities of $w$ crossing the cosmological
constant boundary. As we have  shown, these cases can either lead
to a negative effective sound speed $c_s^2$, or lead to a
divergent dispersion relation which makes the system unstable.

\subsection{Conditions of the $-1$-crossing}\label{sec:qcondi}

Let us close  this section by examining the realization conditions
of  quintom scenario. Dynamically, the necessary condition  can be
expressed as follows: when the EoS is close to the cosmological
constant boundary $w=-1$ we must have $w'|_{w=-1}\neq0$. Under
this condition we additionally require both the background and the
perturbations to be stable and to cross the boundary smoothly.
Therefore, we can achieve a quintom scenario by breaking certain
constraints appearing in the no-go theorem.

Since we have seen that a single fluid or a single scalar field
cannot give rise to quintom, we can introduce an additional degree
of freedom to realize it. Namely, we can construct a model in
terms of two scalars with one being quintessence and the other a
ghost field. For each component separately the EoS does not need
to cross the cosmological constant boundary and so their classical
perturbations are stable. However, the combination of these two
components can lead to a quintom scenario. An alternative way to
introduce an extra degree of freedom is to involve higher
derivative operators in the action. The realization of quintom
scenario has also been discussed within models of modified
gravity, in which we can define an effective EoS to mimic the
dynamical behavior of DE observed today. Furthermore, a few
attempts have been addressed in the frame of string theory,
however, models of this type suffers from  the problem of relating
stringy scale with DE. We will describes these quintom models in
the following sections.

Here we should stress again that in a realistic quintom
construction one ought to consider the perturbation aspects
carefully, since conventionally,  dangerous instabilities do
appear. The concordance cosmology is based on precise
observations,  many of which are tightly connected to the growth
of perturbations, and thus we must ensure their stability. If we
start merely with parameterizations of the scale factor or EoS to
realize a quintom scenario, it will become too arbitrary without
the considerations of perturbations. On the other hand, if we
begin with a scenario described by a concrete action which leads
to an EoS across $w=-1$, we can make a judgement on the model by
considering both its background dynamics and the stability of its
perturbations.



\section{The simplest quintom model with double fields}\label{sec:quintomsimple}

As we proved in the previous section, a single fluid or scalar
field cannot realize a viable quintom model in conventional cases.
Consequently, one must introduce extra degrees of freedom or
introduce the non-minimal couplings or modify the Einstein
gravity. In recent years there has been a large amount of research
in constructing models with $w$ crossing $-1$ \cite{Feng:2004ff,
Guo:2004fq, Zhang:2005eg, Wei:2005nw, Apostolopoulos:2005ff,
McInnes:2005vp, Li:2005fm, Andrianov:2005tm, Cai:2005ie,
Nojiri:2005sr, Capozziello:2005tf, Huang:2005gu, Wei:2005fq,
Wei:2005si, Alimohammadi:2006qi, Lazkoz:2006pa, Zhang:2006ck,
Apostolopoulos:2006si, Chimento:2006xu, Zhao:2006mp,
Aref'eva:2006et, Chimento:2006ac, Alimohammadi:2006tw, Guo:2006pc,
Cai:2006dm, Zhang:2006qu, Setare:2006rf, Zhang:2006at, Wei:2006va,
Cai:2007gs, Lazkoz:2007mx, Alimohammadi:2007jj, Babichev:2007dw,
Saridakis:2007ns, Saridakis:2007wx, Setare:2008pz,
Elizalde:2008yf, Sadatian:2008sv, Sadeghi:2008qp, Xiong:2008ic,
Alimohammadi:2008mh, Cai:2008gk, Zhang:2008ac, Setare:2008dw,
Nozari:2008yv, Nozari:2008gf, Setare:2008si, Nozari:2008ff,
Setare:2008pc, Nozari:2008hz, Setare:2008sf,
Creminelli:2008wc,Chimento:2008ws, Setare:2008mb, Leon:2008aq,
Saridakis:2009uu, Saridakis:2009ej,
Wang:2009av,Leon:2009ce,Saridakis:2009jq}.

\subsection{The model}\label{sec:quintomsimplemodel}

To begin with, let us construct the simple quintom cosmological
paradigm. It requires the simultaneous consideration of two
fields, namely one canonical $\phi$ and one phantom  $\s$, and the
DE is attributed to their combination \cite{Feng:2004ad}. The
action of a universe constituted of a such two fields is
\cite{Guo:2004fq, Zhang:2005eg}:
\begin{eqnarray}
S = \int d^{4}x \sqrt{-g} \bigg[\frac{1}{2\kappa^2} R -
\frac{1}{2}g^{\mu\nu}\partial_{\mu}\phi\partial_{\nu}\phi-\vp +
\frac{1}{2}g^{\mu\nu}\partial_{\mu}\sigma\partial_{\nu}\sigma-\vs
+ \cal{L}_\text{M}\bigg]~, \label{actionquint}
\end{eqnarray}
where we have set $\kappa^2\equiv 8\pi G$ as the gravitational
coupling. The term $\cal{L}_\text{M}$ accounts for the (dark)
matter content of the universe, which energy density $\rho_M$ and
pressure $p_M$ are connected by the EoS $\rho_M=w_M p_M$. Finally,
although we could straightforwardly include baryonic matter and
radiation in the model, for simplicity reasons we neglect them.

In a flat geometry, the Friedmann equations read
\cite{Guo:2004fq, Zhang:2005eg}:
\begin{eqnarray}\label{FR1}
H^{2}&=&\frac{\kappa^{2}}{3}\Big(\rho_{M}+\rho_{\phi}+\rho_{\s}\Big)~,\\
\label{FR2}
\dot{H}&=&-\frac{\kappa^2}{2}\Big(\rho_{M}+p_M+\rho_{\phi}+p_{\phi}+\rho_{\s}+p_{\s}\Big)~.
\end{eqnarray}
The evolution equations for the canonical and the phantom fields
are:
\begin{eqnarray}\label{eom}
\dot{\rho}_\phi+3H(\rho_\phi+p_\phi)&=&0,\\
\dot{\rho}_\s+3H(\rho_\s+p_\s)&=&0,
\end{eqnarray}
where $H=\dot{a}/a$ is the Hubble parameter.

In these expressions, $p_\phi$ and $\rho_{\phi}$ are respectively
the pressure and density of the canonical field, while  $p_\s$ and
$\rho_{\s}$ are the corresponding quantities for the phantom
field. They are given by:
\begin{eqnarray}
\label{rhophi}
 &&\rho_{\phi} = \frac{1}{2}\dot{\phi}^{2} + V_\phi(\phi) ~,~~ p_{\phi} = \frac{1}{2}\dot{\phi}^{2} - V_\phi(\phi)~,\\
\label{rhosigma}
 &&\rho_{\s} = -\frac{1}{2}\dot{\s}^{2} + V_\s(\s) ~,~~ p_{\s} = - \frac{1}{2}\dot{\s}^{2} - V_\s(\s)~,
\end{eqnarray}
where $V_\phi(\phi)$, $V_\s(\s)$ are the potentials for the
canonical and phantom field respectively. Therefore, we can
equivalently write the evolution equations for the two
constituents of the quintom model in field terms:
\begin{eqnarray}
\label{pddot}
 &&\ddot{\phi}+3H\dot{\phi}+\frac{\partial V_\phi(\phi)}{\partial\phi}=0\\
\label{sddot}
 && \ddot{\s}+3H\dot{\s}-\frac{\partial V_\s(\s)}{\partial\s}=0.
\end{eqnarray}
Finally, the equations close by considering the evolution of the
matter density:
\begin{eqnarray}\label{sys3}
\dot{\rho}_M+3H(\rho_M+p_M)=0.
\end{eqnarray}

As we have mentioned, in a double-field quintom model, the DE is
attributed to the combination of the canonical and phantom fields:
\begin{eqnarray}\label{rhoDE}
\rho_{DE}\equiv\rho_\phi+\rho_\s~,~~p_{DE}\equiv p_\phi+p_\s~,
\end{eqnarray}
and its EoS is given by
\begin{eqnarray}
w_{DE}\equiv\frac{p_{DE}}{\rho_{DE}}=\frac{p_\phi+p_\s}{\rho_\phi+\rho_\s}.
\end{eqnarray}
Alternatively, we could introduce the ``total'' energy density
$\rho_{tot}\equiv\rho_M+\rho_\phi+\rho_\s$, obtaining:
\begin{eqnarray}
\label{rhot}
 \dot{\rho}_{tot}+3 H(1+w_{tot})\rho_{tot}=0,
\end{eqnarray}
with
\begin{eqnarray}
 w_{tot}=\frac{p_\phi+p_\s+p_M}{\rho_\phi+\rho_\s+\rho_M}=w_\phi\Omega_\phi+w_\s\Omega_\s+w_M\Omega_M,
\end{eqnarray}
with the individual EoS parameters defined as
\begin{eqnarray}
 w_\phi=\frac{p_\phi}{\rho_\phi} ~,~~ w_\s=\frac{p_\s}{\rho_\s} ~,~~ w_M=\frac{p_M}{\rho_M}~,
\end{eqnarray}
and
\begin{eqnarray}
 \Omega_\phi\equiv\frac{\rho_\phi}{\rho_{tot}}~,~~\Omega_\s\equiv\frac{\rho_\s}{\rho_{tot}}~,~~\Omega_M\equiv\frac{\rho_M}{\rho_{tot}}~,
\end{eqnarray}
are the corresponding individual densities. These constitute:
\begin{eqnarray}
 \Omega_\phi+\Omega_\s\equiv\Omega_{DE},
\end{eqnarray}
and
\begin{eqnarray}
 \Omega_\phi+\Omega_\s+\Omega_M=1.
\end{eqnarray}

\subsection{Phase space analysis}\label{sec:phasespacebasic}

In order to investigate the properties of the constructed simple
quintom model, we proceed to a phase-space analysis. To perform
such a phase-space and stability analysis of the phantom model at
hand, we have to transform the aforementioned dynamical system
into its autonomous form  \cite{Copeland:2006wr,Ferreira:1997au,
Copeland:1997et,
 Chen:2008ft}. This will be achieved by introducing the auxiliary
variables:
\begin{eqnarray}\label{auxiliary0}
 x_\phi \equiv \frac{\kappa \dot{\phi }}{\sqrt{6}H} ~,~~
 x_\sigma \equiv \frac{\kappa \dot{\sigma }}{\sqrt{6}H}~,~~
 y_\phi \equiv \frac{\kappa \sqrt{V_\phi(\phi)}}{\sqrt{3}H} ~,~~
 y_\sigma \equiv \frac{\kappa \sqrt{V_\sigma(\sigma)}}{\sqrt{3}H}~,~~
 z \equiv \frac{\kappa \sqrt{\rho_M}}{\sqrt{3}H}~,
\end{eqnarray}
together with $N=\ln a$. Thus, it is easy to see that for every
quantity $F$ we acquire $\dot{F}=H\frac{dF}{dN}$.

Using these variables we straightforwardly obtain:
\begin{eqnarray}
\label{Omegaphi}
 \Omega_{\phi}&\equiv&\frac{\kappa^{2}\rho_{\phi}}{3H^{2}}=x_\phi^2+y_\phi^2~,\\
\label{Omegas}
 \Omega_{\s}&\equiv&\frac{\kappa^{2}\rho_{\s}}{3H^{2}}=-x_\s^2+y_\s^2~,\\
\label{OmegaDE}
 \Omega_{DE}&\equiv&\frac{\kappa^{2}(\rho_{\phi}+\rho_\s)}{3H^{2}}=x_\phi^2+y_\phi^2-x_\s^2+y_\s^2~,
\end{eqnarray}
and
\begin{eqnarray}
\label{wff}
 &&w_{\phi}=\frac{x_\phi^2-y_\phi^2}{x_\phi^2+y_\phi^2}~,~~
 w_{\s}=\frac{-x_\s^2-y_\s^2}{-x_\s^2+y_\s^2}~,\\
\label{wDE}
 &&w_{DE}=\frac{x_\phi^2-y_\phi^2-x_\s^2-y_\s^2}{x_\phi^2+y_\phi^2-x_\s^2+y_\s^2}~.
\end{eqnarray}
For $w_{tot}$ we acquire:
\begin{eqnarray}\label{wtot}
w_{tot}=x_\phi^2-y_\phi^2-x_\s^2-y_\s^2+(\gamma-1)z^2,
\end{eqnarray}
where we have introduced the barotropic form for the matter EoS,
defining $w_M\equiv \gamma-1$. Finally, the Friedmann constraint
(\ref{FR1}) becomes:
\begin{eqnarray}
\label{AS3} x_{\phi}^2+y_\phi^2-x_{\sigma}^2+y_\sigma^2+z^2=1.
\end{eqnarray}

A final assumption must be made in order to handle the potential
derivatives that are present in  (\ref{pddot}) and (\ref{sddot}).
The usual assumption in the literature is to assume an exponential
potential of the form
\begin{eqnarray}
V_{\phi}&=&V_{\phi_0}\,e^{-\kappa\lambda_\phi \phi}~, \nonumber\\
V_{\sigma}&=&V_{\sigma_0}\,e^{-\kappa\lambda_\s \s}~,
\end{eqnarray}
since exponential potentials are known to be significant in
various cosmological models  \cite{Ferreira:1997au,
Copeland:1997et, Copeland:2006wr}. Note that equivalently, but
more generally, we could consider potentials satisfying
$\lambda_\phi=-\frac{1}{\kappa V_\phi(\phi)}\frac{\partial
V_\phi(\phi)}{\partial\phi}\approx const$ and similarly
$\lambda_\s=-\frac{1}{\kappa V_\s(\s)}\frac{\partial
V_\s(\s)}{\partial\s}\approx const$ (for example this relation is
valid for arbitrary but nearly flat potentials
\cite{Scherrer:2007pu}).

Using the auxiliary variables (\ref{auxiliary0}), the equations of
motion (\ref{FR1}), (\ref{FR2}), (\ref{pddot}), (\ref{sddot}) and
(\ref{sys3}) can be transformed to an autonomous system containing
the variables $x_\phi$, $x_\s$, $y_\phi$, $y_\s$, $z$ and their
derivatives with respect to $N=\ln a$. Thus, we obtain:
\begin{eqnarray}\label{eomscol}
\textbf{X}'=\textbf{f(X)},
\end{eqnarray}
where $\textbf{X}$ is the column vector constituted by the
auxiliary variables, \textbf{f(X)} the corresponding  column
vector of the autonomous equations, and prime denotes derivative
with respect to $N=\ln a$. Then, we can extract its critical
points $\bf{X_c}$  satisfying $\bf{X}'=0$. In order to determine
the stability properties of these critical points, we expand
(\ref{eomscol}) around  $\bf{X_c}$, setting
$\bf{X}=\bf{X_c}+\bf{U}$ with $\textbf{U}$ the perturbations of
the variables considered as a column vector. Thus, for each
critical point we expand the equations for the perturbations up to
the first order as:
\begin{eqnarray}
\label{perturbation} \textbf{U}'={\bf{\Xi}}\cdot \textbf{U},
\end{eqnarray}
where the matrix ${\bf {\Xi}}$ contains the coefficients of the
perturbation equations. Thus, for each critical point, the
eigenvalues of ${\bf {\Xi}}$ determine its type and stability.

In particular, the autonomous form of the cosmological system is\\
\cite{Guo:2004fq, Lazkoz:2006pa, Guo:2006pc, Lazkoz:2007mx,
Alimohammadi:2007jj, Setare:2008pz, Setare:2008dw, Setare:2008si,
Setare:2008sf, Leon:2008aq}:
\begin{eqnarray}
\label{AS1}
 x'_\phi &=& -3x_\phi\left(1+x_{\sigma}^2-x_{\phi}^2-\frac{\gamma}{2} z^2\right)+\lambda_\phi \frac{\sqrt{6}}{2}y_{\phi}^2~, \\
 y'_\phi &=& 3y_\phi\left(-x_{\sigma}^2+x_{\phi}^2+\frac{\gamma}{2} z^2-\lambda_\phi \frac{\sqrt{6}}{6}x_\phi\right)~, \\
 x'_\sigma &=& -3x_\sigma\left(1+x_{\sigma}^2-x_{\phi}^2-\frac{\gamma}{2} z^2\right)-\lambda_\sigma \frac{\sqrt{6}}{2}y_{\sigma}^2~, \\
 y'_\sigma &=& 3y_\sigma\left(-x_{\sigma}^2+x_{\phi}^2+\frac{\gamma}{2} z^2-\lambda_\sigma \frac{\sqrt{6}}{6}x_\sigma\right)~, \\
 z' &=& 3z\left(-x_{\sigma}^2+x_{\phi}^2+\frac{\gamma}{2} z^2-\frac{\gamma}{2}\right)~.
\end{eqnarray}

The real and physically meaningful (i.e corresponding to $y_i>0$
and $0\leq\Omega_i\leq1$) of them are presented in table
\ref{stability1}. In table \ref{stability2} we present the
eigenvalues of the corresponding matrix  ${\bf {\Xi}}$ for these
critical points, which determine their stability properties (a
stable point requires negative real parts of all eigenvalues).

\begingroup

\begin{table*}

\begin{tabular}{c c c c c c c} \hline
Label & $x_{\sigma c}$ & $y_{\sigma c}$ & $x_{\phi c}$ & $y_{\phi
c}$ & $z_{c}$
 & \\ \hline
$A$ & $x_{\phi}^2-x_{\sigma}^2=1$ & 0 & & 0 & 0 & \\
$B$ & $-\frac{\lambda_\sigma}{\sqrt{6}}$
 & $\sqrt{ (1+\frac{\lambda_\sigma^2}{6})}$ & 0 & 0 & 0 &   \\
$C$ & 0 & 0 & $\frac{\lambda_\phi}{\sqrt{6}}$ &
 $\sqrt{ (1-\frac{\lambda_\phi^2}{6})}$ & 0 &   \\
$D$ & 0 & 0 & 0 & 0 & 1 & \\
$E$ & 0 & 0 & $\frac{3\gamma}{\sqrt{6}\lambda_\phi}$
 & $\sqrt{\frac{3\gamma(2-\gamma)}{2\lambda_\phi^2}}$
 & $\sqrt{1-\frac{3\gamma}{\lambda_\phi^2}}$ & \\ \hline
\end{tabular}
\caption[crit]{\label{stability1} The list of the critical points
of the simplest quintom model.}

\end{table*}
\endgroup

\begingroup
\begin{table*}
\begin{tabular}{c c c c c c c} \hline
{\small{Label}} & $m_{1}$ & $m_{2}$ & $m_{3}$ & $m_{4}$ &
{\small{Stability}}
 &  \\ \hline
{\small{$A$}} & {\small{$-6(1-\frac{\gamma}{2})x_{\sigma c}^{2}$}}
&{\small{ $3(1-\frac{\sqrt{6}}{6}\lambda_{\sigma}x_{\sigma c})$}}
&{\small{$6(1-\frac{\gamma}{2})x_{\phi c}^{2}$ }}&{\small{$ 3(1-\frac{\sqrt{6}}{6}\lambda_{\phi}x_{\phi c})$}} & unstable  & \\
{\small{$B$}} &{\small{ $-\frac{\lambda_{\sigma}^{2}}{2}$}}
 &{\small{ $-\frac{1}{2}(6+\lambda_{\sigma}^{2})$}} &{\small{ $-\frac{1}{2}(6+\lambda_{\sigma}^{2})$}} &{\small{ $-3\gamma-
\lambda_{\sigma}^{2}$ }}&{\small{ stable}} &   \\
{\small{$C$}} &{\small{ $\frac{\lambda_{\phi}^{2}}{2}$ }}&{\small{
$-3(1-\frac{\lambda_{\phi}^{2}}{6})$}} &{\small{
$-3(1-\frac{\lambda_{\phi}^{2}}{6})$}} &
 {\small{$-3\gamma+
\lambda_{\phi}^{2}$ }}&{\small{ unstable }}&   \\
{\small{$D$}} &{\small{ $\frac{3\gamma}{2}$ }}&{\small{ $\frac{3\gamma}{2}$ }}&{\small{ $-3(1-\frac{\gamma}{2})$}} &{\small{ $-3(1-\frac{\gamma}{2})$ }}&{\small{ unstable}} &  \\
{\small{$E$}} &{\small{ $\frac{3\gamma}{2}$}} &
{\small{$-(3-\frac{3\gamma}{2})$ }}&
{\tiny{$\frac{3(\gamma-2)}{4}\Big[1+\sqrt{1-\frac{8\gamma
\lambda_{\phi}^{2}-24\gamma^{2}}{2\lambda_{\phi}^{2}-\gamma
\lambda_{\phi}^{2}}}\Big]$}}
 & {\tiny{$\frac{3(\gamma-2)}{4}\Big[1-\sqrt{1-\frac{8\gamma
\lambda_{\phi}^{2}-24\gamma^{2}}{2\lambda_{\phi}^{2}-\gamma
\lambda_{\phi}^{2}}}\Big]$}}
 & {\small{unstable}} & \\ \hline
\end{tabular}
\caption[crit]{\label{stability2} The eigenvalues and stability of
the critical points of the simplest quintom model.}
\end{table*}
\endgroup

From Tables \ref{stability1} and \ref{stability2}, one can see
that the phantom-dominated solution is always a late-time stable
attractor (stable point $B$). In particular, this late-time
solution corresponds to $\Omega_{DE}=\Omega_\s=1$ and
$w_{DE}=w_\s=-1-\lambda_\s^2/3$ (see relations (\ref{OmegaDE}) and
(\ref{wDE})), i.e to a complete DE domination. This is true even
if there exists the interaction between the two fields
\cite{Zhang:2005eg}. Thus, if this coupled system is initially
dominated by the quintessence field, it will eventually evolve
into the phantom-dominated phase and the crossing through the
phantom divide is inevitable. This behavior provides a natural
realization of the quintom scenario, and was the motive of the
present cosmological paradigm. Finally, we mention that during its
motion to the late-time attractor, the dynamical system could
present an oscillating behavior for $w_{DE}$, with observationally
testable effects \cite{Zhang:2005eg}.

\subsection{State-finder diagnosis}\label{sec:statefinder}

In this paragraph we are going to present a way to discriminate
between the various quintom scenarios, in a model independent
manner, following  \cite{Wu:2005apa} and based in the seminal
works \cite{Chiba:1998tc, Sahni:2002fz}. In those works, the
authors proposed a cosmological diagnostic pair $\{r, s\}$ called
statefinder, which are defined as
\begin{eqnarray}\label{state}
 r\equiv \frac{\dddot{a}}{aH^3}~,~~~
s\equiv\frac{r-1}{3(q-1/2)}~,
\end{eqnarray}
to differentiate between different forms of DE. In these
expressions, $q$ is the deceleration parameter $q\equiv -a\ddot
a/\dot a^2= -\ddot a/aH^2$, $r$ forms the next step in the
hierarchy of geometrical cosmological parameters beyond $H$ and
$q$, and $s$ is a linear combination of $r$ and $q$. Apparently,
the statefinder parameters  depend only on  $a$ and its
derivatives, and  thus it is a geometrical diagnostic. Since
different quintom cosmological models exhibit qualitatively
different evolution trajectories in the $s-r$ plane, this
statefinder diagnostic can differentiate between them.

However, let us  use an alternative form of statefinder parameters
defined as \cite{Wu:2005apa}:
\begin{eqnarray}\label{statefinder}
 r=1+\frac{9}{2}\frac{(\rho_{tot}+p_{tot})}{\rho_{tot}}\frac{\dot{p}_{tot}}{\dot{\rho}_{tot}}~,~~
 s=\frac{(\rho_{tot}+p_{tot})}{\rho_{tot}}\frac{\dot{p}_{tot}}{\dot{\rho}_{tot}}~.
\end{eqnarray} Here
$\rho_{tot}\equiv\rho_M+\rho_\phi+\rho_\s+\rho_{rad}$  is the
total energy density (note that in order to be more general we
have also added  a radiation part) and $p_{tot}$ is the
corresponding total pressure in the universe.  Since the total
energy, the DE (quintom) energy and radiation are conserved
separately, we have $\dot{\rho}_{tot}=-3H(\rho_{tot}+p_{tot})$,
$\dot{\rho}_{DE}=-3H(1+w_{DE})\rho_{DE}$ and
$\dot{\rho}_{rad}=-4H\rho_{rad}$, respectively. Thus, we can
obtain
\begin{eqnarray}\label{rs}
 r&=&1-\frac{3}{2}[\dot{w}_{DE}/H-3w_{DE}(1+w_{DE})]\Omega_{DE}+2\Omega_{rad}~,\\
 s&=&\frac{-3[\dot{w}_{DE}/H-3w_{DE}(1+w_{DE})]\Omega_{DE}+4\Omega_{rad}}{9w_{DE}\Omega_{DE}+3\Omega_{rad}}~,
\end{eqnarray}
and
\begin{eqnarray}
 q=\frac{1}{2}(1+3w_{DE}\Omega_{DE}+\Omega_{rad})~,
\end{eqnarray}
where as usual $\Omega_{DE}=\rho_{DE}/\rho_{tot}$ and
$\Omega_{rad}=\rho_{rad}/\rho_{tot}$.

\begin{figure}[tb]
\begin{center}
\includegraphics[
width=4.3in] {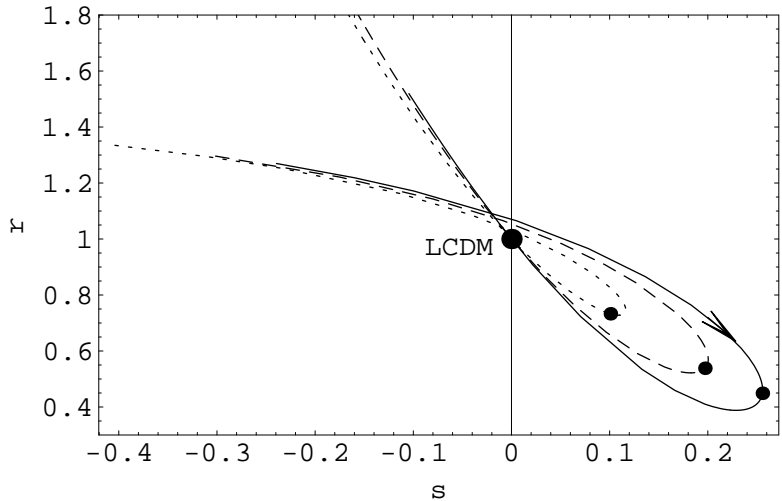} \caption{(Color online) {\it The $r-s$
diagram of no direct coupling exponential potential $V (\sigma,
\phi) = V_{\sigma_0} e^{-\alpha \sigma} +
V_{\phi_0}e^{-\beta\phi}$, with $V_{\sigma_0}=0.3\rho_0$ and
$V_{\phi_0}=0.6\rho_0$, $\beta=3$ and $\alpha=1, ~1.5,~ 2$(solid
line, dashed line and dot-dashed line respectively), and $\rho_0$
the present energy density of our universe. The curves $r(s)$
evolve in the time interval $\frac{t}{t_0}\in[0.5,4]$ where $t_0$
is the present time. Dots locate the current values of the
statefinder parameters. From Ref.  \cite{Wu:2005apa}.} }
\label{fig1statefnder}
\end{center}
\end{figure}

We discuss the statefinder for the quintom scenario, imposing
various potentials $V(\phi,\s)\equiv V_\phi(\phi)+V_\s(\s)$.
Firstly we assume that there is no direct coupling between the
phantom scalar field and the normal scalar field, that is we
assume the simple exponential potentials of the previous
subsection: $V (\sigma, \phi) = V_{\sigma_0} e^{-\alpha \sigma} +
V_{\phi_0}e^{-\beta\phi}$, where $\alpha$ and $\beta$ are
constants. As we have shown, in this case  the universe is moving
towards the phantom dominated, late time attractor ~
\cite{Guo:2004fq, Zhang:2005eg,
Saridakis:2009ej,Saridakis:2009pj}. In Fig.\ref{fig1statefnder} it
is depicted the time evolution of statefinder pair $\{r,s \}$ in
the time interval $\frac{t}{t_0}\in[0.5,4]$ where $t_0$ is the
present time
 \cite{Wu:2005apa}.
 As we can see, in the past and future the
 $r-s$ curve
is almost linear, which means that the deceleration parameter
changes from one constant to another nearly with the increasing of
time, and the parameters will pass the fixed point of LCDM in the
future. These trajectories of  $r(s)$ are different from other DE
models discussed in  \cite{Sahni:2002fz, Alam:2003sc,
Gorini:2002kf, Zimdahl:2003wg, Zhang:2005rj, Zhang:2005yz}.

\begin{figure}[htbp]
\begin{center}
\includegraphics[width=8cm]{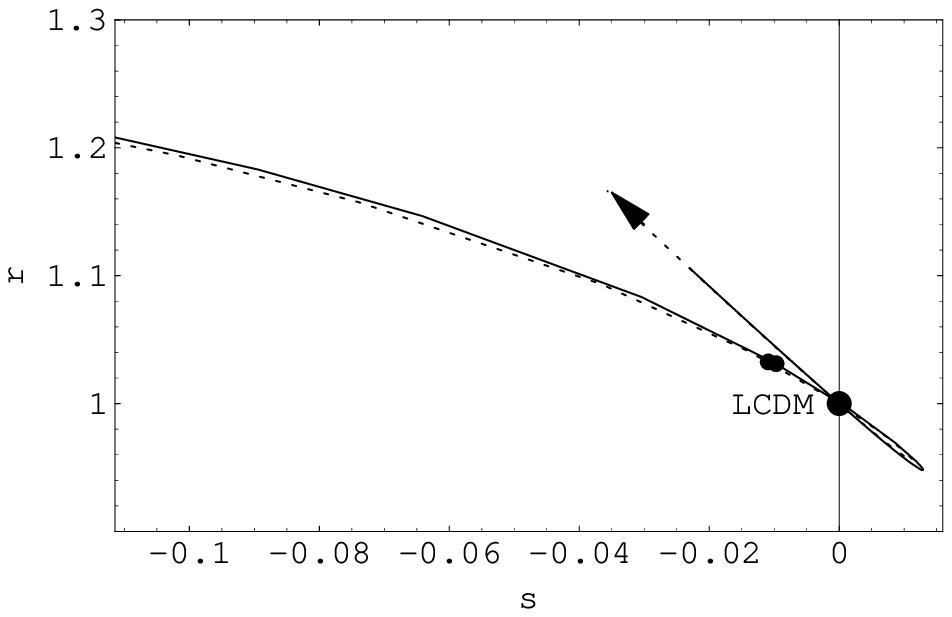}\\ \includegraphics[width=8cm]{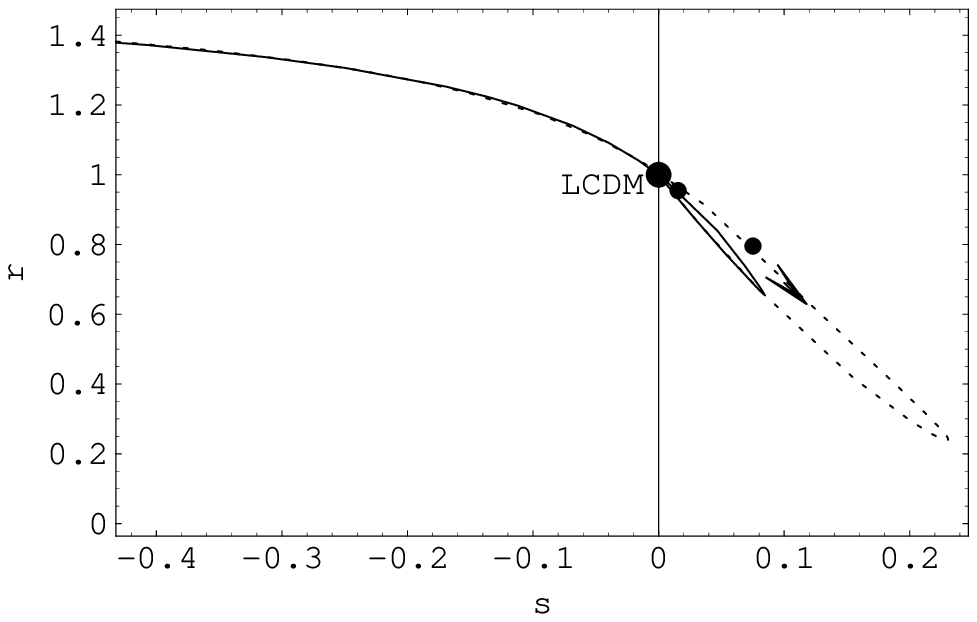}
\caption{ {\it{ In the upper figure curves $r(s)$ evolve  in the
time interval $\frac{t}{t_0}\in[0.5,4]$ where $t_0$ is the present
time for the potential $V (\sigma, \phi) = V_{\sigma_0} e^{-\alpha
\sigma} + V_{\phi_0}e^{-\beta\phi}+V_0e^{-\kappa(\sigma+\phi)}$.
The model parameters are chosen as  $V_{\sigma_0}=0.3\rho_0$,
$V_{\phi_0}=0.6\rho_0$, $V_{0}=0.3\rho_0$, $\alpha=1$, $\beta=1$
and $\kappa=1,~ 2$(solid line and dot-dashed line respectively).
In the lower figure the potential is  $V (\sigma, \phi) =
V_{\sigma_0} e^{-\alpha \sigma^2} + V_{\phi_0}e^{-\beta\phi^2}$.
The model parameter are chosen as $V_{\sigma_0}=0.3\rho_0$,
$V_{\phi_0}=0.6\rho_0$, $\alpha=1$ and $\beta=1,~3$(solid line and
dot-dashed line respectively). Dots locate the current values of
the statefinder parameters. From Ref.  \cite{Wu:2005apa}.}}}
 \label{fig2statefnder}
 \end{center}
\end{figure}

Similarly, in Fig.\ref{fig2statefnder} we can see the  $r(s)$
curves for the potentials $V (\sigma, \phi) = V_{\sigma_0}
e^{-\alpha \sigma} +
V_{\phi_0}e^{-\beta\phi}+V_0e^{-\kappa(\sigma+\phi)}$ and $V
(\sigma, \phi) = V_{\sigma_0} e^{-\alpha \sigma^2} +
V_{\phi_0}e^{-\beta\phi^2}$, which lead to late-time attractors.
The former potential leads to a Big Rip attractor while the latter
one to a de Sitter attractor  \cite{Guo:2004fq, Zhang:2005eg}.
Apparently the upper figure is very similar to
Fig.\ref{fig1statefnder}. This shows that for uncoupling and
coupling exponential potentials the evolutions of our universe are
very similar in the time interval we consider here.  The lower
figure is different from Fig.\ref{fig1statefnder} but has a common
characteristic with the phantom with power-law potential
 \cite{Saridakis:2009pj}, quintessence with inverse power-law
potential and Chaplygin gas model~ \cite{Alam:2003sc} that it
reaches the point of LCDM with the increasing of time.  This is
due to the fact that they all lead to  the same fate of the
universe--de Sitter expansion, but the trajectories to LCDM are
different, therefore they can be differentiated.

\begin{figure}[htbp]
\begin{center}
\includegraphics[width=8cm]{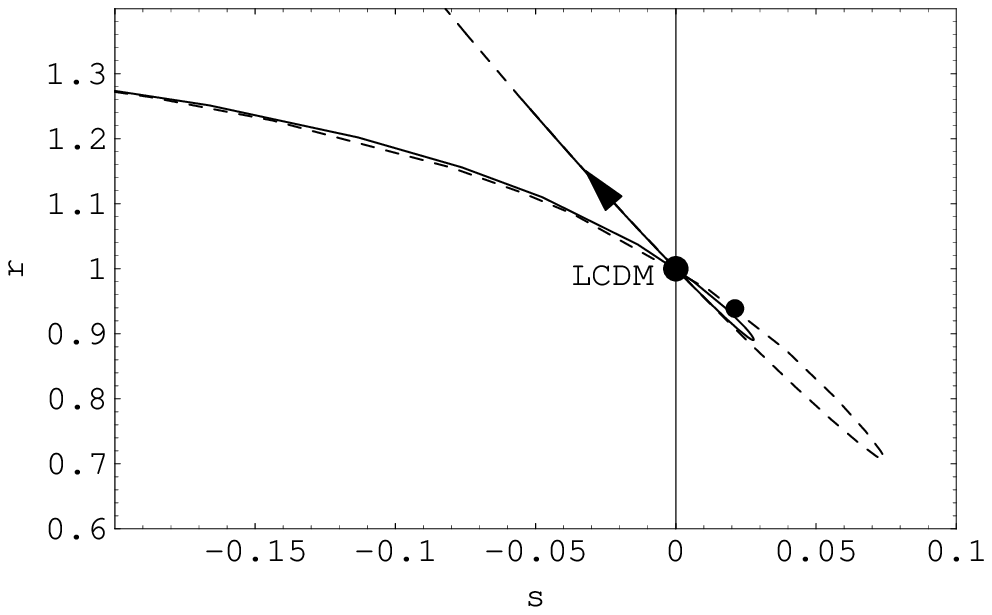}\\
\includegraphics[width=8cm]{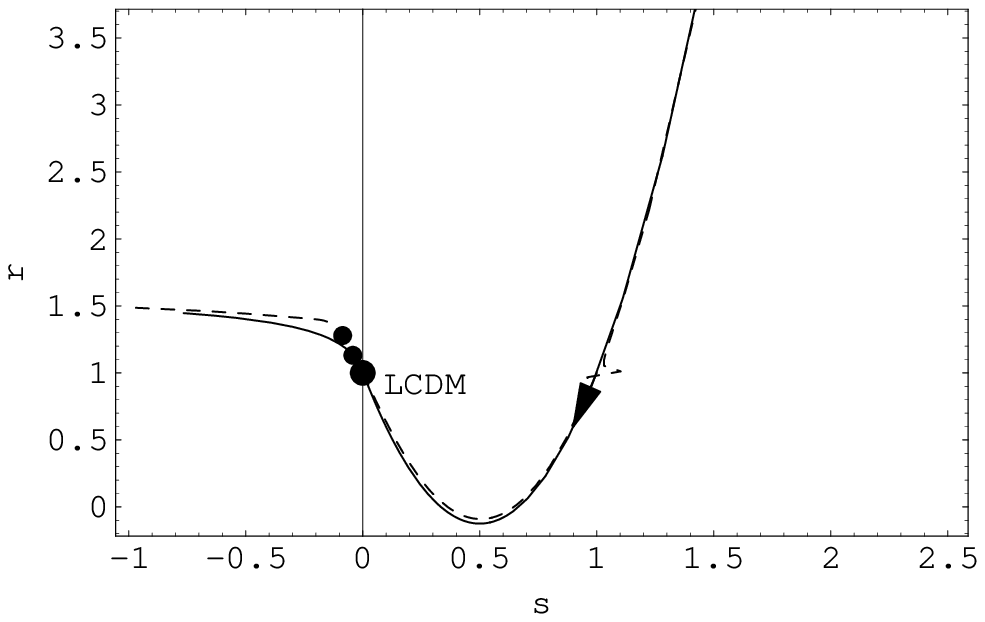}
\caption{ {\it{ The $r-s$ diagrams for the linear potential case
$V (\sigma, \phi) =\kappa( \sigma+\phi) +\lambda\sigma\phi$. The
curves $r(s)$  evolve in the time interval
$\frac{t}{t_0}\in[0.5,4]$. In the upper graph $\kappa=0.8$ and
$\lambda=-0.2,~ -0.8$(solid line and dashed line respectively). In
the lower graph $\kappa=0.8$ and $\lambda=0.2,~ 0.8$(solid line
and dashed line respectively). Dots locate the current values of
the statefinder parameters. From Ref.  \cite{Wu:2005apa}.}}}
 \label{fig3statefnder}
\end{center}
\end{figure}

Let us investigate the  case of linear coupling potential $V
(\sigma, \phi) =\kappa( \sigma+\phi) +\lambda\sigma\phi$, where
$\kappa$ and $\lambda$ are two constants. The scalar field with a
linear potential was first studied in   \cite{Garriga:2003nm} and
it has been argued that such a potential is favored by anthropic
principle considerations  \cite{Garriga:1999bf, Garriga:2002tq,
Garriga:2003hj} and can solve the coincidence problem
 \cite{Avelino:2004vy}. In addition, if the universe is dominated
by quintessence (phantom) with this potential, it ends with a Big
Crunch (Big Rip) \cite{Caldwell:2003vq, Perivolaropoulos:2004yr}.
The time evolutions of statefinder pair $\{r,s \}$ in  the time
interval $\frac{t}{t_0}\in[0.5,4]$ are shown in
Fig.\ref{fig3statefnder}.
We see that in the case of negative coupling the diagram is very
similar to Fig.\ref{fig1statefnder} and the upper graph in
Fig.\ref{fig2statefnder}, which shows that in these cases the
evolutions of our universe are similar in the time interval we
consider here.

In conclusion, the statefinder diagnostic can differentiate the
quintom model with other DE models, but it is not very helpful in
order to differentiate quintom DE models with some different kinds
of potentials which lead to a similar evolution of our unverse in
the time interval we consider.

\subsection{Cosmic duality}\label{sec:duality}

As we have seen in paragraph \ref{sec:quintomsimplemodel} the
simplest quintom model consists of two scalar fields, one
canonical, quintessence-like and one phantom one \cite{Guo:2004fq,
Zhang:2005eg}. In  \cite{Guo:2006pc} it was shown that there
exists two basic categories of such quintom models: one is
quintom-A type, where  the canonical field dominates at early
times while the phantom one dominates at late times and thus the
phantom divide crossing is realized from above to below, and the
other is quintom-B type for which the EoS is arranged to change
from below $-1$ to above $-1$. As we have analyzed in paragraph
\ref{sec:quintomsimplemodel}, the exponential, as well as other
simple potential forms, belong to quintom-A type, which is
realized easily. However, if one desires to construct a specific
model of quintom-B type then he has to use more sophisticated or
fine-tuned potentials, or to add more degrees of freedom
 \cite{Guo:2006pc}. Alternatively he can include higher derivative
terms (see for example  \cite{Li:2005fm}), or add suitably
constructed interactive terms which lead to a transition from
phantom-to-quintessence domination (see  \cite{Zhang:2005eg}).
Generally speaking, both quintom-A and quintom-B types could be
consistent with current observational data.

One question arises naturally: is there a cosmic duality between
these two quintom types? Dualities in field  and string theory
have been widely studied,  predicting many interesting phenomena
\cite{Polchinski:2005}. The authors of \cite{Chimento:2002gb,
Aguirregabiria:2003uh} have considered a possible transformation
with the Hubble parameter and they have studied the relevant
issues with the cosmic duality \cite{Veneziano:1991ek,
Lidsey:1999mc}. Specifically, in \cite{Chimento:2003qy}  a link
between  standard cosmology with quintessence matter and
contracting cosmology with phantom, has been shown. Later on this
duality was generalized into more complicated DE models, where it
has been shown to exist too
\cite{Dabrowski:2003jm,Chimento:2005xa, Chimento:2005au,
Chimento:2006gk, Gupta:2009kk,Dutta:2009yb}. In
\cite{Dabrowski:2006iv, Dabrowski:2006dd} the authors have studied
this cosmic duality and its connection to the fates of the
universe, while in \cite{Singh:2006sg} the author has also
discussed the possibility of realizing the aforementioned duality
in  braneworld cosmological paradigm. A common feature of these
studies is that the EoS parameter does not cross $-1$. Therefore,
it would be interesting to study the implications of this cosmic
duality in quintom models of DE, and in particular between
quintom-A and quintom-B types.

We consider the simple quintom model constructed in
\ref{sec:quintomsimplemodel}. Following
\cite{Cai:2006dm,Chimento:2003qy} we can construct a
form-invariant transformation by defining a group of new
quantities $\bar H$, $\bar\rho$, $\bar p$ and $\bar w$ which keep
Einstein equations invariant:
\begin{eqnarray}
 \bar\rho&=&\bar{\rho}\left(\rho_{DE}\right)~, \\
 \bar{H}&=&-\left(\frac{\bar{\rho}}{\rho_{DE}}\right)^{\frac{1}{2}}H~.
\end{eqnarray}
Under this transformation, we obtain the corresponding changes for
the pressure $p_{DE}$ and the EoS $w_{DE}$,
\begin{eqnarray}
\label{ptrans}
 \bar{p}&=&-\bar\rho-\left(\frac{\rho_{DE}}{\bar\rho}\right)^{\frac{1}{2}}\left(\rho_{DE}+p_{DE}\right)\frac{d\bar\rho}{d\rho_{DE}}~,\\
\label{wtrans}
 \bar{w}&=&-1-\left(\frac{\rho_{DE}}{\bar\rho}\right)^{\frac{3}{2}}\frac{d\bar\rho}{d\rho_{DE}}\left(1+w_{DE}\right)~.
\end{eqnarray}

From relations (\ref{ptrans}) and (\ref{wtrans}) one can see that
for a positive $\frac{d\bar\rho}{d\rho_{DE}}$, one would be able
to establish a connection between the quintom-A and quintom-B
types. Assuming without loss of generality, and as an example for
a detailed discussion, that $\bar\rho=\rho_{DE}$ in (\ref{ptrans})
and (\ref{wtrans}), we can obtain the dual transformation:
\begin{eqnarray}
 \bar H&=&-H~,\\
 \bar p&=&-2\rho_{DE}-p_{DE}~,\\
 \bar w&=&-2-w_{DE}~.
\end{eqnarray}
Consequently, using the canonical and phantom energy density
definitions (\ref{rhophi}) and (\ref{rhosigma}), we can extract
the dual form of the (quintom) DE Lagrangian:
\begin{eqnarray}
\label{Transform of L}
 \bar{\cal L} = \frac{1}{2}\partial_\mu\sigma\partial^\mu\sigma-\frac{1}{2}\partial_\mu \phi\partial^\mu\phi
 - \delta{\cal L}_1(\phi)-\delta{\cal L}_2(\sigma)~,
\end{eqnarray}
where $\delta{\cal L}_1$ and $\delta{\cal L}_2$ are
\begin{eqnarray}
 \delta{\cal L}_1&=&V_\phi(\phi)+{\dot\phi}^2~,\\
 \delta{\cal L}_2&=&V_\sigma(\sigma)-{\dot\sigma}^2~.
\end{eqnarray}
Therefore, we can easily see that if the original Lagrangian is a
quintom-A type then the dual one is a quintom-B type, and vice
versa. Thus, under this duality one expects a general connection
amongst different fates of the universe, and it might be possible
that the early universe is linked to its subsequent epochs.

For a specific discussion let us impose a special form for the
potentials \cite{Cai:2006dm}:
\begin{eqnarray}
 \label{V_1} V_\phi(\phi)&\propto&-3\sqrt{2}\phi+2e^{-\sqrt{2}\phi}~,\\
 \label{V_2} V_\sigma(\sigma)&\propto&\frac{3}{2}{\sigma}^2+4\sigma~.
\end{eqnarray}
Solving explicitly the cosmological equation (\ref{FR1}),
(\ref{FR2}), (\ref{pddot}), (\ref{sddot}), neglecting the matter
density, we study the two periods of the universe evolution. For
early times where $|t|\ll1$ (in Planck mass units), we choose the
initial conditions by fixing $\phi \rightarrow -\infty$ and
$\sigma \rightarrow 0$. With these initial conditions we can see
that the dominant component in DE density   is the exponential
term of $V_\phi(\phi)$ in (\ref{V_1}), and thus the contribution
from the phantom potential in (\ref{V_2}) and the linear part of
quintessence potential in (\ref{V_1}) can be neglected. Therefore,
the universe behaves like being dominated by the quintessence
component $\phi$ and it evolves following the approximate
analytical solution  \cite{Cai:2006dm}:
\begin{eqnarray}\label{result1}
 \phi\sim\sqrt{2}\ln{|t|}~,~~ \sigma\sim\frac{1}{2}t^2~,~~ H\sim\frac{1}{t}~.
\end{eqnarray}
Thus, we see that the scale factor in this period would variate
with respect to time as: $a\propto\pm t$, in which the signal is
determined by the positive definite form of the scale factor.
Therefore, the scale factor here  corresponds to the Big Bang or
Big Crunch of quintessence-dominanated universe.

The dual form of the solution above is a description of a universe
dominated by a phantom component with a Lagrangian given by
(\ref{Transform of L}) and
\begin{eqnarray}
\delta{\cal L}_1+\delta{\cal
L}_2={\dot\phi}^2-{\dot\sigma}^2+V_\phi(\phi)+V_\sigma(\sigma)
\propto\left(-3\sqrt{2}\phi+4e^{-\sqrt{2}\phi}+2\sigma+\frac{3}{2}{\sigma}^2\right)~.
\end{eqnarray}
In addition, the dual Hubble parameter is of the form
$\bar{H}\sim-\frac{1}{t}$ and for the scale factor we acquire
$a\propto\pm\frac{1}{t}$. Accordingly, the scale factor of the
dual form is tending towards infinity in the beginning or the end
of universe. From what we have investigated so far, we can see
that, for the positive branch there is a duality between an
expanding universe with initial singularity at $t=0^+$ and a
contracting one that begins with an infinite scale factor at
$t=0^+$. However, for the negative branch there is a duality
between a contracting universe ending in a big crunch at $t=0^-$
and an expanding one that ends in a final Big Rip at $t=0^-$. The
latter is dominated by a phantom component. Besides, in general,
under phantom domination the contracting solution is not stable,
because the phantom universe will evolve into Big Rip or Big
Sudden or will expand forever approaching to a de Sitter solution,
and thus it will not be able to stay in the contracting phase
forever \cite{Sami:2003xv,Guo:2004ae}. This problem, however, can
be avoided in quintom cosmology since in the dual universe with
quintom-B DE the increase of kinetic energy of phantom during the
contraction can be set off by that of quintessence at late time.

For times where $|t|\gg1$, the phantom component in the first
quintom model will dominate and the universe will expand. For the
specific potentials  (\ref{V_1}) and (\ref{V_2}) the mass term in
$V_\sigma(\sigma)$ will gradually play an important role in the
evolution of DE. Proceeding as above we obtain:
\begin{eqnarray}
 \phi\sim\sqrt{2}\ln{|t|}~,~~ \sigma\sim\sqrt{2}t~,~~ H\sim t,
\end{eqnarray}
where we note that the scale factor is
$a\propto\exp(\frac{t^2}{2})$. Consequently, the scale factor
would increase towards infinity rapidly for the positive branch,
while it would start from infinity for the negative branch. In
this case, the transformed Lagrangian is (\ref{Transform of L})
with
\begin{eqnarray}
\delta{\cal L}_1+\delta{\cal L}_2 \propto\left(
-3\sqrt{2}\phi+4e^{-\sqrt{2}\phi}+4\sigma+\frac{3}{2}{\sigma}^2-2\right).
\end{eqnarray}
The universe is evolving with a Hubble parameter $\bar{H}\sim{-t}$
and a scale factor $a\propto\exp(-\frac{t^2}{2})$, which is close
to singularity related to the origin and the fate of universe.
Finally, the component which dominates the evolution of the
universe is $\sigma$, that is the scalar field resembling
quintessence in (\ref{Transform of L}). Consequently we conclude
that, for the positive branch, there is a dual relation between an
expanding universe with a fate of expanding for ever with
$t\rightarrow+\infty$ and a contracting universe with a destiny of
shrinking for ever with $t\rightarrow+\infty$. Meanwhile, for the
negative one, there is a duality connecting a contracting universe
starting from near infinity with $t\rightarrow-\infty$ and an
expanding universe originating from infinity with
$t\rightarrow-\infty$.

\begin{figure}[htbp]
\begin{center}
\includegraphics[scale=1]{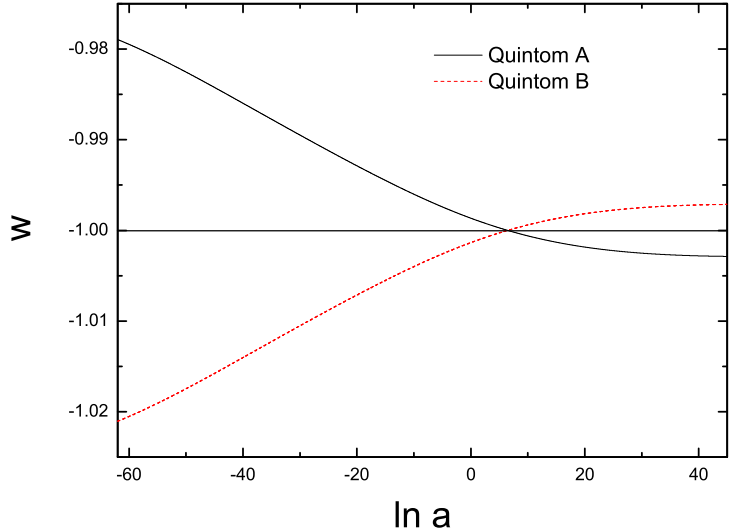}
\caption{(Color Online){\it{Evolution of EoS parameter $w$ of the
quintom model and its dual as a function of the scale factor $\ln
a$ for
$V=-3\sqrt{2}M^3\phi+2M^4e^{-\frac{\sqrt{2}\phi}{M}}+\frac{3}{2}M^2{\sigma}^2+4M^3\sigma$
and $M$ is the Planck mass. From Ref.  \cite{Cai:2006dm}. }}
 \label{Fig:dual S}}
\end{center}
\end{figure}

Having presented the analytical arguments for the duality between
quintom A and quintom B types, we proceed to numerical
investigation. In Fig.\ref{Fig:dual S} we depict the evolution of
the EoS parameter of the quintom model and its dual. One can see
from this figure that under the framework of the duality studied
above, the EoS of the quintom model and its dual are symmetric
around $w=-1$. Accordingly, in this case quintom A  is dual to
quintom B rigorously, which supports our analytical arguments
above.

\begin{figure}[htbp]
\begin{center}
\includegraphics[scale=1]{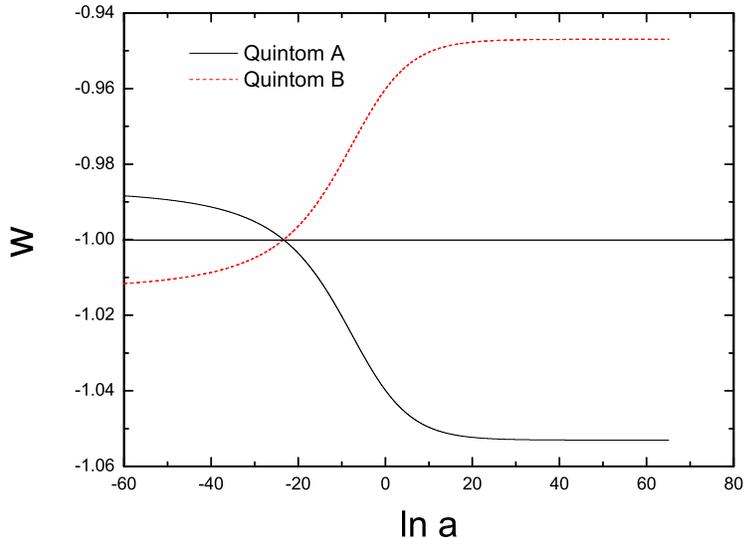}
\caption{(Color Online){\it {Evolution of EoS parameter $w$  of
the quintom model and its dual as a function of the scale factor
$\ln a$ for $V=V_0(e^{-\frac{\phi}{M}}+e^{-\frac{2\sigma}{M}})$
and $M$ is the planck mass. From Ref.  \cite{Cai:2006dm}. }}
 \label{Fig:dualexp}}
\end{center}
\end{figure}

\begin{figure}[htbp]
\begin{center}
\includegraphics[scale=1]{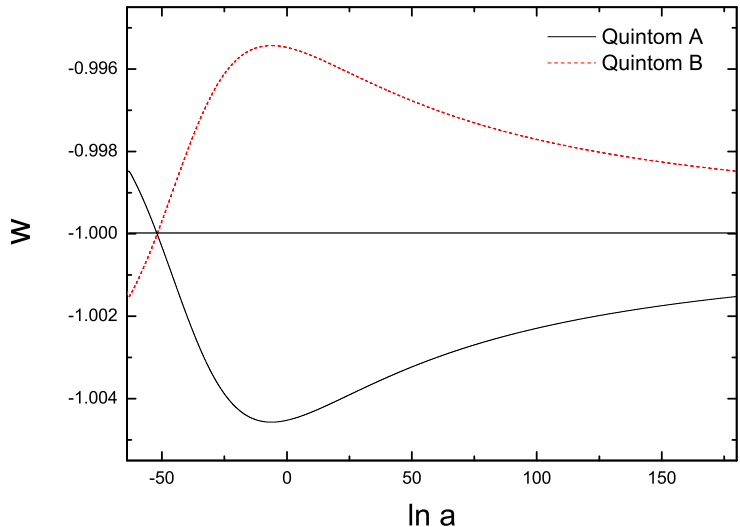}
\caption{(Color Online){\it{Evolution of  EoS parameter $w$  of
the quintom model and its dual as a function of the scale factor
$\ln a$ for $V={m_1}^2{\phi}^2+{m_2}^2{\sigma}^2$ where $m_1$
corresponds to the quintessence component mass and $m_2$ the
phantom component mass. From Ref.  \cite{Cai:2006dm}.}}
 \label{Fig:dualmass}}
\end{center}
\end{figure}

In Fig.\ref{Fig:dualexp} we assume the potentials $V_\phi(\phi)$
and $V_\sigma(\sigma)$ to be exponentials and one can see that the
EoS parameter for quintom A approaches to a fixed value which
corresponds to the attractor solution of this type of model
\cite{Zhang:2005eg}. Through the duality we can see that there
exists a corresponding attractor of the quintom B model dual to
the former one. finally, in Fig.\ref{Fig:dualmass}, we provide
another example for the duality.

In summary, we have shown that the quintom model has its dual
partner, specifically the quintom A model is dual to the quintom B
one, while the cosmological equations are form-invariant. These
two models describe two different behaviors of the universe
evolution with one in the expanding phase and the other in the
contracting one, depending on the imposed initial conditions. The
cosmic duality, which connects the two totally different scenarios
of universe evolution, preserves the energy density of the
universe unchanged  but it transforms the Hubble parameter.


%

\section{Perturbation theory and current observational
constraints}\label{sec:perturbation}

In this section we investigate the perturbations of quintom DE
scenario and the effects of these perturbations on  current
observations. As proved in Section \ref{sec:quintombasic}, it is
forbidden for a single fluid or scalar field to realize a quintom
scenario in conventional cases, and thus one is led to add extra
degrees of freedom. Therefore, it is important to check the
consistency of this extension at the classical level and in
particular to analyze the behavior of perturbations when the EoS
crosses the cosmological constant boundary \cite{Zhao:2005vj}.

\subsection{Analysis of perturbations in quintom
cosmology}\label{sec:quintompert}

As we have argued above, the quintom scenario needs extra degrees
of freedom to the conventional models of a single scalar field,
and the simplest realization of the quintom is a model with two
scalar fields or two ``effective"  scalar fields in the case of
higher-derivative operators. In the following discussions on
quintom perturbations we will restrict ourselves to the
double-field model of quintom with the lagrangian
\begin{equation}\label{quintomlag}
    \mathcal{L}=\mathcal{L}_{Q}+\mathcal{L}_{P},
\end{equation}
 where \begin{equation}\label{qlag}
    \mathcal{L}_{Q}=\frac{1}{2}\partial_{\mu}\phi_{1}\partial^{\mu}\phi_{1}-V_{1}(\phi_{1})
\end{equation}
describes the quintessence component, and
\begin{equation}
\mathcal{L}_{P}=-\frac{1}{2}\partial_{\mu}\phi_{2}\partial^{\mu}\phi_{2}-V_{2}(\phi_{2})
\end{equation}
for the phantom component. The background equations of motion for
the two scalar
 fields $ \phi_i ( i=1, 2) $ are
\begin{equation}\label{phiEOM}
      \ddot{\phi_i} + 2 \mathcal{H} \dot{\phi_i} \pm a^2 \frac{\partial  V_i}{\partial \phi_i}=
      0~,
\end{equation}
where the positive sign is for the quintessence and the minus sign
for the phantom. In general there will be couplings between the
two scalar fields, but for simplicity we neglect them.

For a complete perturbation study, both the fluctuations of the
fields, as well as those of the metric, need to be considered. In
the conformal Newtonian gauge the perturbed metric is given by
\begin{equation}\label{lineelecon}
ds^{2}=a^2(\tau)[(1+2\Psi)d\tau^{2} - (1-2\Phi)dx^{i}dx_{i}]~.
\end{equation}
Using the notations of \cite{Ma:1995ey}, the perturbation
equations satisfied by each component of the two-field quintom
model read:
\begin{eqnarray}
\label{dotdelta}
  \dot\delta_{i}&=&-(1+w_{i})(\theta_{i}-3\dot{\Phi})
  - 3\mathcal{H}\left(\frac{\delta P_{i}}{\delta\rho_{i}}-w_{i}\right)\delta_{i}~, \\
\label{dottheta}
  \dot\theta_{i}&=&-\mathcal{H}(1-3w_{i})\,\theta_{i}-\frac{\dot{w_{i}}}{1+w_{i}}\,\theta_{i}
  + k^{2}\left(\frac{\delta P_{i}/\delta\rho_{i}}{1+w_{i}}\,\delta_{i}-\sigma_{i} +
\Psi\right)~,
\end{eqnarray}
where
\begin{eqnarray}
 \theta_{i}=(k^{2}/\dot{\phi}_{i})\delta\phi_{i}~,~~
 \sigma_{i}=0~,~~
 w_{i}=\frac{P_{i}}{\rho_{i}}~,
\end{eqnarray}
and
\begin{eqnarray}
\label{prho2}
    \delta P_{i}=\delta\rho_{i}-2V_{i}'\delta\phi_i=\delta\rho_{i}+ \frac{\rho_{i}\theta_{i}}
      {k^{2}}\left[3\mathcal{H}(1-w_{i}^{2})+\dot{w_{i}}\right]~.
\end{eqnarray}
Combining Eqs. (\ref{dotdelta}), (\ref{dottheta}) and
(\ref{prho2}), we have
\begin{eqnarray}
    \dot\theta_{i} &=& 2\mathcal{H}\theta_{i}+\frac{k^{2}}{1+w_{i}}\delta_{i}+k^2\Psi~,\label{theta2}\\
    \dot\delta_{i} &=& -(1+w_{i})(\theta_{i}-3\dot{\Phi}) -3\mathcal{H}(1-w_{i})\delta_{i}
    -3\mathcal{H}\left[\frac{\dot w_{i}+3\mathcal{H}(1-w_{i}^{2})}{k^{2}}\right]\theta_{i}~.\ \ \ \ \
\label{delta2}
\end{eqnarray} Since the simple two-field quintom model is essentially a combination
of a quintessence and a phantom field, one obtains the
perturbation equations of quintom by combining the above
equations. The corresponding variables for the quintom system are
\begin{eqnarray}
\label{wq}
    w_{quintom} &=& \frac{\sum_i P_{i}}{\sum_i \rho_{i}}~,\\
\label{delta}
   \delta_{quintom} &=& \frac{\sum_i\rho_{i}\delta_{i}}{\sum_i\rho_{i}}~,\\
\label{theta}
    \theta_{quintom} &=& \frac{\sum_i(\rho_{i}+p_{i})\theta_{i}}{\sum_i(\rho_{i}+P_{i})}~.
\end{eqnarray}
Note that for the quintessence component, $-1\leq w_{1}\leq 1$,
while for the phantom component, $w_{2}\leq-1$.

The two-field quintom model is characterized by the potentials
$V_i$. Let us consider $V_i(\phi_i)=\frac{1}{2} m^2_i \phi^2_i$.
 In general the perturbations of $\phi_i$ today stem from
two origins, the adiabatic and the isocurvature modes. If we use
the gauge invariant variable
$\zeta_i=-\Phi-\mathcal{H}\frac{\delta \rho_i}{\dot{\rho_i}}$
instead of $\delta_i$, and the relation $\Phi=\Psi$ in the
universe without anisotropic stress, the equations (\ref{delta2})
and (\ref{theta2}) can be rewritten as,
\begin{eqnarray}
    \dot\zeta_{i}&=&-\frac{\theta_{i}}{3}-C_i\left(\zeta_i+\Phi+\frac{\mathcal{H}}{k^2}\theta_i\right)~,\label{mdelta2}\\
\dot\theta_{i}&=&2\mathcal{H}\theta_{i}+k^{2}(3\zeta_i+4\Phi)~,\label{mtheta2}
\end{eqnarray}
where
\begin{eqnarray}\label{definec} C_i=\frac{\dot
w_i}{1+w_i}+3\mathcal{H}(1-w_i)=\partial_0[\ln(a^6|\rho_i+p_i|)]~.
\end{eqnarray}
$\zeta_{\alpha}$ is the curvature perturbation on the
uniform-density hypersurfaces for the $\alpha$-component in the
universe  \cite{Wands:2000dp}. Usually, the isocurvature
perturbations of $\phi_i$ are characterized by the differences
between the curvature perturbation of the uniform-$\phi_i$-density
hypersurfaces and that of the uniform-radiation-density
hypersurfaces,
\begin{eqnarray}
 S_{ir}\equiv 3(\zeta_i-\zeta_r)~,
\end{eqnarray}
 where the
subscript $r$ represents radiation. We assume there are no matter
isocurvature perturbations and thus $\zeta_M=\zeta_r$. Eliminating
$\zeta_i$ from equations (\ref{mdelta2}) and (\ref{mtheta2}), we
obtain a second order equation for $\theta_i$, namely
\begin{eqnarray}
\label{2theta} \ddot
\theta_i+(C_i-2\mathcal{H})\dot\theta_i+(C_i{\cal H}-2\dot{\cal
H}+k^2)\theta_i=k^2(4\dot\Phi+C_i\Phi)~.
\end{eqnarray}
This is an inhomogeneous differential equation and its general
solution is the sum of the general solution of its homogeneous
part and a special integration. In the following, we will show
that the special integration corresponds to the adiabatic
perturbation. Before the era of DE domination, the universe is
dominated by some background fluids, for instance, the radiation
or the matter. The perturbation equations of the background fluid
are,
\begin{eqnarray}
 \dot\zeta_f&=&-\theta_f/3~,\nonumber\\
 \dot\theta_f&=&-\mathcal{H}(1-3w_f)\theta_f+k^2[3w_f\zeta_f+(1+3w_f)\Phi]~.
\end{eqnarray}
From the Poisson equation
\begin{eqnarray}
-\frac{k^2}{\mathcal{H}^2}\Phi&=&\frac{9}{2}\sum_{\alpha}\Omega_{\alpha}(1+w_{\alpha})
\left(\zeta_{\alpha}+\Phi+\frac{\mathcal{H}}{k^2}\theta_{\alpha}\right)\nonumber\\
&\simeq&\frac{9}{2}(1+w_{f})
\left(\zeta_{f}+\Phi+\frac{\mathcal{H}}{k^2}\theta_{f}\right)~,
\end{eqnarray}
on large scales we approximately acquire:
\begin{eqnarray}
\Phi\simeq-\zeta_f-\frac{\mathcal{H}}{k^2}\theta_{f}~.
\end{eqnarray}
Combining these equations with ${\cal H}=2/[(1+3w_f)\tau]$,
we obtain (note numerically $\theta_f\sim \mathcal{O}(k^2)\zeta_f$)
\begin{eqnarray}
& &\zeta_f=-\frac{5+3w_f}{3(1+w_f)}\Phi={\rm const.}~,\nonumber\\
& &\theta_f=\frac{k^2 (1+3w_f)}{3(1+w_f)}\Phi\tau~.
\end{eqnarray}
Thus, we can see from Eq. (\ref{2theta}) that there is a special
solution, which is given approximately on large scales by
$\theta_i^{ad}=\theta_f$, and from Eq. (\ref{mtheta2}) we have
$\zeta_i^{ad}=\zeta_f$. This indicates that the special
integration of (\ref{2theta}) corresponds to the adiabatic
perturbation. Hence, concerning the isocurvature perturbations of
$\phi_i$, we can consider only the solution of the homogeneous
part of (\ref{2theta}):
\begin{eqnarray}\label{22theta}
 \ddot \theta_i+(C_i-2\mathcal{H})\dot\theta_i+(C_i\mathcal{H}-2\dot{\cal
H}+k^2)\theta_i=0~.
\end{eqnarray}
These solutions are represented by $\theta_i^{iso}$ and
$\zeta_i^{iso}$ and the relation between them is
\begin{eqnarray}\label{iso}
 \zeta_i^{iso}=\frac{\dot\theta_{i}^{iso}-2\mathcal{H}\theta_{i}^{iso}}{3k^{2}}~.
\end{eqnarray}
Since the general solution of $\zeta_i$ is $
\zeta_i=\zeta_i^{ad}+\zeta_i^{iso}=\zeta_r+\zeta_i^{iso}$, the
isocurvature perturbations are simply $S_{ir}=3\zeta_i^{iso}$.

In order to solve Eq. (\ref{22theta}) we need to know the forms of
$C_i$ and $\mathcal{H}$ as functions of time $\tau$. For this
purpose, we solve the background equations (\ref{phiEOM}). In
radiation dominated period, $a=A\tau~,~\mathcal{H}=1/\tau$ and we
have
\begin{eqnarray}
\label{phi1}
 \phi_1&=&\tau^{-1/2}\bigg[A_{1}J_{1/4}(\frac{A}{2}m_1\tau^2)+A_{2}J_{-1/4}(\frac{A}{2}m_1\tau^2)\bigg]~,\\
\label{phi2}
 \phi_2&=&\tau^{-1/2}\bigg[\tilde{A}_{1}I_{1/4}(\frac{A}{2}m_2\tau^2)+\tilde{A}_{2}I_{-1/4}(\frac{A}{2}m_2\tau^2)\bigg]~,
\end{eqnarray} respectively, where $A$, $A_i$ and $\tilde{A}_i$ are
constants, $J_{\nu}(x)$ is the $\nu$th order  Bessel function and
$I_{\nu}(x)$ is the $\nu$th order  modified Bessel function.
Usually the masses are small in comparison with the expansion rate
in the early universe $m_i\ll \mathcal{H}/a$, and we can
approximate the (modified) Bessel functions as $J_{\nu}(x)\sim
x^{\nu}(c_1+c_2x^2)$ and $I_{\nu}(x)\sim
x^{\nu}(\tilde{c}_1+\tilde{c}_2x^2)$. We note that $J_{-1/4}$ and
$I_{-1/4}$ are divergent when $x\rightarrow 0$.
Given these arguments one can see that this requires large initial
values of $\phi_i$ and $\dot\phi_i$ if $A_2$ and $\tilde{A}_2$ are
not vanished.

If we choose small initial values, which is the natural choice if
the DE fields are assumed to survive after inflation, only $A_1$
and $\tilde{A}_1$ modes exist, so $\dot\phi_i$ will be
proportional to $\tau^3$ in the leading order. Thus, the
parameters $C_i$ in equation (\ref{definec}) will be $C_i=10/\tau$
(we have used $|\rho_i+p_i|=\dot\phi_i^2/a^2$). Therefore, we
obtain the solution of Eq. (\ref{22theta}) as
\begin{eqnarray}
 \theta_i^{iso}=\tau^{-4}\bigg[D_{i1}\cos(k\tau)+D_{i2}\sin(k\tau)\bigg]~.
\end{eqnarray}
That is, $\theta_i^{iso}$  presents an oscillatory behavior,
with an amplitude damping with the expansion of the universe. The
isocurvature perturbations $\zeta_i^{iso}$ decrease rapidly.

On the other hand, if we choose large initial values for $\phi_i$
and $\dot \phi_i$, $A_2$ and $\tilde{A}_2$ modes are present,
$\dot\phi_i$ will be proportional to $\tau^{-2}$ in the leading
order and $C_i=0$. Thus, the solution of (\ref{22theta}) is
\begin{eqnarray}
 \theta_i^{iso}=\tau\bigg[D_{i1}\cos(k\tau)+D_{i2}\sin(k\tau)\bigg]~.
\end{eqnarray}
Therefore, $\theta_i^{iso}$ will oscillate with an increasing
amplitude, so $\zeta_i^{iso}$ remains constant on large scales.

Similarly, during matter dominated era, $a=B\tau^2$,
$\mathcal{H}=2/\tau$, and the solutions for the fields $\phi_i$
read:
\begin{eqnarray}
\label{phi11}
 \phi_1&=&\tau^{-3}\bigg[B_{1}\sin(\frac{B}{3}m_1\tau^3)+B_{2}\cos(\frac{B}{3}m_1\tau^3)\bigg]~,\\
\label{phi22}
 \phi_2&=&\tau^{-3}\bigg[\tilde{B}_{1}\sinh(\frac{B}{3}m_2\tau^3)+\tilde{B}_{2}\cosh(\frac{B}{3}m_2\tau^3)\bigg]~.
\end{eqnarray}
We acquire the same conclusions as those reached by the
aforementioned analysis for the radiation dominated era. If we
choose small initial values at the beginning of the matter
domination we find that the isocurvature perturbations in $\phi_i$
decrease with time. On the contrary, for large initial values the
isocurvature perturbations remain constant at large scales. This
conclusion is expectable. In the case of large initial velocity,
the energy density in the scalar field is dominated by the kinetic
term and it behaves like the fluid with $w=1$. The isocurvature
perturbation in such a fluid remains constant on large scales. In
the opposite case, however, the energy density in the scalar field
will be dominated by the potential energy due to the slow rolling.
It behaves like a cosmological constant, and there are only tiny
isocurvature perturbations in it.

In summary, we have seen that the isocurvature perturbations in
quintessence-like or phantom-like field with quadratic potential,
decrease or remain constant on large scales depending on the
initial velocities. In this sense the isocurvature perturbations
are stable on large scales. The amplitude of these perturbations
will be proportional to the value of Hubble rate evaluated during
the period of inflation $H_{inf}$, if their quantum origins are
from inflation. For a large $H_{inf}$ the isocurvature DE
perturbations may be non-negligible and will contribute to the
observed CMB anisotropy \cite{Kawasaki:2001bq, Moroi:2003pq}. In
the cases discussed here, however, these isocurvature
perturbations are negligible. Firstly, large initial velocities
are not possible if these fields survive after inflation as
mentioned above. Secondly, even if the initial velocities are
large at the beginning of the radiation domination, they will be
reduced to a small value due to the small masses and the damping
effect of Hubble expansion. In general the contributions of DE
isocurvature perturbations are not very large
\cite{Gordon:2005ti}, and here for simplicity we have assumed that
$H_{inf}$ is small enough and thus the isocurvature contributions
are negligible. Therefore, in the next subsection we focus on the
effects of the adiabatic perturbations of the quintom model with
two scalars fields.

\subsection{Signatures of perturbations in quintom DE}\label{sec:quintomsignal}

Based on perturbation equations (\ref{delta}) and (\ref{theta}),
we modify the code of CAMB  \cite{Lewis:1999bs} and we study
preliminarily in this subsection the quintom observational
signatures. For simplicity we impose a flat geometry as a
background, although this is not necessary. Moreover, we assume
the fiducial background parameters to be $\Omega_{b}=0.042,
\Omega_{DM}=0.231, \Omega_{DE}=0.727$, where $b$ stands for
baryons, $DM$ for dark matter and $DE$ for dark energy, while
today's Hubble constant is fixed at $H_{0}=69.255$ km/s
Mpc$^{-2}$. We will calculate the effects of perturbed quintom on
CMB and LSS.

\begin{figure}[tb]
\begin{center}
\includegraphics[
width=6.in] {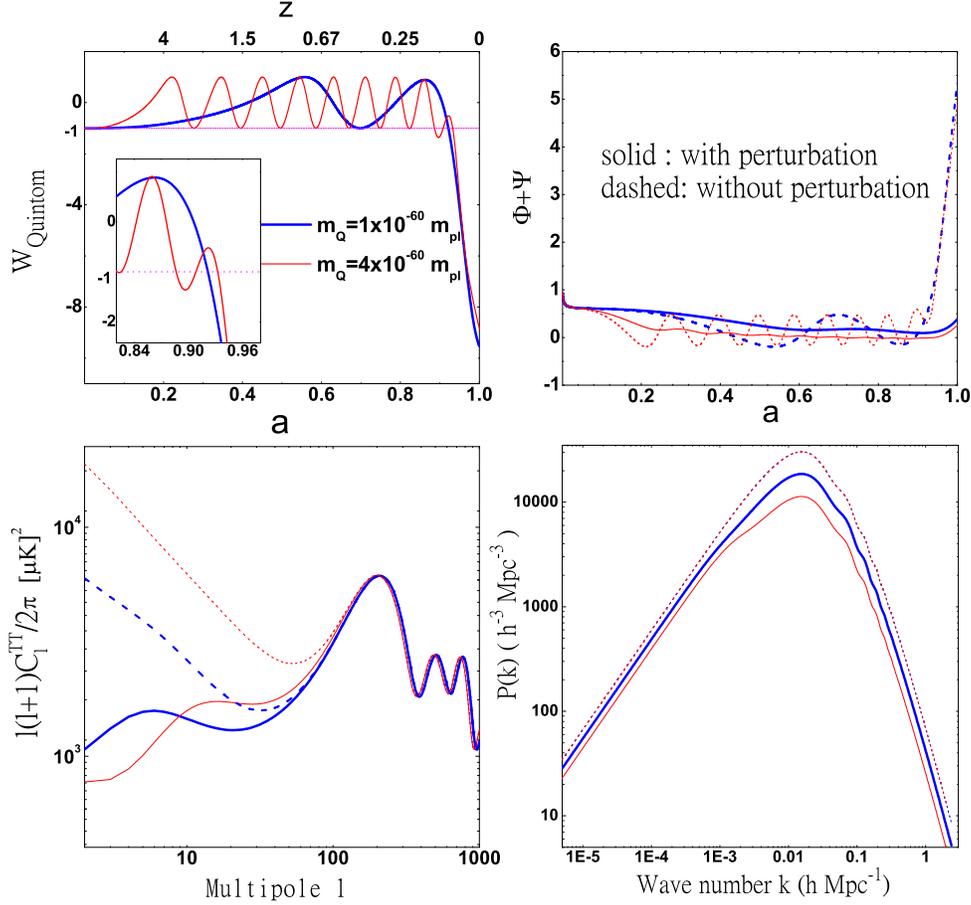} \caption{(Color online) {\it Effects of
the two-field oscillating quintom on the observables. The mass of
the phantom field is fixed at $4\times10^{-61} M_{p}$ and the mass
of the quintessence field are $2\times10^{-61} M_{p}$ (thicker
line) and $8\times10^{-61} M_{p}$ (thinner line) respectively. The
upper right graph illustrates the evolution of the metric
perturbations $\Phi+\Psi$ of the two models, with (solid lines)
and without (dashed lines) DE perturbations. The scale is
$k\sim10^{-3}$ Mpc$^{-1}$. The lower left graph shows the CMB
effects and the lower right graph delineates the effects on the
matter power spectrum, with (solid lines) and without (dashed
lines) DE perturbations. From Ref. \cite{Zhao:2005vj}.} }
\label{ZhaoFig1}
\end{center}
\end{figure}

In the two-field quintom model  there are two parameters, namely
the quintessence and phantom masses. When the quintessence mass is
larger than the Hubble parameter, the field starts to oscillate
and consequently one obtains an oscillating quintom. In the
numerical analysis we fix the phantom mass to be $m_P \sim 4\times
10^{-61} M_{p}$, and we vary the quintessence mass with the
typical values being $m_Q=2\times10^{-61} M_{p}$ and $8 \times
10^{-61} M_{p}$ respectively. In Fig. \ref{ZhaoFig1} we depict the
equation-of-state parameters as a function of the scale factor,
for the aforementioned two parameter-sets, and additionally their
corresponding effects on observations. We clearly observe the
quintom oscillating behavior as the mass of quintessence component
increases. After reaching the $w=-1$ pivot for several times, $w$
crosses $-1$ consequently with the phantom-component domination in
dark energy. As a result, the quintom fields modify the metric
perturbations $\delta g_{00}=2a^{2}\Psi,\delta g_{ii}=2a^{2}\Phi
\delta_{ij}$, and consequently they contribute to the late-time
Integrated Sachs-Wolfe (ISW) effect. The ISW effect is an
integrant of $\dot{\Phi}+\dot{\Psi}$ over conformal time and
wavenumber $k$. The above two specific quintom models yield quite
different evolving $\Phi+\Psi$ as shown in the right graph of
Fig.\ref{ZhaoFig1}, where the scale is $k\sim 10^{-3}$ Mpc$^{-1}$.
As we  can see, the late time ISW effects differ significantly
when DE perturbations are taken into account (solid lines) or not
(dashed lines).

ISW effects play an important role on large angular scales of CMB
and on the matter power spectrum of LSS. For a constant EoS of
phantom the authors of \cite{Weller:2003hw} have shown that the
low multipoles of CMB will be significantly enhanced when DE
perturbations are neglected. On the other hand, for a matter-like
scalar field where the EoS is around zero, perturbations will also
play an important role on the large scales of CMB, as shown in
\cite{Caldwell:1997ii}. The results on CMB and LSS reflect the two
combined effects of phantom and oscillating quintessence. Note
that in the early studies of quintessence effects on CMB, one
usually considers a constant $w_{eff}$ instead
\begin{equation}\label{weff}
 w_{eff}\equiv\frac{\int da \Omega(a) w(a)}{\int da \Omega(a)}~,
\end{equation}
however this is not enough for the study of effects on SNIa, nor
for CMB, when the EoS of DE has a very large variation with
redshift, such as the model of oscillating quintom considered
above.

\begin{figure}[tb]
\begin{center}
\includegraphics[
width=3.7in] {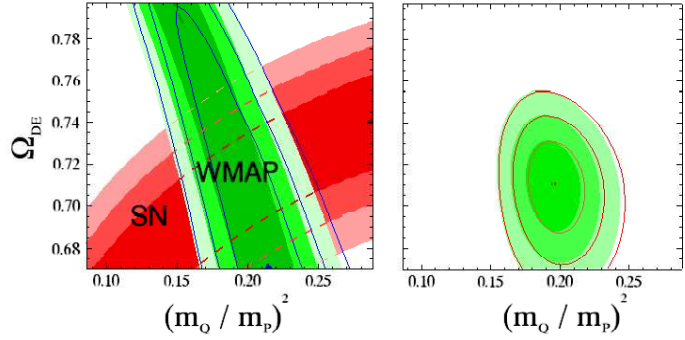} \caption{(Color online) {\it 3$\sigma$ WMAP
and SNIa constraints on two-field quintom model, shown together
with the best fit values. $m_Q$ and $m_P$ denote the quintessence
and phantom  mass respectively. We have fixed $m_P\sim 1.2 \times
10^{-61} M_{p}$ and we have varied the value of $m_Q$. Left graph:
separate WMAP and SNIa constraints. The green (shaded) area marks
the WMAP constraints on models where DE perturbations have been
included, while the blue area (contour with solid lines)
corresponds to the case where DE perturbations have not been taken
into account. Right graph: combined WMAP and SNIa constraints on
the two-field quintom model with perturbations (green/shaded
region) and without perturbations (red region/contour with solid
lines). From Ref.
 \cite{Zhao:2005vj}.} } \label{ZhaoFig2}
\end{center}
\end{figure}

To analyze the oscillating quintom-model under the current
observations, we perform a preliminary fitting to the first year
WMAP TT and the TE temperature--polarization cross-power spectrum,
as well as the recently released 157 ``Gold" SNIa data
\cite{Riess:2004nr}. Following Refs. \cite{Contaldi:2003zv,
Feng:2003zua} in all the fittings below we fix $\tau=0.17$,
$\Omega_M h^2=0.135$ and $\Omega_b h^2=0.022$,  setting the
spectral index as $n_S=0.95$, and using the amplitude of the
primordial spectrum as a continuous parameter. In the fittings of
oscillating quintom we have fixed the phantom-mass to be $m_P\sim
1.2 \times 10^{-61} M_{p}$. Fig.\ref{ZhaoFig2} delineates
3$\sigma$ WMAP and SNIa constraints on the two-field quintom
model, and in addition it shows the corresponding best fit values.
In the left graph of Fig.\ref{ZhaoFig2} we present the separate
WMAP and SNIa constraints. The green (shaded) area marks WMAP
constraints on models where DE perturbations have been included,
while the blue area (contour with solid lines) is the
corresponding area without DE perturbations. The perturbations of
DE have no effects on the geometric constraint of SNIa. The right
graph shows the combined WMAP and SNIa constraints on the
two-field quintom model with perturbations (green/shaded region)
and without perturbations (red region/contour with solid lines).
We conclude that the confidence regions indeed present a large
difference, if the DE perturbations have been taken into account
or not.

\begin{figure}[tb]
\begin{center}
\includegraphics[
width=6.1in] {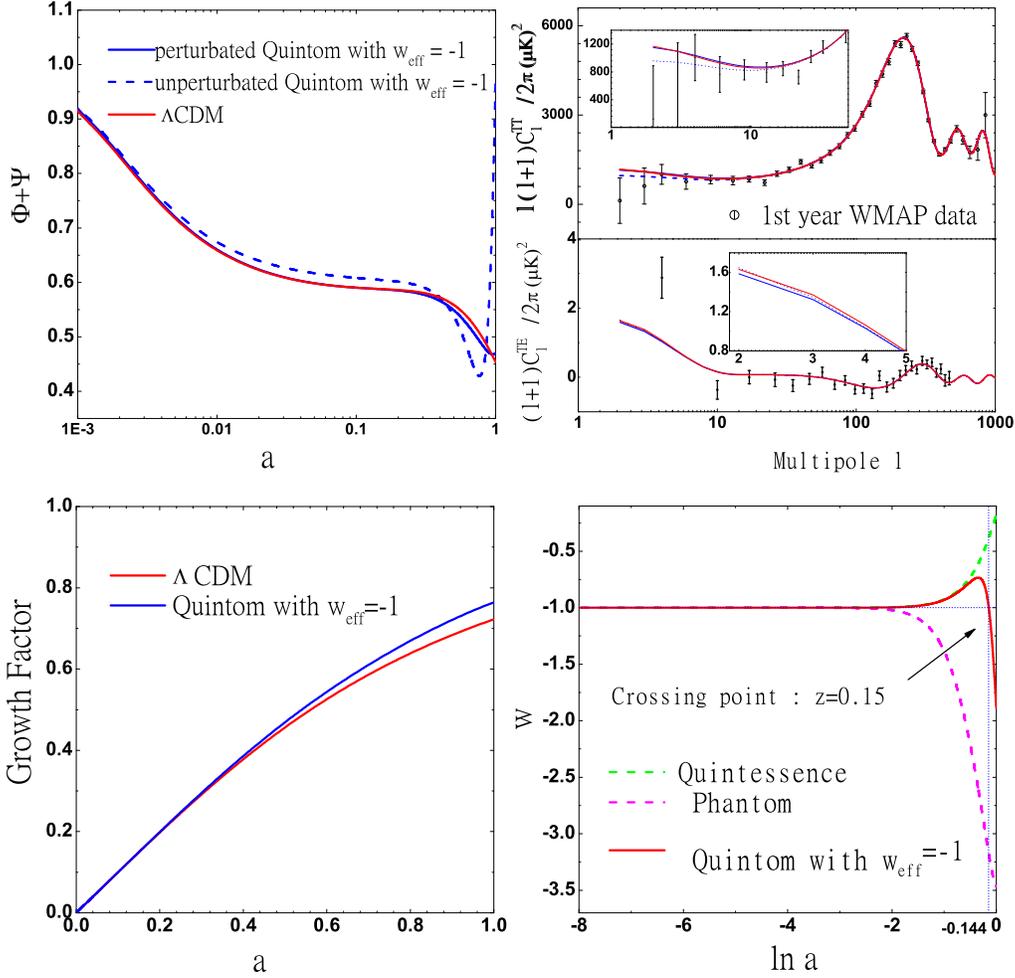} \caption{(Color online) {\it Comparison
of the effects of the two-field quintom model with $w_{eff}=-1$
and of the simple cosmological constant, in CMB (WMAP), the metric
perturbations $\Phi+\Psi$ (the scale is $k\sim10^{-3}$ Mpc$^{-1}$)
and the linear growth factor. The binned error bars in the upper
right graph are WMAP TT and TE data  \cite{Kogut:2003et,
Hinshaw:2003ex}. From Ref.  \cite{Zhao:2005vj}.} }
\label{ZhaoFig3}
\end{center}
\end{figure}

So far we have investigated the imprints of oscillating quintom on
CMB and LSS. Now we consider another example, in which $w$ crosses
$-1$ smoothly without oscillation. It is interesting to study the
effects of this type of  quintom model, with its effective EoS
defined in (\ref{weff}) exactly equal to $-1$, on CMB and matter
power-spectrum. This investigation will help to distinguish the
quintom model from the cosmological constant. We have realized
such a quintom model  in the lower right graph of Fig.
\ref{ZhaoFig3}, which can be easily given in the two-field model
with lighter quintessence mass. In this example we have set
$m_Q\sim 5.2 \times 10^{-62} M_{p}$ and $ m_P\sim 1.2 \times
10^{-61} M_{p}$. Additionally, we assume that there is no initial
kinetic energy. The initial value of the quintessence component is
set to $\phi_{1i}=0.045 M_{p}$, while for the phantom part we
impose $\phi_{2i}=1.32 \times 10^{-3} M_{p}$. We find that the EOS
of quintom crosses $-1$ at $z\sim 0.15$, which is consistent with
the latest SNIa results.

The quintom scenario, which is mildly favored by current SNIa
data, needs to be confronted with other observations in the
framework of concordance cosmology. Since SNIa offer the only
direct detection of DE, this model is the most promising to be
distinguished from the cosmological constant and other dynamical
DE models which do not get across $-1$, by future SNIa projects on
the low redshift (for illustrations see e.g.
\cite{Huterer:2004ch}). This is also the case for quintom scenario
in the full parameter space: it can be most directly tested in low
redshift Type Ia supernova surveys. In the upper left graph of
Fig. \ref{ZhaoFig3} we delineate the different ISW effects among
the cosmological constant (red/light solid) and the quintom model
which exhibits $w_{eff}=-1$, with (blue/dark solid) and without
(blue dashed) perturbations. Similarly to the previous oscillating
case, the difference is big when switching off quintom
perturbations and much smaller when including the perturbations.
In the upper right graph we find that the quintom model cannot be
distinguished from a cosmological constant in light of WMAP. The
two models give almost exactly the same results in CMB TT and TE
power-spectra when including the perturbations. We deduce that the
difference in CMB is hardly distinguishable even by cosmic
variance.

So far we have seen that CMB observations cannot distinguish
between a quintom model with $w_{eff} = -1$ and a cosmological
constant. Thus, in order to acquire distinctive signatures, we
have to rely on other observations. To achieve that we need to
consider the physical observables which can be affected by the
evolving $w$ sensitively. In comparison with the cosmological
constant, such a quintom model exhibits a different evolution of
the universe's expansion history, and in particular it gives rise
to a different epoch of matter-radiation equality. The Hubble
expansion rate $H$ is
\begin{equation}\label{H}
    H \equiv
    \frac{\dot{a}}{a^2}=H_{0}[\Omega_{M}a^{-3}+\Omega_{r}a^{-4}+X]^{1/2},
\end{equation}
where $X$, the energy density ratio of DE between the early times
and today, is quite different between the $quintom$-CDM and
$\Lambda$CDM. In the $\Lambda$CDM scenario $X$ is simply a
constant, while in general for DE models with varying energy
density or EoS,
\begin{equation}\label{omegaquitom}
    X=\Omega_{DE} a^{-3}e^{-3\int w(a)d \ln a}~.
\end{equation}
The two models will give different Hubble expansion rates. This is
also the case between the quintom model with $w_{eff} = -1$ in the
left graph of Fig. \ref{ZhaoFig3} and a cosmological constant.

Finally, we mention that different $H$ leads directly to different
behaviors of the growth factor. In particular, according to the
linear perturbation theory all Fourier modes of the matter density
perturbations grow at the same rate, that is the matter
density perturbations are independent of $k$:
\begin{equation}\label{detamk}
    \ddot{\delta}_k + \mathcal{H} \dot{\delta}_k - 4 \pi G a^2 \rho_{\rm
M} \delta_k=0~.
\end{equation}
The growth factor $D_{1}(a)$ characterizes the growth of the
matter density perturbations: $D_{1}(a)= \delta_k (a)/\delta_k
(a=1)$ and it is normalized to unity today. In the
matter-dominated epoch we have $D_1 (a)=a$. Analytically
$D_{1}(a)$ is often approximated by the Meszaros equation
\cite{Dodelson:2003ft}\footnote{One should notice that the
Meszaros equation is an exact solution to the differential
equation for the linear-theory growth factor only when dark energy
is a cosmological constant or it is absent. However, it is still a
good approximation in the case of dynamical dark energy models on
small length scales of the universe, as in the analysis of Section
3.1 of Ref. \cite{Linder:2003dr}. }:
\begin{equation}\label{d1}
    D_{1}(a)=\frac{5\Omega_{M}H(a)}{2H_{0}}\int^{a}_{0}\frac{da'}{(a'H(a')/H_{0})^{3}}.
\end{equation}
Therefore, we can easily observe the difference between the
quintom and cosmological constant scenarios, due to the different
Hubble expansion rates. More strictly one needs to solve
Eq.(\ref{detamk}) numerically. In the lower left graph of Fig.
\ref{ZhaoFig3} we show the difference of $D_1 (a)$ between the
quintom with $w_{eff} = -1$ and the cosmological constant. The
difference in the linear growth function is considerably large in
the late-time evolution, and thus possibly distinguishable in
future LSS surveys and in weak gravitational lensing (WGL)
observations. WGL has emerged with a direct mapping of cosmic
structures and it has been recently shown that the method of
cosmic magnification tomography can be extremely efficient
\cite{Jain:2003tba, Zhang:2005eb, Zhang:2005pu}, which leaves a
promising future for breaking the degeneracy between quintom and a
cosmological constant.



\section{Quintom models with higher derivative terms}\label{sec:quithd}

As we demonstrated previously, we usually need to introduce new
degrees of freedom into a normal lagrangian in order to obtain a
viable quintom scenario. One approach is to construct a
double-field quintom model introduced in Section IV. In this
section  we provide an alternative possibility of introducing
extra degrees of freedom for the realization of the transition
between quintessence phase and phantom phase
\cite{Stefancic:2005cs}. This model is originally proposed by Lee
and Wick to address on hierarchy problem in particle physics in
the 70's  \cite{Lee:1969fy, Lee:1970iw}, and recently applied in
\cite{Li:2005fm}  involving higher derivative operators in the
Lagrangian.

\subsection{Lee-Wick model}\label{sec:lw}

In this subsection we present a single field model with the
inclusion of higher derivatives, following  \cite{Li:2005fm}. We
consider a Lagrangian of the form
\begin{equation}
 \mathcal{L}=\mathcal{L}\bigg[\phi, X, (\Box\phi)(\Box\phi), (\nabla_{\mu}\nabla_{\nu}\phi)(\nabla^{\mu}\nabla^{\nu}\phi)\bigg]~,
\end{equation}
where $X=1/2 \nabla_{\mu}\phi\nabla^{\mu}\phi$ and $\Box\equiv
\nabla_{\mu}\nabla^{\mu}$ is the d'Alembertian operator. We
mention that this model is just an effective theory and one
assumes that the operators associated with the higher derivatives
can be derived from some fundamental theories, for instance due to
the quantum corrections or the non-local physics in  string theory
\cite{Simon:1990ic, Eliezer:1989cr, Erler:2004hv}. Additionally,
with the inclusion of higher derivative terms to the Einstein
gravity, the theory is shown to be renormalizable
\cite{Stelle:1976gc}. Finally, higher derivative operators have
been shown to be capable of stabilizing the linear fluctuations in
the scenario of ``ghost condensation" \cite{ArkaniHamed:2003uy,
ArkaniHamed:2003uz, Piazza:2004df,
Anisimov:2004sp,ArkaniHamed:2005gu, Mukohyama:2005rw}.

We consider the simple model with
\begin{equation}
\label{lagrangianlih}
 \mathcal{L}=-{1\over 2}\nabla_{\mu}\phi\nabla^{\mu}\phi+{c\over 2M^2}(\Box\phi)^2-V(\phi)~,
\end{equation}
where $M$ is a constant with mass dimension and $c$ is a
dimensionless constant. The energy-momentum tensor reads
\cite{Li:2005fm}:
\begin{eqnarray}
\label{stresslih}
 T^{\mu\nu} =
  g^{\mu\nu} \Big[ \frac{1}{2}\nabla_{\rho}\phi\nabla^{\rho}\phi +\frac{c}{2M^2}(\Box\phi)^2
        +\frac{c}{M^2}\nabla^{\rho}\phi\nabla_{\rho}(\Box\phi) +V\Big]
  - \nabla^{\mu}\phi\nabla^{\nu}\phi -\nonumber\\
 -
 \frac{c}{M^2}\bigg[\nabla^{\nu}\phi\nabla^{\mu}(\Box\phi)+\nabla^{\mu}\phi\nabla^{\nu}(\Box\phi)\bigg]~,\
 \
 \end{eqnarray}
and the equation of motion is given by
\begin{eqnarray}
\label{eqmlih}
 -\Box\phi-\frac{c}{M^2}\Box^2\phi+\frac{dV}{d\phi}=0~.
\end{eqnarray}
However, the energy-momentum tensor (\ref{stresslih}) can be
rewritten as {\small{
\begin{eqnarray} \label{astresslih}
 T^{\mu\nu} = \nabla^{\mu}\chi\nabla^{\nu}\chi-\nabla^{\mu}\psi\nabla^{\nu}\psi
 +\frac{1}{2}\left[\nabla_{\rho}\psi\nabla^{\rho}\psi-\nabla_{\rho}\chi\nabla^{\rho}\chi
 +2V(\psi-\chi)+\frac{M^2}{c}\chi^2\right]g^{\mu\nu}~,
\end{eqnarray}}}
where $\chi$ and $\psi$ are defined by
\begin{eqnarray}\label{changelih}
 \chi =\frac{c}{M^2}\Box\phi~,~~ \psi =\phi+\chi~.
\end{eqnarray}
It is not difficult to see that the energy-momentum tensor
(\ref{astresslih}) can be derived from the following Lagrangian
\begin{eqnarray}
 \label{alagrangianlih}
\mathcal{L}= -{1\over 2}\nabla_{\mu}\psi\nabla^{\mu}\psi+{1\over
2}\nabla_{\mu}\chi\nabla^{\mu}\chi -V(\psi-\chi)-{M^2\over
2c}\chi^2~,
\end{eqnarray}
with $\psi$ and $\chi$ being two independent fields. The
variations of the Lagrangian  (\ref{alagrangianlih}), with respect
to the fields $\psi$ and $\chi$ respectively, give rise to the
equations of motions
\begin{eqnarray}\label{psilih}
& &\Box\psi-V'=0~,\\
& &\Box\chi+\frac{M^2}{c}\chi-V'=0~,
\end{eqnarray} where the
prime denotes the derivative with respect to $\psi-\chi$.

When one imposes $\psi =\phi+\chi$ in (\ref{alagrangianlih}) and
(\ref{psilih}), he recovers the equation of motion (\ref{eqmlih})
of the single field model and the transformation equation
(\ref{changelih}). Thus, one can see that the single field model
(\ref{lagrangianlih}) is equivalent to the one with two fields
(\ref{alagrangianlih}), where one is the canonical scalar field
$\chi$ and the other is the phantom field $\psi$. Therefore,
assuming that $\chi$ dominates the DE sector in the early time,
the EoS parameter  will be larger than $-1$, but when the phantom
mode becomes dominant the EoS parameter will become less than
$-1$, crossing the cosmological constant boundary at an
intermediate redshift.

In general the potential term $V(\psi-\chi)$ (equivalently
$V(\phi)$ in our single field model) should include interactions
between the two fields $\psi$ and $\chi$. However, for some
specific choices of the potential $V$ these two modes could
decouple. As an example we consider $V=(1/2)m^2\phi^2$. Then the
Lagrangian (\ref{alagrangianlih}) can be ``diagonalized" as:
\begin{eqnarray}
\label{aalagrangianlih}
 \mathcal{L}={1\over
2}\nabla_{\mu}\phi_1\nabla^{\mu}\phi_1 -{1\over
2}\nabla_{\mu}\phi_2\nabla^{\mu}\phi_2-{1\over 2}m_1^2\phi_1^2
-{1\over 2}m_2^2\phi_2^2~,
\end{eqnarray}
through the transformation
\begin{eqnarray}
 \left(
\begin{array}{c} \psi \\ \chi\end{array} \right)
=\left(\begin{array} {cc} -a_1 & a_2\\-a_2 & a_1\end{array}\right)
\left( \begin{array}{c} \phi_1\\ \phi_2
\end{array} \right)~,
\end{eqnarray}
where
\begin{eqnarray}
& & a_1={1\over 2} \left(1+\frac{4c m^2}{M^2}\right)^{-1/4}
\left(\sqrt{1+\frac{4c m^2}{M^2}}-1\right)~,\nonumber\\
& & a_2={1\over 2} \left(1+\frac{4c
m^2}{M^2}\right)^{-1/4}\left(\sqrt{1+\frac{4c
m^2}{M^2}}+1\right)~,\ \ \ \ \ \
 \end{eqnarray}
 and
\begin{eqnarray}
\label{masslih}
& &m_1^2=\frac{M^2}{2c}\left(\sqrt{1+\frac{4cm^2}{M^2}}+ 1\right)~,\nonumber\\
& &m_2^2=\frac{M^2}{2c}\left(\sqrt{1+\frac{4cm^2}{M^2}}-
1\right)~.
\end{eqnarray}
Hence, one can see that the model (\ref{lagrangianlih}) with
$V=(1/2)m^2\phi^2$ is equivalent to the uncoupled system
(\ref{aalagrangianlih}). The two modes $\phi_1$ and $\phi_2$
evolve independently in the universe. The positivity of the
parameter $c$ guarantees the positivity of $m^2_1$ ($m^2_2$ is
always positive as long as $1+\frac{4c m^2}{M^2}>0$). In the limit
$m\ll M$, the masses of the two modes are approximately $m_1\simeq
M/\sqrt{c}$ and $m_2\simeq m$. In fact, $\phi_1$ and $\phi_2$ are
the eigenfunctions of the d'Alembertian operator $\Box$, with the
eigenvalues $-m_1^2$ and $m_2^2$ respectively. The solution to the
equation of motion (\ref{eqmlih}) $\phi$, is decomposed by these
eigenfunctions as $\phi=(1+\frac{4c
m^2}{M^2})^{-1/4}(\phi_1+\phi_2)$. Finally, we mention that the
perturbations in the phantom component $\phi_2$ do not present any
unphysical instabilities at the classical level. Thus, the model
is free from the difficulties of the singularity and the
gravitational instabilities of the general k-essence-like
scenarios \cite{Li:2005fm}.

\begin{figure}[tb]
\begin{center}
\includegraphics[
width=4.9in] {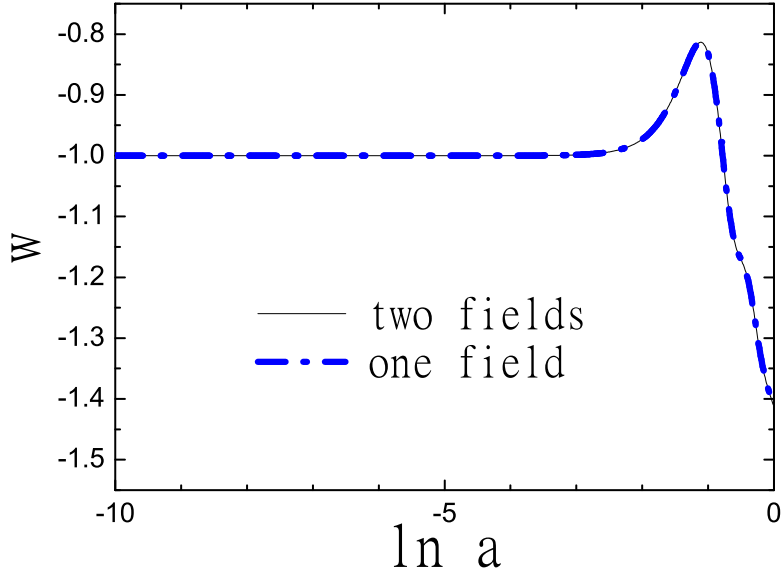} \caption{(Color online) {\it Evolution of
the EoS parameters for the high derivative single-field model and
the equivalent two-field model with one quintessence and one
phantom field.  From Ref. \cite{Li:2005fm}.} }
\label{equivalencelih}
\end{center}
\end{figure}

In Fig. \ref{equivalencelih} we present the EoS parameter of the
single field model as a function of $\ln a$, where $w$ clearly
crosses $-1$ during evolution. On the same figure it is also
depicted the corresponding evolution of EoS parameter of the
equivalent two-field model. Indeed, one can see that the two
models coincide.
Therefore, the high derivative single scalar field model of DE
presents the $-1$-crossing, and it is  equivalent to the usual
two-field quintom paradigm. However, note that when we consider
the interactions between these two fields in the potential, they
may possibly show some different behaviors.

One can easily generalize the aforementioned model considering the
Lagrangian  \cite{Zhang:2006ck}:
\begin{eqnarray}
\label{l1zangh} {\cal
L}=\frac{1}{2}A(\phi)\nabla_\mu\phi\nabla^\mu\phi+\frac{C(\phi)}{2M_{p}^2}(\Box\phi)^2-V(\phi),
\end{eqnarray}
where $ \chi=\frac{C(\phi)}{M_{p}^2}\Box\phi$ and $\phi=\phi(\psi,
\chi)$. Using $A(\phi)=-1,~V(\phi)=0$ this Lagrangian is
simplified as
\begin{eqnarray}
 \label{13lih}
\mathcal{L}=-\frac{1}{2}\nabla_{\mu}\psi\nabla^{\mu}\psi+
\frac{1}{2}\nabla_{\mu}\chi\nabla^{\mu}\chi-\frac{M_{p}^2\chi^2}{2
C[\phi(\psi, \chi)]}~,
\end{eqnarray}
where $\phi=-(\psi+\chi)$. The equations of motion for the two
fields are:
\begin{eqnarray}
 \label{eq1zangh}
 && \Box\psi+\frac{M_{p}^2C'}{2C^2}\chi^2=0\nonumber~,\\
 && \Box\chi-\frac{M_{p}^2C'}{2C^2}\chi^2+\frac{M_{p}^2}{C}\chi=0~,
 \end{eqnarray}
where the prime is the derivative with respect to $\psi+\chi$. In
FRW universe, $\Box =\partial^2/\partial t^2+3H\partial/\partial
t$ with $H$ being the Hubble expansion rate. Furthermore, one can
easily extract the density and the pressure of DE as
\begin{eqnarray}
\label{eq2zangh} \rho &=&
-\frac{\dot\psi^2}{2}+\frac{\dot\chi^2}{2}+
\frac{M_{p}^2\chi^2}{2C}~,\\p &=&
-\frac{\dot\psi^2}{2}+\frac{\dot\chi^2}{2}-\frac{M_{p}^2\chi^2}{2C}~,
\end{eqnarray}
and thus the EoS parameter writes
\begin{eqnarray}
\label{eq3zangh} w ={p\over \rho}
=-1+\frac{2(\dot\chi^2-\dot\psi^2)}
{\dot\chi^2-\dot\psi^2+\frac{M_{p}^2\chi^2}{C}}~,
\end{eqnarray}
where the dot represents the derivative with respect to the cosmic
time. Again, $\chi$ behaves as a canonical field while $\psi$ as a
phantom one. However, in this case they couple to each other
through an effective potential, that is the last term in the right
hand side of equation (\ref{13lih}):
\begin{eqnarray}
V_{eff}=\frac{M_{p}^2\chi^2}{2 C[\phi(\psi, \chi)]}~.
\end{eqnarray}

As an example one can consider \cite{Zhang:2006ck}:
\begin{eqnarray}
C[\phi(\psi, \chi)] =
C_0\left\{\frac{\pi}{2}+\arctan\left[\frac{\lambda(\psi+\chi)}{M_{p}}\right]\right\}~.
\end{eqnarray}
Thus, $C$ is almost constant when $\left|\psi+\chi\right|\gg 0$,
and $\psi$ and $\chi$ are nearly decoupled at these regimes, which
are dubbed ``weak coupling" regimes. By contrast, the two fields
couple tightly in the ``strong coupling" regime where
$\left|\psi+\chi\right|\sim 0$. In the weak coupling regime, as
shown in (\ref{eq1zangh}), the phantom-like field $\psi$ behaves
as a massless scalar field and its energy density
$-(1/2)\dot\psi^2$ dilutes as $a^{-6}$, with $a$ is the scale
factor of the universe. The quintessence-like field $\chi$ has a
mass term with $m_{\chi}=M_{p}/\sqrt{C(\psi+\chi)}$. Its behavior
is determined by the ratio of $m_{\chi}/H$. If $m_{\chi}\ll H$,
$\chi$ is slow-rolling and it behaves like a cosmological
constant. On the other hand, in cases $m_{\chi}\gg H$ the kinetic
term and potential oscillate coherently and their assembly evolves
as $a^{-3}$, just like that of collisionless dust.

\begin{figure}[tb]
\begin{center}
\includegraphics[
width=4.7in] {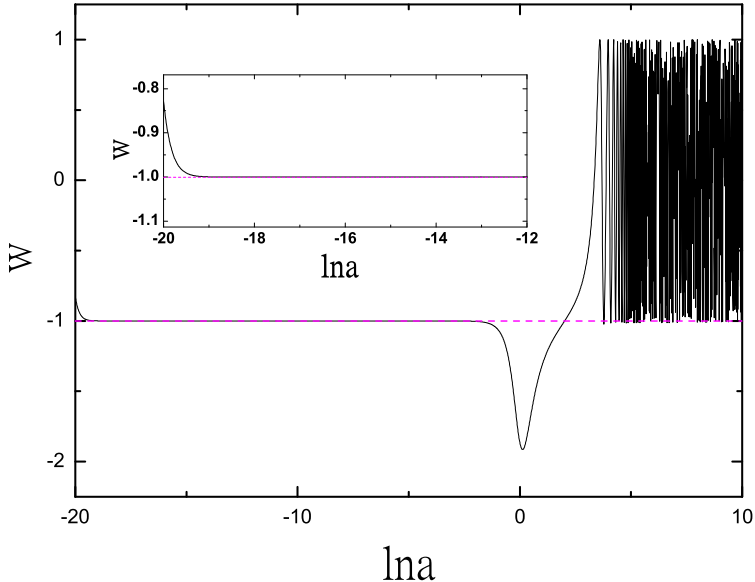} \caption{(Color online) {\it The EoS
parameter $w$ as a function of $\ln a$ for
$C=6.0\times10^{121},~\lambda=25,$ and the initial values are $
\psi_i=-0.052M_{p},~\dot{\psi}_i=1.40\times10^{-63}M_{p}^2,
~\chi_i=0.050M_{p},~\dot{\chi}_i=-1.44\times10^{-63}M_{p}^2,~
with~\Omega_{DE_0}\approx0.73.$  From Ref. \cite{Zhang:2006ck}.} }
\label{wzhanghigh}
\end{center}
\end{figure}

The EoS parameter of such a DE model is depicted in Fig.
\ref{wzhanghigh}.
The initial conditions are chosen in order for the DE to start in
the strong coupling regime with $\chi>0$ at high red-shift. As we
observe,  DE is frozen quickly at  initial times, and then it
becomes significant around redshift $z\sim 1$, and the phantom
kinetic term $\dot\psi^2$ becomes larger than the quintessence one
$\dot\chi^2$. The DE crosses the phantom divide from above to
below. In the future, the system  evolves into the weak coupling
regime, $\dot\psi^2$  dilutes quickly and $\chi$  gets a large
mass. The whole system  behaves as cold matter and its EoS
oscillate around the point of $w=0$. So, the universe will exit
from the phantom phase and end in the state of matter-domination.
This result resembles the late-time behavior of ``B-inflation"
model \cite{Anisimov:2005ne}. Therefore, the present cosmological
scenario differs from the simple quintom DE model, since it leads
to a double crossing of the phantom divide and it can give a nice
exit from phantom phase, in which it stays just for a finite time,
avoiding the realization of a Big Rip
\cite{Caldwell:2003vq,Setare:2008pc, GonzalezDiaz:2003bc,
Kallosh:2003bq, Gibbons:2003yj}. In addition, at late times $w$
oscillates with high frequency around zero, so the universe will
end in an accelerated expansion at some time. Finally, note that
the quintom model based on higher derivative theory possesses also
ghost degrees of freedom, and thus it could face the problem of
quantum instability \cite{Cline:2003gs, Carroll:2003st}. However,
as pointed out in \cite{Hawking:2001yt}, the problem of quantum
instability arises because $\phi$ and $\Box\phi$ are quantized in
canonical way independently. In fact, in the present model both of
them are determined by $\phi$, and a more appropriate quantization
method seems to be possible to avoid the instability
 \cite{Zhang:2006ck}.

\subsection{Quintom DE inspired by string theory}\label{sec:string}

In recent years, there have been active studies on relating
quintom scenario with fundamental physics, and in particular with
string theory. In such a context a rolling tachyon field in the
world volume theory of the open string, stretched between a
D-brane and an anti-D-brane or a non-BPS D-brane, plays the role
of the scalar field in the quintom model  \cite{
Aref'eva:2006et,Aref'eva:2004vw, Aref'eva:2005gg, Aref'eva:2007yr,
Mulryne:2008iq, Barnaby:2008pt, Shi:2008df, Saridakis:2009uk}. The
effective action used in the study of tachyon cosmology consists
of the standard Einstein-Hilbert action plus an effective action
for the tachyon field on unstable D-brane or
(D-brane)-(anti-D-brane) system. Such a single-field effective
action, to the lowest order in $\nabla_\mu\phi\nabla^\mu \phi$
around the top of the tachyon potential, can be obtained by the
stringy computations for either a D3 brane in bosonic theory
\cite{Gerasimov:2000zp, Kutasov:2000qp} or a non-BPS D3 brane in
supersymmetric theory
 \cite{Kutasov:2000aq} (see also \cite{Barnaby:2006hi} for an
application in inflation). What distinguishes the tachyon action
from the standard Klein-Gordon form for a scalar field, is that
the former is non-standard but it is of the ``Dirac-Born-Infeld''
form \cite{Sen:1999md, Bergshoeff:2000dq, Kluson:2000iy}. The
tachyon potential is derived from string theory itself and thus it
has to satisfy some definite properties and requirements, such as
to describe the tachyon condensation.

\subsubsection{A model inspired by string theory}\label{sec:stringmodel}

In this paragraph we construct a specific quintom model inspired
by string theory. One considers an action of the form
\cite{Cai:2007gs}:
\begin{equation}
\label{actionorigincai}
 S=\int
d^4x\sqrt{-g}\left[-V(\phi)\sqrt{1-{\alpha}^\prime
\nabla_{\mu}\phi\nabla^{\mu}\phi+{\beta}^\prime\phi\Box\phi}\right],
\end{equation}
which generalizes the usual``Born-Infeld-type'' action, for the
effective description of tachyon dynamics, by adding a term $\phi
\Box \phi$ to the usual $\nabla_\mu \phi\nabla^\mu \phi $ in the
square root. The two parameters $\alpha'$ and $\beta'$ in
(\ref{actionorigincai}) can be also  made arbitrary when the
background flux is turned on  \cite{Mukhopadhyay:2002en}. Note
that we have defined $\alpha^\prime = \alpha/ M^4$ and
$\beta^\prime = \beta/ M^4$ with $\alpha$ and $\beta$ being the
dimensionless parameters respectively, and $M$ an energy scale
used to make the ``kinetic energy terms" dimensionless. $V(\phi)$
is the potential of scalar field $\phi$ (e.g., a tachyon) with
dimension of ${\rm [mass]}^4$, with a usual behavior in general,
that is bounded and reaching its minimum asymptotically. Note
that,
$\Box=\frac{1}{\sqrt{-g}}\partial_{\mu}\sqrt{-g}g^{\mu\nu}\partial_{\nu}$,
and therefore in (\ref{actionorigincai}) the terms
$\nabla_\mu\phi\nabla^\mu\phi$ and $\phi\Box\phi$ both involve two
fields and two derivatives.

The equation of motion for the scalar field $\phi$ reads:
\begin{equation}
\label{Eqofmbasiccai}
\frac{\beta}{2}\Box\Big(\frac{V\phi}{f}\Big)+
\alpha\nabla_{\mu}\Big(\frac{V\nabla^{\mu}\phi}{f}\Big)+M^4V_{\phi}f+\frac{\beta
V}{2f}\Box\phi=0~,
\end{equation}
where $f=\sqrt{1-{\alpha}^\prime
\nabla_{\mu}\phi\nabla^{\mu}\phi+{\beta}^\prime\phi\Box\phi}$ and
$V_\phi=dV /d\phi$. Following the convention of \cite{Li:2005fm},
the energy-momentum tensor $T^{\mu\nu}$ is given by the standard
definition $\delta_{g_{\mu\nu}}S\equiv-\int d^4 x
\frac{\sqrt{-g}}{2}T^{\mu\nu}\delta g_{\mu\nu}$, thus:
\begin{eqnarray}
\label{emtensorcai}
T^{\mu\nu}=g^{\mu\nu}[V(\phi)f+\nabla_{\rho}(\psi\nabla^{\rho}\phi)]+
\frac{\alpha}{M^4}\frac{V(\phi)}{f}\nabla^\mu\phi\nabla^\nu\phi
-\nabla^\mu\psi\nabla^\nu\phi-\nabla^\nu\psi\nabla^\mu\phi,\ \ \ \
\end{eqnarray}
where $\psi\equiv\frac{\partial{\cal
L}}{\partial\Box\phi}=-\frac{V\beta\phi}{2M^4f}$.

For a flat FRW universe and a homogenous scalar field $\phi$, the
equation of motion  (\ref{Eqofmbasiccai}) can be solved
equivalently by the following two equations
\begin{eqnarray}
\label{EqofmTT} &&\ddot\phi+3H\dot\phi=\frac{\beta
V^2\phi}{4M^4\psi^2}-\frac{M^4}{\beta\phi}+\frac{\alpha}{\beta\phi}\dot\phi^2~,\\
&&\ddot\psi+3H\dot\psi=(2\alpha+\beta)\Big(\frac{M^4\psi}{\beta^2\phi^2}-\frac{V^2}{4M^4\psi}\Big)-
-(2\alpha-\beta)\frac{\alpha\psi}{\beta^2\phi^2}\dot\phi^2-\frac{2\alpha}{\beta\phi}\dot\psi\dot\phi-\frac{\beta
V\phi}{2M^4\psi}V_{\phi}~,\ \ \ \
\end{eqnarray}
where we have made use of $\psi$ as defined above.  Therefore,
(\ref{EqofmTT}) is just the defining equation for $\psi$ in terms
of $\phi$ and its derivatives. Finally, from (\ref{emtensorcai})
we obtain the energy density and pressure:
\begin{eqnarray}
\label{Energydensitycai}
 \rho&=&Vf+\frac{d}{a^3dt}\Big(a^3\psi\dot\phi\Big)+\frac{\alpha}{M^4}\frac{V(\phi)}{f}\dot\phi^2-2\dot\psi\dot\phi~,\\
\label{Pressurecai}
 p&=&-Vf-\frac{d}{a^3dt}(a^3\psi\dot\phi)~.
\end{eqnarray}

\begin{figure}[tb]
\begin{center}
\includegraphics[
width=4.3in] {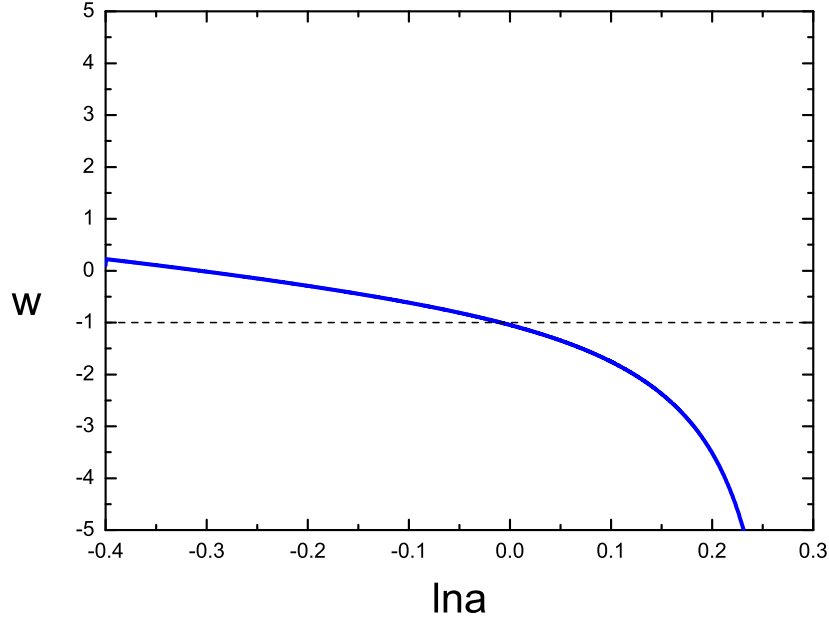} \caption{(Color online) {\it The DE EoS
parameter as a function of $\ln a$ for $V(\phi)=V_0
e^{-\lambda\phi^2}$ with $V_0=0.8$, $\lambda=1$, $\alpha=1$, and
$\beta=-0.8$, and for initial conditions $\phi_i=0.9$,
$\dot\phi_i=0.6$, $(\Box\phi)_i=\frac{d}{dt}(\Box\phi)_i=0$. From
Ref. \cite{Cai:2007gs}.} } \label{fig1bigripcai}
\end{center}
\end{figure}

The numerical exploration of the constructed model has been
performed in  \cite{Cai:2007gs}. In Fig. \ref{fig1bigripcai} one
can see the EoS parameter of DE for the (motivated by string
theory) potential $V(\phi)=V_0 e^{-\lambda\phi^2}$. The model
experiences the phantom-divide crossing, and results to a Big Rip
singularity. Numerically one sees that $\dot \phi = 0$ when $w$
crosses $-1$.

\begin{figure}[tb]
\begin{center}
\includegraphics[
width=4.3in] {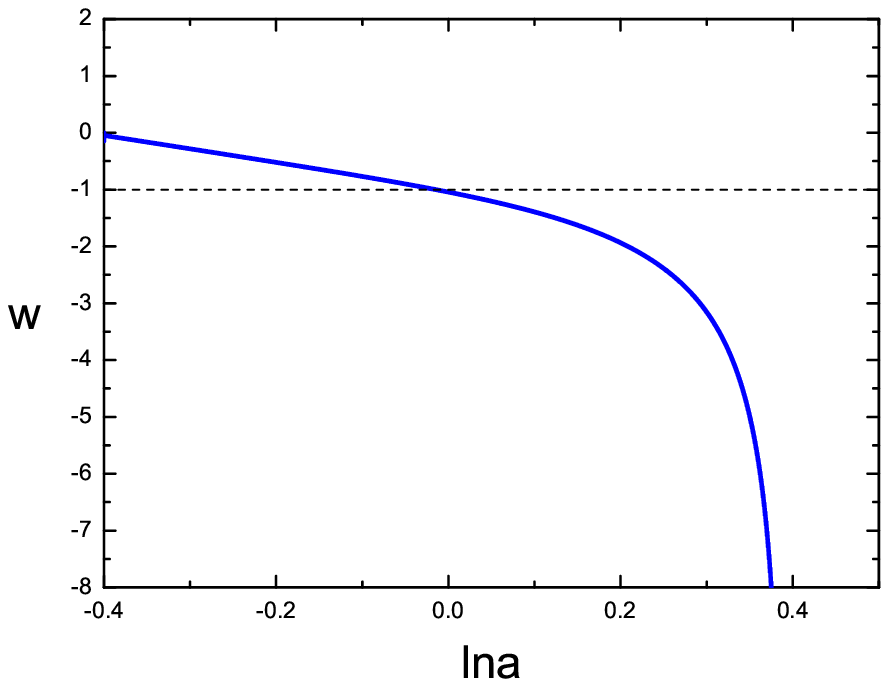} \caption{(Color online) {\it The DE
EoS parameter  as a function of $\ln a$, for
$V(\phi)=\frac{V_0}{e^{\lambda\phi}+e^{-\lambda\phi}}$ with
$V_0=0.5$, $\lambda=1$, $\alpha=1$, and $\beta=-0.8$, and for
initial conditions  $\phi_i=0.9$, $\dot\phi_i=0.6$,
$(\Box\phi)_i=\frac{d}{dt}(\Box\phi)_i=0$. From Ref.
 \cite{Cai:2007gs}.} } \label{fig2bigripcai}
\end{center}
\end{figure}

\begin{figure}[tb]
\begin{center}
\includegraphics[
width=4.3in] {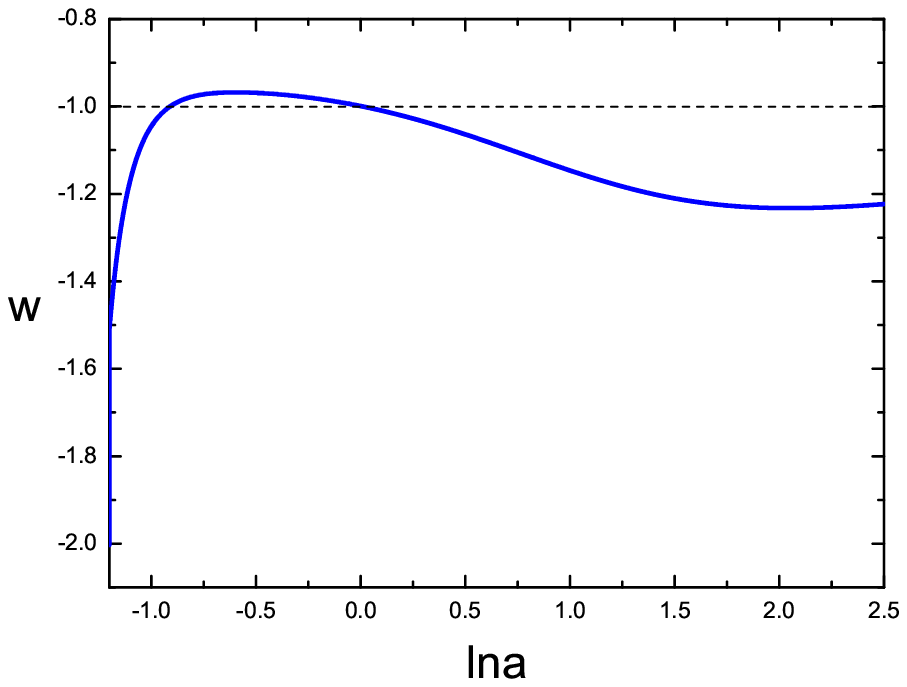} \caption{(Color online) {\it The DE
EoS parameter  as a function of $\ln a$, for
$V(\phi)=\frac{V_0}{e^{\lambda\phi}+e^{-\lambda\phi}}$ with
$V_0=0.5$, $\lambda=1$, $\alpha=1$, and $\beta=0.8$, and for
initial conditions  $\phi_i=0.9$, $\dot\phi_i=0.6$,
$(\Box\phi)_i=\frac{d}{dt}(\Box\phi)_i=0$. From Ref.
 \cite{Cai:2007gs}.} } \label{fig3bigripcai}
\end{center}
\end{figure}

In Fig.\ref{fig2bigripcai} we assume a different potential, namely
$V(\phi)=\frac{V_0}{e^{\lambda\phi}+e^{-\lambda\phi}}$. One
observes again  that the EoS of the model crosses  $-1$ during the
evolution.
Finally, in  Fig.\ref{fig3bigripcai} the potential has been also
chosen as $V(\phi)=\frac{V_0}{e^{\lambda\phi}+e^{-\lambda\phi}}$,
but taking a positive $\beta$ we observe that the EoS parameter
starts with $w<-1$, it crosses  $-1$ from below to above and then
transits again to $w<-1$.

One can generalize the aforementioned model in the case of a
tachyon, non-minimally coupled to gravity
\cite{Chingangbam:2004ng}. Thus, one adds an extra term $T\Box
{T}$ to the usual terms in the square root and the non-minimal
coupling $ Rf(T)$  \cite{Sadeghi:2008qp, Setare:2008qk}:
\begin{eqnarray}
\label{tacset}
 S=\int d^{4}x \sqrt{-g} \left[\frac{M_{p}^{2}}{2}Rf(T) -
AV(T)\sqrt{1-\alpha'g^{\mu\nu}\partial_{\mu}T\partial_{\nu}T+\beta'T
\Box T}\right]~,
\end{eqnarray}
where $M_{p}$ is the reduced Planck mass. Action (\ref{tacset})
can be brought to a simpler form for the derivation of the
equations of motion by performing a conformal transformation
$g_{\mu\nu}\rightarrow f(T) g_{\mu\nu}$: {\small{\begin{equation}
S=\int d^{4}x\sqrt{-g}\
\Bigg[\frac{M_{p}^{2}}{2}\Big(R-\frac{3f'^{2}}{2f^{2}}\partial_{\mu}T\partial^{\mu}T\Big)
-A\tilde{V}(T)\sqrt{1-(\alpha'f(T)-2\beta'f'(T)T)\partial_{\mu}T\partial^{\mu}T+\beta'f(T)T
\Box T}\Bigg]~,
\end{equation}}}
where $\tilde{V}(T)=\frac{V(T)}{f^{2}}$ is the effective potential
of the tachyon. For a flat FRW universe and a homogenous scalar
field $T$, the equations of motion can be solved equivalently by
the following two equations,
\begin{eqnarray}
\label{psiset}
 \ddot{\psi}+3H\dot{\psi}
 &=&\bigg(\frac{2\beta'f'T-\alpha'f}{fT}\bigg)\dot{\psi}\dot{T}-\frac{A^{2}\beta'f\tilde{V}T}{2\psi}\tilde{V}' \nonumber\\
 &&-\frac{3M_{p}^{2}}{2}\Big(\frac{ff'f''-f'^{3}}{f^{3}}\Big)\dot{T}^{2}
 -\left[\frac{(1-\beta')(\alpha'-2\beta')}{\beta'}\frac{f'}{f}-\frac{\alpha'}{T}\right]\frac{\psi\dot{T}^{2}}{T}~,\\
\label{eq4set}
 \ddot{T}+3H\dot{T}
 &=&\frac{2\left[\Big(\frac{ff''+\beta'f'}{f^{2}}\Big)T\dot{T}^{2}-2
 \bigg(\alpha'-2\beta'\frac{f'}{f}T\bigg)H\dot{T}\right]}{1+\frac{2\alpha'}{\beta'}-3\frac{f'}{f}T-
\frac{3M_{p}^{2}}{2}\Big(\frac{f'}{f}\Big)^{2}\frac{T}{\psi}}=\gamma~,
\end{eqnarray}
where
\begin{eqnarray}
\label{pesiset} \psi=\frac{\partial \mathcal{L}}{\partial \Box
T}=-\frac{A\beta'\tilde{V}f T}{2h}~,
\end{eqnarray}
$h=\sqrt{1-(\alpha'f-2\beta'f'T)\partial_{\mu}T\partial^{\mu}T+\beta'fT
\Box T}$, and $\tilde{V}^{'}=\frac{d\tilde{V}}{dT}$. Finally, the
energy density and pressure read
\cite{Sadeghi:2008qp,Setare:2008qk}:
\begin{eqnarray}
\label{rhoset}
 \rho&=&A\tilde{V}h+\frac{d}{a^{3}dt}\Big(a^{3}\psi\dot{T}\Big)+\Big(\alpha'f-2\beta'f'T\Big)
 \frac{A\tilde{V}}{h}\dot{T}^{2}-2\dot{\psi}\dot{T}+\frac{3M_{p}^{2}}{4}\Big(\frac{f'}{f}^{2}\Big)\dot{T}^{2}~, \\
\label{perset}
 p&=&-A\tilde{V}h-\frac{d}{a^{3}dt}\Big(a^{3}\psi\dot{T}\Big)+\frac{3M_{p}^{2}}{4}\Big(\frac{f'}{f}^{2}\Big)\dot{T}^{2}~.
\end{eqnarray}

 \begin{figure}[tb]
\begin{center}
\includegraphics[
width=4.4in] {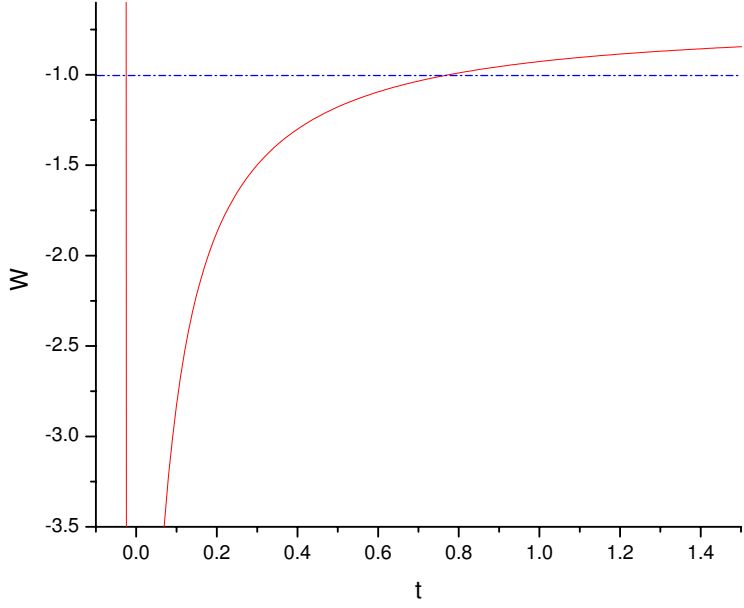} \caption{(Color online) {\it The
plot of EoS parameter for the potential $V(T)=e^{-\lambda T^{2}}$,
for $\alpha=-2$ and $\beta=2.2$. The initial values are $\phi=1$
and $\dot{\phi}=3$. From Ref.  \cite{Sadeghi:2008qp}.} }
\label{IIIB1setnm}
\end{center}
\end{figure}

The numerical exploration of the constructed model has been
performed in \cite{Sadeghi:2008qp}. In Fig. \ref{IIIB1setnm} we
present the EoS parameter of DE for the (motivated by string
theory) potential $V(T)=e^{-\lambda T^{2}}$ and
$f(T)=1+\sum_{i=1}c_{i}T^{2i}$. Thus, the model experiences the
$-1$-crossing, before the tachyon potential reaches asymptotically
to its minimum.

\subsubsection{Analysis on perturbations}\label{sec:stringpert}

Let us examine the stability  of the aforementioned model by
considering  quadratic perturbations. Consider a small
perturbation $\pi$ around the background $\phi$,
\begin{eqnarray}\label{phidecompose}
\phi\rightarrow\phi(t)+\pi(t,{\bf x})~,
\end{eqnarray}
where the background field $\phi$ in the FRW cosmology is
spatially homogenous. After this shift of the field, and a tedious
calculation, we obtain the following terms for the quadratic
perturbations of the action  \cite{Cai:2007gs}:
\begin{eqnarray}
^{(2)}S &\sim&\int
d^4x\sqrt{-g}~\left[\left(\frac{\alpha'}{2}Vf^{-1}+\frac{\alpha'^2}{2}Vf^{-3}\dot\phi^2+\frac{\beta'}{2}Vf^{-1}+\frac{\beta'}{2}\phi
V_\phi f^{-1}-\frac{\beta'^2}{4}\phi\Box\phi
Vf^{-3}\right)\dot\pi^2\right.\nonumber\\
&~&\left.-\left(\frac{\alpha'}{2}Vf^{-1}+\frac{\beta'}{2}Vf^{-1}+\frac{\beta'}{2}\phi
V_\phi f^{-1}-\frac{\beta'^2}{4}\phi\Box\phi
Vf^{-3}\right)(\nabla\pi)^2 +...~\right]~.
\end{eqnarray}

\begin{figure}[tb]
\begin{center}
\includegraphics[
width=6.9in] {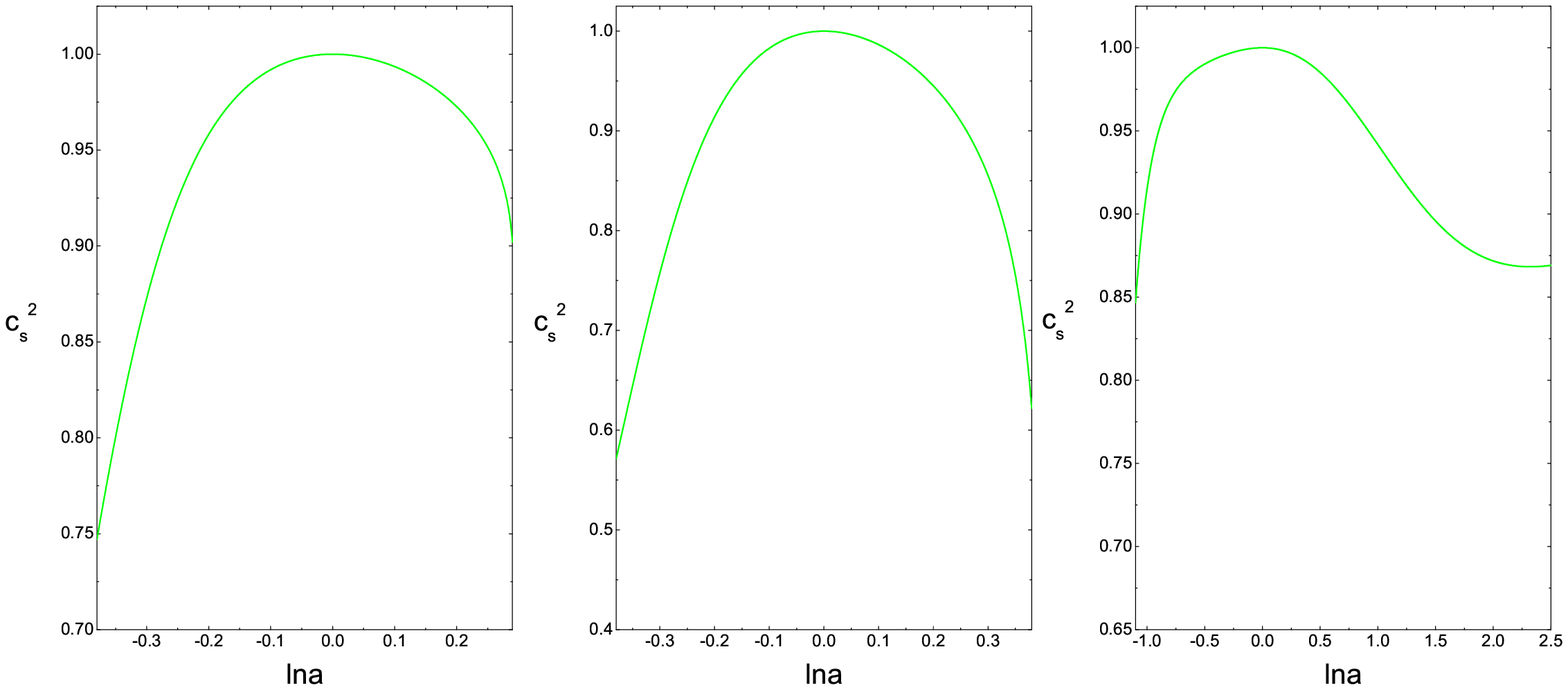} \caption{(Color online) {\it
Evolution of the sound speeds (from left to right) for the three
models considered in the text. From Ref.
 \cite{Cai:2007gs}.} }
\label{fig:cs}
\end{center}
\end{figure}

\begin{figure}[tb]
\begin{center}
\includegraphics[
width=6.9in] {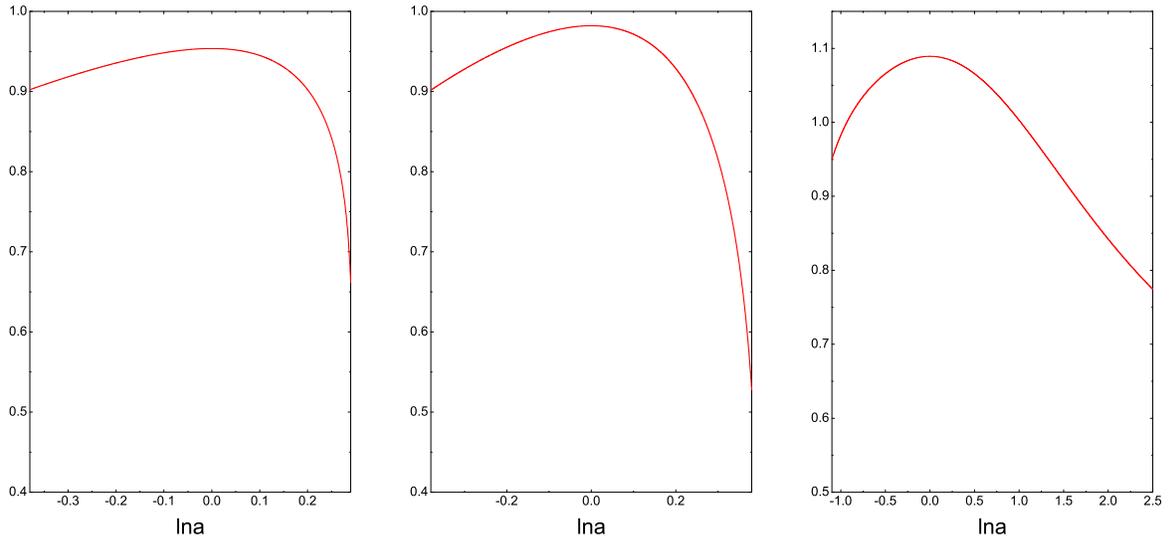} \caption{(Color online) {\it
Evolution of the coefficient of $\dot\pi^2$ (from left to right)
for the three models considered in the text. From Ref.
\cite{Cai:2007gs}.} } \label{fig:kin}
\end{center}
\end{figure}

Interestingly, we notice that due to the positivity of the term
$\frac{\alpha'^2}{2}Vf^{-3}\dot\phi^2$ in this model if the
coefficient of $(\nabla\pi)^2$ is positive, the term in front of
$\dot\pi^2$ is guaranteed to be positively valued. The sound speed
characterizing the stability property of the perturbations is
given by
\begin{equation}
c_s^2=\frac{\frac{\alpha'}{2}Vf^{-1}+\frac{\beta'}{2}Vf^{-1}+\frac{\beta'}{2}\phi
V_\phi f^{-1}-\frac{\beta'^2}{4}\phi\Box\phi
Vf^{-3}}{\frac{\alpha'}{2}Vf^{-1}+\frac{\alpha'^2}{2}Vf^{-3}\dot\phi^2+\frac{\beta'}{2}Vf^{-1}+\frac{\beta'}{2}\phi
V_\phi f^{-1}-\frac{\beta'^2}{4}\phi\Box\phi Vf^{-3}}~.
\end{equation}
If the $c_s^2$ lies in the range between $0$ and $1$ and the
coefficient of $\dot\pi^2$ remains positive, then the model will
be stable. In Fig. \ref{fig:cs} and Fig. \ref{fig:kin} we present
 $c_s^2$ and the coefficients of $\dot\pi^2$ respectively, for
the models of Figs. \ref{fig1bigripcai}, \ref{fig2bigripcai}, and
\ref{fig3bigripcai}. We observe that for these models $c_s^2$ lies
in the range $[0,1]$ and the coefficients of $\dot\pi^2$ are
positive, thus the models are indeed stable.



%

\section{Realizations of quintom scenario in non-scalar
systems}\label{sec:noscalar}

\subsection{Spinor quintom}\label{sec:spinor}

The quintom models described so far have been constructed by
scalar fields, which are able to accommodate a rich variety of
phenomenological behaviors, but the ghost field may lead to
quantum instabilities. To solve this problem one may consider the
linearized perturbations in the quantum-corrected effective field
equation at two-loop order \cite{Onemli:2002hr, Onemli:2004mb,
Brunier:2004sb} and obtain a super-acceleration phase induced by
these quantum effects \cite{Kahya:2006hc, Kahya:2009sz}. However,
there is the alternative possibility where the acceleration of the
universe is driven by a classical homogeneous spinor field. Some
earlier studies on applications of spinor fields in cosmology are
given in Refs. \cite{taub:1937, Brill:1957fx, Parker:1971pt}. In
recent years there are many works on studying spinor fields as
gravitational sources in cosmology. For example see Refs.
\cite{Obukhov:1993fd, ArmendarizPicon:2003qk, Kasuya:2001pr} for
inflation and cyclic universe driven by spinor fields, Refs.
\cite{Saha:2000nk} for spinor matter in Bianchi Type I spacetime,
Refs. \cite{Ribas:2005vr, Chimento:2007fx} for a DE model with
spinor matter etc.

\subsubsection{Algebra of a spinor field in FRW cosmology}

Recently, a delicate quintom model was proposed by making use of
spinor matter \cite{Cai:2008gk}. This scenario can realize the
quintom behavior, with an EoS crossing $-1$, without introducing a
ghost field. Instead, the derivative of its potential with respect
to the scalar bilinear $\bar\psi\psi$, which is defined as the
mass term, becomes negative when the spinor field lies in the
phantom-like phase. If this model can realize its EoS across $-1$
more than one time, the total EoS of the universe can satisfy
$w\geq -1$ during the whole evolution, as it is required by NEC
\cite{Buniy:2005vh, Qiu:2007fd}, and we expect to treat this
process as a phase transition merely existing for a short while.

To begin with, we simply review the dynamics of a spinor field
which is minimally coupled to Einstein gravity (see Refs.
\cite{Weinberg:1972, Birrell:1982, GSW} for a detailed
introduction). Following the general covariance principle, a
connection between the metric $g_{\mu\nu}$ and the vierbein is
given by
\begin{eqnarray}
g_{\mu\nu}e_{a}^{\mu}e_{b}^{\nu}=\eta_{ab}~,
\end{eqnarray}
where $e_{a}^{\mu}$ denotes the vierbein, $g_{\mu\nu}$ is the
space-time metric, and $\eta_{a b}$ is the Minkowski metric with
$\eta_{ab}={\rm diag}(1,-1,-1,-1)$. Note that the Latin indices
represents the local inertial frame, while the Greek indices
represents the space-time frame.

We choose the Dirac-Pauli representation as
\begin{eqnarray}
\gamma^0= \left(\begin{array}{cccc}
1 &   0  \\
0 &   -1
\end{array}\right),~~~
\gamma^{i}=\left(\begin{array}{cccc}
0 &          \sigma_{i} \\
-\sigma_{i}&   0
\end{array}\right),
\end{eqnarray}
where $\sigma_{i}$ is Pauli matrices. One can see that the
$4\times4$ $\gamma^{a}$ satisfy the Clifford
algebra$\{\gamma^{a},\gamma^{b}\}=2\eta_{ab}$. The $\gamma^{a}$
and $e_{a}^{\mu}$ provide the definition of a new set of Gamma
matrices
\begin{eqnarray}
\Gamma^{\mu}=e_{a}^{\mu}\gamma^{a}~,
\end{eqnarray}
which satisfy the algebra
$\{\Gamma^{\mu},\Gamma^{\nu}\}=2g_{\mu\nu}$. The generators of the
Spinor representation of the Lorentz group can be written as
$\Sigma^{ab}=\frac{1}{4}[\gamma^{a},\gamma^{b}]$. Therefore, the
covariant derivatives are given by
\begin{eqnarray}
D_{\mu}\psi&=&(\partial_{\mu}+\Omega_{\mu})\psi\\
D_{\mu}\bar\psi&=&\partial_{\mu}\bar\psi-\bar\psi\Omega_{\mu}~,
\end{eqnarray}
where the Dirac adjoint $\bar\psi$ is defined as $\psi^+\gamma^0$.
The $4\times4$ matrix $\Omega_{\mu}=\frac{1}{2}\omega_{\mu
ab}\Sigma^{ab}$ is the spin connection, where $\omega_{\mu
ab}=e_{a}^{\nu}\nabla_{\mu}e_{\nu b}$ are the Ricci spin
coefficients.

Using the aforementioned algebra we can write the following Dirac
action in a curved space-time background
\cite{ArmendarizPicon:2003qk, Ribas:2005vr, Vakili:2005ya}:
\begin{eqnarray}
\label{action} S_{\psi}=\int d^4
x~e~\left[\frac{i}{2}\left(\bar\psi\Gamma^{\mu}D_{\mu}
\psi-D_{\mu}\bar\psi\Gamma^{\mu}\psi\right)-V\right].
\end{eqnarray}
Here, $e$ is the determinant of the vierbein $e_{\mu}^{a}$ and $V$
stands for the potential of the spinor field $\psi$ and its
adjoint $\bar\psi$. Due to the covariance requirement, the
potential $V$  depends only on the scalar bilinear $\bar\psi\psi$
and on the ``pseudo-scalar" term $\bar\psi\gamma^5\psi$. For
simplicity we neglect the latter term and assume that
$V=V(\bar\psi\psi)$.

Varying the action with respect to the vierbein $e_{a}^{\mu}$, we
obtain the energy-momentum-tensor
\begin{eqnarray}\label{EMT}
T_{\mu\nu}
 &=& \frac{e_{\mu a}}{e}\frac{\delta S_\psi}{\delta e_{a}^{\nu}} \nonumber\\
 &=& \frac{i}{4}\left[\bar\psi\Gamma_{\nu}D_{\mu}\psi+\bar\psi\Gamma_{\mu}D_{\nu}\psi
     -D_{\mu}\bar\psi\Gamma_{\nu}\psi-D_{\nu}\bar\psi\Gamma_{\mu}\psi\right]
     -g_{\mu\nu}{\cal L}_{\psi}~.
\end{eqnarray}
On the other hand, variation of the action with respect to the
fields $\bar\psi$, $\psi$ respectively yields the equation of
motion of the spinor
\begin{eqnarray}
i\Gamma^{\mu}D_{\mu}\psi-\frac{\partial V}{\partial\bar\psi}=0~,~~
iD_{\mu}\bar\psi\Gamma^{\mu}+\frac{\partial V}{\partial\psi}=0~.
\end{eqnarray}

As usual, we deal with the homogeneous and isotropic FRW metric.
Correspondingly, the vierbein are given by
\begin{eqnarray}
e_{0}^{\mu}=\delta_{0}^{\mu}~,~~e_{i}^{\mu}=\frac{1}{a}\delta_{i}^{\mu}~.
\end{eqnarray}
Assuming that the spinor field is space-independent, the equations
of motion read:
\begin{eqnarray}
\label{EoMa}\dot{\psi}+\frac{3}{2}H\psi+i\gamma^{0} V' \psi&=&0\\
\label{EoMb}\dot{\bar\psi}+\frac{3}{2}H\bar\psi-i\gamma^{0}V'
\bar\psi&=&0~,
\end{eqnarray}
where a dot denotes a time derivative `$\frac{d}{dt}$', a prime
denotes a derivative with respect to $\bar\psi\psi$, and $H$ is
the Hubble parameter. Taking a further derivative, we can obtain:
\begin{eqnarray}\label{solution}
\bar\psi\psi=\frac{N}{a^{3}}~,
\end{eqnarray}
where $N$ is a positive time-independent constant, defined as the
value of $\bar\psi\psi$ at present.

From the expression of the energy-momentum tensor in (\ref{EMT}),
we obtain the energy density and  pressure of the spinor field:
\begin{eqnarray}
\label{density}\rho_{\psi}&=&T_{0}^{0}=V\\
\label{pressure}p_{\psi}&=&-T_{i}^{i}=V'\bar\psi\psi-V~,
\end{eqnarray}
where Eqs. (\ref{EoMa}) and (\ref{EoMb}) have been applied. The
EoS of the spinor field, defined as the ratio of its pressure to
energy density, is given by
\begin{eqnarray}\label{eos}
w_{\psi}\equiv\frac{p_{\psi}}{\rho_{\psi}}=-1+\frac{V'\bar\psi\psi}{V}~.
\end{eqnarray}

If we assume the potential to be a power law as $V=V_0
(\frac{\bar\psi\psi}{N})^{\alpha}$, with $\alpha$ as a nonzero
constant, we obtain a constant EoS:
\begin{eqnarray}
w_{\psi}=-1+\alpha~.
\end{eqnarray}
In this case, the spinor matter behaves like a
linear-barotropic-like perfect fluid. For example, if
$\alpha=\frac{4}{3}$, we can acquire $\rho_{\psi}\sim a^{-4}$ and
$w_{\psi}=\frac{1}{3}$, which is the same as radiation. If
$\alpha=1$, then $\rho_{\psi}\sim a^{-3}$ and $w_{\psi}=0$, thus
this component behaves like normal matter.

It is interesting to see that the spinor matter is able to realize
the acceleration of the universe if $\alpha<\frac{2}{3}$. Thus, it
provides  a potential motivation to construct a dynamical DE model
with the spinor matter. In the next subsection we focus our
investigation on constructing the subclass of quintom DE models
using the spinor field, which is called Spinor Quintom.

\subsubsection{Evolutions of Spinor Quintom}

In the model of the previous subsection $V$ is positive in order
for
 the energy density to be positive (see (\ref{density})). Therefore,
from (\ref{eos}) we deduce that   $w_{\psi}>-1$ when $V'>0$ and
$w_{\psi}<-1$ when $V'<0$. The former corresponds to a
quintessence-like phase and the latter to a phantom-like phase.
Therefore, the potential derivative $V'$ is required to change its
sign if one desires to obtain a quintom picture. In terms of the
variation of $V'$ there are three categories of evolutions in spinor quintom:\\
(i)
\begin{eqnarray}
V'>0~~~\rightarrow~~~V'<0 \nonumber~,
\end{eqnarray}
which gives a Quintom-A scenario by describing the universe
evolving from quintessence-like phase with $w_{\psi} > -1$ to
phantom-like phase with $w_{\psi} < -1$; \\
(ii)
\begin{eqnarray}
V'<0~~~\rightarrow~~~V'>0 \nonumber~,
\end{eqnarray}
which gives a Quintom-B scenario for which the EoS is arranged to
change from below $-1$ to above $-1$; \\
(iii), $V'$ changes its sign   more than one time, thus one
obtains a new Quintom scenario with its EoS crossing $-1$ more
than one time, dubbed Quintom-C scenario.\\
In the following, we assume different potentials in order to
realize the three types of evolutions mentioned above
\footnote{Note that we choose the potentials phenomenologically
without any constraints from quantum field theory or other
consensus. From the phenomenological viewpoint it is allowed to
treat the background classically, while dealing with the
perturbations in quantum level, like what it is usual in inflation
theory.}.

\begin{figure}[tb]
\begin{center}
\includegraphics[
width=4.3in] {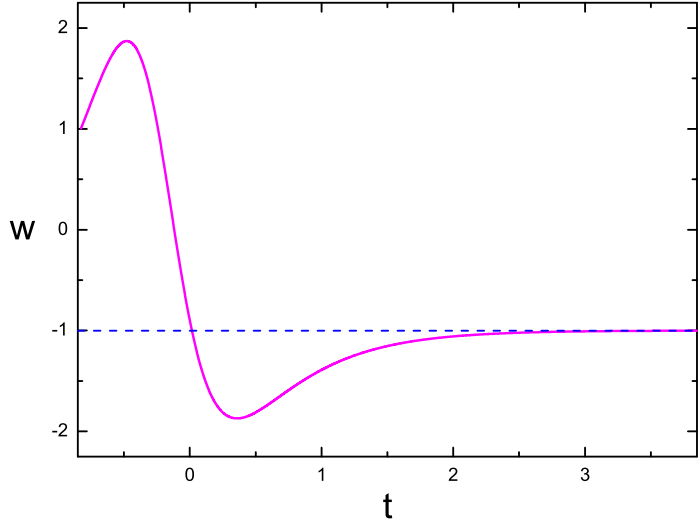} \caption{(Color online) {\it Evolution of
the EoS in Case (i) as a function of time. For numerical
elaboration we consider the potential of the spinor field as $V =
V_{0}[(\bar\psi\psi-b)^{2}+c]$, where we choose
$V_0=1.0909\times10^{-117}$, $b=0.05$ and $c=10^{-3}$ for the
model parameters. For the initial condition we take $N=0.051$.
From Ref.  \cite{Cai:2008gk}.} } \label{Fig:Q-A}
\end{center}
\end{figure}

To begin with, we shall investigate Case (i) and provide a
Quintom-A model. We use the potential
$V=V_{0}[(\bar\psi\psi-b)^{2}+c]$, where $V_{0}$, b, c are
undefined parameters. Then we obtain $V' = 2V_0(\bar\psi\psi-b)$
and the EoS reads:
\begin{eqnarray}
w_{\psi} = \frac { (\bar\psi\psi)^{2} - b^{2} - c} {
(\bar\psi\psi)^{2} - 2b\bar\psi\psi + b^{2} + c} ~.
\end{eqnarray}
According to   (\ref{solution}) one finds that $\bar\psi\psi$ is
decreasing  with   increasing scale factor $a$ during the
expansion of the universe. From the formula of $V'$  we deduce
that at the beginning of the evolution the scale factor $a$ is
very small, so $\bar\psi\psi$ becomes very large and ensures that
initially $V'>0$. Then $\bar\psi\psi$ decreases   with expanding
  $a$. At the moment where $\bar\psi\psi=b$,  one can see that
  $V'=0$,
which results in the EoS $w_{\psi}=-1$. After that $V'$ becomes
less than $0$ and thus the universe enters a phantom-like phase.
Finally, the universe approaches a de-Sitter spacetime. This
behavior is also obtained by  numerical elaboration and it is
shown in Fig. \ref{Fig:Q-A}. From this figure  one observes that
the EoS $w_{\psi}$ starts the evolution from $1$, then mildly
increases to a maximum and then begins to decrease. When
$\bar\psi\psi=b$, it reaches the point $w_{\psi}=-1$ and crosses
$-1$ from above to below smoothly. After that, the EoS
progressively decreases to a minimal value, then increases and
eventually approaches the cosmological constant boundary.

\begin{figure}[tb]
\begin{center}
\includegraphics[
width=4.3in] {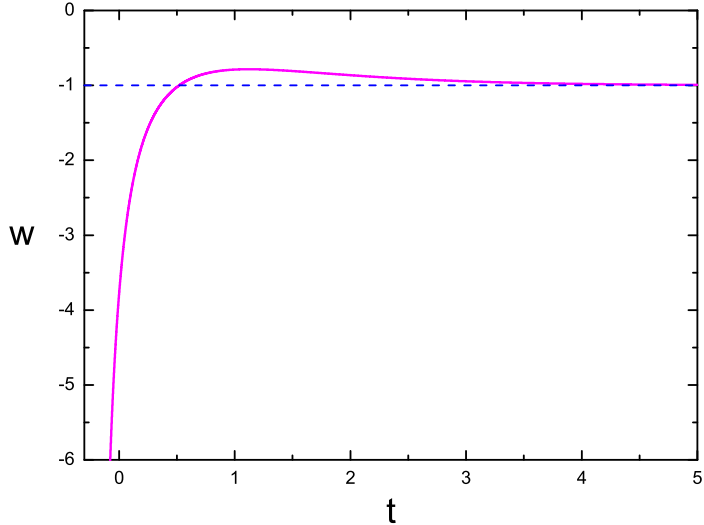} \caption{(Color online) {\it Evolution of
the EoS in Case (ii) as a function of time. For numerical
elaboration we consider the potential of the spinor field as $V =
V_0 [-(\bar\psi\psi-b)\bar\psi\psi+c]$, where we choose
$V_0=1.0909\times10^{-117}$, $b=0.05$ and $c=10^{-3}$ for the
model parameters. For the initial condition we take $N=0.051$.
From Ref.  \cite{Cai:2008gk}.} } \label{Fig:Q-B}
\end{center}
\end{figure}

In Case (ii) we choose the potential as $V = V_0
[-(\bar\psi\psi-b)\bar\psi\psi+c]$. Then one can obtain
$V'=V_0(-2\bar\psi\psi+b)$ and the EoS
\begin{eqnarray}
w_{\psi} =
\frac{-(\bar\psi\psi)^2-c}{-(\bar\psi\psi)^2+b\bar\psi\psi+c}~.
\end{eqnarray}
Initially $V'$ is negative due to the large values of
$\bar\psi\psi$. Then it increases to $0$ when
$\bar\psi\psi=\frac{b}{2}$, whereafter it changes its sign and
becomes larger than $0$, in correspondence with the Case (ii).
From Fig. \ref{Fig:Q-B}, we can see that the EoS evolves from
below $-1$ to above $-1$, and finally it approaches  $-1$.

\begin{figure}[tb]
\begin{center}
\includegraphics[
width=4.5in] {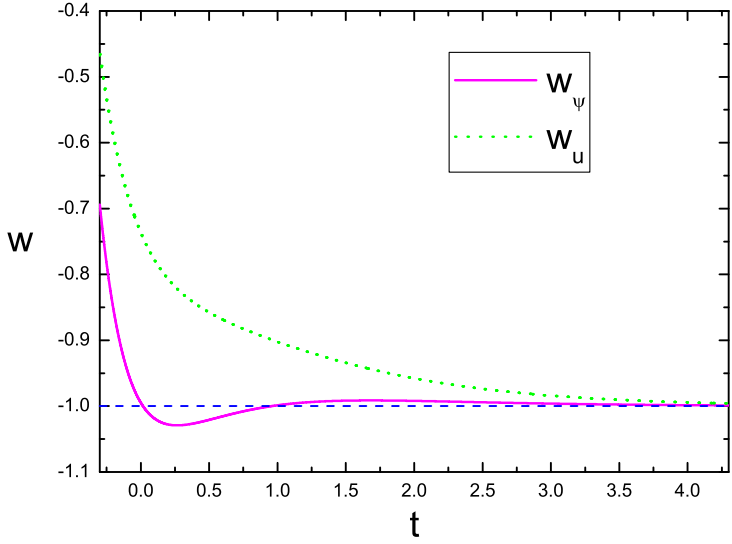} \caption{(Color online) {\it Evolution of
the EoS  in Case (iii) as a function of time. The magenta solid
line stands for the EoS of spinor quintom and the green dot line
stands for that of the whole universe. For numerical elaboration
we consider the potential of the spinor field as
$V=V_{0}[(\bar\psi\psi-b)^{2}\bar\psi\psi+c]$, where we choose
$V_0=1.0909\times10^{-117}$, $b=0.05$ and $c=10^{-3}$ for the
model parameters. For the initial condition we take $N=0.051$.
From Ref.  \cite{Cai:2008gk}.} } \label{Fig:quinc}
\end{center}
\end{figure}

In the third case we explore a quintom scenario which presents the
$-1$-crossing for two times, using the potential $
V=V_{0}[(\bar\psi\psi-b)^{2}\bar\psi\psi+c] $. Thus, we acquire $
V'=V_0(\bar\psi\psi-b)(3\bar\psi\psi-b) $ and the EoS becomes
\begin{eqnarray}\label{eoswsi3}
w_{\psi}=\frac{2(\bar\psi\psi)^3-2b(\bar\psi\psi)^2-c}
{(\bar\psi\psi)^3-2b(\bar\psi\psi)^2+b^2(\bar\psi\psi)+c}~.
\end{eqnarray}
Evidently, the equation $V'=0$ has two solutions which are
$\bar\psi\psi=b$ and $\bar\psi\psi=\frac{b}{3}$, thus $V'$ changes
its sign two times. From the EoS expression (\ref{eoswsi3}), we
find that initially $w_{\psi}>-1$. When the value of
$\bar\psi\psi$ becomes equal to b, it crosses $-1$ for the first
time. After the first crossing, it enters the phantom-like state,
it continuously descends until it passes through its minimum, then
it ascends to $\bar\psi\psi=\frac{b}{3}$ and then it experiences
the second crossing, moving eventually  to the quintessence-like
phase. This behavior is depicted in Fig. \ref{Fig:quinc}. One can
see that in this case the Big Rip can be avoided.

Moreover, considering the dark-matter component with  energy
density $\rho_{M} \propto 1/a^3$   and $w_M=0$, we can see that
the EoS of the universe
$w_{u}=w_{\psi}\frac{\rho_{\psi}}{\rho_{\psi}+\rho_{M}}$ satisfies
the relation $w_u\geq-1$ in this case. As it is argued in Ref.
 \cite{Qiu:2007fd}, NEC might be satisfied in such a model, since
$w_{\psi} < -1$ is only obtained for a short period during the
evolution of the universe.

As mentioned in previous sections,  one important issue of a DE
model is the analysis of its perturbations. Performing such an
investigation we might be able to learn to what degree the system
is stable both at quantum and classical level. Usually systems
with $w<-1$ present instabilities, for example  a phantom universe
suffers from a Big Rip singularity. Although this classical
singularity can be avoided in most quintom models, which usually
enter a de-Sitter expansion in the final epoch, all the scalar
quintom models suffer from  quantum instabilities since there are
negative kinetic modes from ghost fields. Here we would like to
perform the perturbation theory of spinor quintom. Since we do not
introduce any ghost fields in our model, the  phantom divide
crossing is achieved by the spinor field itself and thus it does
not result to any particular instability.

In order to simplify the derivation, we would like to redefine the
spinor as $\psi_N\equiv a^{\frac{3}{2}}\psi$. Then perturbing the
spinor field  we extract  the perturbation equation as
\begin{eqnarray}
\label{perteq}
 \frac{d^2}{d\tau^2}\delta\psi_N-\nabla^2\delta\psi_N+a^2\left[
 V'^2+i\gamma^0 (HV'-3HV''\bar\psi\psi) \right]\delta\psi_N
 =\nonumber\\-2a^2V'V''\delta(\bar\psi\psi)\psi_N-i\gamma^\mu\partial_\mu[a
 V''\delta(\bar\psi\psi)]\psi_N~,
\end{eqnarray}
where $\tau$ is the conformal time defined by $d\tau\equiv dt/a$.
Since the right hand side of the equation decays proportionally to
$a^{-3}$ or even faster, we can neglect those terms during the
late-time evolution of the universe for simplicity.

From the perturbation equation above, we can read that the sound
speed is equal to $1$, and this eliminates the instability of the
system in short wavelength. Furthermore, when the EoS $w$ crosses
$-1$  we have $V'=0$  and thus the eigenfunction of the solution
to Eq. (\ref{perteq}) in momentum space is a Hankel function with
an index $\frac{1}{2}$. Therefore, the perturbations of the spinor
field oscillate inside the hubble radius. This is an interesting
result, because in this way we might be able to establish the
quantum theory of the spinor perturbations, just as what it is
done in inflation theory.

Note that  the above derivation does not mean that spinor quintom
is able to avoid all instabilities. We still have not examined the
effects of the right hand side of Eq. (\ref{perteq}), which could
destroy the system stability under particular occasions. Finally,
another possible instability could arise from  quantum effects, if
this model is non-renormalizable.

\subsubsection{A unified model of Quintom and Chaplygin gas}

As we have analyzed so far, a spinor field with a power-law-like
potential behaves like a perfect fluid with a constant EoS.
However, it is still obscure to establish a concrete scenario to
explain how a universe dominated by matter evolves to the current
stage, that is dominated by DE. The last years, an alternative
interesting perfect fluid, with an exotic EoS $p=-c/\rho$, has
been applied into cosmology \cite{Kamenshchik:2001cp} in the aim
of unifying a matter-dominated phase where $\rho \propto 1/a^3$,
and a de-Sitter phase where $p=-\rho$, which describes the
transition from a universe filled with dust-like matter to an
exponentially expanding universe. This so-called Chaplygin gas
 \cite{Kamenshchik:2001cp} and its generalization
 \cite{GonzalezDiaz:2002hr} have been intensively studied in the
literature.

 Some possibilities of this scenario, motivated by field
theory, are investigated in  \cite{Bilic:2002vm}. Interestingly, a
 Chaplygin gas model can be viewed as an effective fluid
associated with D-branes  \cite{Bordemann:1993ep, Fabris:2001tm},
and it can be also obtained from the Born-Infeld action
\cite{Bento:2002ps, Bento:2003we}. The combination of quintom and
Chaplygin gas has been realized by the interacting Chaplygin gas
scenario
 \cite{Zhang:2005jj}, as well as in the framework of Randall-Sundrum
braneworld  \cite{GarciaCompean:2007vh}. Finally, the Chaplygin
gas model has been thoroughly investigated for its impact on the
cosmic expansion history. A considerable range of models was found
to be consistent with SNIa data
 \cite{Makler:2002jv}, the CMBR  \cite{Bento:2002yx}, the gamma-ray
bursts  \cite{Bertolami:2005aa}, the X-ray gas mass fraction of
clusters  \cite{Cunha:2003vg}, the large-scale structure
 \cite{Bilic:2001cg}, etc.

Here, we propose a new model constructed by spinor quintom, which
combines the property of a Chaplygin gas. The generic expression
of the potential is given by
\begin{eqnarray}
V=\sqrt[1+\beta]{f(\bar\psi\psi)+c}~,
\end{eqnarray}
where $f(\bar\psi\psi)$ is an arbitrary function of
$\bar\psi\psi$. Altering the form of $f(\bar\psi\psi)$, one can
realize both the Chaplygin gas and quintom scenario in a spinor
field.

Firstly, let us show how this model recovers a picture of
generalized Chaplygin gas. Assuming
$f(\bar\psi\psi)=V_0(\bar\psi\psi)^{1+\beta}$ the potential
becomes
\begin{eqnarray}
V=\sqrt[1+\beta]{V_0(\bar\psi\psi)^{1+\beta}+c}~,
\end{eqnarray}
and therefore, the energy density and pressure  of
(\ref{density}),(\ref{pressure}) read
\begin{eqnarray}
\rho_{\psi}&=&\sqrt[1+\beta]{V_0(\bar\psi\psi)^{1+\beta}+c}\\
p_{\psi}&=&-c
[V_0(\bar\psi\psi)^{1+\beta}+c]^{-\frac{\beta}{1+\beta}}~.
\end{eqnarray}
Thus, this spinor model  behaves like a generalized Chaplygin
fluid, satisfying the exotic relation
$p_{\psi}=-\frac{c}{\rho_{\psi}^{\beta}}$. Being more explicit we
assume $\beta=1$, obtaining the expressions of energy density and
pressure as
\begin{eqnarray}
\rho_{\psi}=\sqrt{\frac{N^2}{a^6}+c}~,~~p_{\psi}=-\frac{c}{\rho_{\psi}}~.
\end{eqnarray}
In this case a perfect fluid of Chaplygin gas is given by a spinor
field. Based on the above analysis, we  conclude that this simple
and elegant model is able to mimic different behaviors of a
perfect fluid and therefore it accommodates a large variety of
evolutions phenomenologically.

In succession, we will use this model to realize a combination of
a Chaplygin gas and a Quintom-A model, which is mildly favored by
observations. Choosing $f(\bar\psi\psi)$ to be
$f(\bar\psi\psi)=V_{0}(\bar\psi\psi-b)^{2}$ we acquire the
potential
\begin{eqnarray}\label{case4}
V=\sqrt{V_{0}(\bar\psi\psi-b)^{2}+c}~,
\end{eqnarray}
where $V_{0}$, b, c are undetermined parameters. Its derivative
reads
\begin{eqnarray}
V'=\frac{V_0(\bar\psi\psi-b)}{\sqrt{V_{0}(\bar\psi\psi-b)^{2}+c}}~,
\end{eqnarray}
and thus the EoS becomes
\begin{eqnarray}\label{eosch}
w_{\psi}=-1+\frac{V_0\bar\psi\psi(\bar\psi\psi-b)}{V_0(\bar\psi\psi-b)^2+c}~,
\end{eqnarray}
with the $-1$-crossing  taking place when $\bar\psi\psi=b$. During
the expansion $\bar\psi\psi$ is decreasing following
(\ref{solution}), with $\bar\psi\psi\rightarrow \infty$ at the
beginning of the evolution where $a\rightarrow 0$. Therefore,
$w_{\psi}$ increases from $0$ to a maximum value and then it
starts to decrease. When $\bar\psi\psi=b$ the EoS arrives at the
cosmological constant boundary $w_{\psi}=-1$ and then it crosses
it. Due to the existence of $c$-term, the EoS eventually
approaches the cosmological constant boundary. In this case the
universe finally becomes a de-Sitter space-time.

\begin{figure}[tb]
\begin{center}
\includegraphics[
width=4.3in] {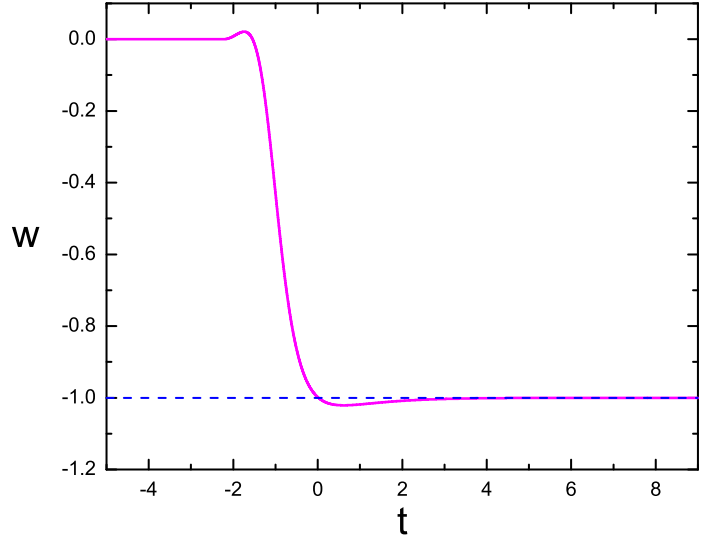} \caption{(Color online) {\it Evolution of
the EoS of the unified model with potential (\ref{case4}) as a
function of time. For the numerical analysis we assume
$V_0=3.0909\times10^{-239}$, $b=0.05$ and $c=9\times10^{-241}$,
while for the initial conditions we take $N=0.051$. From Ref.
\cite{Cai:2008gk}.} } \label{Fig:quindeos}
\end{center}
\end{figure}

\begin{figure}[tb]
\begin{center}
\includegraphics[
width=4.5in] {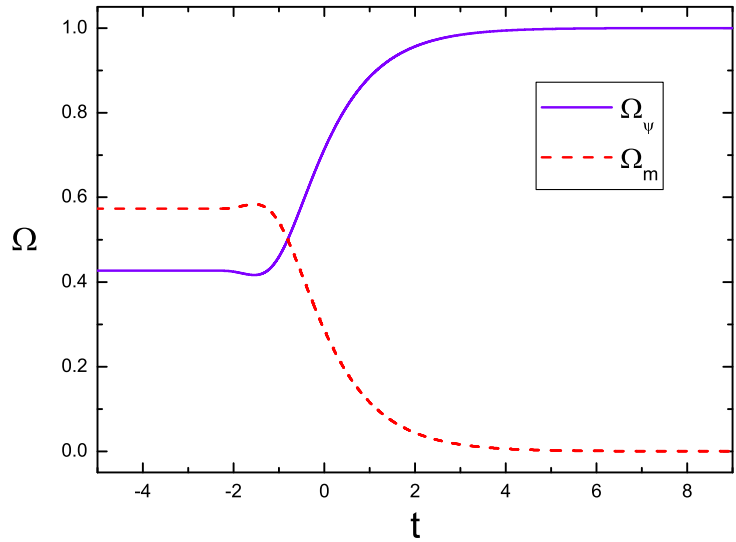} \caption{(Color online) {\it Evolution
of the density parameters
 of DE (violet solid line)
and dark matter (red dash line) as a function of time. From Ref.
\cite{Cai:2008gk}.} } \label{Fig:quindomg}
\end{center}
\end{figure}

Considering a universe filled with dark matter and the
aforementioned spinor quintom matter, we perform a numerical
analysis and we present the evolution of the EoS in Fig.
\ref{Fig:quindeos}. Moreover, in Fig. \ref{Fig:quindomg} we
display the evolution of the density parameters of dark matter
$\Omega_M \equiv \rho_M/(\rho_M+\rho_{\psi})$ and spinor quintom
$\Omega_{\psi} \equiv \rho_{\psi}/(\rho_M+\rho_{\psi})$. It is
evident that the  ratio of these two components from the beginning
of evolution is of order one, in relief of the fine-tuning  and
coincidence problems. During the evolution DE overtakes dark
matter, driving the universe into an accelerating expansion at
present, and eventually it dominates the universe completely,
leading to an asymptotic de-Sitter expansion.


Summarizing the results of this section we see that while a
scalar-type quintom model leads to quantum instabilities (due to
the ghost field with a negative kinetic term), the class of
quintom models in virtue of a spinor field can avoid such a
problem. As shown above, such a scenario could behave as a
generalized Chaplygin gas, and give rise to $-1$-crossing during
the evolution, caused by the sign-flip of the $V'$-term. Compared
with other models  experiencing the $-1$-crossing, this scenario
is also economical, in the sense that it  involves a single spinor
field.

\subsection{Other non-scalar models}\label{sec:noscalarother}

We finish this section mentioning that non-scalar systems may give
rise to a quintom scenario under certain cases. For example,
developed from a vector-like DE model \cite{Novello:2003kh,
Kiselev:2004py}, Ref.
 \cite{ArmendarizPicon:2004pm} has shown that an unconventional
cosmological vector field in absence of gauge symmetry can realize
its EoS across $-1$. This model and its extensions have also been
shown to be possible candidates for DE  \cite{Zhao:2005bu,
Wei:2006tn, Wei:2006gv, Boehmer:2007qa, Beck:2007ru,
Koivisto:2007bp, Bamba:2008xa}, and even solve the coincidence
problem  \cite{Mota:2007sz, Jimenez:2008au, Jimenez:2009py}.
However, see  \cite{Himmetoglu:2008zp} for a different viewpoint.

Finally, recently  it has been noticed that the quintom scenario
can be achieved in a model of non-relativistic gravity
\cite{Saridakis:2009bv, Cai:2009in, Park:2009zr}, but this is
realized at a high energy scale, far away from that of DE. The
possible successes of this model has led to the exploration of the
potential relations among quintom scenario and phenomena of
Lorentz symmetry breaking \cite{Carroll:1989vb, Colladay:1998fq,
Jackiw:1999yp, Kostelecky:2003fs, Lehnert:2006mn}.



\section{Quintom scenario in the braneworld}\label{sec:brane}

An alternative way of explaining the observed acceleration of the
late universe is to modify gravity at large scales. A well-studied
model of modified gravity is the braneworld model. Although the
exciting idea that we live in a fundamentally higher-dimensional
spacetime which is greatly curved by vacuum energy was older
 \cite{Akama:1982jy, Rubakov:1983bb, Antoniadis:1988jn,
Antoniadis:1990ew, Horava:1995qa, Horava:1996ma, Maldacena:1997re,
Lukas:1998yy, Antoniadis:1998ig, ArkaniHamed:1998rs,
ArkaniHamed:1998nn, Binetruy:1999ut}, the new class of "warped"
geometries offered a simple way of localizing the low energy
gravitons on the 4D submanifold  (brane)
\cite{Randall:1999ee,Randall:1999vf, Ida:1999ui}. Motivated by
string/M theory, AdS/CFT correspondence, and the hierarchy problem
of particle physics, braneworld models were studied actively in
recent years
 \cite{Goldberger:1999uk, Kobayashi:2002pw, Brax:2003fv,
Maartens:2003tw, Apostolopoulos:2004rk, Schwindt:2005fm,
Diakonos:2007au}, and have been shown to admit a much wider range
of possibilities for dark energy  \cite{Sahni:2002dx}. A well
studied model is the Dvali-Gabadadze-Porrati (DGP) one
\cite{Dvali:2000hr, Lue:2005ya} where the braneworld is embedded
in the flat bulk with infinite extra dimension. In this model,
gravity appears 4D at short distances but it is altered at
distance larger than some freely adjustable crossover scale $r_c$,
through the slow evaporation of the graviton off our 4D braneworld
universe into the fifth dimension. The inclusion of a graviton
kinetic term on the brane recovers the usual gravitational force
law scaling, $1/r^2$, at short distances, but at large distances
it asymptotes to the 5D scaling, $1/r^3$. The matter particles
cannot freely propagate in the extra dimensions, and are
constrained to live on the (tensionless) brane. In such a model,
late-time self-acceleration solutions appears naturally
\cite{Deffayet:2000uy, Deffayet:2001pu, Deffayet:2002sp}, since
they are driven by the manifestation of the excruciatingly slow
leakage of gravity into the extra dimension.

\subsection{Quintom DE in DGP scenario}\label{sec:DGP}

In this subsection we investigate a simple model of a single
scalar field in the framework of DGP braneworld, following
\cite{Zhang:2006at}. Although the behavior of the effective DE on
a DGP brane with a cosmological constant and dust has been studied
in \cite{Chimento:2006ac, Lazkoz:2006gp}, from the unified
theoretic point of view we can assume that the gravitational
action is not necessarily the Einstein-Hilbert action. In fact,
string theory suggests that the dimensionally reduced effective
action includes not only higher-order curvature terms but also
dilatonic gravitational scalar fields. Thus, at the level of the
low-energy 5D theory (we restrict to 5D braneworld models although
even higher-dimensional are also possible \cite{Gherghetta:2000qi,
Bostock:2003cv, Guendelman:2003ie, Kofinas:2004ae, Erdem:2006qk}),
it is naturally expected that there appears a dilaton-like scalar
field in addition to the Einstein-Hilbert action
\cite{Lukas:1998qs, Lukas:1998tt, Himemoto:2000nd}. Hence, it is
interesting to investigate how such a scalar field in the 5D
theory affects the braneworld \cite{Ellwanger:2000ne,
Carroll:2001zy}. Since we consider a single-field model, the
accelerated expansion of the universe will be a result of the
combined effect of the field evolution and the competition between
4D gravity and 5D gravity.

Let us start from the action of the DGP model
\begin{equation}
\label{totalactionDGP}
 S=S_{\rm bulk}+S_{\rm brane},
\end{equation}
where
\begin{equation}
\label{bulkactionDGP}
  S_{\rm bulk} =\int_{\cal M} d^5X \sqrt{-{}^{(5)}g}
  {1 \over 2 \kappa_5^2} {}^{(5)}R ,
\end{equation}
and
\begin{equation}
\label{braneactionDGP}
 S_{\rm brane}=\int_{M} d^4 x\sqrt{-g} \left[
{1\over\kappa_5^2} K^\pm + L_{\rm brane}(g_{\alpha\beta},\psi)
\right].
\end{equation}
Here $\kappa_5^2$ is the 5-dimensional gravitational constant,
${}^{(5)}R$ is the 5-dimensional curvature scalar, $x^\mu
~(\mu=0,1,2,3)$ are the induced 4D coordinates on the brane and
$K^\pm$ is the trace of extrinsic curvature on either side of the
brane. Finally, $L_{\rm brane}(g_{\alpha\beta},\psi)$ is the
effective 4D Lagrangian, which is given by a generic functional of
the brane metric $g_{\alpha\beta}$ and brane matter fields $\psi$.

We consider a brane Lagrangian consisting of the following terms
\begin{equation}
\label{lbraneDGP}
 L_{\rm brane}=  {\mu^2 \over 2} R  + L_{\rm
m}+L_{\phi},
\end{equation}
where $\mu$ is the 4D reduced Planck mass, $R$ denotes the
curvature scalar on the brane, $L_{\rm m}$ stands for the
Lagrangian of other matters on the brane, and $L_{\phi}$
represents the lagrangian of a scalar confined to the brane.
Assuming a mirror symmetry in the bulk  we obtain the Friedmann
equation on the brane  \cite{Deffayet:2000uy, Deffayet:2001pu,
Deffayet:2002sp, Cai:2005ie}:
\begin{equation}
 H^2+\frac{k}{a^2}=\frac{1}{3\mu^2}\left[\rho+\rho_0+\theta\rho_0
 \Big(1+\frac{2\rho}{\rho_0}\Big)^{1/2}\right],
 \label{friedDGP}
\end{equation}
where $k$ is the spatial curvature of the  three dimensional
maximally symmetric space in the FRW metric on the brane, and
$\theta=\pm 1$ denotes the two branches of DGP model. $\rho$
denotes the total energy density on the brane, including dark
matter and the scalar field  ($\rho=\rho_{\phi}+\rho_{dm}$), and
the term $\rho_0$ relates the   strength of the 5D gravity with
respect to 4D gravity, that is $ \rho_0=\frac{6\mu^2}{r_c^2}$,
where the cross radius is defined as $r_c\triangleq
\kappa_5^2\mu^2$. Comparing the modified Friedmann equation on the
brane with the standard   one:
\begin{equation}
 H^2+\frac{k}{a^2}=\frac{1}{3\mu^2} \Big(\rho_{dm}+\rho_{de}\Big),
 \label{genericFDGP}
\end{equation}
one obtains the density of the effective 4D dark energy as
\begin{equation}
 \rho_{de}=\rho_{\phi}+\rho_0+\theta\rho_0\left[\rho+\rho_0+\theta\rho_0
 \Big(1+\frac{2\rho}{\rho_0}\Big)^{1/2}\right].
 \label{rhodeDGP}
\end{equation}
As usual DE satisfies the continuity equation:
\begin{equation}
 \frac{d\rho_{de}}{dt}+3H(\rho_{de}+p_{eff})=0,
 \label{emDGP}
\end{equation}
where $p_{eff}$ denotes the effective pressure of DE. Then we can
express the EoS of DE as
\begin{equation}
  w_{de}=\frac{p_{eff}}{\rho_{de}}=-1+\frac{1}{3}\frac{d \ln \rho_{de}}{d \ln
  (1+z)},
\end{equation}
where, from (\ref{emDGP})
\begin{equation}
\label{derivationDGP}
 \frac{d \ln \rho_{de}}{d \ln
  (1+z)}= \frac{3}{\rho_{de}}\bigg\{\rho_{de}+p_{de}
  +\theta\bigg[1+2\bigg(\frac{\rho_{de}+\rho_{dm}}{\rho_0}\bigg)\bigg]^{-1/2}\bigg(\rho_{de}+\rho_{dm}+p_{de}\bigg)\bigg\}.
\end{equation}
We mention that  the DE pressure $p_{de}$   is different from
$p_{eff}$.

Clearly, if $\rho_{de}$ decreases and then increases with respect
to redshift, or increases and then decreases, it is implied that
DE experiences a phantom-divide crossing. Equations like
(\ref{friedDGP}) and (\ref{emDGP}) coincide with the subclass of
inhomogeneous EoS of FRW universe  \cite{Nojiri:2005sr} or FRW
universe with general EoS \cite{Nojiri:2004pf}, which can give
rise to the phantom-divide crossing.

\subsubsection{Canonical scalar field}

For a canonical scalar with an exponential potential of the form:
$V=V_0e^{-\lambda\frac{\phi}{\mu}}$ the effective DE evolves as
\cite{Zhang:2006at}:
\begin{eqnarray}
\label{derivation or}
 \frac{d \rho_{de}}{d \ln
  (1+z)}=3\bigg[\dot{\phi}^2+\theta\bigg(1+\frac{\dot{\phi}^2+2V+2\rho_{dm}}{\rho_0}\bigg)^{-1/2}(\dot{\phi}^2+\rho_{dm})\bigg].
\end{eqnarray}
If $\theta=1$  both terms on the RHS are positive and thus it
never goes to zero at finite time, while in the case $\theta=-1$
the two terms of the RHS have opposite sign, therefore it is
possible to obtain the $-1$-crossing. Hence, we restrict our study
to $\theta=-1$.

In order to transform the aforementioned dynamical system into its
autonomous form \cite{Copeland:2006wr,Ferreira:1997au,
Copeland:1997et}, we introduce the auxiliary variables:
\begin{eqnarray}
  x \equiv \frac{\dot{\phi}}{\sqrt{6}\mu H}~,~~
  y \equiv \frac{\sqrt{V}}{\sqrt{3}\mu H}~,~~
  l \equiv \frac{\sqrt{\rho_M}}{\sqrt{3}\mu H}~,~~
  b \equiv \frac{\sqrt{\rho_0}}{\sqrt{3}\mu H}~.
\end{eqnarray}
Thus, the dynamics of the system  can be described by:
\begin{eqnarray}
\label{1DGP}
  x'&=&-\frac{3}{2}\alpha
  x(2x^2+l^2)+3x-\frac{\sqrt{6}}{2}\lambda y^2,\\
\label{2DGP}
  y'&=&-\frac{3}{2}\alpha
  y(2x^2+l^2)+\frac{\sqrt{6}}{2}\lambda xy,\\
\label{3DGP}
  l'&=&-\frac{3}{2}\alpha
  l(2x^2+l^2)+\frac{3}{2}l,\\
\label{4DGP}
  b'&=&-\frac{3}{2}\alpha
  b(2x^2+l^2),
\end{eqnarray}
where $\alpha\equiv 1-(1+2\frac{x^2+y^2+l^2}{b^2})^{-1/2}$, and
here a prime stands for derivation with respect to
$s\equiv-\ln(1+z)$, and we have set $k=0$. One can see that this
system degenerates to a quintessence with dust matter in standard
general relativity. Note that the four equations (\ref{1DGP}),
(\ref{2DGP}), (\ref{3DGP}), (\ref{4DGP}) are not independent,
since the constraint
\begin{eqnarray}
 \label{constraintDGP}
 x^2+y^2+l^2+b^2-b^2\left(1+2\frac{x^2+y^2+l^2}{b^2}\right)^{1/2}=1,
\end{eqnarray}
arising from the Friedmann equation, reduces the number of
independent equations  to three. There are two critical points of
this system, satisfying $x'=y'=l'=b'=0$, appearing at
\begin{eqnarray}
 &x&=y=l=0, ~~b={\rm constant};\\
 &x&=y=l=b=0.
\end{eqnarray}
However, neither of them satisfies the Friedmann constraint
(\ref{constraintDGP}). Hence one proves that there is no (kinetic
energy)-(potential energy) scaling solution or (kinetic
energy)-(potential energy)-(dust matter) scaling solution on a DGP
brane with quintessence and dust matter.

\begin{figure}
\centering
\includegraphics[totalheight=1.9in, angle=0]{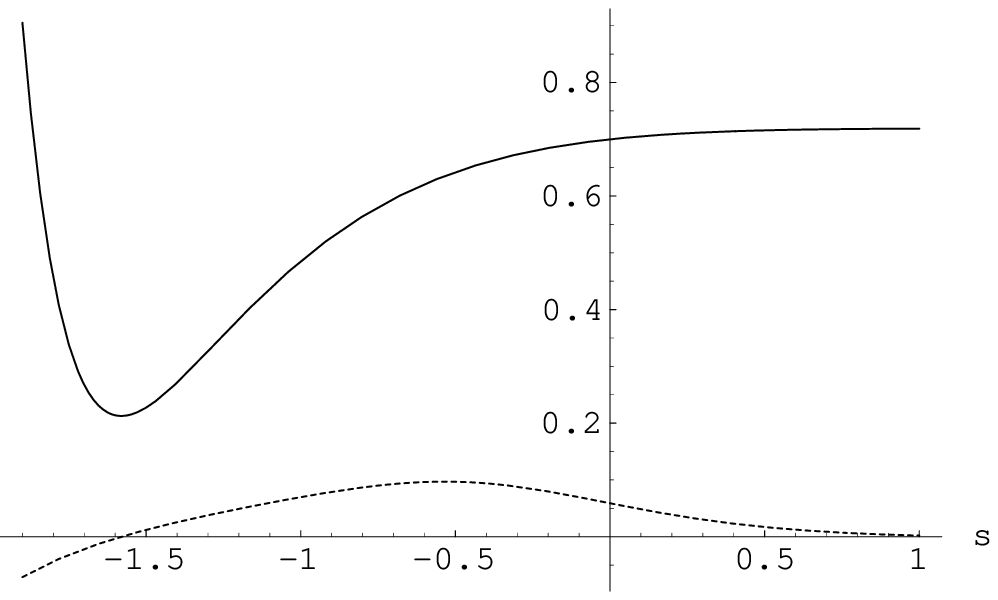}
\includegraphics[totalheight=2.1in, angle=0]{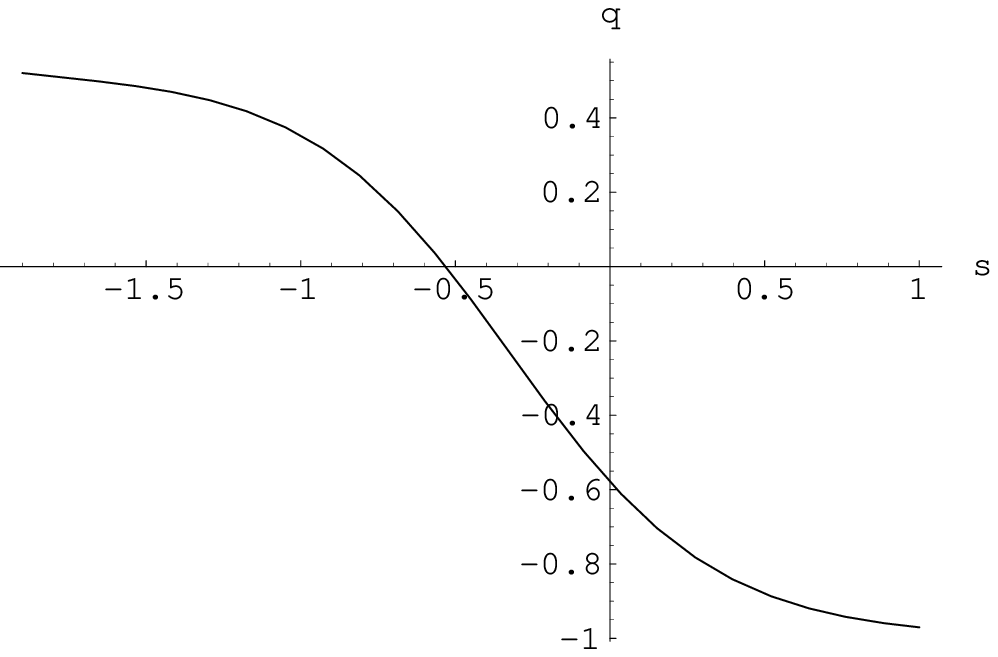}
\caption{ {\it{   { \bf{(a)}} $\beta$ (solid curve) and $\gamma$
(dotted curve) as functions of $s\equiv-\ln(1+z)$ for the
canonical field case of the DGP model. The EoS parameter of DE
crosses $-1$ at about $s=-1.6$, i.e. at $z=3.9$. {\bf{(b)}} The
corresponding deceleration parameter $q$ vs $s$, which crosses 0
at about $s=-0.52$, i.e. at $z=0.68$. For this figure,
$\Omega_{ki}=0.01$, $\Omega_M=0.3$, and
$\Omega_{r_c}\equiv{\rho_0}/{\rho_c}=0.01$, $\lambda=0.05$. From
Ref.  \cite{Zhang:2006at}.}}}
 \label{rho1DGP}
\end{figure}

The most significant parameter from the viewpoint of observations
is the deceleration parameter $q$, which carries the total effects
of cosmic fluids  and it is defined as
$q=-\frac{\ddot{a}a}{\dot{a}^2}=-1+\frac{3}{2}\alpha(2x^2+l^2)$.
In addition, for convenience we introduce the dimensionless
density as
\begin{eqnarray}
\beta=\frac{\rho_{de}}{\rho_c}=\frac{\Omega_{r_c}}{b^2}\left\{x^2+y^2+b^2
-b^2\Big[1+2\Big(\frac{x^2+y^2+l^2}{b^2}\Big)\Big]^{1/2}\right\},
\end{eqnarray}
where $\rho_c$ denotes the present critical density of the
universe, and the rate of change with respect to redshift of DE:
\begin{eqnarray}
 \gamma =
  \frac{1}{\rho_c}\frac{b^2}{\Omega_{r_c}}\frac{d\rho_{de}}{ds}
 =3\left\{\Big[1+2\Big(\frac{x^2+y^2+l^2}{b^2}\Big)\Big]^{-1/2}(2x^2+l^2)-2x^2\right\}.
\end{eqnarray}
In Fig. \ref{rho1DGP}  we present a concrete numerical example of
the $-1$-crossing, depicting $\beta$, $\gamma$ and $q$ as a
function of $s\equiv-\ln(1+z)$. Thus, contrary to the conventional
4D cosmology, where a single canonical field cannot experience the
$-1$-crossing, in the present model this is possible due to the
induced term $\rho_0$ of the ``energy density" of $r_c$. Only a
small component of $\rho_0$, i.e. $\Omega_{r_c}=0.01$, is capable
of making the EoS of DE cross $-1$. At the same time the
deceleration parameter is consistent with observations.

\subsubsection{Phantom scalar field}

\begin{figure}
\centering
\includegraphics[totalheight=1.8in, angle=0]{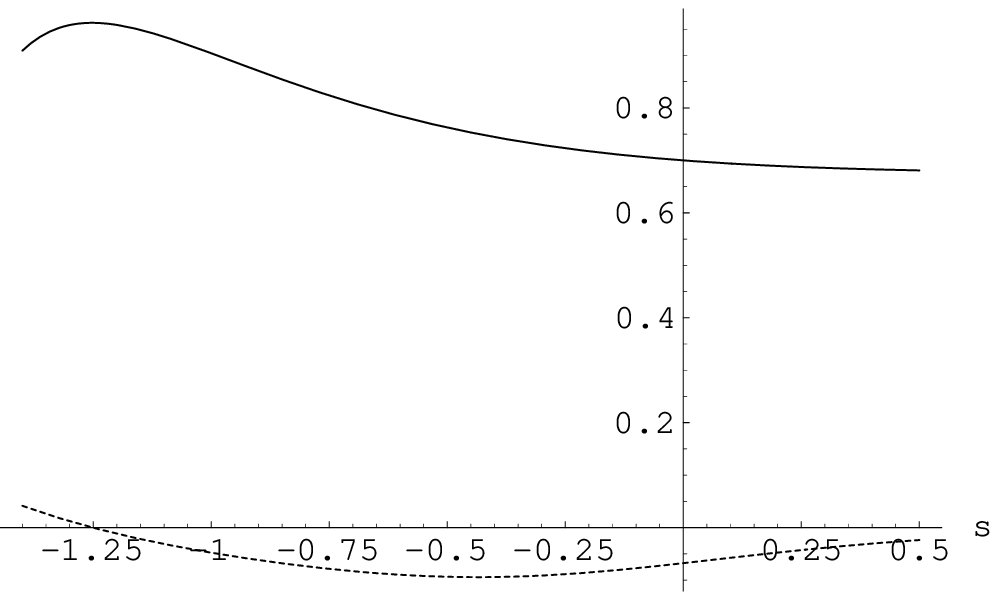}
\includegraphics[totalheight=2in, angle=0]{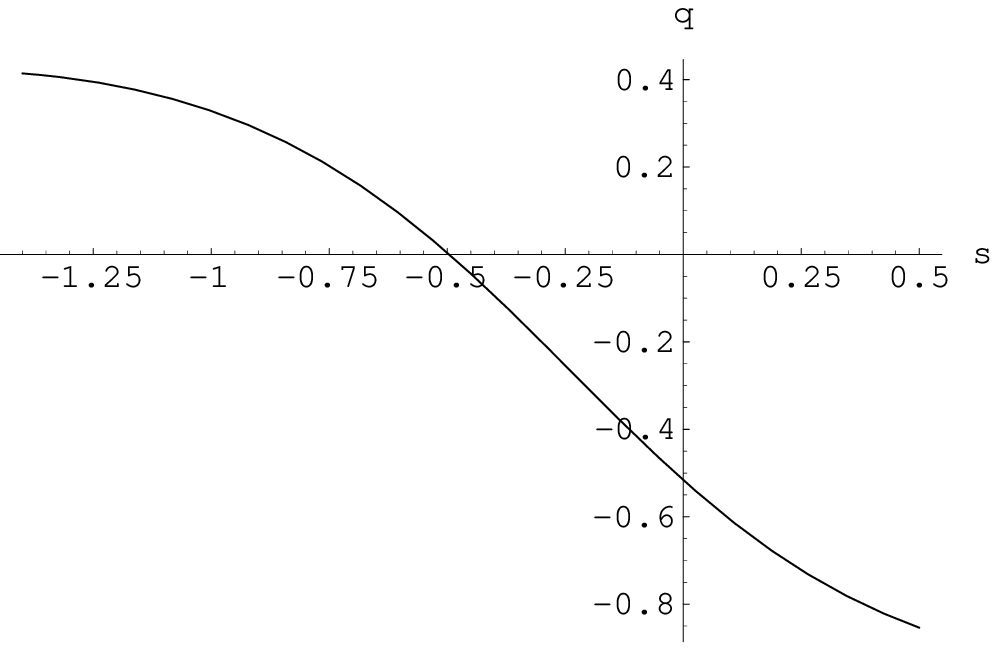}
\caption{ {\it{   { \bf{(a)}} $\beta$ (solid curve) and $\gamma$
(dotted curve) as functions of $s\equiv-\ln(1+z)$ for the phantom
field case of the DGP model. The EoS parameter of DE crosses $-1$
at about $s=-1.25$, i.e. at $z=1.49$. {\bf{(b)}} The corresponding
deceleration parameter $q$ vs $s$, which crosses 0 at about
$s=-0.50$, i.e. at $z=0.65$. For this figure, $\Omega_{ki}=0.01$,
$\Omega_M=0.3$, and $\Omega_{r_c}\equiv{\rho_0}/{\rho_c}=0.01$,
$\lambda=0.01$. From Ref.  \cite{Zhang:2006at}.}}}
 \label{rho4DGP}
\end{figure}

For a phantom scalar with an exponential potential of the form:
$V=V_0e^{-\lambda\frac{\phi}{\mu}}$ the effective DE evolves as
\cite{Zhang:2006at}:
\begin{eqnarray}\label{derivation ph}
 \frac{d \rho_{de}}{d\ln(1+z)} = 3\Big[-\dot{\phi}^2
   + \theta\Big(1+\frac{-\dot{\phi}^2+2V+2\rho_{dm}}{\rho_0}\Big)^{-1/2}(-\dot{\phi}^2+\rho_{dm})\Big].
\end{eqnarray}
If $\theta=-1$, both terms of RHS are negative and it never goes
to zero at finite time. Contrarily, if $\theta=1$  the two terms
of RHS have opposite sign and the phantom-divide crossing is
possible. In the following we consider the branch of $\theta=1$.
The dynamics is described by the following autonomous system:
\begin{eqnarray}
 x'&=&-\frac{3}{2}\alpha'
 x(-2x^2+l^2)+3x-\frac{\sqrt{6}}{2}\lambda y^2,\\
 \label{6}
  y'&=&-\frac{3}{2}\alpha'
 y(-2x^2+l^2)+\frac{\sqrt{6}}{2}\lambda xy,\\
 \label{7}
   l'&=&-\frac{3}{2}\alpha'
 l(-2x^2+l^2)+\frac{3}{2}l,\\
 \label{8}
  b'&=&-\frac{3}{2}\alpha'
 b(-2x^2+l^2),
\end{eqnarray}
where $\alpha'\equiv 1+(1+2\frac{-x^2+y^2+l^2}{b^2})^{-1/2}$, and
we have also adopted the spatial flatness condition. Now the
Friedmann constraint becomes
\begin{eqnarray}
 \label{constraint ph}
 -x^2+y^2+l^2+b^2+b^2\left(1+2\frac{-x^2+y^2+l^2}{b^2}\right)^{1/2}=1,
\end{eqnarray}
with which there are three independent equations left in the
system. Through a similar analysis as in the case of a canonical
scalar field, it can be shown that there is no (kinetic
energy)-(potential energy) scaling solution or (kinetic
energy)-(potential energy)-(dust matter) scaling solution on a DGP
brane with phantom and dust dark metter
 \cite{Zhang:2006at}. The deceleration parameter $q$ becomes
$q=-1+\frac{3}{2}\alpha'(-2x^2+l^2)$, and the dimensionless
density and rate of change with respect to redshift of DE become,
\begin{eqnarray}
 \beta &=& \frac{\Omega_{r_c}}{b^2}\left\{-x^2+y^2+b^2
  +b^2\Big[1+2\Big(\frac{-x^2+y^2+l^2}{b^2}\Big)\Big]^{1/2}\right\},\\
 \gamma &=& 3\left\{-\Big[1+2\Big(\frac{-x^2+y^2+l^2}{b^2}\Big)\Big]^{-1/2}(-2x^2+l^2)+2x^2\right\}.
\end{eqnarray}
In Fig. \ref{rho4DGP}, we depict $\beta$, $\gamma$ and $q$ as a
function of $s\equiv-\ln(1+z)$. As we observe,  the (effective)
EoS parameter of DE crosses $-1$ as expected. This is also in
contrast with conventional 4D cosmology, where a single phantom
field  lies always below $-1$. The 5D gravity plays a critical
role in the realization of the $-1$-crossing, and at the same time
the deceleration parameter is consistent with observations.

Finally, we stress that the most important difference of the
dynamics between an ordinary and a phantom field is that while for
the canonical field the $-1$-crossing takes place from above to
below, in the phantom case it takes place from below to above (see
$\gamma$ vs $s$ in Figs. \ref{rho1DGP} and \ref{rho4DGP}).

\subsection{Quintom DE in the braneworld}\label{sec:hdbrane}

Let us now investigate the effects of the bulk quintom field in
the DGP  braneworld scenario, and in particular whether it is
possible or not to have a late-time accelerated phase on the brane
for i) a tensionless brane with a quintom fluid in the bulk and
ii) a quintom DE fluid on the brane and an empty bulk. We consider
a 5D metric of the form
\begin{eqnarray}
    ds^2=-n^2(t,y)dt^2+a^2(t,y)\gamma_{ij}dx^idx^j+b^2(t,y)dy^2,
\end{eqnarray}
where $y$ is the coordinate of the fifth dimension and
$\gamma_{ij}$ is a maximally symmetric 3-dimensional metric. We
will use $k$ to parameterize the spatial curvature and assume that
the brane is a hypersurface located at $y=0$. We shall be
interested in the model described by the action
\begin{eqnarray}
    S=\int d^5x\sqrt{-g}\Big(\frac{1}{2\kappa^2_5}R^{(5)}-\Lambda+{\mathcal{L}}_B^{mat}\Big)
    +\int
    d^4x\sqrt{-q}(-\delta_b+\mathcal{L}_b^{mat}),
\end{eqnarray}
where $R^{(5)}$ is the curvature scalar of the 5D metric $g_{AB}$,
$\Lambda$ is the bulk cosmological constant, $\delta_b$ is the
brane tension, $\kappa^2_5=8\pi G_5$, and $q_{AB}=g_{AB}-n_An_B$
($n_A $ is the unit vector normal to the brane and
$A,B=0,1,2,3,5$) is the induced metric on the 3-brane. The 5D
Einstein equation can be written as
\begin{eqnarray}
R^{(5)}_{AB}-\frac{1}{2}g_{AB}{R^{(5)}=\kappa^2_5[-\Lambda
g_{AB}+T_{AB}+S_{\mu\nu}\delta^\mu_A\delta^\nu_B\delta(y_b)]}.
\end{eqnarray}
Here $\delta(y_b)=\frac{\delta(y)}{b}$, $T_{AB}$ is the energy
momentum tensor of the bulk matter, and the last term corresponds
to the matter content in the brane
\begin{eqnarray}
    S_{\mu\nu}=-\delta_b g_{\mu\nu}+\tau_{\mu\nu}.
\end{eqnarray}
The non-zero components of the 5D Einstein equation read
\cite{Kiritsis:2002zf, Maeda:2003ar, Diakonos:2004xq,
Umezu:2005dw, Bogdanos:2006pf, Bogdanos:2006km, Bogdanos:2006dt,
Mendes:2000wu, Perivolaropoulos:2003we, Bogdanos:2006qw,
Farakos:2005hz, Farakos:2006tt, BarbosaCendejas:2006hj}
\begin{eqnarray}
\label{3nset}
  &&3\left\{-\frac{\dot{a}}{n^2a}\Big(\frac{\dot{a}}{a}+\frac{\dot{b}}{b}\Big)+\frac{1}{b^2}\Big[\frac{a''}{a}+\frac{a'}{a}
    \Big(\frac{a'}{a}-\frac{b'}{b}\Big)\Big]-\frac{k}{a^2}\right\} =
    \kappa^2_5\Big[-\Lambda+T^0_0+S^0_0\delta(y_b)\Big],\\
\label{3nmset}
  &&\frac{1}{b^2}\delta_{j}^i\Big\{\frac{a'}{a}\Big(\frac{a'}{a}+\frac{n'}{n}\Big)-
    \frac{b'}{ba^2}\Big[\frac{n'}{n}+2\Big(\frac{a'}{a}\Big)\Big]+\frac{a''}{a}+\frac{n''}{n}\Big\}
    +\frac{1}{n^2}\delta_j^i\Big\{\frac{\dot{a}}{a}\Big[-\frac{\dot{a}}{a}+2\Big(\frac{\dot{n}}{n}\Big)\Big]- \nonumber\\
  &&~~~~~~~~~~~~~~ ~~~~
    -2\Big(\frac{\ddot{a}}{a}\Big)+\frac{\dot{b}}{b}
    \Big[-2\Big(\frac{\dot{a}}{a}\Big)+\frac{\dot{n}}{n}\Big]-\frac{\ddot{b}}{b}\Big\}
    -k\delta_j^i = \kappa^2_5[-\Lambda+T^i_j+S^i_j\delta(y_b)],\\
\label{3nbset}
  &&3\Big\{\frac{n'}{n}\frac{\dot{a}}{a}+\frac{a'}{a}\frac{\dot{b}}{b}-\frac{\dot{a'}}{a}\Big\}=
    \kappa^2_5T_{05},\\
\label{3nvset}
  &&3\Big\{\frac{a'}{ab^2}\Big(\frac{a'}{a}+\frac{n'}{n}\Big)
    -\frac{1}{n^2}\Big[\frac{\dot{a}}{a}\Big(\frac{\dot{a}}{a}-\frac{\dot{n}}{n}\Big)
    +\frac{\ddot{a}}{a}\Big]-\frac{k}{a^2}\Big\} = \kappa^2_5(-\Lambda+T^5_5),
\end{eqnarray}
where primes indicate derivatives with respect to $y$ and dots
derivatives with respect to $t$. On the brane we assume a perfect
fluid
\begin{eqnarray}\label{4nset}
    \tau^\mu_\nu=diag(-\rho_b,p_b,p_b,p_b).
\end{eqnarray}
In the bulk space we consider a quintom field, constituted from
the normal scalar field $\phi(t,y)$ and the negative-kinetic
scalar field $\sigma(t,y)$, with a Lagrangian of the form:
\begin{eqnarray}\label{4aset}
    \mathcal{L}^{mat}_B=\frac{1}{2}g^{AB}\Big(\phi_{,A}\phi_{,B}-\sigma_{,A}\sigma_{,B}\Big)
    +V(\phi,\sigma).
\end{eqnarray}
According to this action, the energy momentum tensor of the bulk
quintom field is given by
\begin{eqnarray}\label{5set}
    T_{AB}=\phi_{,A}\phi_{,B}-\sigma_{,A}\sigma_{,B} -g_{AB}\Big[\frac{1}{2}g^{CD}
    (\phi_{,C}\phi_{,D}
    -\sigma_{,C}\sigma_{,D})+V(\phi,\sigma)\Big].
\end{eqnarray}
Finally, the equations of motion of the scalar fields $\phi$ and
$\sigma$ in the bulk space write: {\small{
\begin{eqnarray}
\label{5nset}
  -\ddot{\phi}-3\frac{\dot{a}}{a}\dot{\phi}+\Big(n'+3\frac{a'}{a}\Big)\phi'+\phi''-\frac{\partial V(\phi,\sigma)}
  {\partial\phi} = \frac{\delta\mathcal{L}_b^{mat}}{\delta\phi}\delta(y_b)~,\\
\label{5nsetb}
  -\ddot{\sigma}-3\frac{\dot{a}}{a}\dot{\sigma}+\Big(n'+3\frac{a'}{a}\Big)\sigma'+\sigma''+\frac{\partial V(\phi,\sigma)}{\partial\sigma}=
   -\frac{\delta\mathcal{L}_b^{mat}}{\delta\sigma}\delta(y_b)~.
\end{eqnarray}

We are interested in studying the Einstein equations in the
presence of a bulk quintom field at the location of the brane.
Without loosing generality we choose $b(t,y)=1$ and $n(t,0)=1$,
which can be achieved by scaling the time coordinate. As is well
known, the presence of the brane leads to a singular term
proportional to $\delta$-function in $y$ on the right-hand sides
of the Einstein equations (\ref{3nset}) and (\ref{3nmset}) and the
equation of motions (\ref{5nset}),(\ref{5nsetb}), which have to be
matched by singularity in the second derivatives in $y$ on the
left-hand side. Since all fields under consideration are symmetric
under the orbifold symmetry $Z_2$, these jumps in the first
derivatives in $y$ fix the first derivatives at $y=0$. In our
case, the junction conditions read
\begin{eqnarray}\label{6nset}
  &&\frac{a'}{a}|_{y=0}= -\frac{\kappa^2_5}{6}\Big(\rho_b+\delta_b\Big),\nonumber \\
  &&n'|_{y=0} = \frac{\kappa^2_5}{6}\Big(3p_b+2\rho_b-\delta_b\Big),
\end{eqnarray}
and
\begin{eqnarray}\label{6nlset}
  \phi'|_{y=0} = \frac{1}{2}\frac{\delta \mathcal{L}_b^{mat}}{\delta\phi}~,~~
  \sigma'|_{y=0}=-\frac{1}{2}\frac{\delta \mathcal{L}_b^{mat}}{\delta\sigma}~.
\end{eqnarray}
Using the $00$ and $55$ components of the Einstein equations in
the bulk, one obtains
\begin{eqnarray}\label{6nnset}
    F'&=&\frac{2\kappa^2_5}{3}\Big(\Lambda-T_0^0\Big)a^3a'-\frac{2\kappa^2_5}{3}T_5^0a^3\dot{a}~,\\
\label{6nnnset}
    \dot{F}&=&\frac{2\kappa^2_5}{3}\Big(\Lambda-T_5^5\Big)a^3\dot{a}-\frac{2\kappa^2_5}{3}n^2T_5^0a^3a',
\end{eqnarray}
where $F$ is a function of $t$ and $y$ defined by
\begin{eqnarray}\label{6nmset}
    F(t,y)=\frac{(\dot{a}a)^2}{n^2}-(a'a)^2+ka^2.
\end{eqnarray}
In order to find the solution for $F$, we assume that the quintom
field in the bulk is independent of $y$. Under this assumption the
non-vanishing components of the quintom energy momentum tensor
take the form
\begin{eqnarray}\label{7nset}
 && T^0_0=-\rho_B = -\frac{1}{2}\dot{\phi}^2+\frac{1}{2}\dot{\sigma}^2-V(\phi,\sigma)  \nonumber\\
 && T^i_i=p_B = \frac{1}{2}\dot{\phi}^2-\frac{1}{2}\dot{\sigma}^2-V(\phi,\sigma) \nonumber\\
 && T_{05}=\dot{\phi}\phi'-\dot{\sigma}\sigma'=0  \nonumber\\
 && T^5_5= \frac{1}{2}\dot{\phi}^2-\frac{1}{2}\dot{\sigma}^2-V(\phi,\sigma),
\end{eqnarray}
valid also on the brane. As we observe, in the case of
$y$-independent bulk quintom field, $T_{05}$ vanishes, thus there
is no matter-flow along the fifth dimension. Furthermore, it is
obvious from (\ref{6nlset}) that the quintom field cannot appear
in the matter content on the brane
 \cite{Ellwanger:2000ne,Carroll:2001zy}.

 Solving equation
(\ref{6nnset}) leads to the first integral of the 00 component of
Einstein equation as
\begin{eqnarray}\label{7nmset}
    \frac{\kappa^2_5}{6}\Big(\Lambda+\rho_B\Big)+\frac{\mathcal{C}}{a^4}-\frac{\dot{a}^2}{a^2n^2}+\frac{a'^2}{a^2}-\frac{k}{a^2}=0,
\end{eqnarray}
where $\mathcal{C}$ is a constant of integration usually referred
as dark radiation  \cite{Maartens:2000fg}. Substituting the
junction conditions (\ref{6nset}) into  (\ref{7nmset}), we arrive
at the generalized Friedmann equation on the brane as
\cite{Kiritsis:2002zf, Maeda:2003ar, Diakonos:2004xq,
Umezu:2005dw, Bogdanos:2006pf, Bogdanos:2006km, Bogdanos:2006dt,
Mendes:2000wu, Perivolaropoulos:2003we, Bogdanos:2006qw,
Farakos:2005hz, Farakos:2006tt,
BarbosaCendejas:2006hj,Maartens:2000fg, Kanti:2000rd}:
\begin{eqnarray}\label{8nset}
  H^2+\frac{k}{a^2}=\frac{\kappa^2_5}{6}\Big(\Lambda+\frac{\kappa^2_5}{6}\delta_b^2\Big)+\frac{k^4_5}{18}\delta_b\rho_b+
   \frac{\kappa^2_5}{6}\rho_B+\frac{k^4_5}{36}\rho_b^2+\frac{\mathcal{C}}{a^4}.
\end{eqnarray}
As one can see from (\ref{8nset}), in the absence of the bulk
matter field, cosmological constant and tension, the equation
gives rise to a Friedmann equation of the form $H\propto\rho_b$
instead of $H\propto\sqrt{\rho_b}$, which is inconsistent with
cosmological observations. This problem can be solved by either
considering the cosmological constant and tension on the brane or
considering a matter field in the bulk  \cite{Cline:1999ts,
Kanti:1999sz}. Recalling the junction conditions (\ref{6nset}),
the 05 component of Einstein equations and the field equation
(\ref{5nset}), (\ref{5nsetb}) acquire the following forms on the
brane:
\begin{eqnarray}
 \label{9nset}
  &&\dot{\rho_b}+3H(\rho_b+p_b)=0~, \\
 \label{10nset}
  && \ddot{\phi}+3\Big(\frac{\dot{a}}{a}\Big)\dot{\phi}+\frac{\partial V(\phi,\sigma)}{\partial\phi}= 0~, \\
 \label{11nset}
  && \ddot{\sigma}+3\Big(\frac{\dot{a}}{a}\Big)\dot{\sigma}-\frac{\partial V(\phi,\sigma)}{\partial\sigma}=0~.
\end{eqnarray}
It should be noted that if the scalar fields $\phi$ and $\sigma$
satisfy the field equations (\ref{10nset}) and (\ref{11nset})
respectively, the bulk energy momentum tensor is conserved and we
obtain
\begin{eqnarray}\label{9nnset}
    \dot{\rho_B}+3H(\rho_B+p_B)=0.
\end{eqnarray}

We are interested in studying the acceleration condition for a
universe with quintom field in the bulk. The condition for
acceleration can be obtained from (\ref{8nset}) by using the
conservation equation of the brane and bulk matter ((\ref{9nset})
and (\ref{9nnset}) respectively):
\begin{eqnarray}\label{12nset}
 \frac{\ddot{a}}{a}=\frac{\kappa^2_5}{6}\Big(\Lambda+\frac{\kappa^2_5}{6}\delta_b^2\Big)-\frac{k^4_5}{36}\delta_b\Big(\rho_b+3p_b\Big)
 - \frac{\kappa^2_5}{12}\Big(\rho_B+3p_B\Big)-\frac{k^4_5}{36}\Big(2\rho_b^2+3\rho_bp_b\Big)-\frac{\mathcal{C}}{a^4}.
\end{eqnarray}
In order to study the role of bulk quintom field in the late-time
acceleration phase on the brane, we neglect the effect of tension,
brane matter and dark radiation, and therefore the aforementioned
equation becomes:
\begin{eqnarray}\label{13nset}
 \frac{\ddot{a}}{a}=\frac{\kappa^2_5}{6}\Lambda-\frac{\kappa^2_5}{12}\Big(\rho_B+3p_B\Big).
\end{eqnarray}
Thus, the acceleration condition for a universe with quintom DE in
the bulk reads:
\begin{eqnarray}\label{15set}
    p_B<\frac{1}{3}(\Lambda-\rho_B)\hspace{0.5cm} {\rm or} \hspace{0.5cm}
    \dot{\phi}^2-\dot{\sigma}^2<\Lambda/2+V(\phi,\sigma).
\end{eqnarray}

Now we consider the 55-component of Einstein equations at the
position of the brane, which leads to the Raychaudhuri equation
\begin{eqnarray}
\label{16nset}
 \frac{\ddot{a}}{a}+H^2+\frac{k}{a^2}=\frac{\kappa^2_5}{3}\Big(\Lambda+
 \frac{\kappa^2_5}{6}\delta_b^2\Big)
 -\frac{\kappa^2_5}{36}\Big[\delta_b\Big(3p_b-\rho_b\Big)+\rho_b\Big(3p_b+\rho_b\Big)\Big]-\frac{\kappa^2_5}{3}T^5_5.
\end{eqnarray}
Using (\ref{8nset}) one can rewrite the above equation as
\begin{eqnarray}
\label{17nset}
 \frac{\ddot{a}}{a}=\frac{\kappa^2_5}{6}\Big(\Lambda+\frac{\kappa^2_5}{6}\delta_b^2\Big)
  -\frac{k^4_5}{36}\delta_b\Big(\rho_b+3p_b\Big)-\frac{\kappa^2_5}{6}\rho_B
  -\frac{k^4_5}{36}\Big(2\rho_b^2+3\rho_bp_b\Big)-\frac{\mathcal{C}}{a^4}-\frac{k^4_5}{3}T^5_5.
\end{eqnarray}
Comparing equation (\ref{12nset}) with (\ref{17nset}) we provide a
constraint on the bulk energy momentum tensor:
\begin{eqnarray}
\label{18nset}
    (3p_B-\rho_B)=4T^5_5,
\end{eqnarray}
which for the quintom field with the energy-momentum tensor
(\ref{7nset}) leads to
\begin{eqnarray}\label{18mset}
    \dot{\phi}^2-\dot{\sigma}^2=0.
\end{eqnarray}
Therefore, we deduce that the time-dependent bulk quintom field
influences the brane as a time-dependent cosmological constant,
which can be written in terms of the quintom potential:
$p_B=-\rho_B=-V(\phi)$ \cite{Wetterich:1994bg}. Thus, in order to
acquire an accelerating universe the potential $V$ must be a
positive function of time, and in order to be compatible with
observations it should be decreasing with time. For the particular
solution of (\ref{18mset}) in which $\phi$ and $\sigma$ are
constant on the brane, we arrive at the natural cosmological
constant on the brane, which in this case is induced by
time-dependent bulk fields.

Let us now consider a $y$-independent bulk quintom field with
Lagrangian (\ref{4aset}) \cite{Nozari:2008hz,Setare:2007qu}. The
generalized Friedmann equation in this case reads:
\begin{equation}
\label{24set}
 \Big(1+\frac{\delta_b\kappa^2_5}{6\mu^2}\Big)\Big(H^2+\frac{k}{a^2}\Big)
 -\frac{\kappa^2_5}{6}\Big(\Lambda+\frac{\kappa^2_5}{6}\delta_b^2\Big)-\frac{k^4_5}{18}\delta_b\rho_b
 -\frac{\kappa_5^2}{6}\rho_B+\frac{\mathcal{C}}{a^4}
 =\frac{k^4_5}{36\mu^4}\left[-\mu^2\rho_b+3\Big(H^2+\frac{k}{a^2}\Big)\right]^2,
\end{equation}
where $\mu^2=8\pi G_4$ since $r_c=\frac{\kappa^2_5}{2\mu^2}$. The
brane Friedmann equation (\ref{8nset}) can be derived from
(\ref{24set}) by letting $\mu$ go to infinity. We are interested
in studying the effect of the quintom field on the brane. Ignoring
the matter on the brane and the cosmological constant $\Lambda$,
(\ref{24set}) can straightforwardly be rewritten as:
\begin{eqnarray}\label{25set}
    H^2+\frac{k}{a^2}=\frac{1}{2r^2_c}\left(1+\epsilon\sqrt{1-\frac{2\kappa_5^2}{3}\,\rho_B\,r_c^2}\right),
\end{eqnarray}
where $\rho_B$ is the energy density of the bulk quintom field on
the brane, given by the first relation of (\ref{7nset}). The two
different possible $\epsilon$-values  ($\epsilon=\pm1$),
correspond to two different embeddings of the brane into the bulk
spacetime \cite{Cvetic:1993xe,Gibbons:1993in}. Since the bulk
quintom field satisfies the usual energy momentum conservation law
on the brane (\ref{9nnset}), we have $\rho_B\propto a^{-3(w+1)}$.
Integrating (\ref{25set}) for $k=0$ and $w\geq-1$ (that is
$\dot{\phi}>\dot{\sigma}$), shows that the scale factor $a$
diverges at late times. Thus, the energy density of the bulk
matter goes to zero for late times and thus it reaches a regime
where it is small in comparison with $1/r^2_c$
\cite{Setare:2007qu}. In the case $w<-1$ (that is
$\dot{\phi}<\dot{\sigma}$), integrating (\ref{25set}) indicates a
vanishing scale factor $a$ at late times, therefore the matter
density goes to zero and we can use $\kappa_5^2\rho_B\ll 1/r_c^2$.
In summary, in DGP model with a quintom DE fluid in the bulk
space, one can expand (\ref{25set}) under the condition that
$\kappa_5^2\rho_B\ll1/r^2_c$ for the whole $w$-range.

At zeroth order and for spatially flat metric, two different
results, depending on the value of $\epsilon$, can be derived.
Considering the case $\epsilon=-1$ yields
\begin{eqnarray}
\label{26set}
  H^2=0,
\end{eqnarray}
which describes an asymptotically static universe. Considering
$\epsilon=1$  leads to
\begin{eqnarray}
\label{27set}
 H^2=\frac{1}{r_c^2},~~~{\rm or}~~~a(t)\propto \exp\Big(\frac{t}{r_c}\Big).
\end{eqnarray}
This branch provides the self acceleration solution in the late
universe and it is the most important aspect of the model at hand.
Therefore, the late-time behavior of the universe does not change
even if we ignore the matter field on the brane and consider a
scenario with the bulk quintom fields only.

Let us now consider the ``opposite'' case, and neglect the bulk
matter, considering only   quintom fields confined on the brane,
in DGP framework, with a Lagrangian   as
\begin{eqnarray}\label{1cset}
    \mathcal{L}^{mat}_b=\frac{1}{2}q^{\mu\nu}(\phi_{,\mu}\phi_{,\nu}-\sigma_{,\mu}\sigma_{,\nu})
    +\tilde{V}(\phi,\sigma).
\end{eqnarray}
The energy-momentum tensor of the quintom fields
 on the brane is given by
\begin{eqnarray}\label{2cset}
    \tau_{\mu\nu}=\phi_{,\mu}\phi_{,\nu}-\sigma_{,\mu}\sigma_{,\nu}-q_{\mu\nu}\left[\frac{1}{2}q^{\alpha\beta}
    (\phi_{,\alpha}\phi_{,\beta}
    -\sigma_{,\alpha}\sigma_{,\beta})+\tilde{V}(\phi,\sigma)\right].
\end{eqnarray}
In absence of bulk field and brane tension the first integral of
the 00 component of  (\ref{2cset}) leads to
\cite{Deffayet:2000uy,Abramo:2005be}
\begin{eqnarray}
\label{25bset}
    \sqrt{H^2+\frac{k}{a^2}}=\frac{1}{2r_c}\left(\epsilon+\sqrt{1+\frac{4\mu^2}{3}\rho_br^2_c}\right),
\end{eqnarray}
in which $\rho_b$ is quintom energy density obtained by
(\ref{2cset}). Since the energy-momentum tensor on the brane is
conserved, we could use the aforementioned procedure  to examine
the late-time cosmology on the brane. Integrating (\ref{25bset})
for a flat geometry indicates that for $w\geq-1$ the scale factor
$a$ diverges at late times  \cite{Deffayet:2000uy}, while for
$w<-1$ the scale factor $a$ vanishes at late times \cite{
Nozari:2008hz,Setare:2007qu}. Thus, the energy density of quintom
DE goes to zero for late times and reaches a regime where it is
small in comparison with $1/r^2_c$. Expanding equation
(\ref{25bset}) under the condition $\mu^2\rho_b\ll 1/r_c^2$
provides an asymptotically static universe, $H=0$, in the case
$\epsilon=-1$ and a self accelerated phase, $H=\frac{1}{r_c}$, in
the case $\epsilon=1$.

In summary, the presence of quintom DE on the brane or in the bulk
does not change the late time behavior of the universe. In both
cases and for the whole range of $w$ ($w<-1 $ or $w>-1$), one can
derive the self accelerating universe at late times.

\subsection{Other modified-gravity models}

We close this section mentioning that a quintom scenario can also
be obtained in many other modified gravity models, namely,
gravitational models with Gauss-Bonnet corrections
\cite{Cai:2005ie,Nojiri:2005vv, Sami:2005zc, Koivisto:2006xf,
Tsujikawa:2006ph, Leith:2007bu, Sanyal:2007pu}, $f(R)$ models with
singular $w=-1$ crossing \cite{Amendola:2007nt, Bamba:2008hq,
Bamba:2009ay}, scalar-tensor models with nonminimal gravitational
couplings \cite{Nozari:2008ff,Elizalde:2004mq,
Perivolaropoulos:2005yv, Tsujikawa:2008uc}, gravitational ghost
condenstate models \cite{ArkaniHamed:2003uy, Gripaios:2004ms,
Libanov:2005vu, Rubakov:2006pn, Libanov:2007mq} (see Ref.
\cite{Nojiri:2006ri} for a review). The analysis of the
perturbations of these models can be studied by directly expanding
metric fluctuations \cite{Tsujikawa:2007gd, Tsujikawa:2007tg} or
by the Parameterized Post-Friedmann approach \cite{Hu:2007pj,
Hu:2008zd, Fang:2008sn} under certain cases, when one wants to
check their rationalities. In the framework of these modified
gravity models, the current cosmic acceleration  may be achieved
naturally.



\section{Energy conditions and quintom cosmology in the early
universe}\label{sec:quintomEC}

In this section we investigate the violations of energy conditions
in quintom cosmology, and we study its implications in the early
universe.

\subsection{Null Energy Condition}\label{sec:NEC}

It is well known that energy conditions play an important role in
classical theory of general relativity and thermodynamics
\cite{Hawking:1973uf}. In classical general relativity it is
usually convenient and efficient to restrict a physical system to
satisfy one or more of energy conditions, for example  in the
proof of Hawking-Penrose singularity theorem \cite{Penrose:1964wq,
Hawking:1969sw}, the positive mass theorem \cite{Schon:1981vd,
Witten:1981mf} etc. Furthermore, in thermodynamics the energy
conditions are the bases for obtaining entropy bounds
\cite{Bousso:1999xy, Flanagan:1999jp}. Among those energy
conditions, the null energy condition (NEC) is the weakest one,
and it states that for any null vector $n^\mu$ the stress-energy
tensor $T_{\mu\nu}$ should satisfy the relation
\begin{eqnarray}
T_{\mu\nu} n^\mu n^\nu \geq 0~.
\end{eqnarray}

Usually, the violation of NEC may lead to the breakdown of
causality in general relativity and the violation of the second
law of thermodynamics \cite{ArkaniHamed:2007ky}. These pathologies
suggest that the total stress tensor in a physical spacetime
manifold needs to obey the NEC. In the framework of   standard 4D
FRW  cosmology the NEC implies $\rho+p\geq 0$, which in turn gives
rise to the constraint  $w_{u}\geq -1$ on the EoS of the universe
$w_u$, defined as the ratio of pressure to energy density.

In the epochs of universe evolution in which radiation is dominant
the EoS of the universe $w_{u}$ is approximately equal to $1/3$,
while in the matter-dominated period $w_{u}$ is nearly zero, thus
NEC is always satisfied. However, when the DE component is not
negligible, with the NEC being satisfied, we acquire
\begin{eqnarray}\label{NECd}
w_u=w_M\Omega_M+w_{DE}\Omega_{DE}\geq-1~,
\end{eqnarray}
where the subscripts `$M$' and `$DE$' stand for matter and DE,
respectively. With $w_M = 0$, inequality (\ref{NECd}) becomes
\begin{eqnarray}\label{NECde}
w_{DE}\Omega_{DE}\geq-1~.
\end{eqnarray}
From this inequality  we can deduce that it is impossible for a
universe to satisfy NEC if one of its component does not. In this
case, a quintom scenario could be obtained if NEC is violated for
a short period of time during the evolution of the universe.

\subsection{Quintom Bounce}\label{sec:QB}

A bouncing universe with an initial contraction to a non-vanishing
minimal radius and a subsequent expanding phase, provides a
possible solution to the singularity problem of the standard Big
Bang cosmology. For a successful bounce, it can be shown that
within the framework of  standard 4D FRW cosmology with Einstein
gravity, the NEC is violated for a period of time around the
bouncing point. Moreover, for the universe entering into the hot
Big Bang era after the bouncing, the EoS  must transit from $w<-1$
to $w>-1$.

We start with an examination on the necessary conditions required
for a successful bounce  \cite{Starobinsky:1980te,
Mukhanov:1991zn, Peter:2002cn, Mukherji:2002ft, Tsujikawa:2002qc,
Medved:2003fp, Kanti:2003bx, Foffa:2003gt, Finelli:2003mc,
Hovdebo:2003ug, Piao:2003zm, Piao:2003hh, Setare:2004jx,
Biswas:2005qr, Biswas:2006bs, Creminelli:2007aq, Cai:2007qw,
Wei:2007rp, Zhang:2007an, Saridakis:2007cf} (we refer to
\cite{Novello:2008ra} for a review on bounce cosmology). During
the contracting phase  the scale factor $a(t)$ decreases ($\dot{a}
< 0$), while in the expanding phase it increases ($\dot{a}
> 0$). At the bouncing point, $\dot{a}=0$, and around this point
$\ddot{a}
> 0$ for a period of time. Equivalently, in the bouncing cosmology
the hubble parameter H changes from $H < 0$ to $H > 0$, with $H =
0$ at the bouncing point. A successful bounce requires
\begin{eqnarray} \label{bccong}
\dot{H}=-\frac{1}{2M_{p}^{2}}\Big(\rho+P\Big)=-\frac{1}{2M_{p}^{2}}\rho
\Big(1+w\Big)>0~,
\end{eqnarray}
around this point and thus $w<-1$ in its neighborhood. After the
bounce the universe needs to enter into the hot Big Bang era,
otherwise it will reach the Big Rip singularity similarly to the
phantom DE scenario \cite{Caldwell:2003vq}. This requirement leads
the EoS  to transit from $w<-1$ to $w>-1$, which is exactly a
quintom scenario \cite{Cai:2007qw}.

\subsubsection{A phenomenological analysis}\label{sec:QBpheno}

Let us examine the possibility of obtaining the bouncing solution
in a phenomenological quintom models, in which the  EoS is
described by:
\begin{eqnarray}\label{paratachyon}
w(t)=-r-\frac{s}{t^2}~.
\end{eqnarray}
In (\ref{paratachyon}) $r$ and $s$ are parameters and we require
that $r<1$ and $s>0$. One can see from (\ref{paratachyon}) that
$w$ runs from negative infinity at $t=0$ to the cosmological
constant boundary at $t=\sqrt{\frac{s}{1-r}}$ and then it crosses
this boundary. Assuming that the universe is dominated by the
matter with the EoS given by (\ref{paratachyon}), we solve the
Friedmann equation,  obtaining the corresponding evolution of
hubble parameter $H(t)$ and scale factor $a(t)$ as:
\begin{eqnarray}
H(t)&=&\frac{2}{3}\Big[\frac{t}{(1-r) t^2+s}\Big]\\
a(t)&=&\Big[t^2+\frac{s}{1-r}\Big]^{\frac{1}{3 (1-r)}}.
\end{eqnarray}
Here we choose $t=0$ as the bouncing point and we normalize $a=1$
at this point. Thus, our solution provides a universe evolution
with contracting (for $t<0$),   bouncing (at $t=0$) and expanding
(for $t>0$) phases. In Fig. \ref{fig:1parahw} we depict the
evolution of the EoS, the Hubble parameter and the scale factor,
as it arises from numerical elaboration.
 \begin{figure}[tb]
\begin{center}
\includegraphics[
width=6.7in] {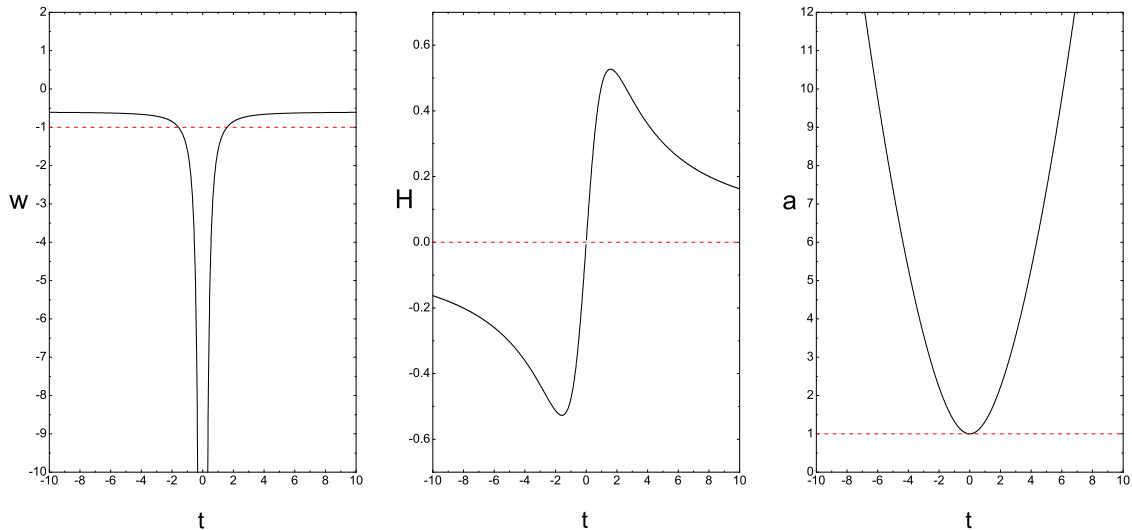} \caption{(Color online) {\it Evolution
of the EoS $w$, the Hubble parameter $H$ and the scale factor $a$
as a function of the cosmic time $t$. For the numerical
elaboration we take $r=0.6$ and $s=1$. From Ref.
\cite{Cai:2007qw}.} } \label{fig:1parahw}
\end{center}
\end{figure}
As we observe, a non-singular bounce is realized  at $t=0$, at a
minimal non-vanishing scale factor $a$, with the Hubble parameter
$H$ running across zero. At the bouncing point $w$ approaches
negative infinity.

\subsubsection{Double-field quintom model}\label{sec:QBdf}

Having presented the bouncing solution with the phenomenological
quintom matter, we now study the bounce in a  two-field quintom
model with the action given by
\begin{eqnarray}
S=\int d^4x \sqrt{-g}
\left[\frac{1}{2}\partial_{\mu}\phi\partial^{\mu}\phi-\frac{1}{2}\partial_{\mu}\sigma\partial_{\mu}\sigma-V(\phi,\sigma)\right],\nonumber
\end{eqnarray}
where   $\phi$ is the canonical and $\sigma$ the phantom field. In
the framework of FRW cosmology  as usual we have  $
\rho=\frac{1}{2}{\dot\phi}^2-\frac{1}{2}{\dot\sigma}^2+V$ and
$p=\frac{1}{2}{\dot\phi}^2-\frac{1}{2}{\dot\sigma}^2-V$ and the
cosmological equations read
\begin{eqnarray}
\label{Einstein1bounce}
&&H^2=\frac{8\pi G}{3}\Big(\frac{1}{2}{\dot\phi}^2-\frac{1}{2}{\dot\sigma}^2+V\Big) \\
\label{Einstein2bounce}
&&\ddot\phi+3H\dot\phi+\frac{dV}{d\phi}=0 \\
\label{Einstein3bounce}
&&\ddot\sigma+3H\dot\sigma-\frac{dV}{d\sigma}=0~.
\end{eqnarray}
Thus, we see that ${\dot\sigma}^2={\dot\phi}^2+2V$ when $H$
crosses zero, and moreover from (\ref{bccong})  we deduce that
${\dot\sigma}^2={\dot\phi}^2$ when $w$ crosses $-1$. These
constraints can be easily satisfied in the parameter space of the
model at hand.

An interesting scenario realized by this model is a universe
originated from a contraction, that has undergone a smooth bounce,
then an inflationary period, and finally it returns to the
standard thermal history. Models of this class have been studied
in Refs. \cite{Piao:2003zm, Cai:2007qw, Cai:2008qb, Cai:2008qw}
and the dynamics of their perturbations have been analyzed in
Refs. \cite{Cai:2007zv, Cai:2008ed, Cai:2009fn, Cai:2009rd,
Cai:2009hc}. In the following we will provide explicitly one
example to illustrate this scenario.

We consider a double-field model in which the potential is only a
function of the field $\phi$ and of Coleman-Weinberg form
\cite{Coleman:1973jx}:
\begin{eqnarray}
V=\frac{1}{4}\lambda\phi^4 \left( \ln\frac{|\phi|}{v}-\frac{1}{4}
\right) + \frac{1}{16}\lambda v^4~,
\end{eqnarray}
which takes its maximum value $\lambda v^4/16$ at $\phi=0$ and
vanishes at the minima where $\phi=\pm v$. Therefore, the scalar
field $\sigma$ affects the evolution only around the bounce but it
decreases  quickly   away from it.

In order to discuss the perturbations explicitly, we first examine
the evolution of the background. In this model a contracting
universe can be driven to reach a minimal size during which the
universe evolves like a matter-dominated one, and then a
quasi-exponential expansion is following. The process to link the
contraction and expansion is a smooth bounce, and the evolution of
the hubble parameter can be treated as a linear function of the
cosmic time approximately.

For the background  initial conditions we assume that $\phi$ lies
at one vacuum (for instance $-v$) when the universe is
contracting, and moreover that $\dot\sigma$ is small enough in
order to be neglected. In this case the field $\phi$ oscillates
around $-v$ leading  the universe EoS to oscillate around $w=0$,
and so this state is in average   similar to a matter-dominated
one. Thus, the useful expressions of the background evolution
write
\begin{eqnarray}\label{relationc1}
a\sim(-\eta)^{2}~,~~{\cal
H}=\frac{2}{\eta}~,~~|\dot\phi|\sim\eta^{-3}~,
\end{eqnarray}
where ${\cal H}\equiv{a'}/{a}$ is the comoving hubble parameter
and the prime denotes the derivative with respect to the comoving
time $\eta$. Moreover, since in the contracting phase the universe
is dominated by the regular field and then we have an
approximation $\phi'^2 \simeq 2({\cal H}^2-{\cal H}')$, we can
obtain another useful relation
\begin{eqnarray}\label{relationc2}
\frac{\phi''}{\phi'}=\frac{2{\cal H}{\cal H}'-{\cal H}''}{2({\cal
H}^2-{\cal H}')}~,
\end{eqnarray}
which will be used to calculate the metric perturbations.

When the universe is contracting, the amplitude of
$\phi$-oscillations increases, while the contribution of the
$\sigma$-field grows rapidly. When the $\phi$-field reaches the
plateau, the bounce starts at the moment $t_{B-}$. During the
bounce  we use the parametrization $H(t)=\alpha(t-t_B)$ around the
bounce point $t_B$, and the coefficient $\alpha$ is a positive
constant determined by numerical elaboration. This parametrization
leads to a scale factor of the form
$a=\frac{a_B}{1-y(\eta-\eta_B)^2/4}$. In the bouncing phase the
kinetic term of $\sigma$ reaches the maximal value, and from the
equation of motion we deduce that
$\ddot\phi/\dot\phi=-3H\dot\sigma^2/(\dot\sigma^2-\frac{\alpha}{4\pi
G})\simeq -3H$ when $\alpha$ is not very large. Finally, we obtain
the approximate relations
\begin{eqnarray}\label{relationb}
{\cal H}\simeq\frac{y}{2}(\eta-\eta_B),~\phi''\simeq-2{\cal
H}\phi',~|\dot\phi|\sim e^{-\frac{3}{4}y(\eta-\eta_B)^2},\ \
\end{eqnarray}
where we have defined $y\equiv8\alpha a_B^2/\pi$.

After the bounce, as the field $\phi$ moves forward slowly along
the plateau, the universe enters into an expanding phase at the
moment $t_{B+}$, and its EoS  is approximately $-1$. Thus, the
universe expands with its scale factor growing almost
exponentially. In this phase we have the well-known relations for
the background evolution
\begin{eqnarray}\label{relationi}
a\sim-\frac{1}{\eta}~,~~H\sim \mathrm{constant}.
\end{eqnarray}
Finally, when the $\phi$-field reaches   the vacuum state $+v$, it
will oscillate again and the EoS of the universe will oscillate
around zero as it happens before the bounce.

In order to present this scenario explicitly, we perform a
numerical elaboration and the results are depicted in Fig.
\ref{fig:background}.
 \begin{figure}[tb]
\begin{center}
\includegraphics[width=4.8in] {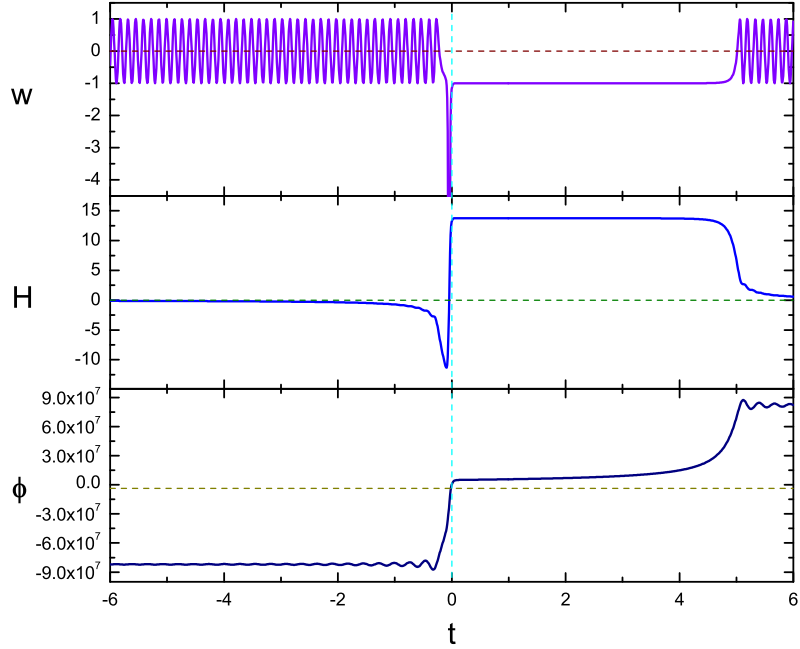} \caption{(Color online) {\it The
evolution of $\phi$-field, of the EoS $w$ and of the hubble
parameter $H$, in the model described in the text. For the
numerical elaboration we choose the parameters
$\lambda=8.0\times10^{-14},~v=0.16M_{p}$, and we impose the
initial conditions $\phi=-0.16M_{p},
~\dot\phi=1.2\times10^{-11}M_{p}^2, ~\sigma=-0.14M_{p},
~\dot\sigma=1.99\times10^{-14}M_{p}^2$. From Ref.
\cite{Cai:2008qb}.} } \label{fig:background}
\end{center}
\end{figure}

Finally, we mention that  if the potential of the $\phi$-field is
not flat enough, the inflationary stage could be very short. An
extreme case of this scenario is that, after a matter-like
contraction, the universe undergoes a smooth bounce and enters a
normal expanding phase directly. This is the so-called ``Matter
Bounce" scenario, which has been intensively studied in the
literature \cite{Starobinsky:1979ty, Wands:1998yp, Finelli:2001sr,
Allen:2004vz, Peter:2008qz}.

\subsection{A cyclic scenario and oscillating universe}\label{sec:cyclic}

The idea of cyclic universe was initially introduced in 1930's by
Richard Tolman  \cite{Tolman:1934}. Since then there have been
various proposals in the literature. The authors of Refs.
 \cite{Khoury:2001wf, Steinhardt:2001vw, Khoury:2001bz,
Buchbinder:2007ad} introduced a cyclic model in high dimensional
string theory with an infinite and flat universe. Within a
modified Friedmann equation the cyclic evolution of the universe
can also be realized  \cite{Brown:2004cs, Baum:2006nz,
Piao:2004hr, Piao:2004me, Piao:2009ku}. In Ref.
\cite{Xiong:2007cn}, it is shown that in the framework of loop
quantum cosmology (LQC) a cyclic universe can be obtained with the
quintom matter. In this section, however, we will study the
solution of oscillating universe in the absence of modifications
of the standard 4D Einstein Gravity, within a flat universe.

To begin with, let us examine in detail the conditions required
for an oscillating solution. The basic picture for the evolution
of the cyclic universe can be shown below:
\begin{equation}\label{btb}
... {\rm bounce} \xrightarrow{expanding} {\rm turn}\texttt{-}{\rm
around} \xrightarrow{contracting} {\rm bounce} ...~.\nonumber
\end{equation}
In the 4D FRW framework the Einstein equations can be written
as:\begin{equation}\label{einstein}
 H^2\equiv \left(\frac{\dot a}{a}\right)^2=\frac{\rho}{3M_p^2}~~~ {\rm and}~~~~
\frac{\ddot a}{a}=-\frac{\rho+3p}{6M_p^2}~,
\end{equation}
where we have defined $M_p^2\equiv\frac{1}{8\pi G}$. $H$ stands
for the Hubble parameter, while $\rho$ and $p$ represent the
energy density and  pressure of the universe respectively. By
definition, for a pivot (bounce or turn-around) process to occur,
one must require that at the pivot point $\dot a=0$ and $\ddot
a>0$ around the bouncing point, while $\ddot a<0$ around the
turn-around point. According to (\ref{einstein}), one obtains
\begin{equation}
\label{conditioncyclic}
\rho=0~,~~p<0\,({\rm or}~p>0)~~{\rm for~the~bounce}~({\rm or}~{\rm
turn}\texttt{-}{\rm around}),
\end{equation}
or equivalently $w\equiv \frac{p}{\rho}\rightarrow -\infty~~({\rm
or}~+\infty)$ at the bounce (or turn-around) point, with the
parameter $w$ being the EoS of the content of the universe. Thus,
when the universe undergoes from bounce to turn-around  $w$
  evolves from $-\infty$ to $+\infty$, while in the
reverse case  $w$ goes from $+\infty$ to $-\infty$. This behavior
 requires   $w$   to cross over the cosmological constant boundary
($w=-1$) in these processes, which interestingly implies the
necessity of the quintom matter in order to achieve  the
realization of the oscillating universe in 4D Einstein Gravity.

An additional interesting feature of an oscillating universe is
that it undergoes accelerations  periodically, avoiding the Big
Rip and Big Crunch, and furthermore we are able to unify inflation
and current acceleration. We mention that the scale factor keeps
increasing from one period to another and thus we are naturally
led to a highly flat universe. This scenario was first proposed in
Ref. \cite{Feng:2004ff}, in which a parameterized quintom model
was used  and   the coincidence problem was argued to be
reconciled.

\subsubsection{A solution of oscillating universe in the double-field quintom
model}

We consider the simplest quintom model consisting of two scalars
with one being quintessence-like and another the phantom-like,
with the usual action
\begin{equation}
S=\int d^{4}x \sqrt{-g}\left [ \frac{1}{2}\partial_{\mu}\phi
\partial^{\mu}\phi-\frac{1}{2}\partial_{\mu}\sigma\partial^{\mu}\sigma-V\left(\phi,\sigma\right)
\right ].\ \ \
\end{equation}
As mentioned in previous sections   the energy density and
pressure are $
\rho=\frac{1}{2}\dot\phi^2-\frac{1}{2}\dot\sigma^2+V(\phi,
\sigma)$ and
$p=\frac{1}{2}\dot\phi^2-\frac{1}{2}\dot\sigma^2-V(\phi, \sigma)
$, and the equations of motion for these two fields write:
$\ddot\phi+3H\dot\phi+V_{,\phi}=0$ and
$\ddot\sigma+3H\dot\sigma-V_{,\sigma}=0$.

Phenomenologically, a general potential form for a renormalizable
model includes operators with dimension 4 or less, including
various powers of the scalar fields. We impose a $Z_2$ symmetry,
that is the potential remains invariant under the simultaneous
transformations $\phi\rightarrow-\phi$ and
$\sigma\rightarrow-\sigma$. Thus, the potential of the model
writes:
\begin{equation}
V(\phi,\sigma)=V_0+\frac{1}{2}m_1^2\phi^2+\frac{1}{2}
m_2^2\sigma^2+\gamma_1\phi^4+\gamma_2\sigma^4
+g_1\phi\sigma+g_2\phi\sigma^3+g_3\phi^2\sigma^2+g_4\phi^3\sigma.
\end{equation}

From condition (\ref{conditioncyclic})  we can see that at both
the bounce and the turn-around point $\dot\sigma^2=\dot\phi^2+2V$.
Similarly, we deduce that $p=-2V$. When the universe undergoes a
bounce the pressure is required to be negative, which implies the
potential to be positive. However, when a turn-around takes place,
the pressure of the universe is required to be positive, and
consequently the potential must be negative. Therefore, the
potential must contain a negative term in order to give rise to an
oscillating scenario.

In the model at hand the two scalar fields dominate the universe
alternately, since the evolution from the bounce to the
turn-around requires the transition from the phantom-dominated
phase into the quintessence-dominated one, or vice versa. However,
as pointed out in \cite{Guo:2004fq, Zhang:2005eg}  this process
does not happen if the two fields are decoupled, and thus
interaction between the two fields is  crucial. For a detailed
quantitative study we consider $$
V\left(\phi,\sigma\right)=\left(\Lambda_0+
\lambda\phi\sigma\right)^{2}+\frac{1}{2}m^{2}\phi^{2}-
\frac{1}{2}m^{2}\sigma^{2}~.$$ This potential acquires a negative
value when $\phi$ is near the origin and $\sigma$ large. However,
due to the interaction  the potential is still bounded from below
and it is positive definite when the fields are both away from
zero. This potential allows for the analytic solution
\begin{equation}\label{ansatz}
\phi=\sqrt{A_{0}}\cos mt~,~~\sigma=\sqrt{A_{0}}\sin mt~
\end{equation}
with $\lambda=\frac{\sqrt{3}m}{2M_p}$, and with the parameter
$A_{0}$ describing the oscillation amplitude. In Fig.
\ref{draftos} we depict the evolution  of the energy density and
scale factor.
 \begin{figure}[tb]
\begin{center}
\includegraphics[
width=4.8in] {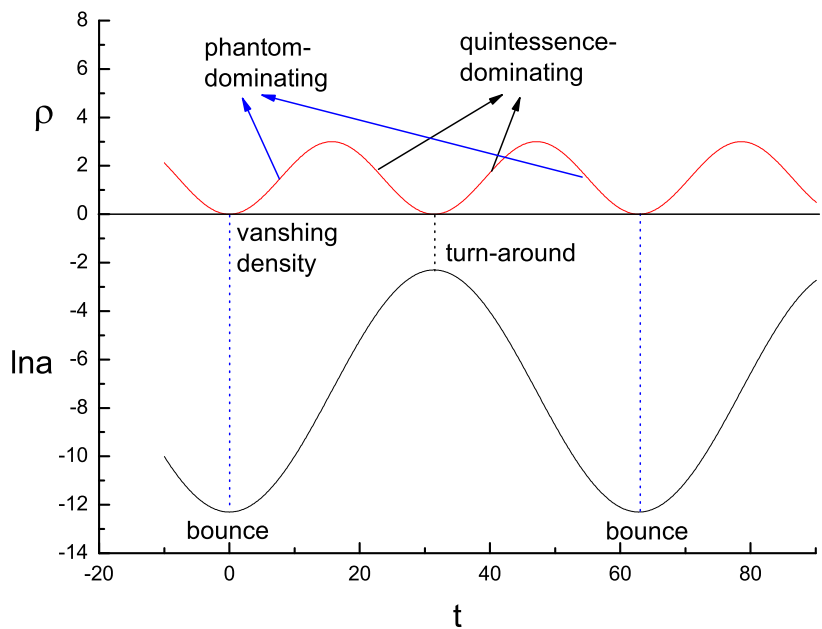} \caption{(Color online) {\it Evolution
of the energy density and scale factor in a cyclic quintom
universe. For each cycle the quintessence-like and phantom-like
components dominate alternately. From Ref. \cite{Xiong:2008ic}.} }
\label{draftos}
\end{center}
\end{figure}

\subsubsection{Classifications of the solutions}

Let us study the detailed cosmological evolutions of the cyclic
quintom model. First of all, from (\ref{einstein}) one acquires
the Hubble parameter
\begin{equation}\label{H}
H=\frac{\sqrt{3}}{3M_p}\left(\Lambda_0+\Lambda_1\sin{2mt}\right)~,
\end{equation}
where  $\Lambda_1\equiv\frac{\sqrt{3}m}{4M_p}A_0$. For different
parameters, the evolution of the universe can be classified into
five cases. Taking $M_p=1$ we have the following:

Case (\uppercase\expandafter{\romannumeral1}): $\Lambda_0=0$. In
this case the Hubble parameter is given by
$H=\frac{mA_{0}}{4M_{p}^{2}}\sin{2mt}$. Thus, the scale factor is
 \begin{equation}
\ln{a}\propto \cos{2mt},
 \end{equation} and therefore   there is
no spacetime singularity. In Fig. \ref{fig1os} we depict the
evolution  of the scale factor, the Hubble parameter, the energy
density and the EoS.
 \begin{figure}[tb]
\begin{center}
\includegraphics[
width=5.2in] {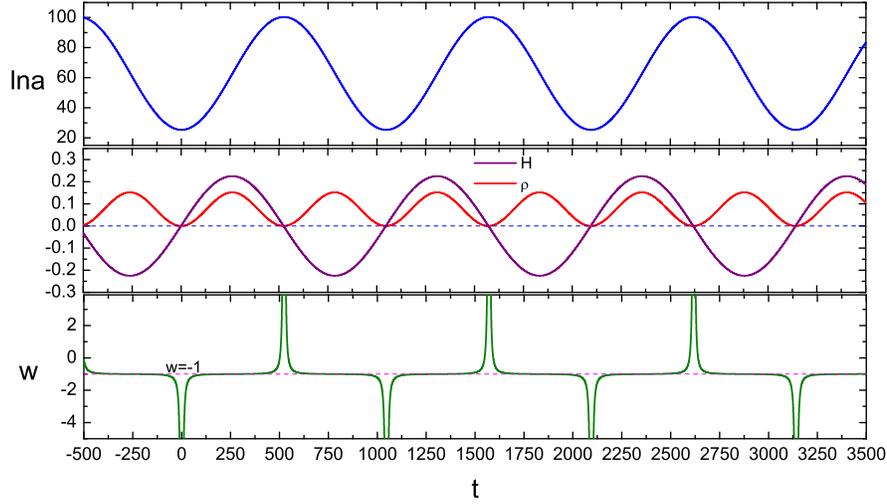} \caption{(Color online) {\it  Evolution
for Case (I). This figure presents an exactly cyclic universe. The
scale factor   oscillates between the minimal and maximal value.
In the numerical elaboration we take $m=3\times 10^{-3}$ and
$A_{0}=300$ or equivalently $\Lambda_1\simeq 0.39$. The Hubble
parameter is depicted by the purple line, while the energy density
by the red line. From Ref.  \cite{Xiong:2008ic}.} } \label{fig1os}
\end{center}
\end{figure}

Case (\uppercase\expandafter{\romannumeral2}):
$0<\Lambda_0<\Lambda_1$. In this case the scale factor is
\begin{equation}
\ln{a}\propto C_{1}t+C_{2}\cos{2mt} \label{a}~,
\end{equation}
where $C_{1}=\Lambda_0/\sqrt{3}M_{p}$, $C_{2}=-A_{0}/8M_{p}^{2}$.
 This solution
also describes a cyclic universe, but both the minimal and maximal
values of the scale factor increase cycle by cycle. Therefore, the
average size of the universe is   growing up gradually without Big
Crunch or Big Rip singularities, although its scale factor
experiences contractions and expansions alternately. Finally, the
backward time evolution cannot lead to a shrinking scale factor
either. These features are presented in Fig. {\ref{fig2os}}.
 \begin{figure}[tb]
\begin{center}
\includegraphics[
width=5.2in] {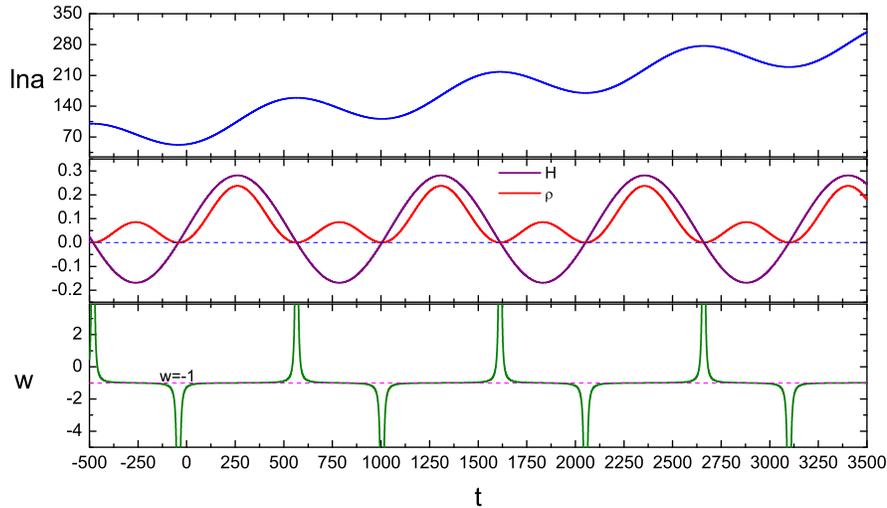} \caption{(Color online) {\it  Evolution
for Case (II).  $m$ and $A_0$ are the same as those used in Fig.
\ref{fig1os}  and $\Lambda_0=0.10$.   From Ref.
\cite{Xiong:2008ic}.} } \label{fig2os}
\end{center}
\end{figure}
 The graph for energy density and Hubble
parameter is similar to Fig.\ref{fig1os}, but the oscillating
amplitude of the energy density in the expanding phase is larger
than that os the contracting one.

Case (\uppercase\expandafter{\romannumeral3}):
$\Lambda_0\geq\Lambda_1$. The solution for the scale factor is the
same as  Case (\uppercase\expandafter{\romannumeral2}),
 however
the evolution of the universe is different. The universe lies in
the expanding period forever and there is no contracting phase,
that is the accelerating expansion   is periodical, without any
bounce or turn-around. Since the EoS $w$ is oscillating around
``$-1$", the Big Rip can be avoided and    the energy density and
Hubble parameter are always positive. Furthermore, for reasonable
parameters one can unify the inflationary period and the late-time
acceleration.
 This behavior is shown
 in Fig. \ref{fig3os}.
  \begin{figure}[tb]
\begin{center}
\includegraphics[
width=5.in] {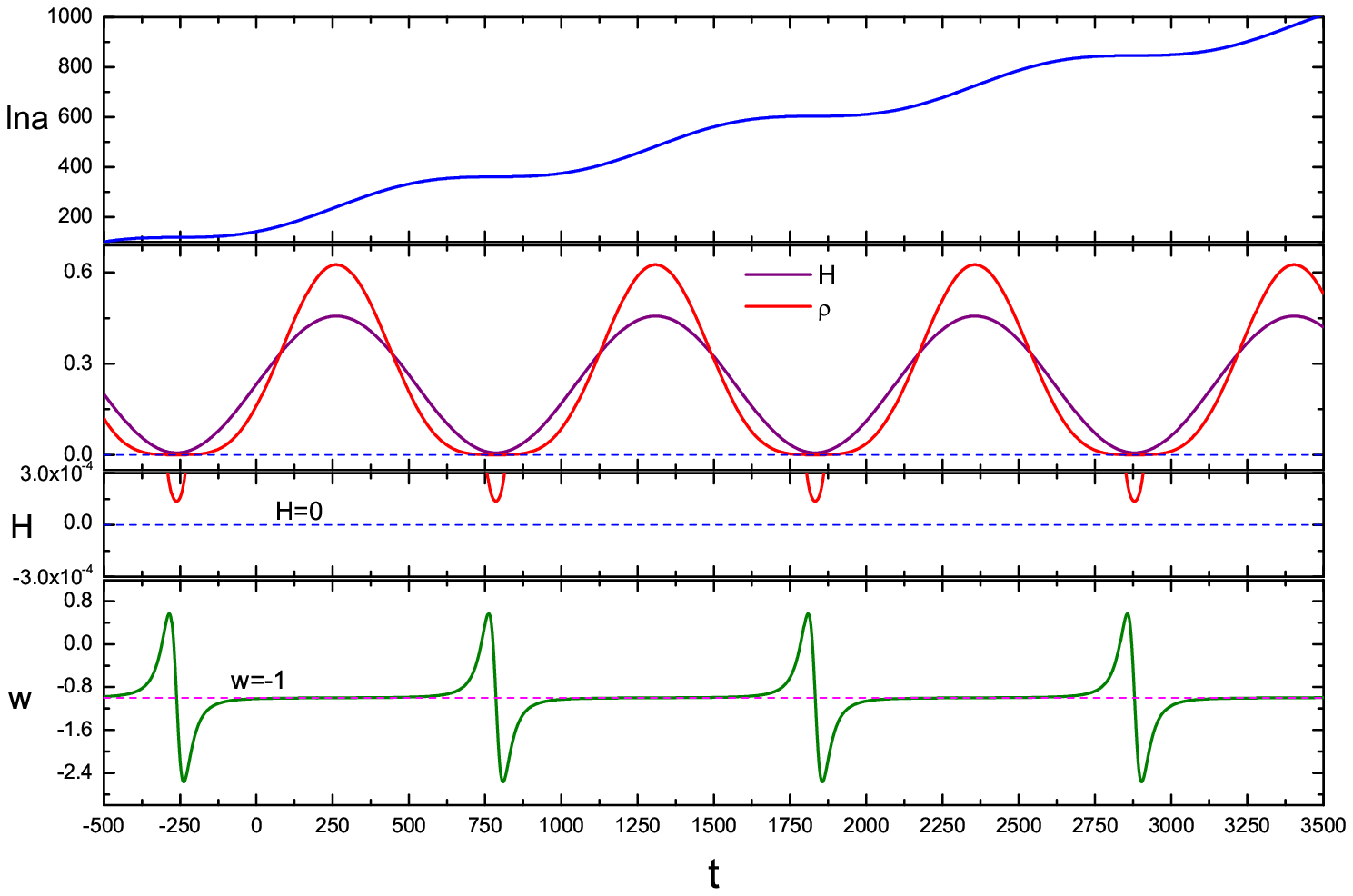} \caption{(Color online) {\it  Evolution
for Case (III). The parameters $m$ and $A_0$ are taken the same as
in Fig. \ref{fig1os}, and $\Lambda_0=0.402$. From Ref.
\cite{Xiong:2008ic}.} } \label{fig3os}
\end{center}
\end{figure}

Case (\uppercase\expandafter{\romannumeral4}):
$-\big|\Lambda_{1}\big| <\Lambda_0<0$. This case corresponds to a
cyclic universe with decreasing minimal and maximal scale factor
for each epoch. Contrary to Case
(\uppercase\expandafter{\romannumeral2}), the total tendency is
that the scale factor decreases cycle by cycle, since the
contracting phase is longer than the expanding phase, but it never
reaches zero. These features are depicted in Fig. \ref{fig4os}
  \begin{figure}[tb]
\begin{center}
\includegraphics[
width=5.3in] {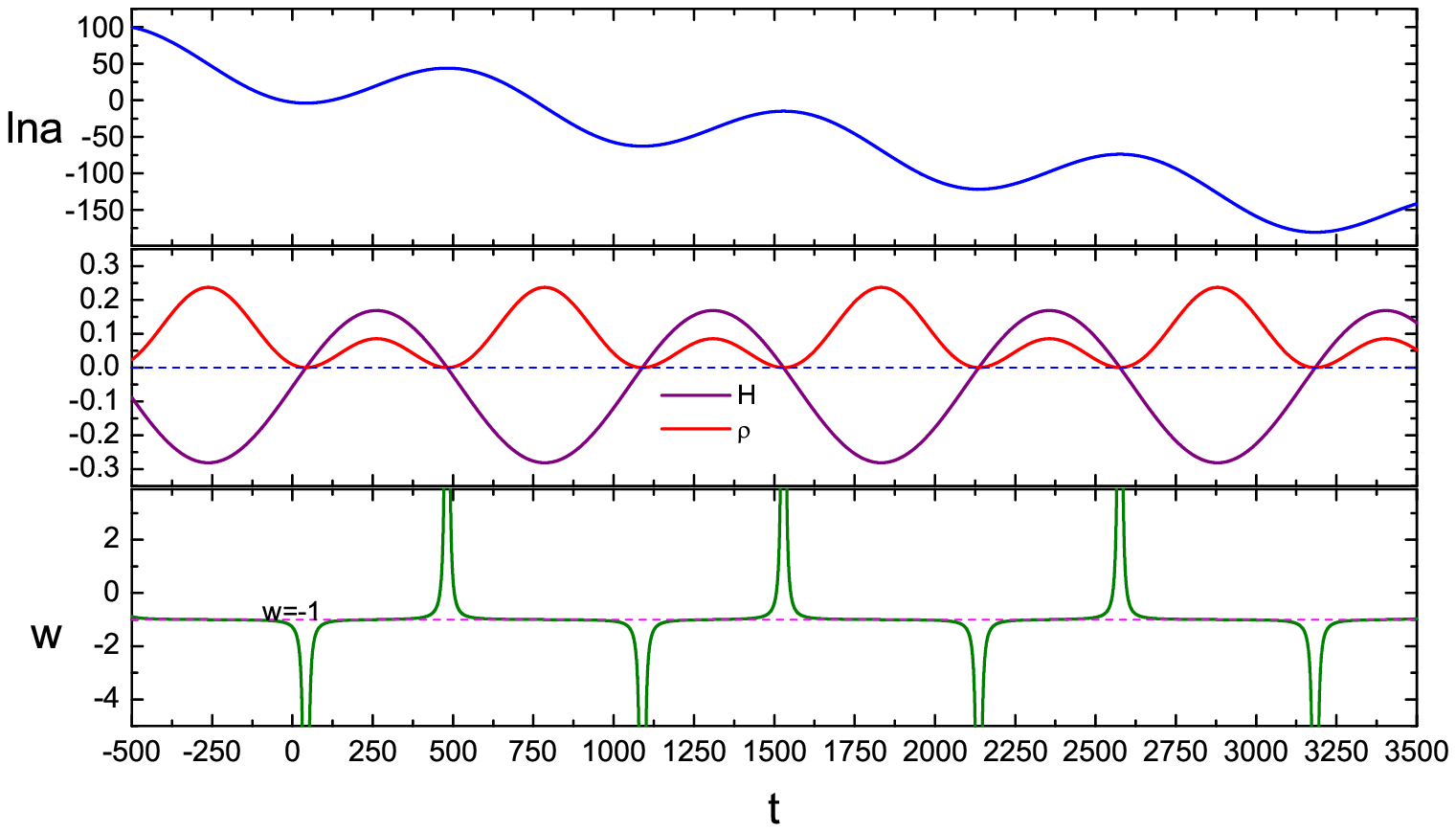} \caption{(Color online) {\it Evolution
for Case (IV). The values of $m$ and $A_0$ are  the same with
Fig.\ref{fig1os}, and $\Lambda_0=-0.10$.  From Ref.
\cite{Xiong:2008ic}.} } \label{fig4os}
\end{center}
\end{figure}

Case (\uppercase\expandafter{\romannumeral5}):
$\Lambda_0<-\big|\Lambda_{1}\big|$. This case describes an
enterally  contracting universe, which corresponds to the reverse
of Case (III), amd thus it is unphysical.

Having presented the possible solutions, we make some final
comments. One may notice that the EoS of the model oscillates
around the cosmological constant boundary, and every time it
approaches $w=-1$ it remains close to that for some time.
Correspondingly, the Hubble parameter evolves near its maximal
value and makes the scale factor expand exponentially. This period
corresponds to an inflationary stage after the bounce. It is a
significant process since it dilutes the relics created in the
last cycle. Furthermore,  entropy can be diluted too, providing a
solution to the problem of enteral entropy increase in cyclic
cosmology. Additionally, some of the primordial perturbations are
able to exit the horizon  and re-enter the horizon when the
inflationary stage ceases. When these perturbations re-enter they
lead to new structures in the next cycle.

Lastly,  let us constrain the model parameters ($m, \Lambda_0,
A_{0}$).  First of all, we require the  oscillation period of the
  Hubble parameter to be no less than 2 times the age of our
universe which is of the order of the present Hubble time. Thus,
from   (\ref{H}) we obtain the period $T=\frac{\pi}{m}\sim {\cal
O}({H_0}^{-1})$, where ${H_0}^{-1}\simeq 10^{60} M_p^{-1}$, and
therefore $m\sim {\cal O}(10^{-60})M_p$. Secondly we require the
maximal value of the Hubble parameter to be able to reach the
inflationary energy scale. The maximal value of $H$ is given by
($\Lambda_0+\Lambda_1$), when ($\sin 2mt$) reaches its maximum.
For case (I), (III) and (IV), $|\Lambda_0|\leq\Lambda_1$, so we
have $(\Lambda_0+\Lambda_1)\sim {\cal O}(\frac{m A_0}{{M_p}^2})$.
If we consider inflation with energy density, for example around
${\cal O}(10^{-20}){M_p}^4$, we find that $A_0$ must be of ${\cal
O}(10^{50}){M_p}^2$. Such a large value of $A_0$ indicates that
the scalar fields $\phi$ and $\sigma$ take values   above the
Planck scale, which makes our effective lagrangian description
invalid. One possibility of solving this problem is to consider a
model with a large number of quintom fields. In this case the
Hubble parameter is amplified by a pre-factor $\sqrt{N}$ with $N$
the number of quintom fields. If $N$ is larger than   $10^{50}$,
$A_0$ can be relaxed to   ${\cal O}(1)M_p^2$ or less.



\section{Concluding remarks}\label{sec:conclu}

Since discovered in 1998, the nature of DE has become one of the
most intriguing puzzles of modern physics and it has been widely
investigated. The simplest candidate of DE is a cosmological
constant, but it suffers from the well-known fine-tuning and
coincidence problems. Alternatively, dynamical DE models have been
proposed, such as quintessence, phantom, and k-essence. Since at
present we know very little about the theoretical aspects of DE,
the cosmological observations play a crucial role in enlightening
our picture. In particular, since astronomical observations have
shown a mild preference for an EoS of DE smaller than $-1$ at
present, a phantom model appeared in the literature.

In these lines, with accumulated observational data  a scenario of
quintom, with DE EoS crossing $-1$ during evolution, has been
constructed in the literature. If such a class of dynamical DE
scenario were verified by future observations, it would be a
challenge to the model-building of DE. This is because of  a No-Go
theorem, which forbids dynamical models with a single scalar field
to lead to an EoS crossing over the cosmological constant boundary
in the frame of Einstein's gravity. The current work aims at
presenting a review of successful examples of quintom models and
the corresponding observational consequences.

We studied in detail the simplest quintom model, which involves
two scalar fields, with one being quintessence and the other being
phantom. We have shown that  these models possess stable critical
points in the phase space, and in addition certain dualities
appear, with form-invariant cosmological equations.

As a consistent consideration  we  analyzed the behavior of
quintom perturbations. In the example of the simplest quintom
model there are two degrees of freedom in its perturbations, which
are finite and stable. It should be noticed that since the quintom
scenario  usually involves extra degrees of freedom, additional
relative pressure perturbations are inevitably produced in order
to seed entropy perturbations.

We also provided alternative approaches to realize the quintom
scenario in a scalar field system within standard Einstein's
gravity, in which higher derivative terms were introduced.
Specifically, we  constructed a Lee-Wick and a string inspired
quintom model. For the latter, the perturbation propagations
 are quite different than usual, with a time-dependent sound
speed due to the non-canonical kinetic term.

Usually, a quintom model constructed by a double-scalar system and
a single scale with higher derivatives, suffers from the problem
of ambiguous quantum behavior inherited from the DE phantom model.
This feature offered a motivation to study the possibilities of
realizing a quintom scenario without a ghost mode. An attempt on
solving this problem is to construct a string-inspired quintom
model, but it is still unclear how to be quantized since it
involves highly non-linear terms.

Furthermore, we considered the model-building  in the framework of
a non-scalar system. A model of spinor quintom is able to evade
the drawbacks of considering a phantom field, since its kinetic
term is well defined, and the sign-change of its potential
derivative  with respect to the scalar bilinear $\bar\psi\psi$ can
realize the $-1$-crossing. However, it is non-renormalizable if we
consider higher order effects at the quantum level.

Another interesting issue is the investigation of a quintom
scenario in  braneworld cosmology, which is motivated by higher
dimensional constructions, closer to fundamental theories.
Braneworld models were studied actively in recent years, offering
a new path to describe DE. In this review we presented the
possibility of realizing a quintom scenario in DGP braneworld
model. In such a construction, solutions of late-time
super-acceleration can be obtained when matter components are
driven by a manifestation of the excruciatingly slow leakage of
gravity into the extra dimension. In particular, we showed that
for a canonical field living on the brane the $-1$-crossing takes
place from above to below, while in the phantom case it takes
place from below to above.

Concerning the theoretical implications of quintom cosmology, an
important issue is the violation of energy conditions. Such
conditions usually play a crucial role in classical theory of
general relativity. An important predication of general relativity
is that a singularity cannot be avoided under assumptions of
certain energy conditions, which was originally proved by Penrose
and Hawking and later developed into cosmological framework by
Borde, Vilenkin and Guth \cite{Borde:2001nh}. Therefore, in a
quintom scenario with NEC violation, one may obtain a bouncing
solution at early times of the universe. This may provide a
possible solution to the singularity problem of standard Big Bang
cosmology. Specifically, we  studied bouncing cosmology realized
by a model of double-field quintom. To extend, we also provided a
scenario of oscillating universe, in which it experiences
expansions and contractions periodically.

As an end, we would like to comment on some unsettled issues in
quintom cosmology. Although the  quintom paradigm  has been
studied intensively in recent years, its nature is still unclear.
In the following, we list three main questions, which are crucial
to the developments of quintom cosmology.
\begin{itemize}
\item (a) how to combine quintom cosmology with the well-known
particle physics or fundamental theories?
\item (b) how to construct a ghost-free quintom scenario with
robust quantum behavior?
\item (c) why the DE EoS crosses the cosmological constant
boundary around today (the so-called second coincidence problem)?
\end{itemize}
Recently, physicists have made many attempts on partially
addressing the above questions in quintom cosmology. For example,
a quintom scenario can be realized by a rolling tachyon in the
cubic string field theory
\cite{Aref'eva:2006et,Aref'eva:2005fu,Vernov:2006dm} due to a
non-local effect \cite{Aref'eva:2003qu}. Interestingly, motivated
by studies on thermodynamics of a cosmological structure
\cite{Cohen:1998zx, Horava:2000tb, Thomas:2002pq, Hsu:2004ri,
Li:2004rb, Huang:2004wt, Wang:2004nqa}, a cosmic holographic bound
(originally suggested by \cite{'tHooft:1993gx, Susskind:1994vu}
and comprehensively reviewed in
 \cite{Bousso:2002ju}) on dark energy dynamics allows a period of
quintom scenario with NEC violation \cite{Elizalde:2005ju,
Izquierdo:2005ku, Wang:2005jx, Zhang:2005hs, Feng:2008kz,
Feng:2009ai}.

However, the three aforementioned questions remain open, requiring
to be faced by the coming theoretical studies on quintom
cosmology. In the meantime, cosmological observations of higher
accuracy are needed in order to determine whether the DE EoS has
crossed the cosmological constant boundary. If this is indeed
true, cosmology will enter in a very interesting and challenging
era.


%


\section*{Acknowledgement}
 We are grateful to Hao Wei, Xin Zhang, Xinmin Zhang, Yang Zhang,
Wen Zhao for giving us very useful comments on the manuscript. We
also thank Hong Li, Mingzhe Li, Jian-Xin Lu, Yun-Song Piao, Taotao
Qiu, Jing Wang, Hua-Hui Xiong, Hong-Sheng Zhang, Xiao-Fei Zhang,
Gongbo Zhao, and Zong-Hong Zhu for permissions to include figures
from their works.

Part of the numerical elaboration was performed on the MagicCube
of Shanghai Supercomputer Center (SSC).  E.N.S wishes to thank
Institut de Physique Th\'eorique, CEA,  for the hospitality during
the preparation of the present work. The work of M.R.S. is
 financially supported by Research Institute for Astronomy and
Astrophysics of Maragha, Iran. The researches of Y.F.C. and J.Q.X.
are partly supported   by National Science Foundation of China
under Grant Nos. 10803001, 10533010, 10821063 and 10675136, and
the 973 program No. 2007CB815401, and by the Chinese Academy of
Science under Grant No. KJCX3-SYW-N2.

\bibliographystyle{elsart-num}

\begin{thebibliography}{99}


\bibitem{Riess:1998cb}
  A.~G.~Riess {\it et al.}  [Supernova Search Team Collaboration],
  Astron.\ J.\  {\bf 116}, 1009 (1998)
  [arXiv:astro-ph/9805201].

\bibitem{Perlmutter:1998np}
  S.~Perlmutter {\it et al.}  [Supernova Cosmology Project Collaboration],
  Astrophys.\ J.\  {\bf 517}, 565 (1999)
  [arXiv:astro-ph/9812133].

\bibitem{Weinberg:1988cp}
  S.~Weinberg,
  Rev.\ Mod.\ Phys.\  {\bf 61}, 1 (1989).

\bibitem{Carroll:1991mt}
  S.~M.~Carroll, W.~H.~Press and E.~L.~Turner,
  Ann.\ Rev.\ Astron.\ Astrophys.\  {\bf 30}, 499 (1992).

\bibitem{Krauss:1995yb}
  L.~M.~Krauss and M.~S.~Turner,
  Gen.\ Rel.\ Grav.\  {\bf 27}, 1137 (1995)
  [arXiv:astro-ph/9504003].

\bibitem{Huey:1998se}
  G.~Huey, L.~M.~Wang, R.~Dave, R.~R.~Caldwell and P.~J.~Steinhardt,
  Phys.\ Rev.\  D {\bf 59}, 063005 (1999)
  [arXiv:astro-ph/9804285].

\bibitem{Peebles:2002gy}
 P.~J.~E.~Peebles and B.~Ratra,
 Rev.\ Mod.\ Phys.\  {\bf 75}, 559 (2003)
 [arXiv:astro-ph/0207347].

\bibitem{Padmanabhan:2002ji}
 T.~Padmanabhan,
 Phys.\ Rept.\  {\bf 380}, 235 (2003)
 [arXiv:hep-th/0212290].

\bibitem{Copeland:2006wr}
 E.~J.~Copeland, M.~Sami and S.~Tsujikawa,
 Int.\ J.\ Mod.\ Phys.\  D {\bf 15}, 1753 (2006)
 [arXiv:hep-th/0603057].

\bibitem{Albrecht:2006um}
 A.~J.~Albrecht {\it et al.},
 arXiv:astro-ph/0609591.

\bibitem{Linder:2008pp}
 E.~V.~Linder,
 Rept.\ Prog.\ Phys.\  {\bf 71}, 056901 (2008)
 [arXiv:0801.2968 [astro-ph]].

\bibitem{Frieman:2008sn}
 J.~Frieman, M.~Turner and D.~Huterer,
 Ann.\ Rev.\ Astron.\ Astrophys.\  {\bf 46}, 385 (2008)
 [arXiv:0803.0982 [astro-ph]].

\bibitem{Caldwell:2009ix}
 R.~R.~Caldwell and M.~Kamionkowski,
 Ann.\ Rev.\ Nucl.\ Part.\ Sci.\  {\bf 59}, 397 (2009)
 [arXiv:0903.0866 [astro-ph.CO]].

\bibitem{Silvestri:2009hh}
 A.~Silvestri and M.~Trodden,
 Rept.\ Prog.\ Phys.\  {\bf 72}, 096901 (2009)
 [arXiv:0904.0024 [astro-ph.CO]].

\bibitem{Ratra:1987rm}
  B.~Ratra and P.~J.~E.~Peebles,
  Phys.\ Rev.\  D {\bf 37}, 3406 (1988).

\bibitem{Wetterich:1987fm}
  C.~Wetterich,
  Nucl.\ Phys.\  B {\bf 302}, 668 (1988).

\bibitem{Caldwell:1999ew}
  R.~R.~Caldwell,
  Phys.\ Lett.\  B {\bf 545}, 23 (2002)
  [arXiv:astro-ph/9908168].

\bibitem{Caldwell:2003vq}
  R.~R.~Caldwell, M.~Kamionkowski and N.~N.~Weinberg,
  Phys.\ Rev.\ Lett.\  {\bf 91}, 071301 (2003)
  [arXiv:astro-ph/0302506].

\bibitem{Feng:2004ad}
  B.~Feng, X.~L.~Wang and X.~M.~Zhang,
  Phys.\ Lett.\  B {\bf 607}, 35 (2005)
  [arXiv:astro-ph/0404224].

\bibitem{Huterer:2004ch}
  D.~Huterer and A.~Cooray,
  Phys.\ Rev.\  D {\bf 71}, 023506 (2005)
  [arXiv:astro-ph/0404062].




\bibitem{Xia:2007km}
  J.~Q.~Xia, Y.~F.~Cai, T.~T.~Qiu, G.~B.~Zhao and X.~Zhang,
  Int.\ J.\ Mod.\ Phys.\  D {\bf 17}, 1229 (2008)
  [arXiv:astro-ph/0703202].

\bibitem{Vikman:2004dc}
  A.~Vikman,
  Phys.\ Rev.\  D {\bf 71}, 023515 (2005)
  [arXiv:astro-ph/0407107].

\bibitem{Hu:2004kh}
  W.~Hu,
  Phys.\ Rev.\  D {\bf 71}, 047301 (2005)
  [arXiv:astro-ph/0410680].

\bibitem{Caldwell:2005ai}
  R.~R.~Caldwell and M.~Doran,
  Phys.\ Rev.\  D {\bf 72}, 043527 (2005)
  [arXiv:astro-ph/0501104].

\bibitem{Zhao:2005vj}
  G.~B.~Zhao, J.~Q.~Xia, M.~Li, B.~Feng and X.~Zhang,
  Phys.\ Rev.\  D {\bf 72}, 123515 (2005)
  [arXiv:astro-ph/0507482].

\bibitem{Kunz:2006wc}
  M.~Kunz and D.~Sapone,
  Phys.\ Rev.\  D {\bf 74}, 123503 (2006)
  [arXiv:astro-ph/0609040].



\bibitem{Perlmutter:2003kf}
  S.~Perlmutter and B.~P.~Schmidt,
  in {\it{ Supernovae and Gamma-Ray
Bursters}}, Lecture Notes in Physics, vol. 598, 195-217 (2003)
  [arXiv:astro-ph/0303428].

\bibitem{Riess:2004nr}
  A.~G.~Riess {\it et al.}  [Supernova Search Team Collaboration],
  Astrophys.\ J.\  {\bf 607}, 665 (2004)
  [arXiv:astro-ph/0402512].

\bibitem{Astier:2005qq}
  P.~Astier {\it et al.}  [The SNLS Collaboration],
  Astron.\ Astrophys.\  {\bf 447}, 31 (2006)
  [arXiv:astro-ph/0510447].

\bibitem{Riess:2006fw}
  A.~G.~Riess {\it et al.},
  Astrophys.\ J.\  {\bf 659}, 98 (2007)
  [arXiv:astro-ph/0611572].

\bibitem{Miknaitis:2007jd}
  G.~Miknaitis {\it et al.},
  Astrophys.\ J.\  {\bf 666}, 674 (2007)
  [arXiv:astro-ph/0701043].

\bibitem{Kowalski:2008ez}
  M.~Kowalski {\it et al.}  [Supernova Cosmology Project Collaboration],
  Astrophys.\ J.\  {\bf 686}, 749 (2008)
  [arXiv:0804.4142 [astro-ph]].

\bibitem{Hicken:2009dk}
  M.~Hicken {\it et al.},
  Astrophys.\ J.\  {\bf 700}, 1097 (2009)
  [arXiv:0901.4804 [astro-ph.CO]].

\bibitem{Komatsu:2008hk}
  E.~Komatsu {\it et al.}  [WMAP Collaboration],
  Astrophys.\ J.\ Suppl.\  {\bf 180}, 330 (2009)
  [arXiv:0803.0547 [astro-ph]].

\bibitem{Tegmark:2003ud}
  M.~Tegmark {\it et al.}  [SDSS Collaboration],
  Phys.\ Rev.\  D {\bf 69}, 103501 (2004)
  [arXiv:astro-ph/0310723].

\bibitem{Tegmark:2006az}
  M.~Tegmark {\it et al.}  [SDSS Collaboration],
  Phys.\ Rev.\  D {\bf 74}, 123507 (2006)
  [arXiv:astro-ph/0608632].

\bibitem{Seljak:2004xh}
  U.~Seljak {\it et al.}  [SDSS Collaboration],
  Phys.\ Rev.\  D {\bf 71}, 103515 (2005)
  [arXiv:astro-ph/0407372].

\bibitem{Freedman:2000cf}
  W.~L.~Freedman {\it et al.}  [HST Collaboration],
  Astrophys.\ J.\  {\bf 553}, 47 (2001)
  [arXiv:astro-ph/0012376].

\bibitem{Jimenez:1996at}
  R.~Jimenez, P.~Thejll, U.~Jorgensen, J.~MacDonald and B.~Pagel,
  Mon.\ Not.\ Roy.\ Astron.\ Soc.\ {\bf 282}, 926 (1996)
  [arXiv:astro-ph/9602132].

\bibitem{Richer:2002tg}
  H.~B.~Richer {\it et al.},
  Astrophys.\ J.\  {\bf 574}, L151 (2002)
  [arXiv:astro-ph/0205086].

\bibitem{Hansen:2002ij}
  B.~M.~S.~Hansen {\it et al.},
  Astrophys.\ J.\  {\bf 574}, L155 (2002)
  [arXiv:astro-ph/0205087].

\bibitem{Krauss:2003em}
  L.~M.~Krauss and B.~Chaboyer,
  Science {\bf 299}, 65 (2003).




\bibitem{Weinberg:2000yb}
  S.~Weinberg,
  arXiv:astro-ph/0005265.

\bibitem{Bousso:2000xa}
  R.~Bousso and J.~Polchinski,
  JHEP {\bf 0006}, 006 (2000)
  [arXiv:hep-th/0004134].

\bibitem{Kachru:2003aw}
  S.~Kachru, R.~Kallosh, A.~Linde and S.~P.~Trivedi,
  Phys.\ Rev.\  D {\bf 68}, 046005 (2003)
  [arXiv:hep-th/0301240].

\bibitem{Weinberg:1987dv}
  S.~Weinberg,
  Phys.\ Rev.\ Lett.\  {\bf 59}, 2607 (1987).




\bibitem{Peebles:1987ek}
  P.~J.~E.~Peebles and B.~Ratra,
  Astrophys.\ J.\  {\bf 325}, L17 (1988).

\bibitem{Wetterich:1994bg}
  C.~Wetterich,
  Astron.\ Astrophys.\  {\bf 301}, 321 (1995)
  [arXiv:hep-th/9408025].

\bibitem{Zlatev:1998tr}
  I.~Zlatev, L.~M.~Wang and P.~J.~Steinhardt,
  Phys.\ Rev.\ Lett.\  {\bf 82}, 896 (1999)
  [arXiv:astro-ph/9807002].


\bibitem{Carroll:2003st}
  S.~M.~Carroll, M.~Hoffman and M.~Trodden,
  Phys.\ Rev.\  D {\bf 68}, 023509 (2003)
  [arXiv:astro-ph/0301273].

\bibitem{Cline:2003gs}
  J.~M.~Cline, S.~Jeon and G.~D.~Moore,
  Phys.\ Rev.\  D {\bf 70}, 043543 (2004)
  [arXiv:hep-ph/0311312].


\bibitem{Chiba:1999ka}
  T.~Chiba, T.~Okabe and M.~Yamaguchi,
  Phys.\ Rev.\  D {\bf 62}, 023511 (2000)
  [arXiv:astro-ph/9912463].

\bibitem{ArmendarizPicon:2000dh}
  C.~Armendariz-Picon, V.~F.~Mukhanov and P.~J.~Steinhardt,
  Phys.\ Rev.\ Lett.\  {\bf 85}, 4438 (2000)
  [arXiv:astro-ph/0004134].

\bibitem{ArmendarizPicon:2000ah}
  C.~Armendariz-Picon, V.~F.~Mukhanov and P.~J.~Steinhardt,
  Phys.\ Rev.\  D {\bf 63}, 103510 (2001)
  [arXiv:astro-ph/0006373].

\bibitem{ArmendarizPicon:1999rj}
  C.~Armendariz-Picon, T.~Damour and V.~F.~Mukhanov,
  Phys.\ Lett.\  B {\bf 458}, 209 (1999)
  [arXiv:hep-th/9904075].

\bibitem{Garriga:1999vw}
  J.~Garriga and V.~F.~Mukhanov,
  Phys.\ Lett.\  B {\bf 458}, 219 (1999)
  [arXiv:hep-th/9904176].

\bibitem{Steinhardt:1999nw}
  P.~J.~Steinhardt, L.~M.~Wang and I.~Zlatev,
  Phys.\ Rev.\  D {\bf 59}, 123504 (1999)
  [arXiv:astro-ph/9812313].




\bibitem{Huterer:2002hy}
  D.~Huterer and G.~Starkman,
  Phys.\ Rev.\ Lett.\  {\bf 90}, 031301 (2003)
  [arXiv:astro-ph/0207517].


\bibitem{Wang:2004py}
  Y.~Wang and M.~Tegmark,
  Phys.\ Rev.\ Lett.\  {\bf 92}, 241302 (2004)
  [arXiv:astro-ph/0403292].

\bibitem{Alam:2004jy}
  U.~Alam, V.~Sahni and A.~A.~Starobinsky,
  JCAP {\bf 0406}, 008 (2004)
  [arXiv:astro-ph/0403687].

\bibitem{Wang:2003gz}
  Y.~Wang and P.~Mukherjee,
  Astrophys.\ J.\  {\bf 606}, 654 (2004)
  [arXiv:astro-ph/0312192].

\bibitem{Alam:2003fg}
  U.~Alam, V.~Sahni, T.~D.~Saini and A.~A.~Starobinsky,
  Mon.\ Not.\ Roy.\ Astron.\ Soc.\  {\bf 354}, 275 (2004)
  [arXiv:astro-ph/0311364].

\bibitem{Padmanabhan:2002vv}
  T.~Padmanabhan and T.~R.~Choudhury,
  Mon.\ Not.\ Roy.\ Astron.\ Soc.\  {\bf 344}, 823 (2003)
  [arXiv:astro-ph/0212573].

\bibitem{Zhu:2004cu}
  Z.~H.~Zhu, M.~K.~Fujimoto and X.~T.~He,
  Astron.\ Astrophys.\  {\bf 417}, 833 (2004)
  [arXiv:astro-ph/0401095].


\bibitem{Chevallier:2000qy}
  M.~Chevallier and D.~Polarski,
  Int.\ J.\ Mod.\ Phys.\  D {\bf 10}, 213 (2001)
  [arXiv:gr-qc/0009008].

\bibitem{Linder:2002et}
  E.~V.~Linder,
  Phys.\ Rev.\ Lett.\  {\bf 90}, 091301 (2003)
  [arXiv:astro-ph/0208512].


\bibitem{Hannestad:2004cb}
  S.~Hannestad and E.~Mortsell,
  JCAP {\bf 0409}, 001 (2004)
  [arXiv:astro-ph/0407259].

\bibitem{Xia:2004rw}
  J.~Q.~Xia, B.~Feng and X.~M.~Zhang,
  Mod.\ Phys.\ Lett.\  A {\bf 20}, 2409 (2005)
  [arXiv:astro-ph/0411501].

\bibitem{Xia:2005ge}
  J.~Q.~Xia, G.~B.~Zhao, B.~Feng, H.~Li and X.~Zhang,
  Phys.\ Rev.\  D {\bf 73}, 063521 (2006)
  [arXiv:astro-ph/0511625].

\bibitem{Xia:2006cr}
  J.~Q.~Xia, G.~B.~Zhao, B.~Feng and X.~Zhang,
  JCAP {\bf 0609}, 015 (2006)
  [arXiv:astro-ph/0603393].

\bibitem{Zhao:2006bt}
  G.~B.~Zhao, J.~Q.~Xia, B.~Feng and X.~Zhang,
  Int.\ J.\ Mod.\ Phys.\  D {\bf 16}, 1229 (2007)
  [arXiv:astro-ph/0603621].

\bibitem{Xia:2006rr}
  J.~Q.~Xia, G.~B.~Zhao, H.~Li, B.~Feng and X.~Zhang,
  Phys.\ Rev.\  D {\bf 74}, 083521 (2006)
  [arXiv:astro-ph/0605366].

\bibitem{Xia:2006wd}
  J.~Q.~Xia, G.~B.~Zhao and X.~Zhang,
  Phys.\ Rev.\  D {\bf 75}, 103505 (2007)
  [arXiv:astro-ph/0609463].

\bibitem{Zhao:2006qg}
  G.~B.~Zhao, J.~Q.~Xia, H.~Li, C.~Tao, J.~M.~Virey, Z.~H.~Zhu and X.~Zhang,
  Phys.\ Lett.\  B {\bf 648}, 8 (2007)
  [arXiv:astro-ph/0612728].



\bibitem{Wang:2007mza}
  Y.~Wang and P.~Mukherjee,
  Phys.\ Rev.\  D {\bf 76}, 103533 (2007)
  [arXiv:astro-ph/0703780].

\bibitem{Wright:2007vr}
  E.~L.~Wright,
  Astrophys.\ J.\  {\bf 664}, 633 (2007)
  [arXiv:astro-ph/0701584].

\bibitem{Li:2008cj}
  H.~Li, J.~Q.~Xia, G.~B.~Zhao, Z.~H.~Fan and X.~Zhang,
  Astrophys.\ J.\  {\bf 683}, L1 (2008)
  [arXiv:0805.1118 [astro-ph]].

\bibitem{Xia:2008ex}
  J.~Q.~Xia, H.~Li, G.~B.~Zhao and X.~Zhang,
  Phys.\ Rev.\  D {\bf 78}, 083524 (2008)
  [arXiv:0807.3878 [astro-ph]].

\bibitem{Li:2008vf}
  H.~Li {\it et al.},
  arXiv:0812.1672 [astro-ph].




\bibitem{Feng:2004ff}
  B.~Feng, M.~Li, Y.~S.~Piao and X.~Zhang,
  Phys.\ Lett.\  B {\bf 634}, 101 (2006)
  [arXiv:astro-ph/0407432].

\bibitem{Guo:2004fq}
  Z.~K.~Guo, Y.~S.~Piao, X.~M.~Zhang and Y.~Z.~Zhang,
  Phys.\ Lett.\  B {\bf 608}, 177 (2005)
  [arXiv:astro-ph/0410654].

\bibitem{Zhang:2005eg}
  X.~F.~Zhang, H.~Li, Y.~S.~Piao and X.~M.~Zhang,
  Mod.\ Phys.\ Lett.\  A {\bf 21}, 231 (2006)
  [arXiv:astro-ph/0501652].

\bibitem{Ma:1995ey}
  C.~P.~Ma and E.~Bertschinger,
  Astrophys.\ J.\  {\bf 455}, 7 (1995)
  [arXiv:astro-ph/9506072].


\bibitem{Wei:2005nw}
  H.~Wei, R.~G.~Cai and D.~F.~Zeng,
  Class.\ Quant.\ Grav.\  {\bf 22}, 3189 (2005)
  [arXiv:hep-th/0501160].

\bibitem{Apostolopoulos:2005ff}
  P.~S.~Apostolopoulos, N.~Brouzakis, E.~N.~Saridakis and N.~Tetradis,
  Phys.\ Rev.\  D {\bf 72}, 044013 (2005)
  [arXiv:hep-th/0502115].

\bibitem{McInnes:2005vp}
  B.~McInnes,
  Nucl.\ Phys.\  B {\bf 718}, 55 (2005)
  [arXiv:hep-th/0502209].

\bibitem{Li:2005fm}
  M.~Z.~Li, B.~Feng and X.~Zhang,
  JCAP {\bf 0512}, 002 (2005)
  [arXiv:hep-ph/0503268].

\bibitem{Andrianov:2005tm}
  A.~A.~Andrianov, F.~Cannata and A.~Y.~Kamenshchik,
  Phys.\ Rev.\  D {\bf 72}, 043531 (2005)
  [arXiv:gr-qc/0505087].

\bibitem{Cai:2005ie}
  R.~G.~Cai, H.~S.~Zhang and A.~Wang,
  Commun.\ Theor.\ Phys.\  {\bf 44}, 948 (2005)
  [arXiv:hep-th/0505186].

\bibitem{Nojiri:2005sr}
  S.~Nojiri and S.~D.~Odintsov,
  Phys.\ Rev.\  D {\bf 72}, 023003 (2005)
  [arXiv:hep-th/0505215].

\bibitem{Capozziello:2005tf}
  S.~Capozziello, S.~Nojiri and S.~D.~Odintsov,
  Phys.\ Lett.\  B {\bf 632}, 597 (2006)
  [arXiv:hep-th/0507182].

\bibitem{Huang:2005gu}
  C.~G.~Huang and H.~Y.~Guo,
  arXiv:astro-ph/0508171.

\bibitem{Wei:2005fq}
  H.~Wei and R.~G.~Cai,
  Phys.\ Rev.\  D {\bf 72}, 123507 (2005)
  [arXiv:astro-ph/0509328].

\bibitem{Wei:2005si}
  H.~Wei and R.~G.~Cai,
  Phys.\ Lett.\  B {\bf 634}, 9 (2006)
  [arXiv:astro-ph/0512018].

\bibitem{Alimohammadi:2006qi}
  M.~Alimohammadi and H.~Mohseni Sadjadi,
  Phys.\ Rev.\  D {\bf 73}, 083527 (2006)
  [arXiv:hep-th/0602268].

\bibitem{Lazkoz:2006pa}
  R.~Lazkoz and G.~Leon,
  Phys.\ Lett.\  B {\bf 638}, 303 (2006)
  [arXiv:astro-ph/0602590].

\bibitem{Zhang:2006ck}
  X.~F.~Zhang and T.~Qiu,
  Phys.\ Lett.\  B {\bf 642}, 187 (2006)
  [arXiv:astro-ph/0603824].

\bibitem{Apostolopoulos:2006si}
  P.~S.~Apostolopoulos and N.~Tetradis,
  Phys.\ Rev.\  D {\bf 74}, 064021 (2006)
  [arXiv:hep-th/0604014].

\bibitem{Chimento:2006xu}
  L.~P.~Chimento and R.~Lazkoz,
  Phys.\ Lett.\  B {\bf 639}, 591 (2006)
  [arXiv:astro-ph/0604090].

\bibitem{Zhao:2006mp}
  W.~Zhao,
  Phys.\ Rev.\  D {\bf 73}, 123509 (2006)
  [arXiv:astro-ph/0604460].

\bibitem{Aref'eva:2006et}
  I.~Y.~Aref'eva and A.~S.~Koshelev,
  JHEP {\bf 0702}, 041 (2007)
  [arXiv:hep-th/0605085].

\bibitem{Chimento:2006ac}
  L.~P.~Chimento, R.~Lazkoz, R.~Maartens and I.~Quiros,
  JCAP {\bf 0609}, 004 (2006)
  [arXiv:astro-ph/0605450].

\bibitem{Alimohammadi:2006tw}
  M.~Alimohammadi and H.~M.~Sadjadi,
  Phys.\ Lett.\  B {\bf 648}, 113 (2007)
  [arXiv:gr-qc/0608016].

\bibitem{Guo:2006pc}
  Z.~K.~Guo, Y.~S.~Piao, X.~Zhang and Y.~Z.~Zhang,
  Phys.\ Rev.\  D {\bf 74}, 127304 (2006)
  [arXiv:astro-ph/0608165].

\bibitem{Cai:2006dm}
  Y.~F.~Cai, H.~Li, Y.~S.~Piao and X.~Zhang,
  Phys.\ Lett.\  B {\bf 646}, 141 (2007)
  [arXiv:gr-qc/0609039].

\bibitem{Zhang:2006qu}
  X.~Zhang,
  Phys.\ Rev.\  D {\bf 74}, 103505 (2006)
  [arXiv:astro-ph/0609699].

\bibitem{Setare:2006rf}
  M.~R.~Setare,
  Phys.\ Lett.\  B {\bf 641}, 130 (2006)
   [arXiv:hep-th/0611165].

\bibitem{Zhang:2006at}
  H.~S.~Zhang and Z.~H.~Zhu,
  Phys.\ Rev.\  D {\bf 75}, 023510 (2007)
  [arXiv:astro-ph/0611834].

\bibitem{Wei:2006va}
  H.~Wei, N.~N.~Tang and S.~N.~Zhang,
  Phys.\ Rev.\  D {\bf 75}, 043009 (2007)
  [arXiv:astro-ph/0612746].

\bibitem{Cai:2007gs}
  Y.~F.~Cai, M.~Z.~Li, J.~X.~Lu, Y.~S.~Piao, T.~Qiu and X.~Zhang,
  Phys.\ Lett.\  B {\bf 651}, 1 (2007)
  [arXiv:hep-th/0701016].

\bibitem{Lazkoz:2007mx}
  R.~Lazkoz, G.~Leon and I.~Quiros,
  Phys.\ Lett.\  B {\bf 649}, 103 (2007)
  [arXiv:astro-ph/0701353].

\bibitem{Alimohammadi:2007jj}
  M.~Alimohammadi,
  Gen.\ Rel.\ Grav.\  {\bf 40}, 107 (2008)
  [arXiv:0706.1360 [gr-qc]].

\bibitem{Babichev:2007dw}
  E.~Babichev, V.~Mukhanov and A.~Vikman,
  JHEP {\bf 0802}, 101 (2008)
  [arXiv:0708.0561 [hep-th]].

\bibitem{Saridakis:2007ns}
  E.~N.~Saridakis,
  JCAP {\bf 0804}, 020 (2008)
  [arXiv:0712.2672 [astro-ph]].

\bibitem{Saridakis:2007wx}
  E.~N.~Saridakis,
  Phys.\ Lett.\  B {\bf 661}, 335 (2008)
  [arXiv:0712.3806 [gr-qc]].

\bibitem{Setare:2008pz}
  M.~R.~Setare and E.~N.~Saridakis,
  Phys.\ Lett.\  B {\bf 668}, 177 (2008)
  [arXiv:0802.2595 [hep-th]].

\bibitem{Elizalde:2008yf}
  E.~Elizalde, S.~Nojiri, S.~D.~Odintsov, D.~Saez-Gomez and V.~Faraoni,
  Phys.\ Rev.\  D {\bf 77}, 106005 (2008)
  [arXiv:0803.1311 [hep-th]].

\bibitem{Sadatian:2008sv}
  S.~D.~Sadatian and K.~Nozari,
  Europhys.\ Lett.\  {\bf 82}, 49001 (2008)
  [arXiv:0803.2398 [gr-qc]].

\bibitem{Sadeghi:2008qp}
  J.~Sadeghi, M.~R.~Setare, A.~Banijamali and F.~Milani,
  Phys.\ Lett.\  B {\bf 662}, 92 (2008)
  [arXiv:0804.0553 [hep-th]].

\bibitem{Xiong:2008ic}
  H.~H.~Xiong, Y.~F.~Cai, T.~Qiu, Y.~S.~Piao and X.~Zhang,
  Phys.\ Lett.\  B {\bf 666}, 212 (2008)
  [arXiv:0805.0413 [astro-ph]].

\bibitem{Alimohammadi:2008mh}
  M.~Alimohammadi and L.~Sadeghian,
  JCAP {\bf 0901}, 035 (2009)
  [arXiv:0806.0141 [gr-qc]].

\bibitem{Cai:2008gk}
  Y.~F.~Cai and J.~Wang,
  Class.\ Quant.\ Grav.\  {\bf 25}, 165014 (2008)
  [arXiv:0806.3890 [hep-th]].

\bibitem{Zhang:2008ac}
  S.~Zhang and B.~Chen,
  Phys.\ Lett.\  B {\bf 669}, 4 (2008)
  [arXiv:0806.4435 [hep-ph]].

\bibitem{Setare:2008dw}
  M.~R.~Setare and E.~N.~Saridakis,
  arXiv:0807.3807 [hep-th].

\bibitem{Nozari:2008yv}
  K.~Nozari, N.~Behrouz and B.~Fazlpour,
  arXiv:0808.0318 [gr-qc].

\bibitem{Nozari:2008gf}
  K.~Nozari and M.~Pourghasemi,
  JCAP {\bf 0810}, 044 (2008)
  [arXiv:0808.3701 [gr-qc]].

\bibitem{Setare:2008si}
  M.~R.~Setare and E.~N.~Saridakis,
  JCAP {\bf 0809}, 026 (2008)
  [arXiv:0809.0114 [hep-th]].

\bibitem{Nozari:2008ff}
  K.~Nozari and S.~D.~Sadatian,
  Eur.\ Phys.\ J.\  C {\bf 58}, 499 (2008)
  [arXiv:0809.4744 [gr-qc]].

\bibitem{Setare:2008pc}
  M.~R.~Setare and E.~N.~Saridakis,
  Phys.\ Lett.\  B {\bf 671}, 331 (2009)
  [arXiv:0810.0645 [hep-th]].

\bibitem{Nozari:2008hz}
  K.~Nozari, M.~R.~Setare, T.~Azizi and N.~Behrouz,
  arXiv:0810.1427 [hep-th].

\bibitem{Setare:2008sf}
  M.~R.~Setare and E.~N.~Saridakis,
  Phys.\ Rev.\  D {\bf 79}, 043005 (2009)
  [arXiv:0810.4775 [astro-ph]].

\bibitem{Creminelli:2008wc}
  P.~Creminelli, G.~D'Amico, J.~Norena and F.~Vernizzi,
  JCAP {\bf 0902}, 018 (2009)
  [arXiv:0811.0827 [astro-ph]].

\bibitem{Chimento:2008ws}
  L.~P.~Chimento, M.~Forte, R.~Lazkoz and M.~G.~Richarte,
  [arXiv:0811.3643 [astro-ph]].

\bibitem{Setare:2008mb}
  M.~R.~Setare and E.~N.~Saridakis,
  JCAP {\bf 0903}, 002 (2009)
  [arXiv:0811.4253 [hep-th]].

\bibitem{Leon:2008aq}
  G.~Leon, R.~Cardenas and J.~L.~Morales,
  [arXiv:0812.0830 [gr-qc]].

\bibitem{Saridakis:2009uu}
  E.~N.~Saridakis, P.~F.~Gonzalez-Diaz and C.~L.~Siguenza,
  arXiv:0901.1213 [astro-ph].

\bibitem{Saridakis:2009ej}
  E.~N.~Saridakis,
  arXiv:0903.3840 [astro-ph.CO].

\bibitem{Wang:2009av}
  T.~Wang,
  arXiv:0908.2477 [hep-th].

\bibitem{Leon:2009ce}
  G.~Leon, Y.~Leyva, E.~N.~Saridakis, O.~Martin and R.~Cardenas,
  arXiv:0912.0542 [gr-qc].

\bibitem{Saridakis:2009jq}
  E.~N.~Saridakis and J.~M.~Weller,
  arXiv:0912.5304 [hep-th].




\bibitem{Ferreira:1997au}
  P.~G.~Ferreira and M.~Joyce,
  Phys.\ Rev.\ Lett.\  {\bf 79}, 4740 (1997)
  [arXiv:astro-ph/9707286].

\bibitem{Copeland:1997et}
  E.~J.~Copeland, A.~R.~Liddle and D.~Wands,
  Phys.\ Rev.\  D {\bf 57}, 4686 (1998)
  [arXiv:gr-qc/9711068].

\bibitem{Chen:2008ft}
  X.~M.~Chen, Y.~G.~Gong and E.~N.~Saridakis,
  JCAP {\bf 0904}, 001 (2009)
  [arXiv:0812.1117 [gr-qc]].

\bibitem{Scherrer:2007pu}
  R.~J.~Scherrer and A.~A.~Sen,
  Phys.\ Rev.\  D {\bf 77}, 083515 (2008)
  [arXiv:0712.3450 [astro-ph]].

\bibitem{Wu:2005apa}
  P.~x.~Wu and H.~w.~Yu,
  Int.\ J.\ Mod.\ Phys.\  D {\bf 14}, 1873 (2005)
  [arXiv:gr-qc/0509036].

\bibitem{Chiba:1998tc}
  T.~Chiba and T.~Nakamura,
  Prog.\ Theor.\ Phys.\  {\bf 100}, 1077 (1998)
  [arXiv:astro-ph/9808022].

\bibitem{Sahni:2002fz}
  V.~Sahni, T.~D.~Saini, A.~A.~Starobinsky and U.~Alam,
  JETP Lett.\  {\bf 77}, 201 (2003)
  [Pisma Zh.\ Eksp.\ Teor.\ Fiz.\  {\bf 77}, 249 (2003)]
  [arXiv:astro-ph/0201498].

\bibitem{Saridakis:2009pj}
  E.~N.~Saridakis,
  Nucl.\ Phys.\  B {\bf 819}, 116 (2009)
  [arXiv:0902.3978 [gr-qc]].

\bibitem{Alam:2003sc}
  U.~Alam, V.~Sahni, T.~D.~Saini and A.~A.~Starobinsky,
  Mon.\ Not.\ Roy.\ Astron.\ Soc.\  {\bf 344}, 1057 (2003)
  [arXiv:astro-ph/0303009].

\bibitem{Gorini:2002kf}
  V.~Gorini, A.~Kamenshchik and U.~Moschella,
  Phys.\ Rev.\  D {\bf 67}, 063509 (2003)
  [arXiv:astro-ph/0209395].

\bibitem{Zimdahl:2003wg}
  W.~Zimdahl and D.~Pavon,
  Gen.\ Rel.\ Grav.\  {\bf 36}, 1483 (2004)
  [arXiv:gr-qc/0311067].

\bibitem{Zhang:2005rj}
  X.~Zhang,
  Phys.\ Lett.\  B {\bf 611}, 1 (2005)
  [arXiv:astro-ph/0503075].

\bibitem{Zhang:2005yz}
  X.~Zhang,
  Int.\ J.\ Mod.\ Phys.\  D {\bf 14}, 1597 (2005)
  [arXiv:astro-ph/0504586].


\bibitem{Garriga:2003nm}
  J.~Garriga, L.~Pogosian and T.~Vachaspati,
  Phys.\ Rev.\  D {\bf 69}, 063511 (2004)
  [arXiv:astro-ph/0311412].

\bibitem{Garriga:1999bf}
  J.~Garriga and A.~Vilenkin,
  Phys.\ Rev.\  D {\bf 61}, 083502 (2000)
  [arXiv:astro-ph/9908115].

\bibitem{Garriga:2002tq}
  J.~Garriga and A.~Vilenkin,
  Phys.\ Rev.\  D {\bf 67}, 043503 (2003)
  [arXiv:astro-ph/0210358].

\bibitem{Garriga:2003hj}
  J.~Garriga, A.~Linde and A.~Vilenkin,
  Phys.\ Rev.\  D {\bf 69}, 063521 (2004)
  [arXiv:hep-th/0310034].

\bibitem{Avelino:2004vy}
  P.~P.~Avelino,
  Phys.\ Lett.\  B {\bf 611}, 15 (2005)
  [arXiv:astro-ph/0411033].

\bibitem{Perivolaropoulos:2004yr}
  L.~Perivolaropoulos,
  Phys.\ Rev.\  D {\bf 71}, 063503 (2005)
  [arXiv:astro-ph/0412308].

\bibitem{Polchinski:2005}
  J. Polchinski,
  \textit{``String Theory,"}
  Cambridge University Press (2005).

\bibitem{Chimento:2002gb}
  L.~P.~Chimento,
  Phys.\ Rev.\  D {\bf 65}, 063517 (2002).

\bibitem{Aguirregabiria:2003uh}
  J.~M.~Aguirregabiria, L.~P.~Chimento, A.~S.~Jakubi and R.~Lazkoz,
  Phys.\ Rev.\  D {\bf 67}, 083518 (2003)
  [arXiv:gr-qc/0303010].

\bibitem{Veneziano:1991ek}
  G.~Veneziano,
  Phys.\ Lett.\  B {\bf 265}, 287 (1991).

\bibitem{Lidsey:1999mc}
  J.~E.~Lidsey, D.~Wands and E.~J.~Copeland,
  Phys.\ Rept.\  {\bf 337}, 343 (2000)
  [arXiv:hep-th/9909061].

\bibitem{Chimento:2003qy}
  L.~P.~Chimento and R.~Lazkoz,
  Phys.\ Rev.\ Lett.\  {\bf 91}, 211301 (2003)
  [arXiv:gr-qc/0307111].

\bibitem{Dabrowski:2003jm}
  M.~P.~Dabrowski, T.~Stachowiak and M.~Szydlowski,
  Phys.\ Rev.\  D {\bf 68}, 103519 (2003)
  [arXiv:hep-th/0307128].

\bibitem{Chimento:2005xa}
  L.~P.~Chimento and D.~Pavon,
  Phys.\ Rev.\  D {\bf 73}, 063511 (2006)
  [arXiv:gr-qc/0505096].

\bibitem{Chimento:2005au}
  L.~P.~Chimento and R.~Lazkoz,
  Class.\ Quant.\ Grav.\  {\bf 23}, 3195 (2006)
  [arXiv:astro-ph/0505254].

\bibitem{Chimento:2006gk}
  L.~P.~Chimento and W.~Zimdahl,
  Int.\ J.\ Mod.\ Phys.\  D {\bf 17}, 2229 (2008)
  [arXiv:gr-qc/0609104].

\bibitem{Gupta:2009kk}
  G.~Gupta, E.~N.~Saridakis and A.~A.~Sen,
  Phys.\ Rev.\  D {\bf 79}, 123013 (2009)
  [arXiv:0905.2348 [astro-ph.CO]].

\bibitem{Dutta:2009yb}
  S.~Dutta, E.~N.~Saridakis and R.~J.~Scherrer,
  Phys.\ Rev.\  D {\bf 79}, 103005 (2009)
  [arXiv:0903.3412 [astro-ph.CO]].

\bibitem{Dabrowski:2006iv}
  M.~P.~Dabrowski,
  Annalen Phys.\  {\bf 15}, 352 (2006)
  [arXiv:astro-ph/0606574].

\bibitem{Dabrowski:2006dd}
  M.~P.~Dabrowski, C.~Kiefer and B.~Sandhofer,
  Phys.\ Rev.\  D {\bf 74}, 044022 (2006)
  [arXiv:hep-th/0605229].

\bibitem{Singh:2006sg}
  P.~Singh,
  Phys.\ Rev.\  D {\bf 73}, 063508 (2006)
  [arXiv:gr-qc/0603043].

\bibitem{Sami:2003xv}
  M.~Sami and A.~Toporensky,
  Mod.\ Phys.\ Lett.\  A {\bf 19}, 1509 (2004)
  [arXiv:gr-qc/0312009].

\bibitem{Guo:2004ae}
  Z.~K.~Guo, Y.~S.~Piao and Y.~Z.~Zhang,
  Phys.\ Lett.\  B {\bf 594}, 247 (2004)
  [arXiv:astro-ph/0404225].

\bibitem{Wands:2000dp}
  D.~Wands, K.~A.~Malik, D.~H.~Lyth and A.~R.~Liddle,
  Phys.\ Rev.\  D {\bf 62}, 043527 (2000)
  [arXiv:astro-ph/0003278].

\bibitem{Kawasaki:2001bq}
  M.~Kawasaki, T.~Moroi and T.~Takahashi,
  Phys.\ Rev.\  D {\bf 64}, 083009 (2001)
  [arXiv:astro-ph/0105161].

\bibitem{Moroi:2003pq}
  T.~Moroi and T.~Takahashi,
  Phys.\ Rev.\ Lett.\  {\bf 92}, 091301 (2004)
  [arXiv:astro-ph/0308208].

\bibitem{Gordon:2005ti}
  C.~Gordon and D.~Wands,
  Phys.\ Rev.\  D {\bf 71}, 123505 (2005)
  [arXiv:astro-ph/0504132].


\bibitem{Lewis:1999bs}
  A.~Lewis, A.~Challinor and A.~Lasenby,
  Astrophys.\ J.\  {\bf 538}, 473 (2000)
  [arXiv:astro-ph/9911177].

\bibitem{Weller:2003hw}
  J.~Weller and A.~M.~Lewis,
  Mon.\ Not.\ Roy.\ Astron.\ Soc.\  {\bf 346}, 987 (2003)
  [arXiv:astro-ph/0307104].

\bibitem{Caldwell:1997ii}
  R.~R.~Caldwell, R.~Dave and P.~J.~Steinhardt,
  Phys.\ Rev.\ Lett.\  {\bf 80}, 1582 (1998)
  [arXiv:astro-ph/9708069].

\bibitem{Contaldi:2003zv}
  C.~R.~Contaldi, M.~Peloso, L.~Kofman and A.~Linde,
  JCAP {\bf 0307}, 002 (2003)
  [arXiv:astro-ph/0303636].

\bibitem{Feng:2003zua}
  B.~Feng and X.~Zhang,
  Phys.\ Lett.\  B {\bf 570}, 145 (2003)
  [arXiv:astro-ph/0305020].

\bibitem{Kogut:2003et}
  A.~Kogut {\it et al.}  [WMAP Collaboration],
  Astrophys.\ J.\ Suppl.\  {\bf 148}, 161 (2003)
  [arXiv:astro-ph/0302213].

\bibitem{Hinshaw:2003ex}
  G.~Hinshaw {\it et al.}  [WMAP Collaboration],
  Astrophys.\ J.\ Suppl.\  {\bf 148}, 135 (2003)
  [arXiv:astro-ph/0302217].

\bibitem{Dodelson:2003ft}
  S.~Dodelson,
  {\it  Amsterdam, Netherlands: Academic Pr. (2003) 440 p}

\bibitem{Linder:2003dr}
  E.~V.~Linder and A.~Jenkins,
  Mon.\ Not.\ Roy.\ Astron.\ Soc.\  {\bf 346}, 573 (2003)
  [arXiv:astro-ph/0305286].

\bibitem{Jain:2003tba}
  B.~Jain and A.~Taylor,
  Phys.\ Rev.\ Lett.\  {\bf 91}, 141302 (2003)
  [arXiv:astro-ph/0306046].

\bibitem{Zhang:2005eb}
  P.~Zhang and U.~L.~Pen,
  Mon.\ Not.\ Roy.\ Astron.\ Soc.\  {\bf 367}, 169 (2006)
  [arXiv:astro-ph/0504551].

\bibitem{Zhang:2005pu}
  P.~Zhang and U.~L.~Pen,
  Phys.\ Rev.\ Lett.\  {\bf 95}, 241302 (2005)
  [arXiv:astro-ph/0506740].

\bibitem{Stefancic:2005cs}
  H.~Stefancic,
  Phys.\ Rev.\  D {\bf 71}, 124036 (2005)
  [arXiv:astro-ph/0504518].

\bibitem{Lee:1969fy}
  T.~D.~Lee and G.~C.~Wick,
  Nucl.\ Phys.\  B {\bf 9}, 209 (1969).

\bibitem{Lee:1970iw}
  T.~D.~Lee and G.~C.~Wick,
  Phys.\ Rev.\  D {\bf 2}, 1033 (1970).

\bibitem{Simon:1990ic}
  J.~Z.~Simon,
  Phys.\ Rev.\  D {\bf 41}, 3720 (1990).

\bibitem{Eliezer:1989cr}
  D.~A.~Eliezer and R.~P.~Woodard,
  Nucl.\ Phys.\  B {\bf 325}, 389 (1989).

\bibitem{Erler:2004hv}
  T.~Erler and D.~J.~Gross,
  arXiv:hep-th/0406199.

\bibitem{Stelle:1976gc}
  K.~S.~Stelle,
  Phys.\ Rev.\  D {\bf 16}, 953 (1977).



\bibitem{ArkaniHamed:2003uy}
  N.~Arkani-Hamed, H.~C.~Cheng, M.~A.~Luty and S.~Mukohyama,
  JHEP {\bf 0405}, 074 (2004)
  [arXiv:hep-th/0312099].

\bibitem{ArkaniHamed:2003uz}
  N.~Arkani-Hamed, P.~Creminelli, S.~Mukohyama and M.~Zaldarriaga,
  JCAP {\bf 0404}, 001 (2004)
  [arXiv:hep-th/0312100].

\bibitem{Piazza:2004df}
  F.~Piazza and S.~Tsujikawa,
  JCAP {\bf 0407}, 004 (2004)
  [arXiv:hep-th/0405054].

\bibitem{Anisimov:2004sp}
  A.~Anisimov and A.~Vikman,
  JCAP {\bf 0504}, 009 (2005)
  [arXiv:hep-ph/0411089].

\bibitem{ArkaniHamed:2005gu}
  N.~Arkani-Hamed, H.~C.~Cheng, M.~A.~Luty, S.~Mukohyama and T.~Wiseman,
  JHEP {\bf 0701}, 036 (2007)
  [arXiv:hep-ph/0507120].

\bibitem{Mukohyama:2005rw}
  S.~Mukohyama,
  Phys.\ Rev.\  D {\bf 71}, 104019 (2005)
  [arXiv:hep-th/0502189].

\bibitem{Anisimov:2005ne}
  A.~Anisimov, E.~Babichev and A.~Vikman,
  JCAP {\bf 0506}, 006 (2005)
  [arXiv:astro-ph/0504560].


\bibitem{GonzalezDiaz:2003bc}
  P.~F.~Gonzalez-Diaz,
  Phys.\ Rev.\  D {\bf 68}, 021303 (2003)
  [arXiv:astro-ph/0305559].

\bibitem{Kallosh:2003bq}
  R.~Kallosh, J.~Kratochvil, A.~Linde, E.~V.~Linder and M.~Shmakova,
  JCAP {\bf 0310}, 015 (2003)
  [arXiv:astro-ph/0307185].

\bibitem{Gibbons:2003yj}
  G.~W.~Gibbons,
  arXiv:hep-th/0302199.

\bibitem{Hawking:2001yt}
  S.~W.~Hawking and T.~Hertog,
  Phys.\ Rev.\  D {\bf 65}, 103515 (2002)
  [arXiv:hep-th/0107088].


\bibitem{Aref'eva:2004vw}
  I.~Y.~Aref'eva, A.~S.~Koshelev and S.~Y.~Vernov,
  Theor.\ Math.\ Phys.\  {\bf 148}, 895 (2006)
  [Teor.\ Mat.\ Fiz.\  {\bf 148}, 23 (2006)]
  [arXiv:astro-ph/0412619].

\bibitem{Aref'eva:2005gg}
  I.~Y.~Aref'eva and L.~V.~Joukovskaya,
  JHEP {\bf 0510}, 087 (2005)
  [arXiv:hep-th/0504200].

\bibitem{Aref'eva:2007yr}
  I.~Y.~Aref'eva, L.~V.~Joukovskaya and S.~Y.~Vernov,
  J.\ Phys.\ A  {\bf 41}, 304003 (2008)
  [arXiv:0711.1364 [hep-th]].

\bibitem{Mulryne:2008iq}
  D.~J.~Mulryne and N.~J.~Nunes,
  Phys.\ Rev.\  D {\bf 78}, 063519 (2008)
  [arXiv:0805.0449 [hep-th]].

\bibitem{Barnaby:2008pt}
  N.~Barnaby, D.~J.~Mulryne, N.~J.~Nunes and P.~Robinson,
  JHEP {\bf 0903}, 018 (2009)
  [arXiv:0811.0608 [hep-th]].

\bibitem{Shi:2008df}
  S.~G.~Shi, Y.~S.~Piao and C.~F.~Qiao,
  JCAP {\bf 0904}, 027 (2009)
  [arXiv:0812.4022 [astro-ph]].

\bibitem{Saridakis:2009uk}
  E.~N.~Saridakis and J.~Ward,
  arXiv:0906.5135 [hep-th].


\bibitem{Gerasimov:2000zp}
  A.~A.~Gerasimov and S.~L.~Shatashvili,
  JHEP {\bf 0010}, 034 (2000)
  [arXiv:hep-th/0009103].

\bibitem{Kutasov:2000qp}
  D.~Kutasov, M.~Marino and G.~W.~Moore,
  JHEP {\bf 0010}, 045 (2000).

\bibitem{Kutasov:2000aq}
  D.~Kutasov, M.~Marino and G.~W.~Moore,
  arXiv:hep-th/0010108.

\bibitem{Barnaby:2006hi}
  N.~Barnaby, T.~Biswas and J.~M.~Cline,
  arXiv:hep-th/0612230.

\bibitem{Sen:1999md}
  A.~Sen,
  JHEP {\bf 9910}, 008 (1999)
  [arXiv:hep-th/9909062].

\bibitem{Bergshoeff:2000dq}
  E.~A.~Bergshoeff, M.~de Roo, T.~C.~de Wit, E.~Eyras and S.~Panda,
  JHEP {\bf 0005}, 009 (2000)
  [arXiv:hep-th/0003221].

\bibitem{Kluson:2000iy}
  J.~Kluson,
  Phys.\ Rev.\  D {\bf 62}, 126003 (2000)
  [arXiv:hep-th/0004106].

\bibitem{Mukhopadhyay:2002en}
  P.~Mukhopadhyay and A.~Sen,
  JHEP {\bf 0211}, 047 (2002)
  [arXiv:hep-th/0208142].

\bibitem{Chingangbam:2004ng}
  P.~Chingangbam, S.~Panda and A.~Deshamukhya,
  JHEP {\bf 0502}, 052 (2005)
  [arXiv:hep-th/0411210].

\bibitem{Setare:2008qk}
  M.~R.~Setare, J.~Sadeghi and A.~R.~Amani,
  arXiv:0811.3343 [hep-th].



\bibitem{Onemli:2002hr}
  V.~K.~Onemli and R.~P.~Woodard,
  Class.\ Quant.\ Grav.\  {\bf 19}, 4607 (2002)
  [arXiv:gr-qc/0204065].

\bibitem{Onemli:2004mb}
  V.~K.~Onemli and R.~P.~Woodard,
  Phys.\ Rev.\  D {\bf 70}, 107301 (2004)
  [arXiv:gr-qc/0406098].

\bibitem{Brunier:2004sb}
  T.~Brunier, V.~K.~Onemli and R.~P.~Woodard,
  Class.\ Quant.\ Grav.\  {\bf 22}, 59 (2005)
  [arXiv:gr-qc/0408080].

\bibitem{Kahya:2006hc}
  E.~O.~Kahya and V.~K.~Onemli,
  Phys.\ Rev.\  D {\bf 76}, 043512 (2007)
  [arXiv:gr-qc/0612026].

\bibitem{Kahya:2009sz}
  E.~O.~Kahya, V.~K.~Onemli and R.~P.~Woodard,
  arXiv:0904.4811 [gr-qc].


\bibitem{taub:1937}
  A.~H.~Taub,
  Phys.\ Rev. {\bf 51}, 512 (1937)

\bibitem{Brill:1957fx}
  D.~R.~Brill and J.~A.~Wheeler,
  Rev.\ Mod.\ Phys.\  {\bf 29}, 465 (1957).

\bibitem{Parker:1971pt}
  L.~Parker,
  Phys.\ Rev.\  D {\bf 3}, 346 (1971)
  [Erratum-ibid.\  D {\bf 3}, 2546 (1971)].

\bibitem{Obukhov:1993fd}
  Y.~N.~Obukhov,
  Phys.\ Lett.\  A {\bf 182}, 214 (1993)
  [arXiv:gr-qc/0008015].

\bibitem{ArmendarizPicon:2003qk}
  C.~Armendariz-Picon and P.~B.~Greene,
  Gen.\ Rel.\ Grav.\  {\bf 35}, 1637 (2003)
  [arXiv:hep-th/0301129].

\bibitem{Kasuya:2001pr}
  S.~Kasuya,
  Phys.\ Lett.\  B {\bf 515}, 121 (2001)
  [arXiv:astro-ph/0105408].

\bibitem{Saha:2000nk}
  B.~Saha,
  Mod.\ Phys.\ Lett.\  A {\bf 16}, 1287 (2001)
  [arXiv:gr-qc/0009002].

\bibitem{Ribas:2005vr}
  M.~O.~Ribas, F.~P.~Devecchi and G.~M.~Kremer,
  Phys.\ Rev.\  D {\bf 72}, 123502 (2005)
  [arXiv:gr-qc/0511099].

\bibitem{Chimento:2007fx}
  L.~P.~Chimento, F.~P.~Devecchi, M.~Forte and G.~M.~Kremer,
  Class.\ Quant.\ Grav.\  {\bf 25}, 085007 (2008)
  [arXiv:0707.4455 [gr-qc]].

\bibitem{Buniy:2005vh}
  R.~V.~Buniy and S.~D.~H.~Hsu,
  Phys.\ Lett.\  B {\bf 632}, 543 (2006)
  [arXiv:hep-th/0502203].

\bibitem{Qiu:2007fd}
  T.~Qiu, Y.~F.~Cai and X.~M.~Zhang,
  Mod.\ Phys.\ Lett.\  A {\bf 23}, 2787 (2008)
  [arXiv:0710.0115 [gr-qc]].

\bibitem{Weinberg:1972}
  S. Weinberg,
  \textit{``Gravitation and Cosmology,"}
  Cambridge University Press (1972).

\bibitem{Birrell:1982}
  N. Birrell and P. Davies,
  \textit{``Quantum Fields in Curved Space,"}
  Cambridge University Press (1982).

\bibitem{GSW}
  M. Green, J.  Schwarz and E. Witten,
  \textit{``Superstring Theory,"} Vol. 2, Chapter 12,
  Cambridge University Press (1987).

\bibitem{Vakili:2005ya}
  B.~Vakili, S.~Jalalzadeh and H.~R.~Sepangi,
  JCAP {\bf 0505}, 006 (2005)
  [arXiv:gr-qc/0502076].

\bibitem{Kamenshchik:2001cp}
  A.~Y.~Kamenshchik, U.~Moschella and V.~Pasquier,
  Phys.\ Lett.\  B {\bf 511}, 265 (2001)
  [arXiv:gr-qc/0103004].

\bibitem{GonzalezDiaz:2002hr}
  P.~F.~Gonzalez-Diaz,
  Phys.\ Lett.\  B {\bf 562}, 1 (2003)
  [arXiv:astro-ph/0212414].

\bibitem{Bilic:2002vm}
  N.~Bilic, G.~B.~Tupper and R.~D.~Viollier,
  arXiv:astro-ph/0207423.

\bibitem{Bordemann:1993ep}
  M.~Bordemann and J.~Hoppe,
  Phys.\ Lett.\  B {\bf 317}, 315 (1993)
  [arXiv:hep-th/9307036].

\bibitem{Fabris:2001tm}
  J.~C.~Fabris, S.~V.~B.~Goncalves and P.~E.~de Souza,
  Gen.\ Rel.\ Grav.\  {\bf 34}, 53 (2002)
  [arXiv:gr-qc/0103083].

\bibitem{Bento:2002ps}
  M.~C.~Bento, O.~Bertolami and A.~A.~Sen,
  Phys.\ Rev.\  D {\bf 66}, 043507 (2002)
  [arXiv:gr-qc/0202064].

\bibitem{Bento:2003we}
  M.~d.~C.~Bento, O.~Bertolami and A.~A.~Sen,
  Phys.\ Lett.\  B {\bf 575}, 172 (2003)
  [arXiv:astro-ph/0303538].

\bibitem{Zhang:2005jj}
  H.~S.~Zhang and Z.~H.~Zhu,
  Phys.\ Rev.\  D {\bf 73}, 043518 (2006)
  [arXiv:astro-ph/0509895].

\bibitem{GarciaCompean:2007vh}
  H.~Garcia-Compean, G.~Garcia-Jimenez, O.~Obregon and C.~Ramirez,
  JCAP {\bf 0807}, 016 (2008)
  [arXiv:0710.4283 [hep-th]].

\bibitem{Makler:2002jv}
  M.~Makler, S.~Quinet de Oliveira and I.~Waga,
  Phys.\ Lett.\  B {\bf 555}, 1 (2003)
  [arXiv:astro-ph/0209486].

\bibitem{Bento:2002yx}
  M.~d.~C.~Bento, O.~Bertolami and A.~A.~Sen,
  Phys.\ Rev.\  D {\bf 67}, 063003 (2003)
  [arXiv:astro-ph/0210468].

\bibitem{Bertolami:2005aa}
  O.~Bertolami and P.~Tavares Silva,
  Mon.\ Not.\ Roy.\ Astron.\ Soc.\  {\bf 365}, 1149 (2006)
  [arXiv:astro-ph/0507192].

\bibitem{Cunha:2003vg}
  J.~V.~Cunha, J.~S.~Alcaniz and J.~A.~S.~Lima,
  Phys.\ Rev.\  D {\bf 69}, 083501 (2004)
  [arXiv:astro-ph/0306319].

\bibitem{Bilic:2001cg}
  N.~Bilic, G.~B.~Tupper and R.~D.~Viollier,
  Phys.\ Lett.\  B {\bf 535}, 17 (2002)
  [arXiv:astro-ph/0111325].

\bibitem{Novello:2003kh}
  M.~Novello, S.~E.~Perez Bergliaffa and J.~Salim,
  Phys.\ Rev.\  D {\bf 69}, 127301 (2004)
  [arXiv:astro-ph/0312093].

\bibitem{Kiselev:2004py}
  V.~V.~Kiselev,
  Class.\ Quant.\ Grav.\  {\bf 21}, 3323 (2004)
  [arXiv:gr-qc/0402095].

\bibitem{ArmendarizPicon:2004pm}
  C.~Armendariz-Picon,
  JCAP {\bf 0407}, 007 (2004)
  [arXiv:astro-ph/0405267].

\bibitem{Zhao:2005bu}
  W.~Zhao and Y.~Zhang,
  Class.\ Quant.\ Grav.\  {\bf 23}, 3405 (2006)
  [arXiv:astro-ph/0510356].

\bibitem{Wei:2006tn}
  H.~Wei and R.~G.~Cai,
  Phys.\ Rev.\  D {\bf 73}, 083002 (2006)
  [arXiv:astro-ph/0603052].

\bibitem{Wei:2006gv}
  H.~Wei and R.~G.~Cai,
  JCAP {\bf 0709}, 015 (2007)
  [arXiv:astro-ph/0607064].

\bibitem{Boehmer:2007qa}
  C.~G.~Boehmer and T.~Harko,
  Eur.\ Phys.\ J.\  C {\bf 50}, 423 (2007)
  [arXiv:gr-qc/0701029].

\bibitem{Beck:2007ru}
  C.~Beck and M.~C.~Mackey,
  Int.\ J.\ Mod.\ Phys.\  D {\bf 17}, 71 (2008)
  [arXiv:astro-ph/0703364].

\bibitem{Koivisto:2007bp}
  T.~Koivisto and D.~F.~Mota,
  Astrophys.\ J.\  {\bf 679}, 1 (2008)
  [arXiv:0707.0279 [astro-ph]].

\bibitem{Bamba:2008xa}
  K.~Bamba, S.~Nojiri and S.~D.~Odintsov,
  Phys.\ Rev.\  D {\bf 77}, 123532 (2008)
  [arXiv:0803.3384 [hep-th]].


\bibitem{Mota:2007sz}
  D.~F.~Mota, J.~R.~Kristiansen, T.~Koivisto and N.~E.~Groeneboom,
  Mon.\ Not.\ Roy.\ Astron.\ Soc.\  {\bf 382}, 793 (2007)
  [arXiv:0708.0830 [astro-ph]].

\bibitem{Jimenez:2008au}
  J.~B.~Jimenez and A.~L.~Maroto,
  Phys.\ Rev.\  D {\bf 78}, 063005 (2008)
  [arXiv:0801.1486 [astro-ph]].

\bibitem{Jimenez:2009py}
  J.~B.~Jimenez, R.~Lazkoz and A.~L.~Maroto,
  Phys.\ Rev.\  D {\bf 80}, 023004 (2009)
  arXiv:0904.0433 [astro-ph.CO].


\bibitem{Himmetoglu:2008zp}
  B.~Himmetoglu, C.~R.~Contaldi and M.~Peloso,
  Phys.\ Rev.\ Lett.\  {\bf 102}, 111301 (2009)
  [arXiv:0809.2779 [astro-ph]].



\bibitem{Saridakis:2009bv}
  E.~N.~Saridakis,
  arXiv:0905.3532 [hep-th].

\bibitem{Cai:2009in}
  Y.~F.~Cai and E.~N.~Saridakis,
  arXiv:0906.1789 [hep-th].

\bibitem{Park:2009zr}
  M.~i.~Park,
  arXiv:0906.4275 [hep-th].

\bibitem{Carroll:1989vb}
  S.~M.~Carroll, G.~B.~Field and R.~Jackiw,
  Phys.\ Rev.\  D {\bf 41}, 1231 (1990).

\bibitem{Colladay:1998fq}
  D.~Colladay and V.~A.~Kostelecky,
  Phys.\ Rev.\  D {\bf 58}, 116002 (1998)
  [arXiv:hep-ph/9809521].

\bibitem{Jackiw:1999yp}
  R.~Jackiw and V.~A.~Kostelecky,
  Phys.\ Rev.\ Lett.\  {\bf 82}, 3572 (1999)
  [arXiv:hep-ph/9901358].

\bibitem{Kostelecky:2003fs}
  V.~A.~Kostelecky,
  Phys.\ Rev.\  D {\bf 69}, 105009 (2004)
  [arXiv:hep-th/0312310].

\bibitem{Lehnert:2006mn}
  R.~Lehnert,
  arXiv:hep-ph/0611177.

\bibitem{Akama:1982jy}
  K.~Akama,
  Lect.\ Notes Phys.\  {\bf 176}, 267 (1982)
  [arXiv:hep-th/0001113].

\bibitem{Rubakov:1983bb}
  V.~A.~Rubakov and M.~E.~Shaposhnikov,
  Phys.\ Lett.\  B {\bf 125}, 136 (1983).

\bibitem{Antoniadis:1988jn}
  I.~Antoniadis, C.~Bachas, D.~C.~Lewellen and T.~N.~Tomaras,
  Phys.\ Lett.\  B {\bf 207}, 441 (1988).

\bibitem{Antoniadis:1990ew}
  I.~Antoniadis,
  Phys.\ Lett.\  B {\bf 246}, 377 (1990).

\bibitem{Horava:1995qa}
  P.~Horava and E.~Witten,
  Nucl.\ Phys.\  B {\bf 460}, 506 (1996)
  [arXiv:hep-th/9510209].

\bibitem{Horava:1996ma}
  P.~Horava and E.~Witten,
  Nucl.\ Phys.\  B {\bf 475}, 94 (1996)
  [arXiv:hep-th/9603142].

\bibitem{Maldacena:1997re}
  J.~M.~Maldacena,
  Adv.\ Theor.\ Math.\ Phys.\  {\bf 2}, 231 (1998)
  [Int.\ J.\ Theor.\ Phys.\  {\bf 38}, 1113 (1999)]
  [arXiv:hep-th/9711200].

\bibitem{Lukas:1998yy}
  A.~Lukas, B.~A.~Ovrut, K.~S.~Stelle and D.~Waldram,
  Phys.\ Rev.\  D {\bf 59}, 086001 (1999)
  [arXiv:hep-th/9803235].

\bibitem{Antoniadis:1998ig}
  I.~Antoniadis, N.~Arkani-Hamed, S.~Dimopoulos and G.~R.~Dvali,
  Phys.\ Lett.\  B {\bf 436}, 257 (1998)
  [arXiv:hep-ph/9804398].

\bibitem{ArkaniHamed:1998rs}
  N.~Arkani-Hamed, S.~Dimopoulos and G.~R.~Dvali,
  Phys.\ Lett.\  B {\bf 429}, 263 (1998)
  [arXiv:hep-ph/9803315].

\bibitem{ArkaniHamed:1998nn}
  N.~Arkani-Hamed, S.~Dimopoulos and G.~R.~Dvali,
  Phys.\ Rev.\  D {\bf 59}, 086004 (1999)
  [arXiv:hep-ph/9807344].

\bibitem{Binetruy:1999ut}
  P.~Binetruy, C.~Deffayet and D.~Langlois,
  Nucl.\ Phys.\  B {\bf 565}, 269 (2000)
  [arXiv:hep-th/9905012].


\bibitem{Randall:1999ee}
  L.~Randall and R.~Sundrum,
  Phys.\ Rev.\ Lett.\  {\bf 83}, 3370 (1999)
  [arXiv:hep-ph/9905221].

\bibitem{Randall:1999vf}
  L.~Randall and R.~Sundrum,
  Phys.\ Rev.\ Lett.\  {\bf 83}, 4690 (1999)
  [arXiv:hep-th/9906064].

\bibitem{Ida:1999ui}
  D.~Ida,
  JHEP {\bf 0009}, 014 (2000)
  [arXiv:gr-qc/9912002].

\bibitem{Goldberger:1999uk}
  W.~D.~Goldberger and M.~B.~Wise,
  Phys.\ Rev.\ Lett.\  {\bf 83}, 4922 (1999)
  [arXiv:hep-ph/9907447].

\bibitem{Kobayashi:2002pw}
  S.~Kobayashi and K.~Koyama,
  JHEP {\bf 0212}, 056 (2002)
  [arXiv:hep-th/0210029].

\bibitem{Brax:2003fv}
  P.~Brax and C.~van de Bruck,
  Class.\ Quant.\ Grav.\  {\bf 20}, R201 (2003)
  [arXiv:hep-th/0303095].

\bibitem{Maartens:2003tw}
  R.~Maartens,
  Living Rev.\ Rel.\  {\bf 7}, 7 (2004)
  [arXiv:gr-qc/0312059].

\bibitem{Apostolopoulos:2004rk}
  P.~S.~Apostolopoulos and N.~Tetradis,
  Class.\ Quant.\ Grav.\  {\bf 21}, 4781 (2004)
  [arXiv:hep-th/0404105].

\bibitem{Schwindt:2005fm}
  J.~M.~Schwindt and C.~Wetterich,
  Nucl.\ Phys.\  B {\bf 726}, 75 (2005)
  [arXiv:hep-th/0501049].

\bibitem{Diakonos:2007au}
  F.~K.~Diakonos and E.~N.~Saridakis,
  JCAP {\bf 0902}, 030 (2009)
  [arXiv:0708.3143 [hep-th]].








\bibitem{Sahni:2002dx}
  V.~Sahni and Y.~Shtanov,
  JCAP {\bf 0311}, 014 (2003)
  [arXiv:astro-ph/0202346].

\bibitem{Dvali:2000hr}
  G.~R.~Dvali, G.~Gabadadze and M.~Porrati,
  Phys.\ Lett.\  B {\bf 485}, 208 (2000)
  [arXiv:hep-th/0005016].

\bibitem{Lue:2005ya}
  A.~Lue,
  Phys.\ Rept.\  {\bf 423}, 1 (2006)
  [arXiv:astro-ph/0510068].

\bibitem{Deffayet:2000uy}
  C.~Deffayet,
  Phys.\ Lett.\  B {\bf 502}, 199 (2001)
  [arXiv:hep-th/0010186].

\bibitem{Deffayet:2001pu}
  C.~Deffayet, G.~R.~Dvali and G.~Gabadadze,
  Phys.\ Rev.\  D {\bf 65}, 044023 (2002)
  [arXiv:astro-ph/0105068].

\bibitem{Deffayet:2002sp}
  C.~Deffayet, S.~J.~Landau, J.~Raux, M.~Zaldarriaga and P.~Astier,
  Phys.\ Rev.\  D {\bf 66}, 024019 (2002)
  [arXiv:astro-ph/0201164].

\bibitem{Lazkoz:2006gp}
  R.~Lazkoz, R.~Maartens and E.~Majerotto,
  Phys.\ Rev.\  D {\bf 74}, 083510 (2006)
  [arXiv:astro-ph/0605701].


\bibitem{Gherghetta:2000qi}
  T.~Gherghetta and M.~E.~Shaposhnikov,
  Phys.\ Rev.\ Lett.\  {\bf 85}, 240 (2000)
  [arXiv:hep-th/0004014].

\bibitem{Bostock:2003cv}
  P.~Bostock, R.~Gregory, I.~Navarro and J.~Santiago,
  Phys.\ Rev.\ Lett.\  {\bf 92}, 221601 (2004)
  [arXiv:hep-th/0311074].

\bibitem{Guendelman:2003ie}
  E.~I.~Guendelman,
  Phys.\ Lett.\  B {\bf 580}, 87 (2004)
  [arXiv:gr-qc/0303048].

\bibitem{Kofinas:2004ae}
  G.~Kofinas,
  Class.\ Quant.\ Grav.\  {\bf 22}, L47 (2005)
  [arXiv:hep-th/0412299].

\bibitem{Erdem:2006qk}
  R.~Erdem,
  Phys.\ Lett.\  B {\bf 639}, 348 (2006)
  [arXiv:gr-qc/0603080].

\bibitem{Lukas:1998qs}
  A.~Lukas, B.~A.~Ovrut and D.~Waldram,
  Phys.\ Rev.\  D {\bf 60}, 086001 (1999)
  [arXiv:hep-th/9806022].

\bibitem{Lukas:1998tt}
  A.~Lukas, B.~A.~Ovrut, K.~S.~Stelle and D.~Waldram,
  Nucl.\ Phys.\  B {\bf 552}, 246 (1999)
  [arXiv:hep-th/9806051].

\bibitem{Himemoto:2000nd}
  Y.~Himemoto and M.~Sasaki,
  Phys.\ Rev.\  D {\bf 63}, 044015 (2001)
  [arXiv:gr-qc/0010035].

\bibitem{Ellwanger:2000ne}
  U.~Ellwanger,
  Eur.\ Phys.\ J.\  C {\bf 25}, 157 (2002)
  [arXiv:hep-th/0001126].


\bibitem{Carroll:2001zy}
  S.~M.~Carroll and L.~Mersini-Houghton,
  Phys.\ Rev.\  D {\bf 64}, 124008 (2001)
  [arXiv:hep-th/0105007].


\bibitem{Nojiri:2004pf}
  S.~Nojiri and S.~D.~Odintsov,
  Phys.\ Rev.\  D {\bf 70}, 103522 (2004)
  [arXiv:hep-th/0408170].


\bibitem{Kiritsis:2002zf}
  E.~Kiritsis, G.~Kofinas, N.~Tetradis, T.~N.~Tomaras and V.~Zarikas,
  JHEP {\bf 0302}, 035 (2003)
  [arXiv:hep-th/0207060].

\bibitem{Maeda:2003ar}
  K.~i.~Maeda, S.~Mizuno and T.~Torii,
  Phys.\ Rev.\  D {\bf 68}, 024033 (2003)
  [arXiv:gr-qc/0303039].


\bibitem{Diakonos:2004xq}
  F.~K.~Diakonos, E.~N.~Saridakis and N.~Tetradis,
  Phys.\ Lett.\  B {\bf 605}, 1 (2005)
  [arXiv:hep-th/0409025].


\bibitem{Umezu:2005dw}
  K.~i.~Umezu, K.~Ichiki, T.~Kajino, G.~J.~Mathews, R.~Nakamura and M.~Yahiro,
  Phys.\ Rev.\  D {\bf 73}, 063527 (2006)
  [arXiv:astro-ph/0507227].

\bibitem{Bogdanos:2006pf}
  C.~Bogdanos and K.~Tamvakis,
  Phys.\ Lett.\  B {\bf 646}, 39 (2007)
  [arXiv:hep-th/0609100].

\bibitem{Bogdanos:2006km}
  C.~Bogdanos, A.~Dimitriadis and K.~Tamvakis,
  Phys.\ Rev.\  D {\bf 75}, 087303 (2007)
  [arXiv:hep-th/0611094].

\bibitem{Bogdanos:2006dt}
  C.~Bogdanos, A.~Dimitriadis and K.~Tamvakis,
  Class.\ Quant.\ Grav.\  {\bf 24}, 3701 (2007)
  [arXiv:hep-th/0611181].

\bibitem{Mendes:2000wu}
  L.~E.~Mendes and A.~Mazumdar,
  Phys.\ Lett.\  B {\bf 501}, 249 (2001)
  [arXiv:gr-qc/0009017].

\bibitem{Perivolaropoulos:2003we}
  L.~Perivolaropoulos,
  Phys.\ Rev.\  D {\bf 67}, 123516 (2003)
  [arXiv:hep-ph/0301237].

\bibitem{Bogdanos:2006qw}
  C.~Bogdanos, A.~Dimitriadis and K.~Tamvakis,
  Phys.\ Rev.\  D {\bf 74}, 045003 (2006)
  [arXiv:hep-th/0604182].


\bibitem{Farakos:2005hz}
  K.~Farakos and P.~Pasipoularides,
  Phys.\ Lett.\  B {\bf 621}, 224 (2005)
  [arXiv:hep-th/0504014].

\bibitem{Farakos:2006tt}
  K.~Farakos and P.~Pasipoularides,
  Phys.\ Rev.\  D {\bf 73}, 084012 (2006)
  [arXiv:hep-th/0602200].


\bibitem{BarbosaCendejas:2006hj}
  N.~Barbosa-Cendejas and A.~Herrera-Aguilar,
  Phys.\ Rev.\  D {\bf 73}, 084022 (2006)
  [Erratum-ibid.\  D {\bf 77}, 049901 (2008)]
  [arXiv:hep-th/0603184].




\bibitem{Maartens:2000fg}
  R.~Maartens,
  Phys.\ Rev.\  D {\bf 62}, 084023 (2000)
  [arXiv:hep-th/0004166].

\bibitem{Kanti:2000rd}
  P.~Kanti, K.~A.~Olive and M.~Pospelov,
  Phys.\ Lett.\  B {\bf 481}, 386 (2000)
  [arXiv:hep-ph/0002229].

\bibitem{Cline:1999ts}
  J.~M.~Cline, C.~Grojean and G.~Servant,
  Phys.\ Rev.\ Lett.\  {\bf 83}, 4245 (1999)
  [arXiv:hep-ph/9906523].

\bibitem{Kanti:1999sz}
  P.~Kanti, I.~I.~Kogan, K.~A.~Olive and M.~Pospelov,
  Phys.\ Lett.\  B {\bf 468}, 31 (1999)
  [arXiv:hep-ph/9909481].

\bibitem{Setare:2007qu}
  M.~R.~Setare, J.~Sadeghi and A.~R.~Amani,
  Phys.\ Lett.\  B {\bf 660}, 299 (2008)
   [arXiv:0712.1873  [hep-th]].

\bibitem{Cvetic:1993xe}
  M.~Cvetic, S.~Griffies and H.~H.~Soleng,
  Phys.\ Rev.\  D {\bf 48}, 2613 (1993)
  [arXiv:gr-qc/9306005].

\bibitem{Gibbons:1993in}
  G.~W.~Gibbons,
  Nucl.\ Phys.\  B {\bf 394}, 3 (1993).

\bibitem{Abramo:2005be}
  L.~R.~Abramo and N.~Pinto-Neto,
  Phys.\ Rev.\  D {\bf 73}, 063522 (2006)
  [arXiv:astro-ph/0511562].


\bibitem{Nojiri:2005vv}
  S.~Nojiri, S.~D.~Odintsov and M.~Sasaki,
  Phys.\ Rev.\  D {\bf 71}, 123509 (2005)
  [arXiv:hep-th/0504052].

\bibitem{Sami:2005zc}
  M.~Sami, A.~Toporensky, P.~V.~Tretjakov and S.~Tsujikawa,
  Phys.\ Lett.\  B {\bf 619}, 193 (2005)
  [arXiv:hep-th/0504154].

\bibitem{Koivisto:2006xf}
  T.~Koivisto and D.~F.~Mota,
  Phys.\ Lett.\  B {\bf 644}, 104 (2007)
  [arXiv:astro-ph/0606078].

\bibitem{Tsujikawa:2006ph}
  S.~Tsujikawa and M.~Sami,
  JCAP {\bf 0701}, 006 (2007)
  [arXiv:hep-th/0608178].

\bibitem{Leith:2007bu}
  B.~M.~Leith and I.~P.~Neupane,
  JCAP {\bf 0705}, 019 (2007)
  [arXiv:hep-th/0702002].

\bibitem{Sanyal:2007pu}
  A.~K.~Sanyal,
  Gen.\ Rel.\ Grav.\  {\bf 41}, 1511 (2009)
  [arXiv:0710.2450 [astro-ph]].


\bibitem{Amendola:2007nt}
  L.~Amendola and S.~Tsujikawa,
  Phys.\ Lett.\  B {\bf 660}, 125 (2008)
  [arXiv:0705.0396 [astro-ph]].

\bibitem{Bamba:2008hq}
  K.~Bamba, C.~Q.~Geng, S.~Nojiri and S.~D.~Odintsov,
  Phys.\ Rev.\  D {\bf 79}, 083014 (2009)
  [arXiv:0810.4296 [hep-th]].

\bibitem{Bamba:2009ay}
  K.~Bamba and C.~Q.~Geng,
  Phys.\ Lett.\  B {\bf 679}, 282 (2009)
  [arXiv:0901.1509 [hep-th]].


\bibitem{Elizalde:2004mq}
  E.~Elizalde, S.~Nojiri and S.~D.~Odintsov,
  Phys.\ Rev.\  D {\bf 70}, 043539 (2004)
  [arXiv:hep-th/0405034].

\bibitem{Perivolaropoulos:2005yv}
  L.~Perivolaropoulos,
  JCAP {\bf 0510}, 001 (2005)
  [arXiv:astro-ph/0504582].

\bibitem{Tsujikawa:2008uc}
  S.~Tsujikawa, K.~Uddin, S.~Mizuno, R.~Tavakol and J.~Yokoyama,
  Phys.\ Rev.\  D {\bf 77}, 103009 (2008)
  [arXiv:0803.1106 [astro-ph]].

\bibitem{Gripaios:2004ms}
  B.~M.~Gripaios,
  JHEP {\bf 0410}, 069 (2004)
  [arXiv:hep-th/0408127].

\bibitem{Libanov:2005vu}
  M.~V.~Libanov and V.~A.~Rubakov,
  JHEP {\bf 0508}, 001 (2005)
  [arXiv:hep-th/0505231].

\bibitem{Rubakov:2006pn}
  V.~A.~Rubakov,
  Theor.\ Math.\ Phys.\  {\bf 149}, 1651 (2006)
  [Teor.\ Mat.\ Fiz.\  {\bf 149}, 409 (2006)]
  [arXiv:hep-th/0604153].

\bibitem{Libanov:2007mq}
  M.~Libanov, V.~Rubakov, E.~Papantonopoulos, M.~Sami and S.~Tsujikawa,
  JCAP {\bf 0708}, 010 (2007)
  [arXiv:0704.1848 [hep-th]].

\bibitem{Nojiri:2006ri}
  S.~Nojiri and S.~D.~Odintsov,
  eConf {\bf C0602061}, 06 (2006)
  [Int.\ J.\ Geom.\ Meth.\ Mod.\ Phys.\  {\bf 4}, 115 (2007)]
  [arXiv:hep-th/0601213].

\bibitem{Tsujikawa:2007gd}
  S.~Tsujikawa,
  Phys.\ Rev.\  D {\bf 76}, 023514 (2007)
  [arXiv:0705.1032 [astro-ph]].

\bibitem{Tsujikawa:2007tg}
  S.~Tsujikawa, K.~Uddin and R.~Tavakol,
  Phys.\ Rev.\  D {\bf 77}, 043007 (2008)
  [arXiv:0712.0082 [astro-ph]].


\bibitem{Hu:2007pj}
  W.~Hu and I.~Sawicki,
  Phys.\ Rev.\  D {\bf 76}, 104043 (2007)
  [arXiv:0708.1190 [astro-ph]].

\bibitem{Hu:2008zd}
  W.~Hu,
  Phys.\ Rev.\  D {\bf 77}, 103524 (2008)
  [arXiv:0801.2433 [astro-ph]].

\bibitem{Fang:2008sn}
  W.~Fang, W.~Hu and A.~Lewis,
  Phys.\ Rev.\  D {\bf 78}, 087303 (2008)
  [arXiv:0808.3125 [astro-ph]].


\bibitem{Hawking:1973uf}
  S.~W.~Hawking and G.~F.~R.~Ellis,
  \textit{``The Large scale structure of space-time,''}
  Cambridge University Press, Cambridge, (1973).

\bibitem{Penrose:1964wq}
  R.~Penrose,
  Phys.\ Rev.\ Lett.\  {\bf 14}, 57 (1965).

\bibitem{Hawking:1969sw}
  S.~W.~Hawking and R.~Penrose,
  Proc.\ Roy.\ Soc.\ Lond.\  A {\bf 314}, 529 (1970).

\bibitem{Schon:1981vd}
  R.~Schon and S.~T.~Yau,
  Commun.\ Math.\ Phys.\  {\bf 79}, 231 (1981).

\bibitem{Witten:1981mf}
  E.~Witten,
  Commun.\ Math.\ Phys.\  {\bf 80}, 381 (1981).

\bibitem{Bousso:1999xy}
  R.~Bousso,
  JHEP {\bf 9907}, 004 (1999)
  [arXiv:hep-th/9905177].

\bibitem{Flanagan:1999jp}
  E.~E.~Flanagan, D.~Marolf and R.~M.~Wald,
  Phys.\ Rev.\  D {\bf 62}, 084035 (2000)
  [arXiv:hep-th/9908070].

\bibitem{ArkaniHamed:2007ky}
  N.~Arkani-Hamed, S.~Dubovsky, A.~Nicolis, E.~Trincherini and G.~Villadoro,
  JHEP {\bf 0705}, 055 (2007)
  [arXiv:0704.1814 [hep-th]].



\bibitem{Starobinsky:1980te}
  A.~A.~Starobinsky,
  Phys.\ Lett.\  B {\bf 91} (1980) 99.

\bibitem{Mukhanov:1991zn}
  V.~F.~Mukhanov and R.~H.~Brandenberger,
  Phys.\ Rev.\ Lett.\  {\bf 68}, 1969 (1992).

\bibitem{Peter:2002cn}
  P.~Peter and N.~Pinto-Neto,
  Phys.\ Rev.\  D {\bf 66}, 063509 (2002)
  [arXiv:hep-th/0203013].

\bibitem{Mukherji:2002ft}
  S.~Mukherji and M.~Peloso,
  Phys.\ Lett.\  B {\bf 547}, 297 (2002)
  [arXiv:hep-th/0205180].

\bibitem{Tsujikawa:2002qc}
  S.~Tsujikawa, R.~Brandenberger and F.~Finelli,
  Phys.\ Rev.\  D {\bf 66}, 083513 (2002)
  [arXiv:hep-th/0207228].

\bibitem{Medved:2003fp}
  A.~J.~M.~Medved,
  JHEP {\bf 0305}, 008 (2003)
  [arXiv:hep-th/0301010].

\bibitem{Kanti:2003bx}
  P.~Kanti and K.~Tamvakis,
  Phys.\ Rev.\  D {\bf 68}, 024014 (2003)
  [arXiv:hep-th/0303073].

\bibitem{Foffa:2003gt}
  S.~Foffa,
  Phys.\ Rev.\  D {\bf 68}, 043511 (2003)
  [arXiv:hep-th/0304004].

\bibitem{Finelli:2003mc}
  F.~Finelli,
  JCAP {\bf 0310}, 011 (2003)
  [arXiv:hep-th/0307068].

\bibitem{Hovdebo:2003ug}
  J.~L.~Hovdebo and R.~C.~Myers,
  JCAP {\bf 0311}, 012 (2003)
  [arXiv:hep-th/0308088].

\bibitem{Piao:2003zm}
  Y.~S.~Piao, B.~Feng and X.~m.~Zhang,
  Phys.\ Rev.\  D {\bf 69}, 103520 (2004)
  [arXiv:hep-th/0310206].

\bibitem{Piao:2003hh}
  Y.~S.~Piao, S.~Tsujikawa and X.~m.~Zhang,
  Class.\ Quant.\ Grav.\  {\bf 21}, 4455 (2004)
  [arXiv:hep-th/0312139].

\bibitem{Setare:2004jx}
  M.~R.~Setare,
  Phys.\ Lett.\  B {\bf 602}, 1 (2004)
  [arXiv:hep-th/0409055].

\bibitem{Biswas:2005qr}
  T.~Biswas, A.~Mazumdar and W.~Siegel,
  JCAP {\bf 0603}, 009 (2006)
  [arXiv:hep-th/0508194].

\bibitem{Biswas:2006bs}
  T.~Biswas, R.~Brandenberger, A.~Mazumdar and W.~Siegel,
  JCAP {\bf 0712}, 011 (2007)
  [arXiv:hep-th/0610274].

\bibitem{Creminelli:2007aq}
  P.~Creminelli and L.~Senatore,
  JCAP {\bf 0711}, 010 (2007)
  [arXiv:hep-th/0702165].

\bibitem{Cai:2007qw}
  Y.~F.~Cai, T.~Qiu, Y.~S.~Piao, M.~Li and X.~Zhang,
  JHEP {\bf 0710}, 071 (2007)
  [arXiv:0704.1090 [gr-qc]].

\bibitem{Wei:2007rp}
  H.~Wei and S.~N.~Zhang,
  Phys.\ Rev.\  D {\bf 76}, 063005 (2007)
  [arXiv:0705.4002 [gr-qc]].

\bibitem{Zhang:2007an}
  J.~f.~Zhang, X.~Zhang and H.~y.~Liu,
  Eur.\ Phys.\ J.\  C {\bf 52}, 693 (2007)
  [arXiv:0708.3121 [hep-th]].

\bibitem{Saridakis:2007cf}
  E.~N.~Saridakis,
  Nucl.\ Phys.\  B {\bf 808}, 224 (2009)
  [arXiv:0710.5269 [hep-th]].


\bibitem{Novello:2008ra}
  M.~Novello and S.~E.~P.~Bergliaffa,
  Phys.\ Rept.\  {\bf 463}, 127 (2008)
  [arXiv:0802.1634 [astro-ph]].

\bibitem{Cai:2008qb}
  Y.~F.~Cai, T.~T.~Qiu, J.~Q.~Xia, H.~Li and X.~Zhang,
  Phys.\ Rev.\  D {\bf 79}, 021303 (2009)
  [arXiv:0808.0819 [astro-ph]].

\bibitem{Cai:2008qw}
  Y.~F.~Cai, T.~Qiu, R.~Brandenberger and X.~Zhang,
  Phys.\ Rev.\  D {\bf 80}, 023511 (2009)
  [arXiv:0810.4677 [hep-th]].


\bibitem{Cai:2007zv}
  Y.~F.~Cai, T.~Qiu, R.~Brandenberger, Y.~S.~Piao and X.~Zhang,
  JCAP {\bf 0803}, 013 (2008)
  [arXiv:0711.2187 [hep-th]].

\bibitem{Cai:2008ed}
  Y.~F.~Cai and X.~Zhang,
  JCAP {\bf 0906}, 003 (2009)
  [arXiv:0808.2551 [astro-ph]].

\bibitem{Cai:2009fn}
  Y.~F.~Cai, W.~Xue, R.~Brandenberger and X.~Zhang,
  JCAP {\bf 0905}, 011 (2009)
  [arXiv:0903.0631 [astro-ph.CO]].

\bibitem{Cai:2009rd}
  Y.~F.~Cai, W.~Xue, R.~Brandenberger and X.~m.~Zhang,
  JCAP {\bf 0906}, 037 (2009)
  [arXiv:0903.4938 [hep-th]].

\bibitem{Cai:2009hc}
  Y.~F.~Cai and X.~Zhang,
  Phys.\ Rev.\ D {\bf 80}, 043520 (2009)
  arXiv:0906.3341 [astro-ph.CO].


\bibitem{Coleman:1973jx}
  S.~R.~Coleman and E.~J.~Weinberg,
  Phys.\ Rev.\  D {\bf 7}, 1888 (1973).

\bibitem{Starobinsky:1979ty}
  A.~A.~Starobinsky,
  JETP Lett.\  {\bf 30} (1979) 682
  [Pisma Zh.\ Eksp.\ Teor.\ Fiz.\  {\bf 30} (1979) 719].

\bibitem{Wands:1998yp}
  D.~Wands,
  Phys.\ Rev.\  D {\bf 60}, 023507 (1999)
  [arXiv:gr-qc/9809062].

\bibitem{Finelli:2001sr}
  F.~Finelli and R.~Brandenberger,
  Phys.\ Rev.\  D {\bf 65}, 103522 (2002)
  [arXiv:hep-th/0112249].

\bibitem{Allen:2004vz}
  L.~E.~Allen and D.~Wands,
  Phys.\ Rev.\  D {\bf 70}, 063515 (2004)
  [arXiv:astro-ph/0404441].

\bibitem{Peter:2008qz}
  P.~Peter and N.~Pinto-Neto,
  Phys.\ Rev.\  D {\bf 78}, 063506 (2008)
  [arXiv:0809.2022 [gr-qc]].


\bibitem{Tolman:1934}
  R. C. Tolman,
  \textit{``Relativity, Thermodynamics and Cosmology,"}
  Oxford U. Press, Clarendon Press, (1934).

\bibitem{Khoury:2001wf}
  J.~Khoury, B.~A.~Ovrut, P.~J.~Steinhardt and N.~Turok,
  Phys.\ Rev.\  D {\bf 64}, 123522 (2001)
  [arXiv:hep-th/0103239].

\bibitem{Steinhardt:2001vw}
  P.~J.~Steinhardt and N.~Turok,
  arXiv:hep-th/0111030.

\bibitem{Khoury:2001bz}
  J.~Khoury, B.~A.~Ovrut, N.~Seiberg, P.~J.~Steinhardt and N.~Turok,
  Phys.\ Rev.\  D {\bf 65}, 086007 (2002)
  [arXiv:hep-th/0108187].

\bibitem{Buchbinder:2007ad}
  E.~I.~Buchbinder, J.~Khoury and B.~A.~Ovrut,
  Phys.\ Rev.\  D {\bf 76}, 123503 (2007)
  [arXiv:hep-th/0702154].






\bibitem{Brown:2004cs}
  M.~G.~Brown, K.~Freese and W.~H.~Kinney,
  JCAP {\bf 0803}, 002 (2008)
  [arXiv:astro-ph/0405353].

\bibitem{Baum:2006nz}
  L.~Baum and P.~H.~Frampton,
  Phys.\ Rev.\ Lett.\  {\bf 98}, 071301 (2007)
  [arXiv:hep-th/0610213].

\bibitem{Piao:2004hr}
  Y.~S.~Piao and Y.~Z.~Zhang,
  Nucl.\ Phys.\  B {\bf 725}, 265 (2005)
  [arXiv:gr-qc/0407027].

\bibitem{Piao:2004me}
  Y.~S.~Piao,
  Phys.\ Rev.\  D {\bf 70}, 101302 (2004)
  [arXiv:hep-th/0407258].

\bibitem{Piao:2009ku}
  Y.~S.~Piao,
  Phys.\ Lett.\  B {\bf 677}, 1 (2009)
  [arXiv:0901.2644 [gr-qc]].


\bibitem{Xiong:2007cn}
  H.~H.~Xiong, T.~Qiu, Y.~F.~Cai and X.~Zhang,
  arXiv:0711.4469 [hep-th].

\bibitem{Borde:2001nh}
  A.~Borde, A.~H.~Guth and A.~Vilenkin,
  Phys.\ Rev.\ Lett.\  {\bf 90}, 151301 (2003)
  [arXiv:gr-qc/0110012].


\bibitem{Aref'eva:2005fu}
  I.~Y.~Aref'eva, A.~S.~Koshelev and S.~Y.~Vernov,
  Phys.\ Rev.\  D {\bf 72}, 064017 (2005)
  [arXiv:astro-ph/0507067].

\bibitem{Vernov:2006dm}
  S.~Y.~Vernov,
  Teor.\ Mat.\ Fiz.\  {\bf 155}, 47 (2008)
  [Theor.\ Math.\ Phys.\  {\bf 155}, 544 (2008)]
  [arXiv:astro-ph/0612487].

\bibitem{Aref'eva:2003qu}
  I.~Y.~Aref'eva, L.~V.~Joukovskaya and A.~S.~Koshelev,
  JHEP {\bf 0309}, 012 (2003)
  [arXiv:hep-th/0301137].


\bibitem{Cohen:1998zx}
  A.~G.~Cohen, D.~B.~Kaplan and A.~E.~Nelson,
  Phys.\ Rev.\ Lett.\  {\bf 82}, 4971 (1999)
  [arXiv:hep-th/9803132].

\bibitem{Horava:2000tb}
  P.~Horava and D.~Minic,
  Phys.\ Rev.\ Lett.\  {\bf 85}, 1610 (2000)
  [arXiv:hep-th/0001145].

\bibitem{Thomas:2002pq}
  S.~D.~Thomas,
  Phys.\ Rev.\ Lett.\  {\bf 89}, 081301 (2002).

\bibitem{Hsu:2004ri}
  S.~D.~H.~Hsu,
  Phys.\ Lett.\  B {\bf 594}, 13 (2004)
  [arXiv:hep-th/0403052].

\bibitem{Li:2004rb}
  M.~Li,
  Phys.\ Lett.\  B {\bf 603}, 1 (2004)
  [arXiv:hep-th/0403127].

\bibitem{Huang:2004wt}
  Q.~G.~Huang and Y.~G.~Gong,
  JCAP {\bf 0408}, 006 (2004)
  [arXiv:astro-ph/0403590].

\bibitem{Wang:2004nqa}
  B.~Wang, E.~Abdalla and R.~K.~Su,
  Phys.\ Lett.\  B {\bf 611}, 21 (2005)
  [arXiv:hep-th/0404057].


\bibitem{'tHooft:1993gx}
  G.~'t Hooft,
  arXiv:gr-qc/9310026.

\bibitem{Susskind:1994vu}
  L.~Susskind,
  J.\ Math.\ Phys.\  {\bf 36}, 6377 (1995)
  [arXiv:hep-th/9409089].

\bibitem{Bousso:2002ju}
  R.~Bousso,
  Rev.\ Mod.\ Phys.\  {\bf 74}, 825 (2002)
  [arXiv:hep-th/0203101].

\bibitem{Elizalde:2005ju}
  E.~Elizalde, S.~Nojiri, S.~D.~Odintsov and P.~Wang,
  Phys.\ Rev.\  D {\bf 71}, 103504 (2005)
  [arXiv:hep-th/0502082].

\bibitem{Izquierdo:2005ku}
  G.~Izquierdo and D.~Pavon,
  Phys.\ Lett.\  B {\bf 633}, 420 (2006)
  [arXiv:astro-ph/0505601].

\bibitem{Wang:2005jx}
  B.~Wang, Y.~g.~Gong and E.~Abdalla,
  Phys.\ Lett.\  B {\bf 624}, 141 (2005)
  [arXiv:hep-th/0506069].

\bibitem{Zhang:2005hs}
  X.~Zhang and F.~Q.~Wu,
  Phys.\ Rev.\  D {\bf 72}, 043524 (2005)
  [arXiv:astro-ph/0506310].

\bibitem{Feng:2008kz}
  C.~J.~Feng,
  Phys.\ Lett.\  B {\bf 672}, 94 (2009)
  [arXiv:0810.2594 [hep-th]].

\bibitem{Feng:2009ai}
  C.~J.~Feng and X.~Z.~Li,
  Phys.\ Lett.\  B {\bf 679}, 151 (2009)
  [arXiv:0904.2976 [hep-th]].



\end{thebibliography}


\end{document}